\pdfoutput=1

\documentclass[11pt,a4paper]{article}
\usepackage{placeins}

\usepackage{jheppub}
\usepackage{amsmath}
\usepackage{multicol,bbm}
\usepackage{enumerate}
\usepackage{bibentry}

\usepackage{amssymb}
\usepackage{bm}
\usepackage{multirow}
\usepackage{caption}
\usepackage{tabularx}
\usepackage{enumitem}
\usepackage{extarrows}
\usepackage{tikz}
\usepackage{verbatim} 
\usepackage{float} 

\def\be{\begin{equation}}
\def\ee{\end{equation}}
\def\ba{\begin{eqnarray}}
\def\ea{\end{eqnarray}}

\def\CP1{\mathbb{CP}^1}
\def\SL2C{\mathrm{SL}(2,\mathbb{C})}

\def\Z2{\mathbb{Z}_2}

\def\su2{{SU(2)}}

\def\[{\left[}
\def\]{\right]}

\def\({\left(}
\def\){\right)}
\def\[{\left[}
\def\]{\right]}

\def\<{\langle}
\def\>{\rangle}

\def\i2{\frac{i}{2}}

\def\2F1{\,_2{\rm F}_1}

\usepackage{amsthm}

\usepackage{todonotes}
\begin{document}


\title{Leading singularities and chambers of Correlahedron}


\date{\today}
\author[a,b,c]{Song He}
\author[d,e,f]{Yu-tin Huang}
\author[g]{Chia-Kai Kuo}


\affiliation[a]{New Cornerstone Science Laboratory, Institute of Theoretical Physics, Chinese Academy of Sciences, Beijing 100190, China}
\affiliation[b]{
School of Fundamental Physics and Mathematical Sciences,
Hangzhou Institute for Advanced Study; ICTP-AP International Centre for Theoretical Physics Asia-Pacific, UCAS, Hangzhou 310024, China}
\affiliation[c]
{Peng Huanwu Center for Fundamental Theory, Hefei, Anhui 230026, P. R. China}
\affiliation[d]{Department of Physics and Center for Theoretical Physics, National Taiwan University, Taipei 10617, Taiwan}
\affiliation[e]{Physics Division, National Center for Theoretical Sciences, Taipei 10617, Taiwan}
\affiliation[f]{Max Planck{-}IAS{-}NTU Center for Particle Physics, Cosmology and Geometry, Taipei 10617, Taiwan}
\affiliation[g]{Max–Planck–Institut für Physik, Werner–Heisenberg–Institut, D–85748 Garching bei München, Germany}


\emailAdd{songhe@itp.ac.cn}
\emailAdd{yutinyt@gmail.com}
\emailAdd{chia-kai.kuo@mpp.mpg.de}

\preprint{ \begin{flushright} MPP-2025-80\end{flushright}}

\abstract{In this paper, we explore the chamber dissection of the loop-geometry of Correlahedron, which encodes the loop integrand of four-point stress-energy correlators in planar $\mathcal{N}=4$ super Yang-Mills. We demonstrate that at four loops, continuing the pattern of lower loops, the integrand of the four-point correlation function can be written as a sum over products of chamber-forms and local loop integrands. The chambers and their associated forms are identical to those of three loops, indicating that the dissection may be complete to all loop orders. Furthermore, this suggests that the leading singularities at all loops are simply linear combinations of these chamber forms. This is especially intriguing at four loops since it contains elliptic functions. Interestingly, each elliptic function appears in a subset of chambers. Our geometric approach motivates us to ``diagonalize" the representation, where the local integrals only possess a single leading singularity or elliptic cut. In such a representation, all integrands must evaluate to pure functions, including a single pure elliptic integrand. Inspired by this picture, we also present a simplified form of the three-loop correlator in terms of two independent pure functions (weight-$6$ single-valued multiple polylogarithms), which are directly computed from local integrands with unit leading singularities, multiplied by the leading singularities from chamber forms. 
}


\maketitle

\section{Introduction}

In~\cite{Eden:2017fow}, a geometric object called the {\it Correlahedron} was introduced as a certain ``off-shell" generalization of the {\it Amplituhedron}~\cite{Arkani-Hamed:2013jha, Arkani-Hamed:2013kca, Arkani-Hamed:2017vfh} in planar $\mathcal{N}=4$ super Yang-Mills. This generalization yields the correlation functions of the stress-tensor multiplet in $\mathcal{N}=4$ super Yang-Mills.
In our previous work~\cite{He:2024xed}, we proposed a reformulation of the Correlahedron and initiated a systematic study of the resulting positive geometries. Just as the canonical forms~\cite{Arkani-Hamed:2017mur} of the Amplituhedron are conjectured to yield all-loop integrands for scattering amplitudes in planar SYM, we provided strong evidence that the canonical forms of these geometries give the loop integrands of four-point stress-energy correlators. These correlators form an important class of observables of great interest for the study of the AdS/CFT correspondence, the conformal bootstrap, and scattering amplitudes (c.f. a recent review~\cite{Heslop:2022xgp} and references therein). Since the correlator reduces to the square of amplitudes (or equivalently Wilson loops) in lightlike limits~\cite{Eden:2010ce, Eden:2011ku, Alday:2010zy, Eden:2010zz, Eden:2011yp, Adamo:2011dq}, the Correlahedron in such limits is expected to reduce to the square of the Amplituhedron~\cite{Dian:2021idl}.

Our main focus will be the integrand of the four-point stress-energy correlator, which is known up to twelve loops~\cite{Bourjaily:2015bpz, Bourjaily:2016evz,  He:2024cej, Bourjaily:2025iad}. This has been largely facilitated by the discovery of a hidden permutation symmetry, which reduces the basis of integrands with certain conformal properties, the so-called $f$-graphs~\cite{Eden:2011we}. An efficient graphical method has been developed to obtain the coefficients of these graphs and hence the correlator integrand~\cite{ Bourjaily:2025iad}. As we will see, not only can the integrand be interpreted as canonical forms of a positive geometry, the geometric formulation in turn implies a hidden simplicity of the integrated result.

As we will review shortly, similar to the Correlahedron, we define our geometry in bosonized twistor space, where a set of positivity conditions are imposed on four lines representing the positions of the operators. One novel feature of our reformulation is the interpretation of the geometry as a \textit{fibration}: we characterize a positive subspace, the tree region, to which an $L$-loop geometry is attached. The loop geometry then consists of $L$ lines in twistor space satisfying positivity conditions among themselves and with respect to the base. As the boundary of the loop-geometry simply corresponds to propagators going on-shell (in the context of correlators the propagators are becoming light-like), the associated canonical form is naturally given in terms of local integrands. This is drastically different from previous triangulations of the loop Amplituhedron which yields non-local representations~\cite{ Kojima:2018qzz, Kojima:2020tjf}. These ideas have played a crucial role in recent studies for both the ${\cal N}=4$ SYM loop Amplituhedron as well as a new, closely-related geometry known as the ABJM Amplituhedron~\cite{He:2023rou, He:2022cup}; it has been conjectured to give all-loop integrands in the three-dimensional ABJM theory~\cite{Aharony:2008ug, Hosomichi:2008jb}.

Importantly, this approach naturally induces the notion of \textit{chambers}, which are dissections of the tree geometry such that within each chamber the loop-geometry is homogeneous. That is we have a unique loop-form in each chamber, and the forms are distinct between different chambers. The full observable is then given as sum over the product of chamber- and loop-form, across all chambers. Such chamber dissection was previously discussed in the context of $\mathcal{N}=4$ SYM and ABJM Amplituhedron~\cite{He:2023rou}. (see~\cite{Lukowski:2023nnf, Ferro:2023qdp,Ferro:2024vwn} for further discussion). The presence of chambers is a reflection of the fact that for a given point in the tree region, not all boundaries of the loop-geometry are accessible! To illustrate this point, let us consider a simple example, the six-point one-loop NMHV Amplituhedron. The loop-geometry is defined on the line $(AB)$ in twistor space satisfying $\langle ABii{+}1\rangle>0$ for all $i$ and $\langle AB1i\rangle$ having 3 sign flips~\cite{Arkani-Hamed:2017vfh}. We can parameterize the line as 
\begin{equation}
(AB)=(Z_1+x Z_2 -w Z_4,\,y Z_2+Z_3+ z Z_4)\,.
\end{equation}
The tree-region is then characterized by $\langle ii{+}1jj{+}1\rangle>0$ and $\langle 123 i\rangle$ having $1$ sign flip. Now at each point in this tree-region we can analyze the loop-geometry which is enclosed by the codimension one boundaries $\langle ABii{+}1\rangle=0$. However, some higher codimension boundaries might not be present. Take the codimension four boundary $\langle AB12\rangle=\langle AB23\rangle=\langle AB34\rangle=\langle AB61\rangle=0$, which fixes $(A,B)=(1,3)$. This implies that $\langle AB45\rangle=\langle 1345\rangle$ and $\langle AB56\rangle=\langle 1356\rangle$. Clearly if $\langle 1345\rangle<0$ or $\langle 1356\rangle<0$, this codimension-four boundary is inaccessible and should be absent from its loop-form. More importantly since the boundary structure is different, the loop-form should be distinct on either side $\langle 1345\rangle=0$ and $\langle 1356\rangle=0$, therefore this represents a boundary in the tree region where the fibration changes as one passes through.

From the above discussion, one sees that each chamber is determined by the subset of boundaries that are accessible. To determine the full chamber structure, one needs to determine the kinematic constraints for all possible boundaries, which naively can be an unbounded problem as one increases the loop order. In the Amplituhedron, where the tree-geometry is understood as a map of positive Grassmannian $G(k,n)$ to $G(k,k+4)$, these chambers can be understood as the intersection of the images of $4k$-dimensional positroid cells under the Amplituhedron map. Thus for $n$-point N$^k$MHV at fixed $k$ and $n$, the number of chambers are finite as is the number of positroid cells.

In our previous work~\cite{He:2024xed}, it was shown that up to three loops, the chambers are simply organized by the six orderings of $x^2_{1,2}x^2_{3,4}\equiv s$, $x^2_{1,4}x^2_{2,3}\equiv t$ and $x^2_{1,3}x^2_{2,4}\equiv u$. An immediate implication is that since each chamber is attached to a $4L$-canonical form, the leading singularities of the correlation functions~\cite{Drummond:2013nda} must simply be given as linear combinations of the chamber forms. We have explicitly verified that this is the case in~\cite{He:2024xed}. Note that an understanding of the Correlahedron as a map of positroid cells is lacking, and thus it is unclear how the chamber structure might change as one progresses to higher loop order.

In this paper, we extend our analysis to four-loops, where we find that the chamber structure is identical to that at three-loops, {\it i.e.} no new chamber boundaries appear. We construct the local integrand representation of the loop-form for each chamber. We have explicitly verified that all known four-loop leading singularities~\cite{He:2025vqt} can also be written as linear combinations of the chamber forms. This fascinating fact becomes even more striking given that, at four loops, the loop form already includes elliptic cuts.  This leads us to conjecture that to all loops there are only six chambers, given by the ordering of $s,t,u$. Note that this is saying that the only kinematic condition that determines whether or not a given boundary is accessible can only be the ordering of $s,t,u$, along with the universal requirement $\Delta^2>0$ defined in eq.~\eqref{eq: DeltaDef}. Interestingly, these chamber boundaries not only encapsulate the conditions of when a cut is positive, they also capture the different orientations of the cut. 

Since our chamber structure reflects the positivity condition of cuts, the associated loop forms could be organized to the form which encode the cut's positivity condition: 
\begin{equation}
\Omega^{(L)\pm}_{r_i}=2 \Delta^2 A_{\sigma_3} \pm2\Delta\big(B+ (\sigma_2-\sigma_1)C_{\sigma_2,\sigma_1} + (\sigma_3-\sigma_1)C_{\sigma_3,\sigma_1} + (\sigma_3-\sigma_2) C_{\sigma_3,\sigma_2} \big)\,,   
\end{equation}
where the $\sigma_i$ correspond to the Mandelstam-like variables in the inequality $r_i$ of eq.~\eqref{eq:L=3_chambers}, {\it i.e.} $\sigma_1<\sigma_2<\sigma_3$ (see Subsection~\ref{sec: three-loop chamber analysis} for details).
This structure naturally emerges when tailoring a basis of integrands that realize a single leading singularity or elliptic cut. In particular, while such integrands may admit multiple distinct cuts, their residues are always proportional to the same function—$\Delta$, $\Delta^2$, or $\Delta(\sigma_i - \sigma_j)$. We refer to the process of identifying local integrands with this property as ``diagonalization" in the space of leading singularities. Note that this immediately implies that our loop forms are directly pure functions. We demonstrate that this is indeed the case at three-loops.

 At four-loops, we are able to isolate a single scalar integrand (and its two permutation images) that is elliptic,

\begin{eqnarray}
&&\vcenter{\hbox{\scalebox{1}{
\begin{tikzpicture}[x=0.75pt,y=0.75pt,yscale=-1,xscale=1]


\draw  [line width=0.75] (-30,140) node [anchor=north west][inner sep=0.75pt]  [align=center] {$ \begin{tikzpicture}[x=0.75pt,y=0.75pt,yscale=-1,xscale=1]

\draw  [line width=0.75]  (135.27,102.72) -- (178.71,102.72) -- (178.71,146.16) -- (135.27,146.16) -- cycle ;
\draw  [line width=0.75]  (178.71,146.16) -- (222.16,146.16) -- (222.16,189.61) -- (178.71,189.61) -- cycle ;
\draw  [line width=0.75]   (123.3,90.75) -- (135.27,102.72) ;
\draw   [line width=0.75]  (222.16,189.61) -- (234.13,201.58) ;
\draw   [line width=0.75]  (178.8,85.8) -- (178.71,102.72) ;
\draw   [line width=0.75]  (135.27,189.61) -- (121.13,202.13) ;
\draw   [line width=0.75]  (178.71,189.61) -- (178.47,206.8) ;
\draw  [line width=0.75]  (178.71,102.72) -- (222.16,102.72) -- (222.16,146.16) -- (178.71,146.16) -- cycle ;
\draw  [line width=0.75]  (135.27,146.16) -- (178.71,146.16) -- (178.71,189.61) -- (135.27,189.61) -- cycle ;
\draw   [line width=0.75]  (236.29,90.19) -- (222.16,102.72) ;
\draw (154.69,119) node [anchor=north west][inner sep=0.75pt]    {$a$};
\draw (195.58,119.45) node [anchor=north west][inner sep=0.75pt]    {$b$};
\draw (154.06,161) node [anchor=north west][inner sep=0.75pt]   [align=left] {$\displaystyle c$};
\draw (194.08,158) node [anchor=north west][inner sep=0.75pt]   [align=left] {$\displaystyle d$};
\draw (151.39,79.98) node [anchor=north west][inner sep=0.75pt]    {$1$};
\draw (195.07,80.72) node [anchor=north west][inner sep=0.75pt]    {$2$};
\draw (194.89,195.14) node [anchor=north west][inner sep=0.75pt]    {$1$};
\draw (115.14,139.37) node [anchor=north west][inner sep=0.75pt]    {$3$};
\draw (231.75,139.13) node [anchor=north west][inner sep=0.75pt]    {$4$};
\draw (152.4,193.72) node [anchor=north west][inner sep=0.75pt]    {$2$};

\end{tikzpicture}
$};

\draw (-90,200) node [anchor=north west][inner sep=0.75pt]    {$I_G^{12;34}:$};

\end{tikzpicture}

}
}}
\end{eqnarray}
 while the remaining are pure functions. Interestingly, a given elliptic integrand is only present in a subset of chambers. Note that while the definition of canonical forms for such ``elliptic boundaries" is lacking at the moment
 , the kinematic coefficient of this integrand can be determined by aligning it with the normalization of the chamber. The latter is fixed by the leading singularities of other non-elliptic leading singularities. This allows us to conclude that the kinematic coefficient of the elliptic integrand is normalized as
\begin{equation}
     \Delta^2 I_G^{12;34}\, .
\end{equation}
In this way, geometry provides the natural normalization for the elliptic integrands.

This paper is organized as follows. In section~\ref{sec2}, we review the  four-point correlation function of half-BPS stress-tensor multiplet in ${\cal N}=4$ SYM and its loop integrand. In section~\ref{sec3}, we present the definition and construction of the four-point Correlahedron, including its tree region, the loop geometry and importantly the {\it chambers}, which lead to the computation of loop forms based on such a dissection. In section~\ref{sec4}, we elaborate the results up to three loops: no chamber dissection is needed for one- and two-loop cases but six chambers appear for the three-loop geometry, which we carefully study and relate these regions to positivity of corresponding cuts; furthermore, we present a chamber-inspired rewriting of the final result for the three-loop correlator (after loop integration), which is given by leading singularities (LS) multiplied by pure functions computed from integrals with unit LS. In section~\ref{sec5}, we move to four loops, where it turns out the six chambers still suffice to organize the four-loop geometry. 
We ``diagonalize" all the ordinary leading singularities such that the corresponding loop forms again have unit LS and must evaluate to pure functions; we also discuss the elliptic cuts in detail and compare both elliptic and ordinary LS with those from the $f$-graph construction. Appendix~\ref{sec:3loop original form} collects the chamber loop-forms at three loops using the original (non-pure) basis, which serves as a comparison with the representation in terms of pure integrals. 
   Appendix~\ref{sec:four-loop integral basis} records the set of pure integrands used in presenting the four-loop chamber integrands.

\section{The four-point correlator}~\label{sec2}
We will consider four-point correlation functions of the half-BPS stress-tensor multiplet in planar ${\cal N}=4$ SYM. The multiplet contains a scalar operator that transforms in the $\mathbf{20'}$ representation of the SO(6) R-symmetry ${\cal O}_{20'}^{IJ}:=\mathrm{tr}(\Phi^I \Phi^J)-\frac{1}{6}\, \delta^{IJ} \mathrm{tr}(\Phi^K \Phi^K)$. The correlator is naturally given in analytic  superspace which includes coordinates 
\begin{equation}
(x^{\alpha\dot{\alpha}}, \theta_\alpha^{+ a},  \bar{\theta}^{\dot{\alpha}}_{- a'}, u^{+a}_{A}) ,\quad \theta_\alpha^{+ a}=\theta^{A}_\alpha u^{+a}_{A},\quad \bar{\theta}^{\dot{\alpha}}_{-a'}=\bar{\theta}_A^{\dot{\alpha}}u^A_{- a'}\,.
\end{equation}
where $(\alpha, \dot{\alpha}) \in$ SL(2,$\mathbb{C}$), $A\in$ SU(4) and $a,a'\in$ SU(2)$\times$SU(2)$\subset$ SU(4). The bosonic variables $u^{+a}_{A}$ and $u^A_{- a'}$ parameterize the coset SU(4)/SU(2)$\times$SU(2)$\times$U(1).  The chiral part of stress-tensor multiplet is given by setting $\bar{\theta}^{\dot{\alpha}}_{- a'}=0$,  having an expansion
\begin{equation}\label{eq: StressTensor}
\mathcal{T}(x,\theta^+, 0, u)=\mathcal{O}_{20'}^{+4}(x)+\cdots +(\theta^+)^4\mathcal{L}(x)
\end{equation}
where in harmonic superspace, the SU(4) antisymmetric field $\Phi^{AB}$ (or $\Phi^I$ in SO(6)) projects as
$\Phi^{++} = \Phi^{AB}u_A^{+a}u_{B a}^{+}$.

The $n$-point correlator is given as a sum over a collection of homogeneous polynomials of degree $4k$ in $\theta^+$, with $k=0,\ldots,n-4$ (see~\cite{Eden:2011we, Chicherin:2014uca} for details). At four points, there are no $\theta^+$ dependence, and one simply has:
\begin{equation}
G_4=\langle {\cal O}(x_1, y_1) \cdots {\cal O}(x_4, y_4)\rangle=\sum_{\ell=0}^\infty a^{\ell} G_4^{(\ell)},
\end{equation}
where $y^I\rightarrow y_{AB}=u_A^{+a}u_{Ba}^{+}$ and $a=g^2 N_c/(4\pi^2)$ is the t' Hooft coupling. The tree correlator $G_4^{(0)}$ is given by 
\begin{equation}
G_4^{(0)}=\frac{(N_c^2-1)^2}{(4\pi^2)^4}(d_{12}^2 d_{34}^2 + 2~{\rm perms.}) + \frac{N_c^2-1}{(4\pi^2)^2}(d_{12} d_{23} d_{34} d_{41} + 2~{\rm perms.})
\end{equation}
where $d_{i,j}:=\frac{y_{i,j}^2}{x_{i,j}^2}$ with $y_{i,j}=(y_i-y_j)^2$ from harmonics. Starting $L\geq 1$, the correlator is given by a universal prefactor, $R(1,2,3,4)$, times the $L$-loop function $F^{(L)}$:
\begin{equation}\label{eq: CorrelatorLoop}
G_4^{(L)}=\frac{2 (N_c^2-1)}{(4\pi^2)^4}\,R(1,2,3,4)\times F^{(L)}(x_1, x_2, x_3, x_4)\qquad {\rm for}~L\geq 1 \,,
\end{equation}
where the prefactor is given by 
\begin{equation}
R(1,2,3,4):= \left(d_{1,2} d_{2,3} d_{3,4} d_{4,1} (1{-}U{-}V){+}d_{1,3}^2 d_{2,4}^2\right) x_{1,3}^2 x_{2,4}^2  
+(1\leftrightarrow 2) + (1\leftrightarrow 4)  \\  \nonumber  \,.
\end{equation}
with $U=\frac{x^2_{1,2}x^2_{3,4}}{x^2_{1,3}x^2_{2,4}}$, $V=\frac{x^2_{1,4}x^2_{2,3}}{x^2_{1,3}x^2_{2,4}}$.
Note that $R(1,2,3,4)$ is distinct from $G_4^{(0)}$. We will be interested in its {\it integrand} which also depends on the loop variables $x_a\equiv x_5, x_b\equiv x_6, \ldots, x_{4+L}$:
\begin{equation}\label{eq: Fgraph}
F^{(L)}:=\frac{\prod_{1\leq i<j\leq 4} x_{i,j}^2}{L!} \int \prod_{a=1}^L d^4 x_{4{+}a} \mathcal{F}^{(L)}(x_1, \ldots, x_{4{+}L})\,.
\end{equation}
As discussed in~\cite{ Eden:2011we}, $\mathcal{F}^{(L)}$ integrand enjoys permutation invariance among all $4+L$ points. Thus the four-point $L$-loop integrand is generally given in terms of a basis of $f$-graphs, which are rational functions of $x^2_{i,j}$ with conformal weight $-4$ on each site:
\begin{equation}
\mathcal{F}^{(L)}=\sum_i c_i f_i^{(L)}\,.
\end{equation}
The coefficients $c_i$ have been determined up to 12 loops~\cite{ Bourjaily:2016evz, He:2024cej,  Bourjaily:2025iad}.

In~\cite{Chicherin:2014uca}, twistorial methods based on chiral superspace was introduced, where the stress-tensor multiplet in eq.~\eqref{eq: StressTensor} is related to the interacting Lagrangian in chiral superspace
\begin{equation}
\mathcal{T}(x,\theta^+, 0, u)=\int d^4 \theta^- L_{\rm int} (x,\theta)\,.
\end{equation}
Thus the correlation function of $\mathcal{T}$ can be extracted from the correlation function of $L_{\rm int} (x,\theta)$ in Minkowski chiral-superspace:
\begin{eqnarray}
\langle \mathcal{T}(x_1,\theta_1^+, 0, u_1)\cdots \mathcal{T}(x_n,\theta_n^+, 0, u_n)\rangle=\int\prod_{i=1}^n d^4\theta^-_i\langle L_{\rm int}(x_1,\theta_1)\cdots L_{\rm int}(x_n,\theta_n)\rangle\,, 
\end{eqnarray}
 At four-points we have:  
\begin{equation}
    \begin{split}
        \langle \mathcal{T}(x_1, u_1)\cdots \mathcal{T}(x_4, u_4)\rangle&=\left[\prod_{i=1}^4 D^4_i \right]\langle  L_{\rm int}(x_1,\theta_1)\cdots L_{\rm int}(x_4,\theta_4)\rangle\\
&\equiv \,\left[\prod_{i=1}^4 D^4_i \right] \mathcal{G} (x,\theta)
    \end{split}
\end{equation}
where the fermionic integral is replaced with the differential operator $D^4_i$ defined as,
\begin{equation}\label{eq: D4Def}
D^4_i=y_{i}^{AB} y_{i}^{CD} \frac{\partial}{\partial \theta_i^{A\alpha}}\frac{\partial}{\partial \theta_i^{B\beta}}\frac{\partial}{\partial \theta_{i\alpha}^C}\frac{\partial}{\partial \theta_{i\beta}^D}\,.
\end{equation}
Note that since the fermionic derivative yields a conformal weight of 2 at each point, the requisite weight of the BPS operators, the function $\mathcal{G} (x,\theta)$ is a conformal invariant function.

In~\cite{Eden:2017fow} a geometry was proposed for  the potential $\mathcal{G} (x,\theta)$ in twistor space. One begins with the embedding space formalism where the four-dimensional space is mapped to the projective null plane in $\mathbb{R}^{2,4}$. Each point is represented as a skew $4\times4$ matrix $X_i^{IJ}=-X_i^{JI}$, where $I,J\in \operatorname{SU}(4)$, and the null condition is given as:
\begin{equation}\label{eq:self zero}
\langle i, i \rangle\equiv\epsilon_{IJKL}X_i^{IJ}X_i^{KL}=0\,.
\end{equation}
This suggests that a given point is parameterized by a pair of twistors $\{Z_{i\alpha}\}$, $X^{IJ}_i=Z^I_{i,[1}Z^{J}_{i,2]}$. Since the $X_i$s are defined projectively, invariants are given by ratios of brackets $\langle i,j\rangle$ such that the  projective weight at each point cancels. For example, we can identify
\begin{equation}\label{eq: DisDef}
x^2_{i,j}\equiv(x_i-x_j)^2=\frac{\langle i,j\rangle}{\langle i, I\rangle\langle j, I\rangle}
\end{equation}
where $I$ is the infinity twistor that breaks  the SU(4) conformal symmetry. The fermionic variables of the chiral superspace can also be parametrized by a pair of fermionic twistors $\{\theta_{i,\alpha}^A\}$. To define the geometry, one bosonizes the fermionic variables~\cite{Arkani-Hamed:2013jha}. More precisely, at four-points we introduce $4\times 4$ fermionic auxiliary variables $\phi_{A,a}$ such that we can bosonize the bi-twistors:\footnote{The inner product is with respect to the SU(4) index $A$}
\begin{equation}
    \textbf{X}_i=
\begin{pmatrix}
Z_{i,\alpha} \\
\theta_i^\alpha \cdot \phi_1\\
\vdots\\
\theta_i^\alpha \cdot \phi_{4}
\end{pmatrix}\,.
\end{equation}
Each bi-twistors $\textbf{X}_i$ is a $(4{+}4)\times 2$ matrix. A recurring object is the determinant of the $8\times 8$ matrix constructed from collecting the four set of bi-twistors:
\begin{equation}
\langle \! \langle \mathbf{X}_1, \mathbf{X}_2, \mathbf{X}_3, \mathbf{X}_4\rangle \! \rangle\equiv {\rm Det}\left(\mathbf{X}_1, \mathbf{X}_2, \mathbf{X}_3, \mathbf{X}_4\right)\,.
\end{equation}
From now on, single and double brackets represent the determinants of $4\times 4$ and $8\times 8$ matrices, respectively. 

The potential is then identified with the geometry in the auxiliary space  $Y_a^{\mathcal{I}}\in$ Gr($4,8$), which is 16-dimensional. The canonical form, denoted as 
\begin{equation}
\mathcal{G}(\mathbf{X},Y)\prod_{a=1}^4 \langle\!\langle Yd^4Y_{a}\rangle\!\rangle\,.
\end{equation}
The potential is then extracted from the ``volume form" $\mathcal{G}(\mathbf{X},Y)$ by projecting $\langle \! \langle Y, \mathbf{X}_i, \mathbf{X}_j\rangle \! \rangle\rightarrow x^2_{i,j}$ and 
integrate out the auxiliary variables $\phi$, 
\begin{equation}
\mathcal{G}(x,\theta)=\int d^{16}\phi\;\left[\left.\mathcal{G}(\mathbf{X},Y)\right|_{\langle \! \langle Y, \mathbf{X}_i, \mathbf{X}_j\rangle \! \rangle\rightarrow x^2_{i,j}}\right]
\end{equation}
As we will see in the following sections,  the only $\phi$ dependence in $\mathcal{G}(\mathbf{X},Y)$ after projection is through the factor $\langle \! \langle \mathbf{X}_1, \mathbf{X}_2, \mathbf{X}_3, \mathbf{X}_4\rangle \! \rangle^4$. Direct integration and applying the fermionic derivatives gives,
\begin{eqnarray}\label{eq: FermionicPro}
    &&\left[\prod_{i=1}^4 D^4_i \right]\int d^{16}\phi\, \langle \! \langle \mathbf{X}_1, \mathbf{X}_2, \mathbf{X}_3, \mathbf{X}_4\rangle \! \rangle^4 =R(1,2,3,4)\prod_{i<j} x_{i,j}^2\,,
\end{eqnarray}
where we have obtained the desired universal prefactor of the loop-level correlator in eq.~\eqref{eq: CorrelatorLoop}.

\section{The four-point Correlahedron}~\label{sec3}
\begin{figure}[H]
\includegraphics[width=5.5cm]{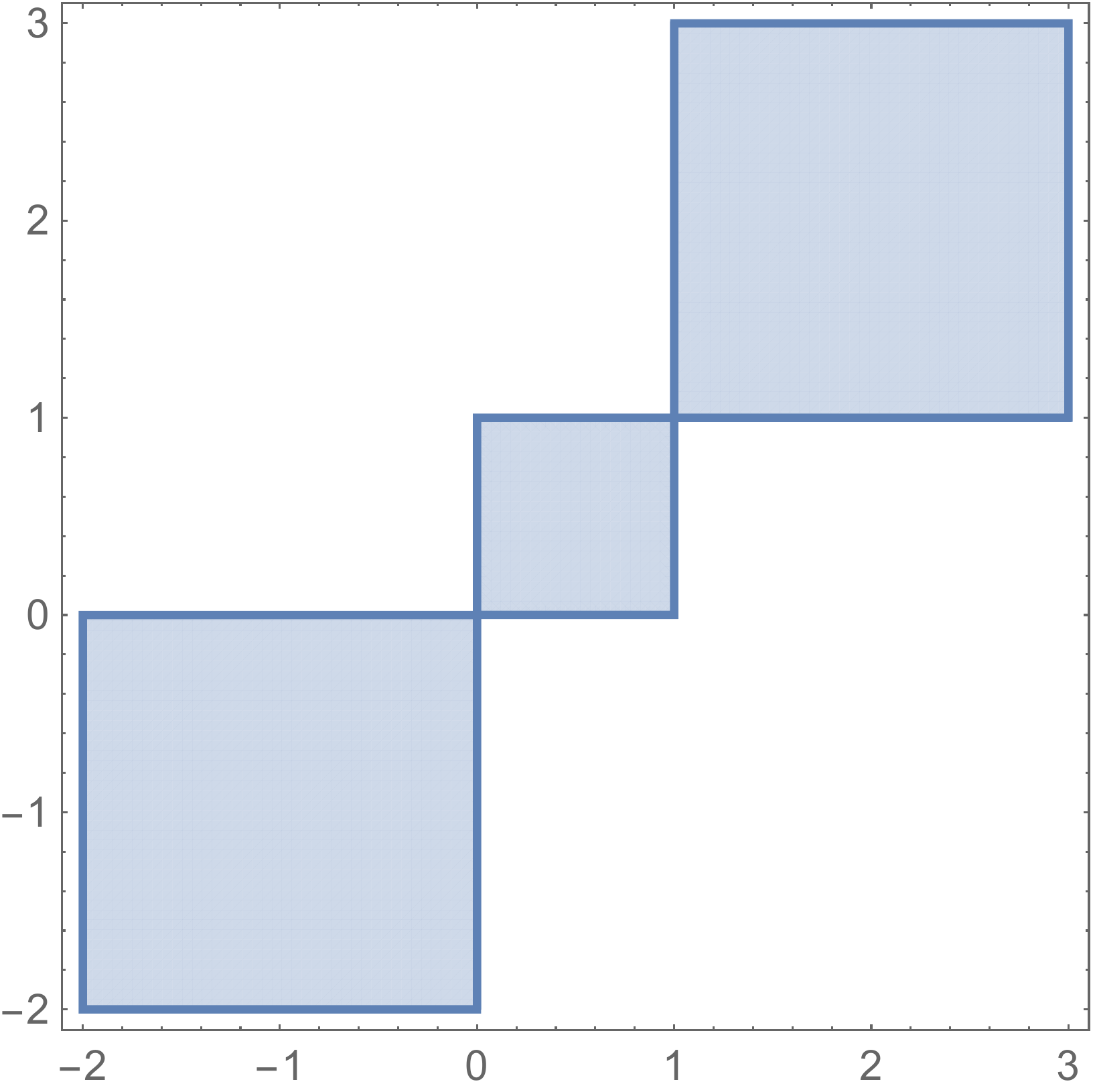} 
\centering
\caption{The region in $c_1, c_2$ plane bounded by eq.~\eqref{eq: TreeRegion0}.}
\label{fig:c1c2}
\end{figure}

Let us begin with the definition of the geometry. We first define the \textit{tree region}, on which our loop-geometry is built. Consider the sub-region defined by the following non-negative condition 
\begin{equation}\label{eq: TreeRegion0}
 \langle i,j\rangle\equiv \frac{1}{4!}\epsilon_{IJKL}X_i^{IJ}X_j^{KL}>0\,.
\end{equation}
This constraint is invariant under the group  $\mathfrak{g}=$  GL(4)$_\text{left}\times$(GL(2)$_\text{right}$)$^4$ action on the bi-twistors. It is convenient to use this freedom to fix the gauge as: 
\begin{equation}\label{eq: GaugeFix}
\{X_1, X_2, X_3,X_4\}=\begin{pmatrix}
\mathbb{I}_{2\times 2} & 0 & \mathbb{I}_{2\times 2} & \mathbb{I}_{2\times 2}\\
0 & \mathbb{I}_{2\times 2} & \mathbb{I}_{2\times 2} & \begin{matrix}
c_1 & 0 \\
0 & c_2
\end{matrix}
\end{pmatrix}\,.
\end{equation} 
The remaining parameters $c_1, c_2$ are closely related to the cross-ratios, via  $(1-c_1)(1-c_2)=\frac{\langle 1,2\rangle \langle 3,4\rangle}{\langle 1,3\rangle \langle 2,4\rangle}=U$  and $c_1 c_2=\frac{\langle 1,4\rangle \langle 2,3\rangle }{\langle 1,3\rangle \langle 2,4\rangle}=V$.\footnote{The 
remaining parameters $c_1$, $c_2$ are naturally related to $z, \bar{z}$, which are commonly used in the literature. These are defined via $z \bar{z} \equiv U$, $(1-z) (1-\bar{z}) \equiv V$. } We further impose the constraint 
\begin{equation}\label{eq: DeltaDef}
\Delta^2=\det (\langle i,j\rangle)\big|_{i,j= 1,\ldots,4 }=s^2{+}t^2{+}u^2{-}2(st{+}tu{+}us)>0\,
\end{equation}
where for convenience we've defined the Mandelstam-like variables $s\equiv \langle 1,2\rangle\langle 3,4\rangle$, $t\equiv \langle 1,4\rangle\langle 2,3\rangle$ and $u\equiv \langle 1,3\rangle\langle 2,4\rangle$. Note that from the Levi-Cevita expression for $\Delta^2$, it is understood that $\Delta$ attains a minus sign under permutation. This $\Delta$ is precisely the square-root factor present in the symbol of the four-mass box one-loop scalar integral. Since in the frame eq.~\eqref{eq: GaugeFix} $\Delta^2=(c_1{-}c_2)^2$, this is tantamount to requiring that $c_{1,2}$ are real. The tree-region is then defined as  
\begin{equation}\label{eq: TreeRegion}
\mathbb{T}_4: \quad \langle i,j\rangle>0 \cap \Delta^2>0\,.
\end{equation}
In $c_1,c_2$ plane $\mathbb{T}_4$ is represented in Figure~\ref{fig:c1c2}. Note that $c_1, c_2$ is a double cover of $\mathbb{T}_4$. To see this,  consider a kinematic point inside the tree region
\begin{equation}
    \{X\}=\left(
\begin{array}{cccccccc}
 \frac{218}{23} & \frac{86}{9} & \frac{138}{7} & \frac{137}{12} & \frac{4700}{161} & \frac{755}{36} & \frac{36440}{161} & \frac{583}{18} \\
 \frac{15}{4} & \frac{155}{11} & \frac{209}{12} & \frac{84}{11} & \frac{127}{6} & \frac{239}{11} & \frac{586}{3} & \frac{323}{11} \\
 \frac{37}{8} & \frac{130}{11} & \frac{190}{11} & \frac{166}{9} & \frac{1927}{88} & \frac{2996}{99} & \frac{1557}{8} & \frac{4822}{99} \\
 \frac{239}{12} & \frac{4}{9} & \frac{221}{12} & \frac{127}{11} & \frac{115}{3} & \frac{1187}{99} & \frac{445}{2} & \frac{2330}{99} \\
\end{array}
\right)
\end{equation}
Through a series of GL(2) transformations together with a  GL(4) action, one can convert this to the standard form in eq.~\eqref{eq: GaugeFix}. There are two solutions: $(c_1, c_2)=(11, 2)$ and $(2, 11)$. Thus we see that the two \textbf{branches} $c_1 > c_2$ and $c_1 < c_2$ correspond to the same point in $\mathbb{T}_4$. Importantly, since $\Delta=(c_1-c_2)$, we see that $\Delta$ for a given point in $\mathbb{T}_4$ takes opposite signs on the two branches. This will be important in constructing the loop-form.

Before moving on to the loop-geometry, we first consider the canonical form for $\mathbb{T}_4$. Since $\mathbb{T}_4$ is two-dimensional, using the parameterization in eq.~\eqref{eq: GaugeFix}, one constructs a two-form \begin{equation}
f(c_1, c_2) dc_1dc_2
\end{equation}
where $f(c_1,c_2)$ develops simple poles at the boundary for $\mathbb{T}_4$.

We should compute the form in each branch separately and then combine. For the branch $c_1>c_2$, the associated form is given as, 
\begin{equation}
f^+(c_1,c_2)=\frac{(1 {-}c_2 {+} c_1 c_2)dc_1 dc_2}{(c_1{-}1)(c_2{-}1)c_1 c_2},
\end{equation}
where the superscript $^+$ denotes the branch. The form for the branch $c_2>c_1$ is given as $f^-(c_1,c_2)=f^+(c_1,c_2)|_{c_1\leftrightarrow c_2}$. To obtain the $\mathfrak{g}$ invariant form, we first apply (GL(2))$^4$ to restore eq.~\eqref{eq: GaugeFix} into an element of  $Y_a^{\mathcal{I}}\in$ Gr($4,8$), which is 16-dimensional. The final GL(4) invariant form is obtained by identifying the Jacobian factor from the gauge fixing is simply $\Delta^2$:  
\begin{equation}\label{eq:tree-form}
f^\pm(c_1,c_2)\,dc_1dc_2\;\rightarrow\;\left( \frac{s{+}t{+}u\pm \Delta}{stu}\right)\frac{\langle \! \langle \mathbf{X}_1, \mathbf{X}_2, \mathbf{X}_3, \mathbf{X}_4\rangle \! \rangle^4}{\Delta^2}\prod_{a=1}^4\langle\! \langle Y d^4 Y_a\rangle\!\rangle\equiv \omega^\pm(Y,\mathbf{X})\,d\mu_Y\,,
\end{equation}
where we use the shorthand notation $d\mu_Y=\prod_{a=1}^4\langle\! \langle Y d^4 Y_a\rangle\!\rangle$ and  $Y_a$ is the 8 component vector that is the row of Gr($4,8$). The arrow $\rightarrow$ indicates that the two-form in $(c_1,c_2)$  is \emph{uplifted} to the fully $\mathrm{GL}(4)$-invariant 16-form on $\mathrm{Gr}(4,8)$. In the above one identifies 
\begin{equation}
\langle i,j\rangle=\langle\! \langle Y, \mathbf{X}_i, \mathbf{X}_j \rangle\!\rangle\,.
\end{equation}
In other words, the bi-twistors $Z$ lives in the subspace of the bosonized twistors $\mathbf{X}$ defined by the complement of $Y$. Note that the projective weights of $Y$ cancel in eq.~\eqref{eq:tree-form} and the factor $\langle\! \langle \mathbf{X}_1, \mathbf{X}_2, \mathbf{X}_3, \mathbf{X}_4\rangle\!\rangle^4$ is necessary to balance the conformal weight at each point.

We now consider the loop-geometry. Given a point in 
 $\mathbb{T}_4$, we consider loop-fibers which is parameterized by a pair of twistors, one for each loop $(Z_{A_{\ell_i}}, Z_{B_{\ell_i}})$ with $i=1,\ldots, L$ at $L$ loops. The loop geometry is defined via the following inequalities, 
\begin{equation}\label{eq: LoopRegion0} \langle(A_{\ell_i},B_{\ell_i}),X_j\rangle>0, \quad \langle (A_{\ell_i},B_{\ell_i}),(A_{\ell_j},B_{\ell_j})\rangle>0\,.
\end{equation}
This definition is manifestly SL(2)- but not GL(2)-invariant. This can be easily rectified by defining the geometry through ratios: \begin{equation}\label{eq: LoopRegion} \frac{\langle(A_{\ell_i},B_{\ell_i}),X_j\rangle}{\langle(A_{\ell_i},B_{\ell_i}),X_i\rangle}>0, \quad \frac{\langle (A_{\ell_i},B_{\ell_i}),(A_{\ell_j},B_{\ell_j})\rangle}{\langle (A_{\ell_i},B_{\ell_i}),X_\ell)\rangle\langle (A_{\ell_j},B_{\ell_j}),X_k\rangle}>0\,.
\end{equation}
The two definitions are equivalent up to subtleties associated with local coordinate charts for  $(A_{\ell},B_{\ell})$ that do not cover the full geometry. This subtlety will be discussed in detail in Subsection \ref{sec: Loopfiber}. We will use eq.~\eqref{eq: LoopRegion} as our primary definition as it is free of such ambiguities. In practice, once the sign of any $\langle(A_{\ell_i},B_{\ell_i}),X_j\rangle$ is given, the remaining signs of four brackets $\langle(A_{\ell_i},B_{\ell_i}),\cdots\rangle$ are fixed. Thus at $L$-loops we have a total of $2^L$ sectors for the positive region. The loop-integrand is then the canonical form of this geometry, which has degree $2+4L$.

The duality between the correlator in the light-like limit~\cite{Eden:2010ce, Eden:2011ku, Alday:2010zy, Eden:2010zz, Eden:2011yp, Adamo:2011dq} and the square of the amplitude is manifest from the geometry. Taking  $\langle X_i,X_{i+1}\rangle \rightarrow 0$ to form a null 4-gon, bi-twistors are then identified as  $X^{IJ}_i=Z^{[I}_i Z^{J]}_{i{+}1}$ with $Z_5=Z_1$.  The inequality~\eqref{eq: LoopRegion0} then reduces precisely to that defining the squared Amplituhedron, and the resulting region reproduces the square of the amplitude.

Importantly, while the geometric form naturally captures the squared amplitude, the correlator is conventionally normalized with an extra factor of 2. Thus, to properly match the correlator normalization, we include an overall factor of 
$\frac{1}{2}$ when expressing the potential in terms of the geometric form in eq.~\eqref{eq: corr potential}.

\subsection{Chambers}

\begin{figure}
    \centering
    \includegraphics[width=0.4\linewidth]{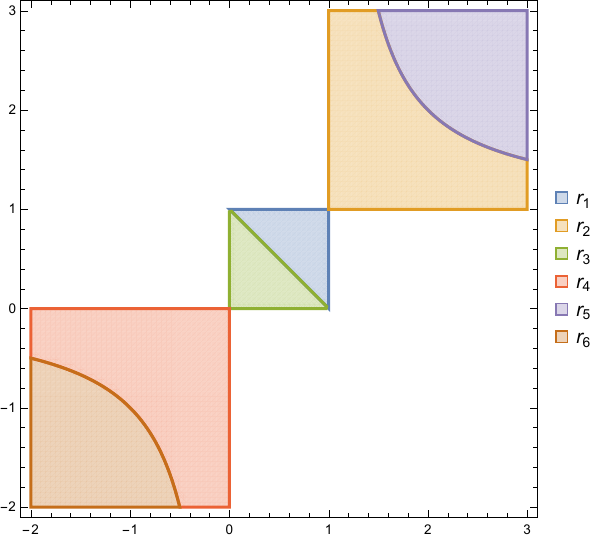}
    \caption{The six chambers of $\mathbb{T}_4$ in the  parametrization of eq.~\eqref{eq: GaugeFix}.  }
    \label{fig: c1c2 plane}
\end{figure}

As we traverse $\mathbb{T}_4$, the geometry defined in eq.~\eqref{eq: LoopRegion} might change; that is, the canonical form can differ between different points in $\mathbb{T}_4$. This suggests that the tree-region is further dissected into \textit{chambers}, 
\begin{equation}
\mathbb{T}_4=\sum_{\alpha}\, \mathbb{T}_{4,\alpha}
\end{equation}
where the canonical form for eq.~\eqref{eq: LoopRegion} is uniform within each chamber $\mathbb{T}_{4,\alpha}$. This phenomenon was observed in recent analysis of Amplituhedron in~\cite{He:2023rou, He:2024xed}, where the chambers are characterized by the collection of maximal cuts being positive. Indeed, as we will see, starting at three-loops, the positivity of certain cuts depends on the ordering of $s,t,u$. For example cuts maybe positive only if $s>t$ or $s$ being the largest (or not the largest) among the three. This suggests us that $\mathbb{T}_4$ should be decomposed into 6 chambers, 
\begin{equation}\label{eq:L=3_chambers}
    \begin{split}
        r_1:&\,s<t<u,\quad r_2:\,s<u<t,\quad r_3:\,t<s<u,\\
        r_4:&\,t<u<s,\quad r_5:\,u<s<t,\quad r_6:\,u<t<s\,.
    \end{split}
\end{equation}
In terms of parametrization $(c_1,c_2)$ the dissection for each chamber with $c_1>c_2$ is given as follows:
\begin{eqnarray}\label{eq: Chamberc}
  r_1(c_1>c_2):&&  \Big(0<c_2\leq \frac{1}{2}\; \land \;1{-}c_2<c_1<1\Big) \lor \Big(\frac{1}{2}<c_2<1 \;\land \;c_2<c_1<1 \Big) ,\nonumber\\
  r_2(c_1>c_2):&&  \Big(1<c_2<2\;\land \;c_2<c_1<\frac{c_2}{c_2{-}1}\Big) ,\nonumber\\
  r_3(c_1>c_2):&&  \Big(0<c_2<\frac{1}{2}\;\land \;c_2<c_1<1{-}c_2\Big) ,\nonumber\\
  r_4(c_1>c_2):&&  \Big(c_2\leq {-}1\; \land \;\frac{1}{c_2}<c_1<0\Big) \lor \Big({-}1<c_2<0 \;\land \;c_2<c_1<0 \Big) ,\nonumber\\
  r_5(c_1>c_2):&&  \Big(1<c_2\leq2\; \land \;c_1>\frac{c_2}{c_2-1}\Big) \lor \Big(c_2>2 \;\land \;c_1>c_2 \Big) ,\nonumber\\
  r_6(c_1>c_2):&&  \Big(c_2<{-}1\; \land \;c_2<c_1<\frac{1}{c_2}\Big)\,.
\end{eqnarray}
The chambers for $c_1 < c_2$ are obtained by interchanging $c_1 \leftrightarrow c_2$. The full chamber structure is illustrated in Figure~\ref{fig: c1c2 plane}.
For each chamber, we construct its chamber form $\omega^\pm_\sigma (Y,\mathbf{X})$. Explicitly, they are given as:
\begin{equation}\label{eq: ChamberForm}
 \omega_{r_1}^\pm=\frac{\langle \! \langle \mathbf{X}_1, \mathbf{X}_2, \mathbf{X}_3, \mathbf{X}_4\rangle\!\rangle^4}{\Delta^2}\times\left( \frac{1}{s (t-s)} {\pm} \frac{\Delta}{s(t-s)(u-t)}\right)\,.
\end{equation}
The poles of this form reflect the chamber boundaries  $s=0$, $s=t$, $\Delta^2=0$ and $t=u$. The forms for the other chambers are given by interchanging $s,t,u$ on $\omega_{r_1}^\pm$. Summing over the six chambers, we can recover the tree-form in eq.~\eqref{eq:tree-form}, {\it i.e.} 
\begin{equation}
\sum_{\sigma } \;\omega^\pm_\sigma (Y,\mathbf{X})=\omega^\pm (Y,\mathbf{X})=\left( \frac{s{+}t{+}u\pm \Delta}{stu}\right)\frac{\langle \! \langle \mathbf{X}_1, \mathbf{X}_2, \mathbf{X}_3, \mathbf{X}_4\rangle \! \rangle^4}{\Delta^2} d\mu_Y\,.
\end{equation}

\subsection{Loop fibration and loop-forms}\label{sec: Loopfiber}
The potential for the correlation function is then identified as $\mathcal{G}(\mathbf{X},Y)$ through
\begin{equation}\label{eq: corr potential}
\mathcal{G}^{(L)}(\mathbf{X},Y)=\frac{1}{2}\sum_{\sigma, \pm} \;\omega^\pm_\sigma (Y,\mathbf{X})\,\Omega^{(L) \pm}_\sigma \,.
\end{equation}
We divide the geometric form by 2 to properly match the correlator normalization.

Note that eq.~\eqref{eq: TreeRegion0} combined with eq.~\eqref{eq: LoopRegion0} enjoys a $P_{4{+}L}$ permutation invariance. This symmetry is broken by our separation into tree and loop forms. However as we will see, in the end the resulting integrand indeed enjoys the full $P_{4{+}L}$ permutation invariance. To construct the form we expand $(A_{\ell_i}, B_{\ell_i})$ on any four external twistors $Z$s and compute the $4L$-form for eq.~\eqref{eq: LoopRegion}, which we denote as $\Omega^{(L) \pm}$.

To explicitly construct the loop geometry, at each loop  we introduce a pair of twistors $(Z_{A_{\ell}}, Z_{B_{\ell}})$, with $\ell=1,\cdots, L$, expanded on any set of four twistors. For convenience we will choose two  bi-twistors $\{X_{i}, X_{j}\}$:
\begin{equation}\label{eq: LoopPara}
\begin{pmatrix}
Z_{A_\ell} \\ Z_{B_\ell}
\end{pmatrix}=\begin{pmatrix}
1& x_{\ell}& 0 & -w_{\ell} \\ 0& y_{\ell}& 1& z_{\ell}
\end{pmatrix}\begin{pmatrix}
Z_{1,i} \\ Z_{2,i}\\ Z_{1,j}\\ Z_{2,j}\,
\end{pmatrix}
\end{equation}
Then for any given point in $\mathbb{T}_4$ we construct a canonical form (4-form at one-loop) from the region defined in eq.~\eqref{eq: LoopRegion0}. An important subtlety is that local charts such as eq.~\eqref{eq: LoopPara} in general cannot  cover the whole geometry. For example, if one chooses $i = 1, j = 2$ in \eqref{eq: LoopPara}, then  $\langle (A,B), Z_{2,2}  , Z_{2,1}\rangle$ becomes positive definite. This limitation manifests itself as the presence of spurious boundaries in the cylindrical decomposition eq.~\eqref{eq: oneLoopRegion}, since in this chart, $\langle (A,B), Z_{2,2}  , Z_{2,1}\rangle=0$ is a boundary of the image of eq.~\eqref{eq: LoopPara}.  This feature was already discussed in great detail in Subsection 4.4 of~\cite{Dian:2021idl}. One can compute the canonical form using another non-overlapping chart, say
\begin{equation}\label{eq: LoopPara2}
\begin{pmatrix}
Z_{A_\ell} \\ Z_{B_\ell}
\end{pmatrix}=\begin{pmatrix}
1& x_{\ell}& 0 & -w_{\ell} \\ 0& y_{\ell}& -1& z_{\ell}
\end{pmatrix}\begin{pmatrix}
Z_{1,i} \\ Z_{2,i}\\ Z_{1,j}\\ Z_{2,j}\,
\end{pmatrix}
\end{equation}
and add the two forms together. Alternatively, one defines the geometry in a manifest gauge invariant manner, as in our eq.~\eqref{eq: LoopRegion}, and sum over its $2^L$ distinct sectors. Each sector is non-overlapping, as can be confirmed from the fact that the coordinate charts eqs.~\eqref{eq: LoopPara} and \eqref{eq: LoopPara2} do not intersect. While the two approaches are equivalent, the latter approach allows us to use a single chart to compute the form.

Next we perform cylindrical decomposition of the inequality eq.~\eqref{eq: LoopRegion} under the parameterization of $(c_1, c_2)$ in eq.~\eqref{eq: LoopPara}. Since the boundary of the loop-region corresponds to propagator becoming null, we expect the canonical form can be expanded on local integrands. As mentioned previously, starting at three loops~\cite{He:2024xed} the loop-forms of the different chambers become distinct. The loop-form for a given chamber can be further separated into sub-blocks, reflecting the fact that for a given cut, the criteria of positivity maybe compatible with multiple chambers. For example if a cut is positive  when $s$ is the greatest, then integrands that reproduce this cut must be present both in $r_4$ and $r_6$. Turning this statement around, for a given chamber, say $r_1$, the positive cuts can be organized into subsets whose positivity only rely on a subset of the chamber condition. This suggests us to further decompose each chamber into 
\begin{equation}\label{eq: ChamberBlocks}
\Omega^{(L)\pm}_{r_i}= 2\Delta^2 A_{\sigma_3} \pm2\Delta\big(B+ (\sigma_2-\sigma_1)C_{\sigma_2,\sigma_1} + (\sigma_3-\sigma_1) C_{\sigma_3,\sigma_1} + (\sigma_3-\sigma_2) C_{\sigma_3,\sigma_2} \big)\,,   
\end{equation}
where the $\sigma_i$'s correspond to the Mandelstam variables in the inequality $r_i$ in eq.~\eqref{eq:L=3_chambers}, {\it i.e.} $\sigma_1<\sigma_2<\sigma_3$. Thus $A_{\sigma_3}$ for each chamber is labeled by the largest $s,t,u$. In the above, $A_{\sigma_3}$ contains cuts whose positivity only requires $\Delta^2>0$ and $\sigma_3$ to be the greatest, while cuts in $C_{\sigma_3,\sigma_1}$ require $\sigma_3>\sigma_1$ along with $\Delta^2>0$. In contrast, cuts in $B$ only require $\Delta^2>0$. Thus one can conclude that the block $B$ is universal, being present in all chambers.

Finally, recall that upon integrating over the auxiliary fermionic variables as well as taking the fermionic derivatives, the factor $\langle\! \langle \mathbf{X}_1, \mathbf{X}_2, \mathbf{X}_3, \mathbf{X}_4\rangle\!\rangle^4$ reduces to $R(1,2,3,4)$ as shown in eq.~\eqref{eq: FermionicPro}. Since the chamber forms in eq.~\eqref{eq: ChamberForm} are proportional to $\langle\! \langle \mathbf{X}_1, \mathbf{X}_2, \mathbf{X}_3, \mathbf{X}_4\rangle\!\rangle^4$, the integrand constructed from the geometry makes manifest the non-renormalization of the four-point correlator~\cite{Eden:2000bk,Eden:2011we}.

\section{Results up to three-loops}\label{sec4}
In the previous section, we presented a general overview of the properties and structure of the four-point Correlahedron. We now examine the explicit one- to three-loop results computed in~\cite{He:2024xed}, in order to illustrate and highlight the details and subtleties of the geometry.  

\subsection{One-loop}
At one-loop, the loop-form is uniform across $\mathbb{T}_4$. There are two sectors arising from eq.~\eqref{eq: LoopRegion},
\begin{equation}\label{eq: oneLoopRegion}
    \langle (A,B), X_j\rangle>0 \quad \text{or} \quad \langle (A,B), X_j\rangle<0, \quad j=1,2,3,4\,.
\end{equation}
We use the local chart in eq.~\eqref{eq: LoopPara} with $i = 1, j = 2$. Within each sector of eq.~\eqref{eq: oneLoopRegion}, the resulting form will contain the spurious boundary due to the fact that the local chart does not cover the whole geometry. In our case the spurious boundary takes the form,
\begin{equation}
x^2_{a,*}=\langle (A,B), Z_{2,2}  , Z_{2,1}\rangle\,.
\end{equation}
Note that we have freely transitioned between $\langle i, j\rangle$ and $x^2_{i,j}$, since the conformal invariant nature of the geometry ensures that all dependence on the infinity twistors, appearing in eq.~\eqref{eq: DisDef}, cancels in the end. Indeed, the explicit forms of the two sectors are then given by 
\begin{equation}
    \begin{split}
\Omega_\pm^{(1)+}=&\frac{\Delta}{x_{a,1}^2x_{a,2}^2x_{a,3}^2x_{a,4}^2}\pm \frac{x_{2,3}^2 x_{4,*}^2{+}x_{3,4}^2 x_{2,*}^2{-} x_{2,4}^2 x_{3,*}^2 }{x_{a,2}^2 x_{a,3}^2 x_{a,4}^2 x_{a,*}^2 } {\pm} \frac{x_{3,4}^2 x_{1,*}^2{+}x_{1,4}^2 x_{3,*}^2{-}x_{1,3}^2 x_{4,*}^2}{x_{a,1}^2 x_{a,3}^2 x_{a,4}^2 x_{a,*}^2}\\
        &\phantom{yyy}\mp \frac{x_{1,2}^2 x_{4,*}^2{+}{x_{1,4}^2 x_{2,*}^2}{-} x_{2,4}^2 x_{1,*}^2 }{x_{a,1}^2 x_{a,2}^2 x_{a,4}^2 x_{a,*}^2 } \mp \frac{x_{1,2}^2 x_{3,*}^2{+}x_{2,3}^2 x_{1,*}^2{-} x_{1,3}^2 x_{2,*}^2 }{x_{a,1}^2 x_{a,2}^2 x_{a,3}^2 x_{a,*}^2 }\,,
    \end{split}
\end{equation}
where we have used the subscript $\pm$ to denote the two sectors in eq.~\eqref{eq: oneLoopRegion}.
Since the sum of the two sectors corresponds to the gauge invariant definition~\eqref{eq: LoopRegion}, the spurious boundaries should cancel. Indeed, one sees that $\Omega^{(1)+}=\Omega_+^{(1)+}+\Omega_-^{(1)+}$ is free of spurious boundaries.

Summing over the sectors gives rise to the form for one branch $(c_1>c_2)$, namely $\Omega^{(1)+}$. For the other branch, $\Omega^{(1)-}$, one simply reverses the sign in front of $\Delta$. Thus we have,
\begin{equation}\label{eq: one-loop form}
\Omega^{(1)\pm}=\frac{\pm2\Delta}{x_{a,1}^2 x_{a,2}^2 x_{a,3}^2 x_{a,4}^2}~d^4 x_a=\pm 2 \Delta\,g(1,2,3,4)\,.
\end{equation}
The factor of 2 ensures that, in the light-like limit, the geometric form reduces precisely to the squared tree amplitude—twice the value of the scalar box integrand.

Combining the forms for the two branches,  one obtains: 
\begin{equation}
\mathcal{G}^{(1)}(\mathbf{X},Y)=\frac{1}{2}\sum_{\pm}\;\omega^\pm\Omega^{(1)\pm} =\frac{\langle \! \langle \mathbf{X}_1, \mathbf{X}_2, \mathbf{X}_3, \mathbf{X}_4\rangle \! \rangle^4}{stu\Delta} \left[\Delta\,g(1,2,3,4)\right] \,,
\end{equation}
where $\omega^\pm$ is defined in eq.~\eqref{eq:tree-form}.
The $\langle \! \langle \mathbf{X}_1, \mathbf{X}_2, \mathbf{X}_3, \mathbf{X}_4\rangle \! \rangle^4$ factor eventually yields the universal prefactor via eq.~\eqref{eq: FermionicPro}, and after canceling the infinity-twistor dependence, we arrive at precisely the one-loop result:
\begin{equation}
G_4^{(1)}=R(1,2,3,4)\left(\prod_{i<j}x^2_{ij}\right)\frac{g(1,2,3,4)}{\prod_{i<j}x^2_{ij}}\,.
\end{equation}
We deliberately separate out the combination $\frac{g(1,2,3,4)}{\prod_{i<j}x^2_{ij}}$, which is precisely $\mathcal{F}^{(1)}$ in eq.~\eqref{eq: Fgraph}, and enjoys permutation invariant with respect to all five points $x_i$ with $i=1,2,3,4,a$. Before moving on to two loops, we note that the sign difference for the two branches exactly conspires to cancel the $1/\Delta^2$ in the tree-form.

\subsection{Two-loops}

Starting from two loop, direct triangulation of the geometry becomes challenging due to the higher dimensionality of the form. A practical approach is to construct an ansatz for the canonical form and determine its coefficients by matching to lower-dimensional boundaries — specifically, multi-cuts~(see Supplemental Material of \cite{He:2022cup} for details). These boundary forms can be computed using cylindrical decomposition of the defining inequalities, evaluated on the cut solutions where the geometry will be simplified. 

At two-loops the geometry is split into four sectors 
\begin{equation}
    \begin{split}
        \ &\langle (AB) X_i\rangle>0,\quad \langle (CD) X_i\rangle>0, \quad  \langle (AB) (CD)\rangle>0,\\
        \ &\langle (AB) X_i\rangle<0,\quad \langle (CD) X_i\rangle>0, \quad  \langle (AB) (CD)\rangle<0,\\
        \ &\langle (AB) X_i\rangle>0,\quad \langle (CD) X_i\rangle<0, \quad  \langle (AB) (CD)\rangle<0,\\
        \ &\langle (AB) X_i\rangle<0,\quad \langle (CD) X_i\rangle<0, \quad  \langle (AB) (CD)\rangle>0.
    \end{split}
\end{equation}
When determining the relative signs for combining the four sectors, special care must be taken with the orientation. For example, consider the regions $x\geq a$ versus $x\leq a$. The oriented canonical forms for these two regions coincide up to an overall sign:
 \begin{equation}
  x\geq a:\quad   \frac{dx}{x-a}, \quad x\leq a:\quad -\frac{dx}{x-a}\,.
 \end{equation} 
The relative sign simply encodes the direction from which the boundary is approached, as illustrated:
\begin{eqnarray}\label{eq: orientation}
&&\vcenter{\hbox{\scalebox{1}{
\begin{tikzpicture}[x=0.75pt,y=0.75pt,yscale=-1,xscale=1]

\draw  [line width=0.75] (-30,140) node [anchor=north west][inner sep=0.75pt]  [align=center] {$ \begin{tikzpicture}[x=0.75pt,y=0.75pt,yscale=-1,xscale=1]

\draw [line width=0.75]    (210.8,136.6) -- (328,136.6) ;
\draw    (281.65,136.35) -- (302,136.35) ;
\draw [shift={(305,136.35)}, rotate = 180] [fill={rgb, 255:red, 0; green, 0; blue, 0 }  ][line width=0.08]  [draw opacity=0] (8.93,-4.29) -- (0,0) -- (8.93,4.29) -- cycle    ;
\draw  [fill={rgb, 255:red, 0; green, 0; blue, 0 }  ,fill opacity=1 ] (266.28,136.6) .. controls (266.28,134.87) and (267.67,133.48) .. (269.4,133.48) .. controls (271.13,133.48) and (272.53,134.87) .. (272.53,136.6) .. controls (272.53,138.33) and (271.13,139.73) .. (269.4,139.73) .. controls (267.67,139.73) and (266.28,138.33) .. (266.28,136.6) -- cycle ;
\draw [line width=1.5]    (269.4,136.6) -- (328.5,136.6) ;

\draw (250,161) node [anchor=north west][inner sep=0.75pt]   [align=left] {$\displaystyle x >a$};
\draw (264.5,120) node [anchor=north west][inner sep=0.75pt]   [align=left] {$\displaystyle a$};

\end{tikzpicture}
$};

\draw  [line width=0.75] (180,140) node [anchor=north west][inner sep=0.75pt]  [align=center] {$ \begin{tikzpicture}[x=0.75pt,y=0.75pt,yscale=-1,xscale=1]

\draw [line width=0.75]    (210.8,136.6) -- (328,136.6) ;
\draw    (228.65,136.35) -- (249,136.35) ;
\draw [shift={(252,136.35)}, rotate = 180] [fill={rgb, 255:red, 0; green, 0; blue, 0 }  ][line width=0.08]  [draw opacity=0] (8.93,-4.29) -- (0,0) -- (8.93,4.29) -- cycle    ;
\draw  [fill={rgb, 255:red, 0; green, 0; blue, 0 }  ,fill opacity=1 ] (266.28,136.6) .. controls (266.28,134.87) and (267.67,133.48) .. (269.4,133.48) .. controls (271.13,133.48) and (272.53,134.87) .. (272.53,136.6) .. controls (272.53,138.33) and (271.13,139.73) .. (269.4,139.73) .. controls (267.67,139.73) and (266.28,138.33) .. (266.28,136.6) -- cycle ;
\draw [line width=1.5]    (211.4,136.6) -- (270.5,136.6) ;

\draw (250,161) node [anchor=north west][inner sep=0.75pt]   [align=left] {$\displaystyle x< a$};
\draw (264.5,120) node [anchor=north west][inner sep=0.75pt]   [align=left] {$\displaystyle a$};

\end{tikzpicture}
$};

\end{tikzpicture}

}
}}
\end{eqnarray}
Here, the arrows indicate the direction in which $x - a$ increases. For the region $x > a$, the arrow points outward, meaning the boundary is approached from above with positive orientation (assigned $+1$). Conversely, for $x < a$, the arrow points inward, corresponding to an approach from below with negative orientation (assigned $-1$). As we will see in Subsection~\ref{sec: three-loop chamber analysis}, this graphical convention provides a convenient way to visualize orientation associated with a given cut.

For each cut, the orientation can therefore be tracked by whether the boundary is approached from above or below. The \textit{net orientation} of multi-cuts is defined as the product of their individual orientations. We refer to configurations with a positive product as having \textit{net-positive orientation}, and those with a negative product as having \textit{net-negative orientation}.


Returning to the Correlahedron, a useful boundary for determining the canonical form is the two-cut geometry defined by $\langle AB X_1\rangle = \langle CD X_3\rangle = 0$. Parameterizing the loop lines as: $(A,B)=(Z_{1,1} + x_1 Z_{2,1}-w_1 Z_{2,2}, y_1 Z_{2,2}+Z_{1,2}+z_1 Z_{2,2})$ and $(C,D)=(Z_{1,3} + x_2 Z_{2,3}-w_2 Z_{2,4}, y_2 Z_{2,3}+Z_{1,4}+z_2 Z_{2,4})$, the cut condition $\langle AB X_1\rangle= \langle CD X_3 \rangle=0$ corresponds to approaching $w_1=w_2=0$ since
\begin{equation}
    \langle AB X_1\rangle= w_1,\quad \langle CD X_3\rangle= (1-c_1)(1-c_2) w_2\,. 
\end{equation}
Note that Jacobian factors—here $(1-c_1)(1-c_2)$—must be taken into account when determining the correct orientation. In this example, the Jacobian factor is strictly positive and one can read off the orientation directly for each sector. As we can see for the first (fourth) sector, we approach both boundaries $\langle(AB) X_1\rangle$ and  $\langle(CD) X_3\rangle$ from above (below) so resulting in a net-positive orientation. While for the second and third sector, one boundary is approached from above and the other from below, thus yielding a net-negative orientation. The full contribution is therefore the oriented sum:
\begin{equation}
\Omega^{(2)+}=    \Omega^{(2)+}_{+,+}-\Omega^{(2)+}_{+,-}-\Omega^{(2)+}_{-,+}+\Omega^{(2)+}_{-,-}\,.
\end{equation}
The loop-forms for the two branches are given by 
\begin{equation}
\Omega^{(2)\pm}=2\,\Delta^2 g(1,2,3,4)^2 \pm 2\, \Delta \left(h(1,2; 3,4) + 5~{\rm perm.}\right)\,, 
\end{equation}
where it is expanded in terms of the square of the one-loop box and the double-box integrands
\begin{eqnarray}\label{eq: IntDef}
g(1,2,3,4)^2:=& \frac{d^4 x_a d^4 x_b}{2 x_{a,1}^2 x_{a,2}^2 x_{a,3}^2 x_{a,4}^2 x_{b,1}^2 x_{b,2}^2 x_{b,3}^2 x_{b,4}^2 
} + (a\leftrightarrow b)\,,\nonumber\\   
h(1,2;3,4):=& \frac{d^4 x_a d^4 x_b x_{3,4}^2}{x_{a,1}^2 x_{a,3}^2 x_{a,4}^2 x_{a,b}^2 x_{b,2}^2 x_{b,3}^2 x_{b,4}^2} + (a\leftrightarrow b)\,. 
\end{eqnarray}

Summing over both branches and dividing by 2 again produces the correct integrand for two-loop correlators (with $\Delta$ canceled): 
\begin{eqnarray}
&&\mathcal{G}^{(2)}(\mathbf{X},Y)=\frac{\langle \! \langle \mathbf{X}_1, \mathbf{X}_2, \mathbf{X}_3, \mathbf{X}_4\rangle\!\rangle^4}{stu}\left((s{+}t{+}u)g(1,2,3,4)^2+ \left(h(1,2; 3,4) + 5~{\rm perm.}\right) \right).\nonumber\\
\end{eqnarray}
Once again, the integrand is proportional to $\langle\! \langle \textbf{X}_1, \textbf{X}_2, \textbf{X}_3, \textbf{X}_4\rangle\!\rangle^4$ and the combination in the bracket together with  $s t u$ in the  denominator is in fact permutation invariant with respect to six points $x_i$ with $i=1,2,3,4,a,b$.

\subsection{Three-loops}

As mentioned previously, starting at three loops the loop-forms depend on the chambers. The three-loop form can be organized into the following pattern:
\begin{equation}\label{eq: 3Loops}
\Omega^{(3)\pm}_{r_i}=2 \Delta^2 A_{\sigma_3} \pm2\Delta\big(B+ (\sigma_2-\sigma_1)C_{\sigma_2,\sigma_1} + (\sigma_3-\sigma_1)C_{\sigma_3,\sigma_1} + (\sigma_3-\sigma_2) C_{\sigma_3,\sigma_2} \big)\,,   
\end{equation}
where the $\sigma_i$ correspond to the Mandelstam-like variables in the inequality $r_i$ of eq.~\eqref{eq:L=3_chambers}. Thus $A_{\sigma_3}$ for each chamber is labeled by the largest among $s,t,u$, and $B$ is universal. For instance, the loop-form for the chamber $r_1$ ($s<t<u$) is given by
\begin{equation}
    \Omega^{(3)\pm}_{r_1} =2\Delta^2 A_{u} \pm 2\Delta \big(B + (t-s)\,C_{s,t}+ (u-s)\,C_{s,u} +(u-t)C_{t,u}\big)\,.
\end{equation}
Each block consists of loop-integrands whose maximal cuts——where the 12 loop variables are localized by the propagators——are positive. That is, it lies within the loop-geometry defined in eq.~\eqref{eq: LoopRegion}. Importantly, not only are the cuts positive, but their leading singularities become pure numbers once combined with the prefactor. For example,
\begin{eqnarray}\label{eq: LS}
&&{\rm LS}[A_{\sigma_i}]\propto \frac{1}{\Delta^2},\quad {\rm LS}[B]\propto \frac{1}{\Delta},\nonumber\\
&&{\rm LS}[C_{s,t}]\propto \frac{1}{\Delta(t-s)},\quad {\rm LS}[C_{s,u}]\propto \frac{1}{\Delta(u-s)}, \quad {\rm LS}[C_{t,u}]\propto \frac{1}{\Delta(u-t)}\,,
\end{eqnarray}
where the proportionality factor is a pure number and possibly zero. Note that, since a given integrand can possess multiple leading singularities, the above statement applies to \textit{all} such singularities within each block.

The blocks are given explicitly as follows:
\begin{equation}\label{eq: three-loop block}
    \begin{split}
          B &:= T^{12;34}+E^{12;34} + 11\, \text{perms.} +  L^{12;34}  + 5\, \text{perms.} , \\
           A_s &:=\big[  H^{14;23}+ (1,4){\leftrightarrow}(2,3) \big]+ (3{\leftrightarrow} 4)+    gh^{12;34} +  gh^{34;12}  ,\\
          &  C_{t,u}:= 2 (      {E^{\prime}}^{12;34} + {E^{\prime}}^{34;12})   \,.
    \end{split}
\end{equation}
As can be read off from the labels, the cuts in $B$ are positive everywhere in $\mathbb{T}_4$. The explicit integrands include the following topologies, 
\begin{eqnarray}
&&\vcenter{\hbox{\scalebox{1}{
\begin{tikzpicture}[x=0.75pt,y=0.75pt,yscale=-1,xscale=1]

\draw  [line width=0.75] (100,100) node [anchor=north west][inner sep=0.75pt]  [align=center] {$ \begin{tikzpicture}[x=0.75pt,y=0.75pt,yscale=-1,xscale=1]

\draw  [line width=0.75]  (196.78,234.67) -- (205.92,243.81) ;
\draw  [line width=0.75]  (127.59,235.14) -- (118.45,244.28) ;
\draw  [line width=0.75] (127.3,192.45) -- (127.06,157.87) -- (161.64,157.63) -- (161.87,192.21) -- cycle ;
\draw  [line width=0.75] (161.87,192.21) -- (161.64,157.63) -- (196.22,157.4) -- (196.45,191.98) -- cycle ;
\draw [line width=0.75]  (127.59,235.14) -- (127.3,192.45) -- (196.49,191.98) -- (196.78,234.67) -- cycle ;
\draw [line width=0.75]   (117.93,148.73) -- (127.06,157.87) ;
\draw   [line width=0.75] (205.35,148.26) -- (196.22,157.4) ;

\draw (155.1,237.54) node [anchor=north west][inner sep=0.75pt]   [align=left] {$\displaystyle j$};
\draw (156.24,209) node [anchor=north west][inner sep=0.75pt]   [align=left] {$\displaystyle a$};
\draw (139.67,166.4) node [anchor=north west][inner sep=0.75pt]   [align=left] {$\displaystyle b$};
\draw (173.74,169.6) node [anchor=north west][inner sep=0.75pt]   [align=left] {$\displaystyle c$};
\draw (157.11,139.18) node [anchor=north west][inner sep=0.75pt]   [align=left] {$\displaystyle i$};
\draw (109.66,184.25) node [anchor=north west][inner sep=0.75pt]   [align=left] {$\displaystyle k$};
\draw (203.95,184.62) node [anchor=north west][inner sep=0.75pt]   [align=left] {$\displaystyle l$};

\end{tikzpicture}
$};

\draw  [line width=0.75] (240,110) node [anchor=north west][inner sep=0.75pt]  [align=center] {$ \begin{tikzpicture}[x=0.75pt,y=0.75pt,yscale=-1,xscale=1]

\draw  [line width=0.75] (310.59,211.36) -- (301.45,183.43) -- (325.19,166.11) -- (349.01,183.33) -- (339.98,211.3) -- cycle ;
\draw  [line width=0.75] (310.6,211.03) -- (283,220.18) -- (273.85,192.58) -- (301.45,183.43) -- cycle ;
\draw  [line width=0.75] (367.62,220.35) -- (339.98,211.3) -- (349.03,183.67) -- (376.67,192.71) -- cycle ;
\draw  [line width=0.75]  (273.85,192.58) -- (262.31,186.77) ;
\draw  [line width=0.75]  (277.2,231.72) -- (283,220.18) ;
\draw  [line width=0.75]  (376.67,192.71) -- (388.19,186.87) ;
\draw  [line width=0.75]  (373.46,231.87) -- (367.62,220.35) ;
\draw  [line width=0.75]  (325.19,154.45) -- (325.19,166.11) ;

\draw (287.57,195.14) node [anchor=north west][inner sep=0.75pt]   [align=left] {$\displaystyle a$};
\draw (319.67,185.74) node [anchor=north west][inner sep=0.75pt]   [align=left] {$\displaystyle b$};
\draw (353.74,195.14) node [anchor=north west][inner sep=0.75pt]   [align=left] {$\displaystyle c$};
\draw (287.77,162.21) node [anchor=north west][inner sep=0.75pt]   [align=left] {$\displaystyle i$};
\draw (353.66,162.25) node [anchor=north west][inner sep=0.75pt]   [align=left] {$\displaystyle j$};
\draw (260.62,200.62) node [anchor=north west][inner sep=0.75pt]   [align=left] {$\displaystyle k$};
\draw (377.95,201.28) node [anchor=north west][inner sep=0.75pt]   [align=left] {$\displaystyle k$};
\draw (322.44,220.84) node [anchor=north west][inner sep=0.75pt]   [align=left] {$\displaystyle l$};

\end{tikzpicture}
$};

\draw  [line width=0.75] (400,120) node [anchor=north west][inner sep=0.75pt]  [align=center] {$ \begin{tikzpicture}[x=0.75pt,y=0.75pt,yscale=-1,xscale=1]

\draw  [line width=0.75] (276.68,187.07) -- (276.93,216.46) -- (249.06,225.78) -- (231.58,202.15) -- (248.65,178.23) -- cycle ;
\draw  [line width=0.75] (276.88,216.46) -- (276.68,187.27) -- (305.87,187.07) -- (306.07,216.26) -- cycle ;
\draw  [line width=0.75] (334.23,225.31) -- (306.21,216.45) -- (305.98,187.06) -- (333.86,177.76) -- (351.32,201.4) -- cycle ;

\draw  [line width=0.75]  (243.67,166) -- (248.65,178.23) ;
\draw   [line width=0.75] (249.06,225.78) -- (241.67,238) ;
\draw  [line width=0.75]  (217,202) -- (231.58,202.15) ;
\draw  [line width=0.75]  (351.32,201.4) -- (365.9,201.56) ;
\draw  [line width=0.75]  (341.25,165.54) -- (333.86,177.76) ;
\draw  [line width=0.75]  (334.23,225.31) -- (339.22,237.54) ;

\draw (252.57,198.14) node [anchor=north west][inner sep=0.75pt]   [align=left] {$\displaystyle a$};
\draw (286.67,195.4) node [anchor=north west][inner sep=0.75pt]   [align=left] {$\displaystyle b$};
\draw (321.41,198.19) node [anchor=north west][inner sep=0.75pt]   [align=left] {$\displaystyle c$};
\draw (226.1,175.87) node [anchor=north west][inner sep=0.75pt]   [align=left] {$\displaystyle k$};
\draw (223.32,211.25) node [anchor=north west][inner sep=0.75pt]   [align=left] {$\displaystyle l$};
\draw (350.1,174.54) node [anchor=north west][inner sep=0.75pt]   [align=left] {$\displaystyle k$};
\draw (348.66,212.59) node [anchor=north west][inner sep=0.75pt]   [align=left] {$\displaystyle l$};
\draw (286.29,167.62) node [anchor=north west][inner sep=0.75pt]   [align=left] {$\displaystyle i$};
\draw (286.78,221.84) node [anchor=north west][inner sep=0.75pt]   [align=left] {$\displaystyle j$};

\end{tikzpicture}
$};

\draw  [line width=0.75] (170,250) node [anchor=north west][inner sep=0.75pt]  [align=center] {$ \begin{tikzpicture}[x=0.75pt,y=0.75pt,yscale=-1,xscale=1]

\draw [line width=0.75]  (214.75,204.12) -- (214.54,173.37) -- (245.3,173.16) -- (245.51,203.91) -- cycle ;
\draw  [line width=0.75] (245.51,203.91) -- (245.3,173.16) -- (276.05,172.95) -- (276.26,203.71) -- cycle ;
\draw [line width=0.75]  (276.26,203.71) -- (276.05,172.95) -- (306.81,172.74) -- (307.02,203.5) -- cycle ;
\draw  [line width=0.75]  (205.4,164.23) -- (214.54,173.37) ;
\draw  [line width=0.75]  (214.75,204.12) -- (205.61,213.26) ;
\draw  [line width=0.75]  (315.95,163.6) -- (306.81,172.74) ;
\draw  [line width=0.75]  (307.02,203.5) -- (316.16,212.63) ;

\draw (225.24,183.48) node [anchor=north west][inner sep=0.75pt]   [align=left] {$\displaystyle a$};
\draw (256,180.4) node [anchor=north west][inner sep=0.75pt]   [align=left] {$\displaystyle b$};
\draw (286.07,183.19) node [anchor=north west][inner sep=0.75pt]   [align=left] {$\displaystyle c$};
\draw (202.1,181.54) node [anchor=north west][inner sep=0.75pt]   [align=left] {$\displaystyle i$};
\draw (310.66,179.59) node [anchor=north west][inner sep=0.75pt]   [align=left] {$\displaystyle j$};
\draw (254.95,153.28) node [anchor=north west][inner sep=0.75pt]   [align=left] {$\displaystyle k$};
\draw (254.11,208.51) node [anchor=north west][inner sep=0.75pt]   [align=left] {$\displaystyle l$};

\end{tikzpicture}
$};

\draw  [line width=0.75] (320,250) node [anchor=north west][inner sep=0.75pt]  [align=center] {$\begin{tikzpicture}[x=0.75pt,y=0.75pt,yscale=-1,xscale=1]

\draw  [line width=0.75] (245.95,221.19) -- (245.74,190.43) -- (276.5,190.22) -- (276.71,220.98) -- cycle ;
\draw  [line width=0.75] (276.71,220.98) -- (276.5,190.22) -- (307.25,190.01) -- (307.46,220.77) -- cycle ;
\draw  [line width=0.75] (357.46,220.77) -- (357.25,190.01) -- (388.01,189.81) -- (388.22,220.56) -- cycle ;
\draw  [line width=0.75]  (307.46,220.77) -- (316.6,229.91) ;
\draw   [line width=0.75] (316.39,180.88) -- (307.25,190.01) ;
\draw  [line width=0.75]  (236.6,181.3) -- (245.74,190.43) ;
\draw   [line width=0.75] (245.95,221.19) -- (236.81,230.33) ;
\draw  [line width=0.75]  (388.22,220.56) -- (397.36,229.7) ;
\draw  [line width=0.75]  (397.15,180.67) -- (388.01,189.81) ;
\draw  [line width=0.75]  (348.12,180.88) -- (357.25,190.01) ;
\draw  [line width=0.75]  (357.46,220.77) -- (348.33,229.91) ;

\draw (326.81,197.14) node [anchor=north west][inner sep=0.75pt]  [font=\normalsize]  {$\times $};
\draw (256.91,199.14) node [anchor=north west][inner sep=0.75pt]   [align=left] {$\displaystyle a$};
\draw (286.33,196.74) node [anchor=north west][inner sep=0.75pt]   [align=left] {$\displaystyle b$};
\draw (368.74,199.85) node [anchor=north west][inner sep=0.75pt]   [align=left] {$\displaystyle c$};
\draw (233.1,197.21) node [anchor=north west][inner sep=0.75pt]   [align=left] {$\displaystyle i$};
\draw (310.99,196.59) node [anchor=north west][inner sep=0.75pt]   [align=left] {$\displaystyle j$};
\draw (271.95,170.95) node [anchor=north west][inner sep=0.75pt]   [align=left] {$\displaystyle k$};
\draw (273.11,227.51) node [anchor=north west][inner sep=0.75pt]   [align=left] {$\displaystyle l$};
\draw (344.1,197.87) node [anchor=north west][inner sep=0.75pt]   [align=left] {$\displaystyle 1$};
\draw (392.66,197.92) node [anchor=north west][inner sep=0.75pt]   [align=left] {$\displaystyle 2$};
\draw (366.95,170.28) node [anchor=north west][inner sep=0.75pt]   [align=left] {$\displaystyle 3$};
\draw (367.44,227.51) node [anchor=north west][inner sep=0.75pt]   [align=left] {$\displaystyle 4$};

\end{tikzpicture}
$};

\draw (140.5,215) node [anchor=north west][inner sep=0.75pt]    {$T^{ij,kl}$};
\draw (298,215) node [anchor=north west][inner sep=0.75pt]    {$E^{ij,kl}$};
\draw (470,215) node [anchor=north west][inner sep=0.75pt]    {$H^{ij,kl}$};
\draw (222,330) node [anchor=north west][inner sep=0.75pt]    {$L^{ij,kl}$};
\draw (390,330) node [anchor=north west][inner sep=0.75pt]    {$gh^{ij,kl}$};

\end{tikzpicture}

}
}}
\end{eqnarray}
In the above, we've used the dual graph to represent the integrand, with the position space points corresponding to regions in the graph, and the edge between the regions represent propagators. Each graph is dressed with the following numerators: 
\begin{equation}
    \begin{split}
        &n_T:=x_{a,i}^2 x_{k,l}^2,.\phantom{my} \quad n_E:=x_{b,k}^2 x_{i,l}^2 x_{j,l}^2{-} x_{b,l}^2 (x_{i,l}^2 x_{j,k}^2{+} x_{i,k}^2 x_{j,l}^2)/2 , \quad n_E^\prime:=x_{b,l}^2,\\
        &n_L:=x_{k,l}^4  ,\phantom{auuuuuuu}  n_H:=\frac{1}{2} x_{a,c}^2 x_{i,j}^2 {-} x_{a,j}^2 x_{c,i}^2, \phantom{auuukkkkkkuuuu}    n_{gh}:=x_{k,l}^2.
    \end{split}
\end{equation}
Note that $E$ and $E'$ have the same topology but are dressed with different numerators.

The easy and hard integrands differ from those of the standard conventions in the literature.\footnote{In literature, the easy and the hard integrands are defined as 
\begin{equation*}
    \begin{split}
        &E_\text{c}^{12;34}:=\frac{x_{b,3}^2x_{1,4}^2 x_{2,4}^2 }{x_{a,1}^2 x_{a,3}^2 x_{a,4}^2 x_{a,b}^2 x_{b,1}^2 x_{b,2}^2 x_{b,4}^2 x_{b,c}^2 x_{c,2}^2  x_{c,3}^2  x_{c,4}^2 },\\
        &H_\text{c}^{12;34}:=\frac{x_{a,c}^2 x_{1,2}^2}{2\, x_{a,1}^2 x_{a,2}^2 x_{a,3}^2 x_{a,4}^2 x_{a,b}^2 x_{b,1}^2 x_{b,2}^2 x_{b,c}^2 x_{c,1}^2 x_{c,2}^2 x_{c,3}^2 x_{c,4}^2}.
    \end{split}
\end{equation*}}
They are related by 
\begin{equation}
    \begin{split}
        &E_{\text{c}}^{12;34}=E^{12;34}+ \frac{1}{2} (x_{1,4}^2 x_{2,3}^2+x_{1,3}^2 x_{2,4}^2) E^{\prime12;34}\\
        &H_{\text{c}}^{12;34}=H^{12;34}+ E^{\prime 12;34}\,.
    \end{split}
\end{equation}
 We have engineered these numerators such that the leading singularities are of the form shown in eq.~\eqref{eq: LS}. As an example, the easy integrand $E^{ 12;34}$ contributes to the following set of maximal cuts:
\begin{eqnarray}
&&\vcenter{\hbox{\scalebox{1}{
\begin{tikzpicture}[x=0.75pt,y=0.75pt,yscale=-1,xscale=1]

\draw  [line width=0.75] (150,150) node [anchor=north west][inner sep=0.75pt]  [align=center] {$ \begin{tikzpicture}[x=0.75pt,y=0.75pt,yscale=-1,xscale=1]

\draw    (148,130.12) -- (212.34,104.12) ;
\draw [shift={(215.12,103)}, rotate = 158] [fill={rgb, 255:red, 0; green, 0; blue, 0 }  ][line width=0.08]  [draw opacity=0] (8.93,-4.29) -- (0,0) -- (8.93,4.29) -- cycle    ;

\end{tikzpicture}
$};

\draw  [line width=0.75] (150,220) node [anchor=north west][inner sep=0.75pt]  [align=center] {$ \begin{tikzpicture}[x=0.75pt,y=0.75pt,yscale=-1,xscale=1]

\draw    (148,130.12) -- (213.21,155.88) ;
\draw [shift={(216,156.98)}, rotate = 201.55] [fill={rgb, 255:red, 0; green, 0; blue, 0 }  ][line width=0.08]  [draw opacity=0] (8.93,-4.29) -- (0,0) -- (8.93,4.29) -- cycle    ;

\end{tikzpicture}
$};

\draw  [line width=0.75] (10,150) node [anchor=north west][inner sep=0.75pt]  [align=center] {$ \begin{tikzpicture}[x=0.75pt,y=0.75pt,yscale=-1,xscale=1]

\draw  [line width=0.75] (310.59,211.36) -- (301.45,183.43) -- (325.19,166.11) -- (349.01,183.33) -- (339.98,211.3) -- cycle ;
\draw  [line width=0.75] (310.6,211.03) -- (283,220.18) -- (273.85,192.58) -- (301.45,183.43) -- cycle ;
\draw  [line width=0.75] (367.62,220.35) -- (339.98,211.3) -- (349.03,183.67) -- (376.67,192.71) -- cycle ;
\draw  [line width=0.75]  (273.85,192.58) -- (262.31,186.77) ;
\draw  [line width=0.75]  (277.2,231.72) -- (283,220.18) ;
\draw  [line width=0.75]  (376.67,192.71) -- (388.19,186.87) ;
\draw  [line width=0.75]  (373.46,231.87) -- (367.62,220.35) ;
\draw  [line width=0.75]  (325.19,154.45) -- (325.19,166.11) ;

\draw (287.57,195.14) node [anchor=north west][inner sep=0.75pt]   [align=left] {$\displaystyle a$};
\draw (319.67,185.74) node [anchor=north west][inner sep=0.75pt]   [align=left] {$\displaystyle b$};
\draw (353.74,195.14) node [anchor=north west][inner sep=0.75pt]   [align=left] {$\displaystyle c$};
\draw (287.77,162.21) node [anchor=north west][inner sep=0.75pt]   [align=left] {$\displaystyle 1$};
\draw (353.66,162.25) node [anchor=north west][inner sep=0.75pt]   [align=left] {$\displaystyle 2$};
\draw (260.62,200.62) node [anchor=north west][inner sep=0.75pt]   [align=left] {$\displaystyle 3$};
\draw (377.95,201.28) node [anchor=north west][inner sep=0.75pt]   [align=left] {$\displaystyle 3$};
\draw (322.44,220.84) node [anchor=north west][inner sep=0.75pt]   [align=left] {$\displaystyle 4$};

\end{tikzpicture}
$};

\draw  [line width=0.75] (240,100) node [anchor=north west][inner sep=0.75pt]  [align=center] {$ \begin{tikzpicture}[x=0.75pt,y=0.75pt,yscale=-1,xscale=1]

\draw  [line width=0.75] (310.59,211.36) -- (301.45,183.43) -- (325.19,166.11) -- (349.01,183.33) -- (339.98,211.3) -- cycle ;
\draw  [line width=0.75] (310.6,211.03) -- (283,220.18) -- (273.85,192.58) -- (301.45,183.43) -- cycle ;
\draw  [line width=0.75] (367.62,220.35) -- (339.98,211.3) -- (349.03,183.67) -- (376.67,192.71) -- cycle ;
\draw  [line width=0.75]  (273.85,192.58) -- (262.31,186.77) ;
\draw  [line width=0.75]  (277.2,231.72) -- (283,220.18) ;
\draw  [line width=0.75]  (376.67,192.71) -- (388.19,186.87) ;
\draw  [line width=0.75]  (373.46,231.87) -- (367.62,220.35) ;
\draw  [line width=0.75]  (325.19,154.45) -- (325.19,166.11) ;

\draw [line width=0.75] [color=red  ,draw opacity=1 ]   (325.15,206.56) -- (325.15,215.28) ;
\draw [line width=0.75] [color=red  ,draw opacity=1 ]   (353.68,212.01) -- (350.98,220.3) ;
\draw [line width=0.75] [color=red  ,draw opacity=1 ]   (364.68,184.01) -- (361.98,192.3) ;
\draw [line width=0.75] [color=red ,draw opacity=1 ]   (377.48,208.5) -- (369.18,205.81) ;
\draw [line width=0.75] [color=red ,draw opacity=1 ]   (348.48,201.5) -- (340.18,198.81) ;
\draw [color=red  ,draw opacity=1 ]   (295.8,211.77) -- (298.5,220.06) ;
\draw[line width=0.75] [color=red  ,draw opacity=1 ]   (288.8,182.77) -- (291.5,191.06) ;
\draw [line width=0.75] [color=red  ,draw opacity=1 ]   (282.3,206.07) -- (274,208.76) ;
\draw [line width=0.75] [color=red  ,draw opacity=1 ]   (340.3,172.07) -- (335,179) ;
\draw [line width=0.75][color=red  ,draw opacity=1 ]   (311.3,199.07) -- (303,201.76) ;
\draw [line width=0.75][color=red ,draw opacity=1 ]   (349.48,198.5) -- (341.18,195.81) ;
\draw [line width=0.75][color=red  ,draw opacity=1 ]   (314.68,179.47) -- (309.62,172.37) ;

\draw (287.57,195.14) node [anchor=north west][inner sep=0.75pt]   [align=left] {$\displaystyle a$};
\draw (319.67,185.74) node [anchor=north west][inner sep=0.75pt]   [align=left] {$\displaystyle b$};
\draw (353.74,195.14) node [anchor=north west][inner sep=0.75pt]   [align=left] {$\displaystyle c$};
\draw (287.77,162.21) node [anchor=north west][inner sep=0.75pt]   [align=left] {$\displaystyle 1$};
\draw (353.66,162.25) node [anchor=north west][inner sep=0.75pt]   [align=left] {$\displaystyle 2$};
\draw (260.62,200.62) node [anchor=north west][inner sep=0.75pt]   [align=left] {$\displaystyle 3$};
\draw (377.95,201.28) node [anchor=north west][inner sep=0.75pt]   [align=left] {$\displaystyle 3$};
\draw (322.44,220.84) node [anchor=north west][inner sep=0.75pt]   [align=left] {$\displaystyle 4$};

\end{tikzpicture}
$};

\draw  [line width=0.75] (250,200) node [anchor=north west][inner sep=0.75pt]  [align=center] {$ \begin{tikzpicture}[x=0.75pt,y=0.75pt,yscale=-1,xscale=1]

\draw  [line width=0.75] (233.13,220.95) -- (254.88,199.2) -- (276.63,220.95) -- (254.88,242.7) -- cycle ;
\draw  [line width=0.75] (254.88,199.2) -- (276.63,177.45) -- (298.38,199.2) -- (276.63,220.95) -- cycle ;
\draw   [line width=0.75] (320.28,220.65) -- (333.2,220.65) ;
\draw   [line width=0.75] (276.48,164.53) -- (276.48,177.45) ;
\draw  [line width=0.75]  (220.36,221.25) -- (233.28,221.25) ;
\draw  [line width=0.75]  (255.18,242.85) -- (255.18,255.77) ;
\draw  [line width=0.75]  (298.68,242.55) -- (298.68,255.47) ;
\draw  [line width=0.75] (276.78,220.8) -- (298.53,199.05) -- (320.28,220.8) -- (298.53,242.55) -- cycle ;
\draw [line width=0.75][color=red  ,draw opacity=1 ]   (241.17,206.33) -- (247.33,212.5) ;
\draw [line width=0.75][color=red  ,draw opacity=1 ]   (262.5,228.67) -- (265.33,231.5) -- (268.67,234.83) ;
\draw [line width=0.75][color=red  ,draw opacity=1 ]   (262.17,185.33) -- (268.33,191.5) ;
\draw [line width=0.75][color=red  ,draw opacity=1 ]   (284.17,207.33) -- (290.33,213.5) ;
\draw [line width=0.75][color=red  ,draw opacity=1 ]   (286.17,205.33) -- (292.33,211.5) ;

\draw [line width=0.75][color=red  ,draw opacity=1 ]   (306.17,229.33) -- (312.33,235.5) ;
\draw [line width=0.75][color=red  ,draw opacity=1 ]   (247.33,229) -- (241.17,235.17) ;
\draw [line width=0.75][color=red  ,draw opacity=1 ]   (290.33,185) -- (284.17,191.17) ;
\draw [line width=0.75][color=red  ,draw opacity=1 ]   (312.33,207) -- (306.17,213.17) ;
\draw [line width=0.75][color=red  ,draw opacity=1 ]   (290.33,228) -- (284.17,234.17) ;
\draw [line width=0.75][color=red  ,draw opacity=1 ]   (268,206.67) -- (261.83,212.83) ;
\draw [line width=0.75][color=red  ,draw opacity=1 ]   (270,208.67) -- (263.83,214.83) ;

\draw (226.94,235.54) node [anchor=north west][inner sep=0.75pt]   [align=left] {$\displaystyle 3$};
\draw (316.44,238.04) node [anchor=north west][inner sep=0.75pt]   [align=left] {$\displaystyle 3$};
\draw (272.94,241.04) node [anchor=north west][inner sep=0.75pt]   [align=left] {$\displaystyle 4$};
\draw (233.94,181.54) node [anchor=north west][inner sep=0.75pt]   [align=left] {$\displaystyle 1$};
\draw (307.94,185.54) node [anchor=north west][inner sep=0.75pt]   [align=left] {$\displaystyle 2$};
\draw (248.57,211.64) node [anchor=north west][inner sep=0.75pt]   [align=left] {$\displaystyle a$};
\draw (271.67,190.54) node [anchor=north west][inner sep=0.75pt]   [align=left] {$\displaystyle b$};
\draw (294.47,212.14) node [anchor=north west][inner sep=0.75pt]   [align=left] {$\displaystyle c$};

\end{tikzpicture}
$};

\draw (165,140) node [anchor=north west][inner sep=0.75pt]    {$\alpha\text{-cut}$};

\draw (165,245) node [anchor=north west][inner sep=0.75pt]    {$\beta\text{-cut}$};

\end{tikzpicture}

}
}}
\end{eqnarray}
On the $\alpha$-cut, the leading singularity is $1/\Delta$, whereas on the $\beta$-cut it becomes $\pm 1/(2\Delta)$, with the sign depending on the order in which one takes the second of the double cut for $x_{a,b}^2$ and $x_{b,c}^2$. On the other hand, $E'(1,2;3,4)$ only contributes to the $\beta$-cut, with leading singularity  $\frac{1}{\Delta(t-u)}$. Note that even though both $E$ and $E'$ integrands contribute to the $\beta$-cut, when summed over permutations $E^{12;34}$ and $E^{12;43}$ cancel each other out:
\begin{equation*}
   \Delta\, E^{12;34}\big|_\beta+\Delta\,E^{12;43}\big|_\beta=\frac{1}{2 }-\frac{1}{2 }=0\,.
\end{equation*}
Similarly, the hard integrand $H^{12;34}$ contributes to the following cuts: 
\begin{eqnarray}
&&\vcenter{\hbox{\scalebox{1}{
\begin{tikzpicture}[x=0.75pt,y=0.75pt,yscale=-1,xscale=1]

\draw  [line width=0.75] (150,150) node [anchor=north west][inner sep=0.75pt]  [align=center] {$ \begin{tikzpicture}[x=0.75pt,y=0.75pt,yscale=-1,xscale=1]

\draw    (148,130.12) -- (212.34,104.12) ;
\draw [shift={(215.12,103)}, rotate = 158] [fill={rgb, 255:red, 0; green, 0; blue, 0 }  ][line width=0.08]  [draw opacity=0] (8.93,-4.29) -- (0,0) -- (8.93,4.29) -- cycle    ;

\end{tikzpicture}
$};

\draw  [line width=0.75] (150,220) node [anchor=north west][inner sep=0.75pt]  [align=center] {$ \begin{tikzpicture}[x=0.75pt,y=0.75pt,yscale=-1,xscale=1]

\draw    (148,130.12) -- (213.21,155.88) ;
\draw [shift={(216,156.98)}, rotate = 201.55] [fill={rgb, 255:red, 0; green, 0; blue, 0 }  ][line width=0.08]  [draw opacity=0] (8.93,-4.29) -- (0,0) -- (8.93,4.29) -- cycle    ;

\end{tikzpicture}
$};

\draw  [line width=0.75] (-20,150) node [anchor=north west][inner sep=0.75pt]  [align=center] {$ \begin{tikzpicture}[x=0.75pt,y=0.75pt,yscale=-1,xscale=1]

\draw  [line width=0.75] (276.68,187.07) -- (276.93,216.46) -- (249.06,225.78) -- (231.58,202.15) -- (248.65,178.23) -- cycle ;
\draw  [line width=0.75] (276.88,216.46) -- (276.68,187.27) -- (305.87,187.07) -- (306.07,216.26) -- cycle ;
\draw  [line width=0.75] (334.23,225.31) -- (306.21,216.45) -- (305.98,187.06) -- (333.86,177.76) -- (351.32,201.4) -- cycle ;

\draw  [line width=0.75]  (243.67,166) -- (248.65,178.23) ;
\draw   [line width=0.75] (249.06,225.78) -- (241.67,238) ;
\draw  [line width=0.75]  (217,202) -- (231.58,202.15) ;
\draw  [line width=0.75]  (351.32,201.4) -- (365.9,201.56) ;
\draw  [line width=0.75]  (341.25,165.54) -- (333.86,177.76) ;
\draw  [line width=0.75]  (334.23,225.31) -- (339.22,237.54) ;

\draw (252.57,198.14) node [anchor=north west][inner sep=0.75pt]   [align=left] {$\displaystyle a$};
\draw (286.67,195.4) node [anchor=north west][inner sep=0.75pt]   [align=left] {$\displaystyle b$};
\draw (321.41,198.19) node [anchor=north west][inner sep=0.75pt]   [align=left] {$\displaystyle c$};
\draw (226.1,175.87) node [anchor=north west][inner sep=0.75pt]   [align=left] {$\displaystyle 3$};
\draw (223.32,211.25) node [anchor=north west][inner sep=0.75pt]   [align=left] {$\displaystyle 4$};
\draw (350.1,174.54) node [anchor=north west][inner sep=0.75pt]   [align=left] {$\displaystyle 3$};
\draw (348.66,212.59) node [anchor=north west][inner sep=0.75pt]   [align=left] {$\displaystyle 4$};
\draw (286.29,167.62) node [anchor=north west][inner sep=0.75pt]   [align=left] {$\displaystyle 1$};
\draw (286.78,221.84) node [anchor=north west][inner sep=0.75pt]   [align=left] {$\displaystyle 2$};

\end{tikzpicture}
$};

\draw  [line width=0.75] (240,100) node [anchor=north west][inner sep=0.75pt]  [align=center] {$ \begin{tikzpicture}[x=0.75pt,y=0.75pt,yscale=-1,xscale=1]

\draw  [line width=0.75] (276.68,187.07) -- (276.93,216.46) -- (249.06,225.78) -- (231.58,202.15) -- (248.65,178.23) -- cycle ;
\draw  [line width=0.75] (276.88,216.46) -- (276.68,187.27) -- (305.87,187.07) -- (306.07,216.26) -- cycle ;
\draw  [line width=0.75] (334.23,225.31) -- (306.21,216.45) -- (305.98,187.06) -- (333.86,177.76) -- (351.32,201.4) -- cycle ;

\draw  [line width=0.75]  (243.67,166) -- (248.65,178.23) ;
\draw   [line width=0.75] (249.06,225.78) -- (241.67,238) ;
\draw  [line width=0.75]  (217,202) -- (231.58,202.15) ;
\draw  [line width=0.75]  (351.32,201.4) -- (365.9,201.56) ;
\draw  [line width=0.75]  (341.25,165.54) -- (333.86,177.76) ;
\draw  [line width=0.75]  (334.23,225.31) -- (339.22,237.54) ;

\draw  [line width=0.75][color=red  ,draw opacity=1 ]   (293.15,182.56) -- (293.15,191.28) ;
\draw  [line width=0.75][color=red  ,draw opacity=1 ]   (294.15,212.56) -- (294.15,221.28) ;
\draw  [line width=0.75][color=red  ,draw opacity=1 ]   (311.51,200.92) -- (302.79,200.92) ;
\draw  [line width=0.75][color=red  ,draw opacity=1 ]   (281.51,200.92) -- (272.79,200.92) ;
\draw  [line width=0.75][color=red  ,draw opacity=1 ]   (266.5,225.06) -- (263.8,216.77) ;
\draw  [line width=0.75][color=red  ,draw opacity=1 ]   (238.12,216.98) -- (245.18,211.85) ;
\draw  [line width=0.75][color=red  ,draw opacity=1 ]   (237.12,186.85) -- (244.18,191.98) ;
\draw  [line width=0.75][color=red  ,draw opacity=1 ]   (263,178.27) -- (260.3,186.56) ;
\draw  [line width=0.75][color=red  ,draw opacity=1 ]   (322.5,186.06) -- (319.8,177.77) ;
\draw  [line width=0.75][color=red  ,draw opacity=1 ]   (338.62,192.48) -- (345.68,187.35) ;
\draw  [line width=0.75][color=red  ,draw opacity=1 ]   (340.62,210.35) -- (347.68,215.48) ;
\draw  [line width=0.75][color=red  ,draw opacity=1 ]   (322.5,216.77) -- (319.8,225.06) ;

\draw (252.57,198.14) node [anchor=north west][inner sep=0.75pt]   [align=left] {$\displaystyle a$};
\draw (286.67,195.4) node [anchor=north west][inner sep=0.75pt]   [align=left] {$\displaystyle b$};
\draw (321.41,198.19) node [anchor=north west][inner sep=0.75pt]   [align=left] {$\displaystyle c$};
\draw (226.1,175.87) node [anchor=north west][inner sep=0.75pt]   [align=left] {$\displaystyle 3$};
\draw (223.32,211.25) node [anchor=north west][inner sep=0.75pt]   [align=left] {$\displaystyle 4$};
\draw (350.1,174.54) node [anchor=north west][inner sep=0.75pt]   [align=left] {$\displaystyle 3$};
\draw (348.66,212.59) node [anchor=north west][inner sep=0.75pt]   [align=left] {$\displaystyle 4$};
\draw (286.29,167.62) node [anchor=north west][inner sep=0.75pt]   [align=left] {$\displaystyle 1$};
\draw (286.78,221.84) node [anchor=north west][inner sep=0.75pt]   [align=left] {$\displaystyle 2$};

\end{tikzpicture}
$};

\draw  [line width=0.75] (260,200) node [anchor=north west][inner sep=0.75pt]  [align=center] {$ \begin{tikzpicture}[x=0.75pt,y=0.75pt,yscale=-1,xscale=1]

\draw  [line width=0.75] (233.13,220.95) -- (254.88,199.2) -- (276.63,220.95) -- (254.88,242.7) -- cycle ;
\draw  [line width=0.75] (254.88,199.2) -- (276.63,177.45) -- (298.38,199.2) -- (276.63,220.95) -- cycle ;
\draw   [line width=0.75] (320.28,220.65) -- (333.2,220.65) ;
\draw   [line width=0.75] (276.48,164.53) -- (276.48,177.45) ;
\draw  [line width=0.75]  (220.36,221.25) -- (233.28,221.25) ;
\draw  [line width=0.75]  (255.18,242.85) -- (255.18,255.77) ;
\draw  [line width=0.75]  (298.68,242.55) -- (298.68,255.47) ;
\draw  [line width=0.75] (276.78,220.8) -- (298.53,199.05) -- (320.28,220.8) -- (298.53,242.55) -- cycle ;
\draw [line width=0.75][color=red  ,draw opacity=1 ]   (241.17,206.33) -- (247.33,212.5) ;
\draw [line width=0.75][color=red  ,draw opacity=1 ]   (262.5,228.67) -- (265.33,231.5) -- (268.67,234.83) ;
\draw [line width=0.75][color=red  ,draw opacity=1 ]   (262.17,185.33) -- (268.33,191.5) ;
\draw [line width=0.75][color=red  ,draw opacity=1 ]   (284.17,207.33) -- (290.33,213.5) ;
\draw [line width=0.75][color=red  ,draw opacity=1 ]   (286.17,205.33) -- (292.33,211.5) ;

\draw [line width=0.75][color=red  ,draw opacity=1 ]   (306.17,229.33) -- (312.33,235.5) ;
\draw [line width=0.75][color=red  ,draw opacity=1 ]   (247.33,229) -- (241.17,235.17) ;
\draw [line width=0.75][color=red  ,draw opacity=1 ]   (290.33,185) -- (284.17,191.17) ;
\draw [line width=0.75][color=red  ,draw opacity=1 ]   (312.33,207) -- (306.17,213.17) ;
\draw [line width=0.75][color=red  ,draw opacity=1 ]   (290.33,228) -- (284.17,234.17) ;
\draw [line width=0.75][color=red  ,draw opacity=1 ]   (268,206.67) -- (261.83,212.83) ;
\draw [line width=0.75][color=red  ,draw opacity=1 ]   (270,208.67) -- (263.83,214.83) ;

\draw (226.94,235.54) node [anchor=north west][inner sep=0.75pt]   [align=left] {$\displaystyle 3$};
\draw (316.44,238.04) node [anchor=north west][inner sep=0.75pt]   [align=left] {$\displaystyle 3$};
\draw (272.94,241.04) node [anchor=north west][inner sep=0.75pt]   [align=left] {$\displaystyle 4$};
\draw (233.94,181.54) node [anchor=north west][inner sep=0.75pt]   [align=left] {$\displaystyle 1$};
\draw (307.94,185.54) node [anchor=north west][inner sep=0.75pt]   [align=left] {$\displaystyle 2$};
\draw (248.57,211.64) node [anchor=north west][inner sep=0.75pt]   [align=left] {$\displaystyle a$};
\draw (271.67,190.54) node [anchor=north west][inner sep=0.75pt]   [align=left] {$\displaystyle b$};
\draw (294.47,212.14) node [anchor=north west][inner sep=0.75pt]   [align=left] {$\displaystyle c$};

\end{tikzpicture}
$};

\draw (165,140) node [anchor=north west][inner sep=0.75pt]    {$\gamma\text{-cut}$};

\draw (165,245) node [anchor=north west][inner sep=0.75pt]    {$\beta\text{-cut}$};

\end{tikzpicture}

}
}}
\end{eqnarray}
The $\gamma$-cut yields a leading singularity $1/\Delta^2$, while the $\beta$-cut vanishes in this case. Thus we see that the leading singularity associated with $\alpha$, $\beta$ and $\gamma$ cut is reproduced by $E^{12;34}$, $E^{\prime 12;34}$ and $H^{12;34}$ respectively.

In contrast, the conventional integrands exhibit multiple leading singularities. As analyzed in~\cite{Drummond:2013nda}, the standard integrand $E_{\text{c}}^{12;34}$ carries three leading singularities:
\begin{equation}
    \frac{1}{\Delta} \quad \text{on the $\alpha$-cut}, \quad \frac{t}{(t-u)\Delta},\ \frac{u}{(t-u)\Delta} \quad  \text{on the $\beta$-cut,}
\end{equation}
while the standard hard integrand $H_{\text{c}}^{12;34}$ contributes with
\begin{equation}
    \frac{1}{\Delta^2} \quad \text{on the $\gamma$-cut}, \quad \frac{1}{(t-u)\Delta} \quad  \text{on the $\beta$-cut.}
\end{equation}
Thus, in our newly defined basis, we have systematically added and subtracted subsector integrands to ensure that all appearing integrands are manifestly pure. The leading singularities are effectively \textit{diagonalized}, such that each integrand exhibits only one. Remarkably, the underlying geometry naturally supplies the required subtopologies. In Appendix~\ref{sec:3loop original form}, we present the chamber loop-form for the ordering $u < t < s$ in the conventional off-shell integrand basis. There, one observes that many of the $E^\prime$ integrands carry nontrivial coefficients—precisely due to the fact that the standard $E_{\text{c}}$ and $H_{\text{c}}$ are not pure.

 Finally, one finds that, after summing over all $6$ chambers, $\mathcal{G}^{(3)}(\mathbf{X},Y)=\frac{1}{2}\sum_{i,\pm}\;\omega_i^\pm \Omega_{r_i}^{(3)\pm}$; the result exactly reproduces the three-loop answer in~\cite{EDEN2012450}.

\subsubsection{Three-loop chambers and positivity of the cuts}\label{sec: three-loop chamber analysis}

Since the refined chamber structure of eq.~\eqref{eq:L=3_chambers} first emerges at three loops, it is instructive to examine how the corresponding chamber boundaries and the block structure of eq.~\eqref{eq: 3Loops} arise from the positivity of the cuts. We focus on integrands $E^{12;34}$, $E^{\prime 12;34}$, and $H^{12;34}$ to facilitate our analysis since (1) each is associated with a distinct block in eq.~\eqref{eq: 3Loops}; and (2) each is responsible for a different leading singularity, namely the $(\alpha,\beta,\gamma)$-cuts.

Let us begin with a warm-up example, the one-dimensional eleven-cut geometry:

\begin{eqnarray}
&&\vcenter{\hbox{\scalebox{1}{
\begin{tikzpicture}[x=0.75pt,y=0.75pt,yscale=-1,xscale=1]

\draw  [line width=0.75] (320,250) node [anchor=north west][inner sep=0.75pt]  [align=center] {$\begin{tikzpicture}[x=0.75pt,y=0.75pt,yscale=-1,xscale=1]

\draw  [line width=0.75] (245.95,221.19) -- (245.74,190.43) -- (276.5,190.22) -- (276.71,220.98) -- cycle ;
\draw  [line width=0.75] (276.71,220.98) -- (276.5,190.22) -- (307.25,190.01) -- (307.46,220.77) -- cycle ;
\draw  [line width=0.75] (357.46,220.77) -- (357.25,190.01) -- (388.01,189.81) -- (388.22,220.56) -- cycle ;
\draw  [line width=0.75]  (307.46,220.77) -- (316.6,229.91) ;
\draw   [line width=0.75] (316.39,180.88) -- (307.25,190.01) ;
\draw  [line width=0.75]  (236.6,181.3) -- (245.74,190.43) ;
\draw   [line width=0.75] (245.95,221.19) -- (236.81,230.33) ;
\draw  [line width=0.75]  (388.22,220.56) -- (397.36,229.7) ;
\draw  [line width=0.75]  (397.15,180.67) -- (388.01,189.81) ;
\draw  [line width=0.75]  (348.12,180.88) -- (357.25,190.01) ;
\draw  [line width=0.75]  (357.46,220.77) -- (348.33,229.91) ;

\draw [line width=0.75][color=red  ,draw opacity=1 ]   (260.05,185.46) -- (260.05,194.18) ;
\draw [line width=0.75][color=red  ,draw opacity=1 ]   (260.05,217.46) -- (260.05,226.18) ;

\draw [line width=0.75][color=red  ,draw opacity=1 ]   (294.05,185.46) -- (294.05,194.18) ;
\draw [line width=0.75][color=red  ,draw opacity=1 ]   (294.05,217.46) -- (294.05,226.18) ;

\draw [line width=0.75][color=red  ,draw opacity=1 ]   (374.05,185.46) -- (374.05,194.18) ;
\draw [line width=0.75][color=red  ,draw opacity=1 ]   (374.05,217.46) -- (374.05,226.18) ;

\draw [line width=0.75][color=red  ,draw opacity=1 ]   (393.16,203.57) -- (384.44,203.57) ;
\draw [line width=0.75][color=red  ,draw opacity=1 ]   (362.16,203.57) -- (353.44,203.57) ;

\draw [line width=0.75][color=red  ,draw opacity=1 ]   (311.16,203.57) -- (302.44,203.57) ;
\draw [line width=0.75][color=red  ,draw opacity=1 ]   (281.16,203.57) -- (272.44,203.57) ;
\draw [line width=0.75][color=red  ,draw opacity=1 ]   (251.16,203.57) -- (242.44,203.57) ;

\draw (326.81,197.14) node [anchor=north west][inner sep=0.75pt]  [font=\normalsize]  {$\times $};
\draw (256.91,199.14) node [anchor=north west][inner sep=0.75pt]   [align=left] {$\displaystyle a$};
\draw (286.33,196.74) node [anchor=north west][inner sep=0.75pt]   [align=left] {$\displaystyle b$};
\draw (368.74,199.85) node [anchor=north west][inner sep=0.75pt]   [align=left] {$\displaystyle c$};
\draw (233.1,197.21) node [anchor=north west][inner sep=0.75pt]   [align=left] {$\displaystyle 1$};
\draw (310.99,196.59) node [anchor=north west][inner sep=0.75pt]   [align=left] {$\displaystyle 2$};
\draw (271.95,170.95) node [anchor=north west][inner sep=0.75pt]   [align=left] {$\displaystyle 3$};
\draw (273.11,227.51) node [anchor=north west][inner sep=0.75pt]   [align=left] {$\displaystyle 4$};
\draw (344.1,197.87) node [anchor=north west][inner sep=0.75pt]   [align=left] {$\displaystyle 1$};
\draw (392.66,197.92) node [anchor=north west][inner sep=0.75pt]   [align=left] {$\displaystyle 2$};
\draw (366.95,170.28) node [anchor=north west][inner sep=0.75pt]   [align=left] {$\displaystyle 3$};
\draw (367.44,227.51) node [anchor=north west][inner sep=0.75pt]   [align=left] {$\displaystyle 4$};

\end{tikzpicture}
$};

\draw  [line width=0.75] (344,255) node [anchor=north west][inner sep=0.75pt]  [align=center] {$ \hbox{\scalebox{0.6}{\begin{tikzpicture}[x=0.75pt,y=0.75pt,yscale=-1,xscale=1]

\draw (93.5,68) node [anchor=north west][inner sep=0.75pt]  [font=\large,color=blue  ,opacity=1 ] [align=left] {$\displaystyle 1$};

\end{tikzpicture}

}}
$};

\draw  [line width=0.75] (344,310) node [anchor=north west][inner sep=0.75pt]  [align=center] {$ \hbox{\scalebox{0.6}{\begin{tikzpicture}[x=0.75pt,y=0.75pt,yscale=-1,xscale=1]

\draw (93.5,68) node [anchor=north west][inner sep=0.75pt]  [font=\large,color=blue  ,opacity=1 ] [align=left] {$\displaystyle 2$};

\end{tikzpicture}

}}
$};

\draw  [line width=0.75] (377,255) node [anchor=north west][inner sep=0.75pt]  [align=center] {$ \hbox{\scalebox{0.6}{\begin{tikzpicture}[x=0.75pt,y=0.75pt,yscale=-1,xscale=1]

\draw (93.5,68) node [anchor=north west][inner sep=0.75pt]  [font=\large,color=blue  ,opacity=1 ] [align=left] {$\displaystyle 3$};

\end{tikzpicture}

}}
$};

\draw  [line width=0.75] (380,310) node [anchor=north west][inner sep=0.75pt]  [align=center] {$ \hbox{\scalebox{0.6}{\begin{tikzpicture}[x=0.75pt,y=0.75pt,yscale=-1,xscale=1]

\draw (93.5,68) node [anchor=north west][inner sep=0.75pt]  [font=\large,color=blue  ,opacity=1 ] [align=left] {$\displaystyle 4$};

\end{tikzpicture}

}}
$};

\draw  [line width=0.75] (322,290) node [anchor=north west][inner sep=0.75pt]  [align=center] {$ \hbox{\scalebox{0.6}{\begin{tikzpicture}[x=0.75pt,y=0.75pt,yscale=-1,xscale=1]

\draw (93.5,68) node [anchor=north west][inner sep=0.75pt]  [font=\large,color=blue  ,opacity=1 ] [align=left] {$\displaystyle 5$};

\end{tikzpicture}

}}
$};

\draw  [line width=0.75] (400,290) node [anchor=north west][inner sep=0.75pt]  [align=center] {$ \hbox{\scalebox{0.6}{\begin{tikzpicture}[x=0.75pt,y=0.75pt,yscale=-1,xscale=1]

\draw (93.5,68) node [anchor=north west][inner sep=0.75pt]  [font=\large,color=blue  ,opacity=1 ] [align=left] {$\displaystyle 6$};

\end{tikzpicture}

}}
$};

\draw  [line width=0.75] (355,286) node [anchor=north west][inner sep=0.75pt]  [align=center] {$ \hbox{\scalebox{0.6}{\begin{tikzpicture}[x=0.75pt,y=0.75pt,yscale=-1,xscale=1]

\draw (93.5,68) node [anchor=north west][inner sep=0.75pt]  [font=\large,color=blue  ,opacity=1 ] [align=left] {$\displaystyle 7$};

\end{tikzpicture}

}}
$};

\draw  [line width=0.75] (433,290) node [anchor=north west][inner sep=0.75pt]  [align=center] {$ \hbox{\scalebox{0.6}{\begin{tikzpicture}[x=0.75pt,y=0.75pt,yscale=-1,xscale=1]

\draw (93.5,68) node [anchor=north west][inner sep=0.75pt]  [font=\large,color=blue  ,opacity=1 ] [align=left] {$\displaystyle 8$};

\end{tikzpicture}

}}
$};

\draw  [line width=0.75] (482,290) node [anchor=north west][inner sep=0.75pt]  [align=center] {$ \hbox{\scalebox{0.6}{\begin{tikzpicture}[x=0.75pt,y=0.75pt,yscale=-1,xscale=1]

\draw (93.5,68) node [anchor=north west][inner sep=0.75pt]  [font=\large,color=blue  ,opacity=1 ] [align=left] {$\displaystyle 9$};

\end{tikzpicture}

}}
$};

\draw  [line width=0.75] (445,257) node [anchor=north west][inner sep=0.75pt]  [align=center] {$ \hbox{\scalebox{0.6}{\begin{tikzpicture}[x=0.75pt,y=0.75pt,yscale=-1,xscale=1]

\draw (93.5,68) node [anchor=north west][inner sep=0.75pt]  [font=\large,color=blue  ,opacity=1 ] [align=left] {$\displaystyle 10$};

\end{tikzpicture}

}}
$};

\draw  [line width=0.75] (445,308) node [anchor=north west][inner sep=0.75pt]  [align=center] {$ \hbox{\scalebox{0.6}{\begin{tikzpicture}[x=0.75pt,y=0.75pt,yscale=-1,xscale=1]

\draw (93.5,68) node [anchor=north west][inner sep=0.75pt]  [font=\large,color=blue  ,opacity=1 ] [align=left] {$\displaystyle 11$};

\end{tikzpicture}

}}
$};

\end{tikzpicture}

}
}}
\end{eqnarray}
We solve the cut using the parameterization
\begin{equation}
    \begin{split}
        (A,B) &= \left(Z_{1,3} + x_1 Z_{2,3} - w_1 Z_{2,4},\quad y_1 Z_{2,3} + Z_{1,4} + z_1 Z_{2,4} \right),\\ \ (C,D) &= \left(Z_{1,3} + x_2 Z_{2,3} - w_2 Z_{2,4},\quad y_2 Z_{2,3} + Z_{1,4} + z_2 Z_{2,4} \right),\\ \ (E,F) &= \left(Z_{1,1} + x_3 Z_{2,1} - w_3 Z_{2,2},\quad y_3 Z_{2,1} + Z_{1,2} + z_3 Z_{2,2} \right). 
    \end{split}
\end{equation}
and solve the cuts sequentially in the following order
\begin{equation}
    (  w_1,y_1, w_2,y_2,x_1,x_2,z_2, w_3,y_3, x_3,z_3)\,.
\end{equation}
The resulting one-dimensional geometry is captured by inequalities involving $z_1$. As it turns out, for the net-positive-oriented sector, eq.~\eqref{eq: LoopRegion} simply reduces to
\begin{equation}
    z_1<0, \quad \text{for max$(s,t,u)=s$,}
\end{equation}
while in the net-negative-oriented sector
\begin{equation}
    z_1>0, \quad \text{for max$(s,t,u)=s$.}
\end{equation}
As mentioned previously, chamber boundaries correspond to kinematic conditions for which certain boundaries of the Correlahedron become inaccessible. Thus, we see that this codimension-11 boundary is  accessible only when  $\text{max}(s,t,u)=s$, {\it i.e.} chambers $r_4$ and $r_6$.

We now turn to the two-dimensional ten-cut geometry of:  
\begin{eqnarray}
&&\vcenter{\hbox{\scalebox{1}{
\begin{tikzpicture}[x=0.75pt,y=0.75pt,yscale=-1,xscale=1]

\draw  [line width=0.75] (-10,140) node [anchor=north west][inner sep=0.75pt]  [align=center] {$ \begin{tikzpicture}[x=0.75pt,y=0.75pt,yscale=-1,xscale=1]

\draw  [line width=0.75] (233.13,220.95) -- (254.88,199.2) -- (276.63,220.95) -- (254.88,242.7) -- cycle ;
\draw  [line width=0.75] (254.88,199.2) -- (276.63,177.45) -- (298.38,199.2) -- (276.63,220.95) -- cycle ;
\draw   [line width=0.75] (320.28,220.65) -- (333.2,220.65) ;
\draw   [line width=0.75] (276.48,164.53) -- (276.48,177.45) ;
\draw  [line width=0.75]  (220.36,221.25) -- (233.28,221.25) ;
\draw  [line width=0.75]  (255.18,242.85) -- (255.18,255.77) ;
\draw  [line width=0.75]  (298.68,242.55) -- (298.68,255.47) ;
\draw  [line width=0.75] (276.78,220.8) -- (298.53,199.05) -- (320.28,220.8) -- (298.53,242.55) -- cycle ;
\draw [line width=0.75][color=red  ,draw opacity=1 ]   (241.17,206.33) -- (247.33,212.5) ;
\draw [line width=0.75][color=red  ,draw opacity=1 ]   (262.5,228.67) -- (265.33,231.5) -- (268.67,234.83) ;
\draw [line width=0.75][color=red  ,draw opacity=1 ]   (262.17,185.33) -- (268.33,191.5) ;
\draw [line width=0.75][color=red  ,draw opacity=1 ]   (284.17,207.33) -- (290.33,213.5) ;

\draw [line width=0.75][color=red  ,draw opacity=1 ]   (306.17,229.33) -- (312.33,235.5) ;
\draw [line width=0.75][color=red  ,draw opacity=1 ]   (247.33,229) -- (241.17,235.17) ;
\draw [line width=0.75][color=red  ,draw opacity=1 ]   (290.33,185) -- (284.17,191.17) ;
\draw [line width=0.75][color=red  ,draw opacity=1 ]   (312.33,207) -- (306.17,213.17) ;
\draw [line width=0.75][color=red  ,draw opacity=1 ]   (290.33,228) -- (284.17,234.17) ;
\draw [line width=0.75][color=red  ,draw opacity=1 ]   (270,208.67) -- (263.83,214.83) ;

\draw (226.94,235.54) node [anchor=north west][inner sep=0.75pt]   [align=left] {$\displaystyle 3$};
\draw (316.44,238.04) node [anchor=north west][inner sep=0.75pt]   [align=left] {$\displaystyle 3$};
\draw (272.94,241.04) node [anchor=north west][inner sep=0.75pt]   [align=left] {$\displaystyle 4$};
\draw (233.94,181.54) node [anchor=north west][inner sep=0.75pt]   [align=left] {$\displaystyle 1$};
\draw (307.94,185.54) node [anchor=north west][inner sep=0.75pt]   [align=left] {$\displaystyle 2$};
\draw (248.57,211.64) node [anchor=north west][inner sep=0.75pt]   [align=left] {$\displaystyle a$};
\draw (271.67,190.54) node [anchor=north west][inner sep=0.75pt]   [align=left] {$\displaystyle b$};
\draw (294.47,212.14) node [anchor=north west][inner sep=0.75pt]   [align=left] {$\displaystyle c$};
\draw (350,215) node [anchor=north west][inner sep=0.75pt]  [font=\LARGE] [align=left] {$\displaystyle : $};

\end{tikzpicture}
$};

\draw  [line width=0.75] (160,104) node [anchor=north west][inner sep=0.75pt]  [align=center] {$ \hbox{\scalebox{0.75}{\begin{tikzpicture}[x=0.75pt,y=0.75pt,yscale=-1,xscale=1]

\draw  [line width=0.75] (233.13,220.95) -- (254.88,199.2) -- (276.63,220.95) -- (254.88,242.7) -- cycle ;
\draw  [line width=0.75] (254.88,199.2) -- (276.63,177.45) -- (298.38,199.2) -- (276.63,220.95) -- cycle ;
\draw   [line width=0.75] (320.28,220.65) -- (333.2,220.65) ;
\draw   [line width=0.75] (276.48,164.53) -- (276.48,177.45) ;
\draw  [line width=0.75]  (220.36,221.25) -- (233.28,221.25) ;
\draw  [line width=0.75]  (255.18,242.85) -- (255.18,255.77) ;
\draw  [line width=0.75]  (298.68,242.55) -- (298.68,255.47) ;
\draw  [line width=0.75] (276.78,220.8) -- (298.53,199.05) -- (320.28,220.8) -- (298.53,242.55) -- cycle ;

\draw (226.94,235.54) node [anchor=north west][inner sep=0.75pt]   [align=left] {$\displaystyle 3$};
\draw (316.44,238.04) node [anchor=north west][inner sep=0.75pt]   [align=left] {$\displaystyle 3$};
\draw (272.94,241.04) node [anchor=north west][inner sep=0.75pt]   [align=left] {$\displaystyle 4$};
\draw (233.94,181.54) node [anchor=north west][inner sep=0.75pt]   [align=left] {$\displaystyle 1$};
\draw (307.94,185.54) node [anchor=north west][inner sep=0.75pt]   [align=left] {$\displaystyle 2$};
\draw (248.57,211.64) node [anchor=north west][inner sep=0.75pt]   [align=left] {$\displaystyle a$};
\draw (271.67,190.54) node [anchor=north west][inner sep=0.75pt]   [align=left] {$\displaystyle b$};
\draw (294.47,212.14) node [anchor=north west][inner sep=0.75pt]   [align=left] {$\displaystyle c$};

\draw (260,265) node [anchor=north west][inner sep=0.75pt]    {${E^\prime}^{12;34}$};

\end{tikzpicture}}}
$};

\draw  [line width=0.75] (270,118) node [anchor=north west][inner sep=0.75pt]  [align=center] {$ \hbox{\scalebox{0.75}{\begin{tikzpicture}[x=0.75pt,y=0.75pt,yscale=-1,xscale=1]

\draw  [line width=0.75] (276.68,187.07) -- (276.93,216.46) -- (249.06,225.78) -- (231.58,202.15) -- (248.65,178.23) -- cycle ;
\draw  [line width=0.75] (276.88,216.46) -- (276.68,187.27) -- (305.87,187.07) -- (306.07,216.26) -- cycle ;
\draw  [line width=0.75] (334.23,225.31) -- (306.21,216.45) -- (305.98,187.06) -- (333.86,177.76) -- (351.32,201.4) -- cycle ;

\draw  [line width=0.75]  (243.67,166) -- (248.65,178.23) ;
\draw   [line width=0.75] (249.06,225.78) -- (241.67,238) ;
\draw  [line width=0.75]  (217,202) -- (231.58,202.15) ;
\draw  [line width=0.75]  (351.32,201.4) -- (365.9,201.56) ;
\draw  [line width=0.75]  (341.25,165.54) -- (333.86,177.76) ;
\draw  [line width=0.75]  (334.23,225.31) -- (339.22,237.54) ;

\draw (252.57,198.14) node [anchor=north west][inner sep=0.75pt]   [align=left] {$\displaystyle a$};
\draw (286.67,195.4) node [anchor=north west][inner sep=0.75pt]   [align=left] {$\displaystyle b$};
\draw (321.41,198.19) node [anchor=north west][inner sep=0.75pt]   [align=left] {$\displaystyle c$};
\draw (226.1,175.87) node [anchor=north west][inner sep=0.75pt]   [align=left] {$\displaystyle 3$};
\draw (223.32,211.25) node [anchor=north west][inner sep=0.75pt]   [align=left] {$\displaystyle 4$};
\draw (350.1,174.54) node [anchor=north west][inner sep=0.75pt]   [align=left] {$\displaystyle 3$};
\draw (348.66,212.59) node [anchor=north west][inner sep=0.75pt]   [align=left] {$\displaystyle 4$};
\draw (286.29,167.62) node [anchor=north west][inner sep=0.75pt]   [align=left] {$\displaystyle 1$};
\draw (286.78,221.84) node [anchor=north west][inner sep=0.75pt]   [align=left] {$\displaystyle 2$};

\draw (285,250) node [anchor=north west][inner sep=0.75pt]    {$H^{12,34}$};

\end{tikzpicture}}}
$};

\draw  [line width=0.75] (155,198) node [anchor=north west][inner sep=0.75pt]  [align=center] {$ \hbox{\scalebox{0.75}{\begin{tikzpicture}[x=0.75pt,y=0.75pt,yscale=-1,xscale=1]

\draw  [line width=0.75] (310.59,211.36) -- (301.45,183.43) -- (325.19,166.11) -- (349.01,183.33) -- (339.98,211.3) -- cycle ;
\draw  [line width=0.75] (310.6,211.03) -- (283,220.18) -- (273.85,192.58) -- (301.45,183.43) -- cycle ;
\draw  [line width=0.75] (367.62,220.35) -- (339.98,211.3) -- (349.03,183.67) -- (376.67,192.71) -- cycle ;
\draw  [line width=0.75]  (273.85,192.58) -- (262.31,186.77) ;
\draw  [line width=0.75]  (277.2,231.72) -- (283,220.18) ;
\draw  [line width=0.75]  (376.67,192.71) -- (388.19,186.87) ;
\draw  [line width=0.75]  (373.46,231.87) -- (367.62,220.35) ;
\draw  [line width=0.75]  (325.19,154.45) -- (325.19,166.11) ;

\draw (287.57,195.14) node [anchor=north west][inner sep=0.75pt]   [align=left] {$\displaystyle a$};
\draw (319.67,185.74) node [anchor=north west][inner sep=0.75pt]   [align=left] {$\displaystyle b$};
\draw (353.74,195.14) node [anchor=north west][inner sep=0.75pt]   [align=left] {$\displaystyle c$};
\draw (287.77,162.21) node [anchor=north west][inner sep=0.75pt]   [align=left] {$\displaystyle 1$};
\draw (353.66,162.25) node [anchor=north west][inner sep=0.75pt]   [align=left] {$\displaystyle 2$};
\draw (260.62,200.62) node [anchor=north west][inner sep=0.75pt]   [align=left] {$\displaystyle 3$};
\draw (377.95,201.28) node [anchor=north west][inner sep=0.75pt]   [align=left] {$\displaystyle 3$};
\draw (322.44,220.84) node [anchor=north west][inner sep=0.75pt]   [align=left] {$\displaystyle 4$};

\draw (320,249.6) node [anchor=north west][inner sep=0.75pt]    {$E^{12,34}$};

\end{tikzpicture}}}
$};

\draw  [line width=0.75] (283,198) node [anchor=north west][inner sep=0.75pt]  [align=center] {$ \hbox{\scalebox{0.75}{\begin{tikzpicture}[x=0.75pt,y=0.75pt,yscale=-1,xscale=1]

\draw  [line width=0.75] (310.59,211.36) -- (301.45,183.43) -- (325.19,166.11) -- (349.01,183.33) -- (339.98,211.3) -- cycle ;
\draw  [line width=0.75] (310.6,211.03) -- (283,220.18) -- (273.85,192.58) -- (301.45,183.43) -- cycle ;
\draw  [line width=0.75] (367.62,220.35) -- (339.98,211.3) -- (349.03,183.67) -- (376.67,192.71) -- cycle ;
\draw  [line width=0.75]  (273.85,192.58) -- (262.31,186.77) ;
\draw  [line width=0.75]  (277.2,231.72) -- (283,220.18) ;
\draw  [line width=0.75]  (376.67,192.71) -- (388.19,186.87) ;
\draw  [line width=0.75]  (373.46,231.87) -- (367.62,220.35) ;
\draw  [line width=0.75]  (325.19,154.45) -- (325.19,166.11) ;

\draw (287.57,195.14) node [anchor=north west][inner sep=0.75pt]   [align=left] {$\displaystyle a$};
\draw (319.67,185.74) node [anchor=north west][inner sep=0.75pt]   [align=left] {$\displaystyle b$};
\draw (353.74,195.14) node [anchor=north west][inner sep=0.75pt]   [align=left] {$\displaystyle c$};
\draw (287.77,162.21) node [anchor=north west][inner sep=0.75pt]   [align=left] {$\displaystyle 1$};
\draw (353.66,162.25) node [anchor=north west][inner sep=0.75pt]   [align=left] {$\displaystyle 2$};
\draw (260.62,200.62) node [anchor=north west][inner sep=0.75pt]   [align=left] {$\displaystyle 4$};
\draw (377.95,201.28) node [anchor=north west][inner sep=0.75pt]   [align=left] {$\displaystyle 4$};
\draw (322.44,220.84) node [anchor=north west][inner sep=0.75pt]   [align=left] {$\displaystyle 3$};
\draw (320,249.6) node [anchor=north west][inner sep=0.75pt]    {$E^{12,43}$};

\end{tikzpicture}}}
$};

\draw  [line width=0.75] (5,175) node [anchor=north west][inner sep=0.75pt]  [align=center] {$ \hbox{\scalebox{0.6}{\begin{tikzpicture}[x=0.75pt,y=0.75pt,yscale=-1,xscale=1]

\draw (93.5,68) node [anchor=north west][inner sep=0.75pt]  [font=\large,color=blue  ,opacity=1 ] [align=left] {$\displaystyle 1$};

\end{tikzpicture}

}}
$};

\draw  [line width=0.75] (5,214) node [anchor=north west][inner sep=0.75pt]  [align=center] {$ \hbox{\scalebox{0.6}{\begin{tikzpicture}[x=0.75pt,y=0.75pt,yscale=-1,xscale=1]

\draw (93.5,68) node [anchor=north west][inner sep=0.75pt]  [font=\large,color=blue  ,opacity=1 ] [align=left] {$\displaystyle 2$};

\end{tikzpicture}

}}
$};

\draw  [line width=0.75] (24,155) node [anchor=north west][inner sep=0.75pt]  [align=center] {$ \hbox{\scalebox{0.6}{\begin{tikzpicture}[x=0.75pt,y=0.75pt,yscale=-1,xscale=1]

\draw (93.5,68) node [anchor=north west][inner sep=0.75pt]  [font=\large,color=blue  ,opacity=1 ] [align=left] {$\displaystyle 3$};

\end{tikzpicture}

}}
$};

\draw  [line width=0.75] (62,155) node [anchor=north west][inner sep=0.75pt]  [align=center] {$ \hbox{\scalebox{0.6}{\begin{tikzpicture}[x=0.75pt,y=0.75pt,yscale=-1,xscale=1]

\draw (93.5,68) node [anchor=north west][inner sep=0.75pt]  [font=\large,color=blue  ,opacity=1 ] [align=left] {$\displaystyle 4$};

\end{tikzpicture}

}}
$};

\draw  [line width=0.75] (82,176) node [anchor=north west][inner sep=0.75pt]  [align=center] {$ \hbox{\scalebox{0.6}{\begin{tikzpicture}[x=0.75pt,y=0.75pt,yscale=-1,xscale=1]

\draw (93.5,68) node [anchor=north west][inner sep=0.75pt]  [font=\large,color=blue  ,opacity=1 ] [align=left] {$\displaystyle 5$};

\end{tikzpicture}

}}
$};

\draw  [line width=0.75] (82,206) node [anchor=north west][inner sep=0.75pt]  [align=center] {$ \hbox{\scalebox{0.6}{\begin{tikzpicture}[x=0.75pt,y=0.75pt,yscale=-1,xscale=1]

\draw (93.5,68) node [anchor=north west][inner sep=0.75pt]  [font=\large,color=blue  ,opacity=1 ] [align=left] {$\displaystyle 6$};

\end{tikzpicture}

}}
$};

\draw  [line width=0.75] (36,212) node [anchor=north west][inner sep=0.75pt]  [align=center] {$ \hbox{\scalebox{0.6}{\begin{tikzpicture}[x=0.75pt,y=0.75pt,yscale=-1,xscale=1]

\draw (93.5,68) node [anchor=north west][inner sep=0.75pt]  [font=\large,color=blue  ,opacity=1 ] [align=left] {$\displaystyle 7$};

\end{tikzpicture}

}}
$};

\draw  [line width=0.75] (54,212) node [anchor=north west][inner sep=0.75pt]  [align=center] {$ \hbox{\scalebox{0.6}{\begin{tikzpicture}[x=0.75pt,y=0.75pt,yscale=-1,xscale=1]

\draw (93.5,68) node [anchor=north west][inner sep=0.75pt]  [font=\large,color=blue  ,opacity=1 ] [align=left] {$\displaystyle 8$};

\end{tikzpicture}

}}
$};

\draw  [line width=0.75] (32,190) node [anchor=north west][inner sep=0.75pt]  [align=center] {$ \hbox{\scalebox{0.6}{\begin{tikzpicture}[x=0.75pt,y=0.75pt,yscale=-1,xscale=1]

\draw (93.5,68) node [anchor=north west][inner sep=0.75pt]  [font=\large,color=blue  ,opacity=1 ] [align=left] {$\displaystyle 9$};

\end{tikzpicture}

}}
$};

\draw  [line width=0.75] (54,190) node [anchor=north west][inner sep=0.75pt]  [align=center] {$ \hbox{\scalebox{0.6}{\begin{tikzpicture}[x=0.75pt,y=0.75pt,yscale=-1,xscale=1]

\draw (93.5,68) node [anchor=north west][inner sep=0.75pt]  [font=\large,color=blue  ,opacity=1 ] [align=left] {$\displaystyle 10$};

\end{tikzpicture}

}}
$};

\end{tikzpicture}

}
}}
\end{eqnarray}
This cut-geometry is of particular interest, as it serves as the ``parent" of the $(\alpha, \beta, \gamma)$-maximal cut discussed earlier, and thus receives contributions from the local integrands $E^{12;34}$, $E^{12;43}$, ${E^\prime}^{12;34}$ and $H^{12;34}$ respectively. We choose the following loop parameterization, 
\begin{equation}
    \begin{split}
         \ (A,B) &= \left(Z_{1,1} + x_1 Z_{2,1} - w_1 Z_{2,3},\quad y_1 Z_{2,1} + Z_{1,3} + z_1 Z_{2,3} \right),\\ 
         \ (C,D) &= \left(Z_{1,1} + x_2 Z_{2,1} - w_2 Z_{2,2},\quad y_2 Z_{2,1} + Z_{1,2} + z_2 Z_{2,2} \right),\\
         \ (E,F) &= \left(Z_{1,2} + x_3 Z_{2,2} - w_3 Z_{2,3},\quad y_3 Z_{2,2} + Z_{1,3} + z_3 Z_{2,3} \right).
    \end{split}
\end{equation}
and solve the cuts sequentially in the following order
\begin{equation}
    (w_1,y_1, w_2,y_2,w_3,y_3,x_1,x_3,x_2,z_2 )\,.
\end{equation}

Focusing on the $\Delta>0$ branch and summing over the net-positive sector of the cut-geometry, we find that the resulting form defined through the inequalities on $z_1, z_3$ depends on the relative magnitude of $(t, u)$:
\begin{equation}\label{eq: NetPositive}
    \begin{split}
       t<u:&\quad \left(z_3-z_1\right)\left(z_3-\frac{(1-c_1)z_1}{1-c_2}\right)<0 \ \lor \  \left(z_3-\frac{c_1 z_1}{c_2}\right)\left(z_3-\frac{c_1(1-c_2)z_1}{c_2(1-c_1)}\right)<0\\
      t>u:&\quad  \left(z_3-\frac{c_1 z_1}{c_2}\right)\left(z_3-\frac{(1-c_1)z_1}{1-c_2}\right)<0 \ \lor \  \left(z_3-z_1\right)\left(z_3-\frac{c_1(1-c_2)z_1}{c_2(1-c_1)}\right)<0,
    \end{split}
\end{equation}
The geometry for the sum of the net-negative sector is even more refined. Besides the relative magnitude of $(t,u)$, it also depends on $\text{max}(s,t,u)$. We find
\begin{equation}\label{eq: NetNegative}
    \begin{split}
       \text{max}(s,t,u)\neq s,\;t<u:&\quad  z_3 \left(z_3-\frac{(1-c_1)z_1}{1-c_2}\right)<0 \lor\ z_1  \left(z_3-\frac{c_1(1-c_2)z_1}{c_2(1-c_1)}\right) >0\\
       \text{max}(s,t,u)\neq s,\;t>u:&\quad  z_3 \left(z_3- \frac{c_1(1-c_2)z_1}{c_2(1-c_1)}\right)<0 \lor\ z_1  \left(z_3-\frac{(1-c_1)z_1}{1-c_2} \right) >0\\
     \text{max}(s,t,u)= s:&\quad  \left(z_3- \frac{c_1(1-c_2)z_1}{c_2(1-c_1)}\right)  \left(z_3-\frac{(1-c_1)z_1}{1-c_2} \right) <0 \,.
    \end{split}
\end{equation}
Once again, this ten-cut is present in all kinematics, albeit the geometry is distinct.

\begin{figure}[H] 
\begin{center}
\includegraphics[width=13.5cm]{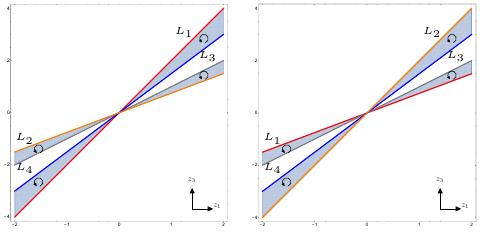} 
\end{center}
\caption{Ten-cut geometry in the net-positive sector (counterclockwise orientation). Left: $t < u$ with $(c_1, c_2) = \left(\tfrac{1}{2}, \tfrac{1}{3}\right)$; Right: $t > u$ with $(c_1, c_2) = (3, 2)$. For the $\alpha$-cut, the approach proceeds via $L_4$ toward the origin; in both cases, the inward-pointing arrow along $L_4$ indicates negative orientation. For the $\beta$-cut, one approaches $L_2$ first, then the origin. The arrow directions along $L_2$ differ between the two cases, reflecting opposite orientations.}
\label{fig:neg positve plot}
\end{figure}

\begin{figure}[H] 
\begin{center}
\includegraphics[width=13.5cm]{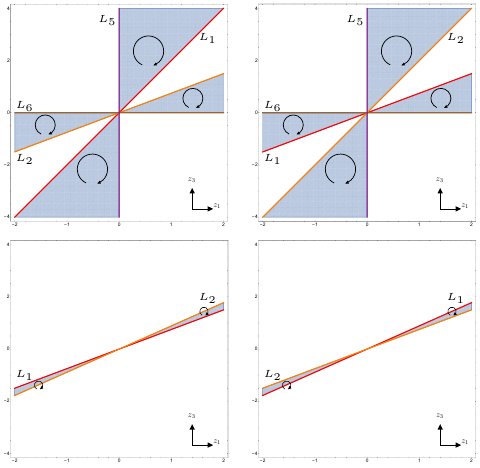} 
\end{center}
\caption{Ten-cut geometry in the net-negative sector (clockwise orientation). Upper left: $\text{max}(s,t,u) \neq s$ and $t < u$ with $(c_1, c_2) = \left(\tfrac{1}{2}, \tfrac{1}{3}\right)$; Upper right: $\text{max}(s,t,u) \neq s$ and $t > u$ with $(c_1, c_2) = (3, 2)$; Lower left: $\text{max}(s,t,u) = s$ and $t < u$ with $(c_1, c_2) = \left( -\tfrac{1}{3},-\tfrac{1}{2}\right)$; Lower right: $\text{max}(s,t,u) = s$ and $t > u$ with $(c_1, c_2) = (-2, -3)$. For the $\gamma$-cut, the approach proceeds along $L_6$ toward the origin. The boundary is accessible only when $\max(s, t, u) \neq s$. In these cases, the inward-pointing arrow along $L_6$ indicates a negative orientation, consistent with eq.~\eqref{eq: gamma-cut approach ways} due to the overall orientation flip in the net-negative sector. For the $\beta$-cut, the orientations for $t < u$ and $t > u$ agree with those in the net-positive sector.}
\label{fig:neg negative plot}
\end{figure}

We can now study the $(\alpha, \beta,\gamma)$ maximal cuts, which appear as the codimension-two boundaries of this 10-cut geometry.  
\begin{itemize}
\item The $\alpha$-cut—obtained by imposing $z_3-z_1=0$ and $z_1=0$ in  eq.~\eqref{eq: NetPositive}—is accessible in all kinematic regions of the net-positive sector but absent in the net-negative one.

In the net-positive sector, the boundary can be approached in two distinct ways:
\begin{eqnarray}
t<u:&& \quad z_3\rightarrow z_1^{\pm}, \quad z_1\frac{c_1-c_2}{1-c_2}\rightarrow 0^\mp,\nonumber\\
t>u:&& \quad z_3\rightarrow z_1^{\pm}, \quad z_1\frac{c_2-c_1}{(1-c_1)c_2}\rightarrow 0^\mp.
\end{eqnarray}
Since we are on the positive branch $c_1>c_2$, and from eq.~\eqref{eq: Chamberc} we have $1-c_2>0$ for $t<u$ and $c_2(1-c_1)<0$ for $t>u$, the relations simplify to
\begin{eqnarray}
t<u:&& \quad z_3\rightarrow z_1^{\pm}, \quad z_1\rightarrow 0^\mp,\nonumber\\
t>u:&& \quad z_3\rightarrow z_1^{\pm}, \quad z_1\rightarrow 0^\mp.
\end{eqnarray}
The fact that the boundary can be approached from two distinct directions indicates that the geometry is disjoint. Furthermore, the orientations of both regions are net-negative from the perspective of the codimension-two boundary ($\alpha$-cut), and thus the residues add up, acquiring an overall multiplicative factor of 2. This situation is known as an ``internal boundary"~\cite{Dian:2022tpf}, where both orientations contribute an overall multiplicative factor to the associated form.  In this case, the orientations for $t>u$ and $t<u$ are the same. These descriptions are illustrated in Figure~\ref{fig:neg positve plot}.

\item The $\gamma$-cut—obtained by imposing $z_1=0$ and $z_3=0$—is present only when $\text{max}(s,t,u) \neq s$ in the net-negative sector. Thus, the $\gamma$-cut boundary is accessible only for $r_1, r_2, r_3$, and $r_5$. Taking into account eq.~\eqref{eq: Chamberc}, the boundary is approached as
\begin{eqnarray}\label{eq: gamma-cut approach ways}
    t < u:&& \quad z_3 \rightarrow 0^\pm, \quad z_1 \rightarrow 0^\pm,\nonumber\\
t > u:&& \quad z_3 \rightarrow 0^\pm, \quad z_1 \rightarrow 0^\pm.
\end{eqnarray}
Once again, we find that the geometry is disjoint but carries the same orientation, and hence the contributions add up; the orientation is also consistent between $t<u$ and $t>u$. These details are depicted in Figure~\ref{fig:neg negative plot}.

\item The $\beta$-cut—obtained by setting $z_3 = (\tfrac{1 - c_1}{1 - c_2} z_1, \tfrac{c_1(1 - c_2)}{c_2(1 - c_1)} z_1)$ and $z_1 = 0$—is  present in both the net-positive and net-negative sectors. A novel feature of this cut is that its orientation changes depending on the  relative magnitude of $t$ and $u$.

Let us consider in more detail the orientation with respect to the boundary $z_3=\tfrac{1 - c_1}{1 - c_2} z_1$ and then $z_1=0$. In the net-positive sector, described by eq.~\eqref{eq: NetPositive}, the boundary is approached for $t < u$ as: 
\begin{equation}
    \begin{split}
 z_3\rightarrow \left(\tfrac{1 - c_1}{1 - c_2} z_1\right)^{\pm}, \quad      \tfrac{(c_1-c_2)}{c_2-1} z_1\rightarrow 0^{\mp} 
    \end{split}
\end{equation}
From eq.~\eqref{eq: Chamberc}, the ratio $\frac{(c_1-c_2)}{c_2-1}$ is strictly negative for $t<u$. Thus, the $\beta$-cut is approached as:
\begin{equation}
  z_3\rightarrow \left(\tfrac{1 - c_1}{1 - c_2} z_1\right)^{\pm}, \quad   z_1 \rightarrow 0^{\pm}.
\end{equation}
Again, the two disjoint geometries carry the same orientation.  Doing the same for $t>u$, one finds:
\begin{equation}
  z_3\rightarrow \left(\tfrac{1 - c_1}{1 - c_2} z_1\right)^{\pm}, \quad   z_1 \rightarrow 0^{\mp}\,.
\end{equation}
Although both disjoint regions are negatively oriented, their orientation is opposite to that for $t<u$! Thus, due to the change in orientation, the integrand responsible for the $\beta$-cut changes sign as one crosses the $t=u$ boundary.  The net-negative sector in eq.~\eqref{eq: NetNegative} yields a similar result. Refer to Figures~\ref{fig:neg positve plot} and \ref{fig:neg negative plot} for these descriptions.

\end{itemize}

In summary, the above discussion is encapsulated in Figures~\ref{fig:neg positve plot} and~\ref{fig:neg negative plot}, where the relevant boundaries are defined by
$L_1: z_3 = \frac{c_1(1 - c_2)}{c_2(1 - c_1)} z_1$, $L_2: z_3 = \frac{1 - c_2}{1 - c_1} z_1$, $L_3: z_3 = \frac{c_1}{c_2} z_1$,  $L_4: z_3 = z_1$, $L_5: z_1=0$, and $L_6: z_3=0$.
The $\alpha$-cut is associated with $L_4$, the $\beta$-cut involves $L_1$ and $L_2$, while the $\gamma$-cut corresponds to $L_5$ and $L_6$. 

We can now use the discussion above to understand the block structure of each three-loop chamber~\eqref{eq: 3Loops}. Each block — $A_{\sigma_i}$, $B$, and $C_{\sigma_i,\sigma_j}$ contains distinct sets of cuts that are accessible within a given chamber. Specifically $A_{\sigma_i}$ corresponds to cuts that are accessible only when $\text{max}(s,t,u) = \sigma_i$ or $\text{max}(s,t,u) \neq \sigma_i$. The blocks  $B$ and $C$ are universally accessible, where $C_{\sigma_i,\sigma_j}$ further depends on the relative magnitude of  $\sigma_i, \sigma_j$. We have seen that the $\alpha$-cut is accessible across all regions. In terms of local integrands, this cut is realized by $ E^{12;34}$, which is why it is contained in $B$. The $\gamma$-cut, on the other hand, only contributes when $\text{max}(s,t,u) \neq s$, and is realized by the integrand $ H^{12;34}$. More interestingly, the $\beta$-cut, realized by ${E^\prime}^{12;34}$ with an orientation change at $t-u=0$, is reflected in the block $C_{t,u}$, where the multiplying factor $\sigma_i - \sigma_j$ depends on the region, with $\sigma_i$ and $\sigma_j$ interchanging accordingly. Finally, each of these maximal cuts is associated with disjoint geometries — {\it i.e.} they are internal boundaries —leading to a multiplicative factor of 2 in the integrand. Furthermore, since the $\beta$-cut is present in both the net-positive and net-negative sectors, there is an additional factor of 2 in the overall normalization, as seen in eq.~\eqref{eq: three-loop block}.

\subsubsection{Chamber-forms and leading singularities}
As the loop-form for each chamber is constructed as a canonical form of positive geometry, this suggests that the integrated result of eq.~\eqref{eq: 3Loops} consists of pure functions. Indeed, as shown above, we find that all leading singularities are pure numbers. The leading singularities of the full correlator were computed up to three-loops~\cite{Drummond:2013nda}. Writing the correlator as 
\begin{equation}
G_4^{(L)}=R(1,2,3,4)stu F^{(L)},
\end{equation}
the leading singularities for $F^{(1,2)}$ are simply $\frac{1}{stu\Delta}$. This indeed matches our result, where the tree-form — sans the $\langle \! \langle \mathbf{X}_1, \mathbf{X}_2, \mathbf{X}_3, \mathbf{X}_4\rangle \! \rangle^4$ factor that yields $R(1,2,3,4)stu$ — is given by $\frac{1}{stu\Delta}$. At three loops, there are several distinct leading singularities. For completeness, we list them as follows:
\begin{align}
&\frac{1}{s t u \Delta},\, \frac{1}{s t\Delta^2},\, \frac{1}{u t\Delta^2},\, \frac{1}{s u\Delta^2},\nonumber\\\, \frac{1}{t s\Delta(t{-}u)},\, \frac{1}{u s\Delta(t{-}u)}&,\, \frac{1}{t s\Delta(u{-}s)},  \, \frac{1}{t u\Delta(u{-}s)}, \, \frac{1}{t u\Delta(s{-}t)},\, \frac{1}{s u\Delta(s{-}t)}\,.    
\end{align}

Since the leading singularities of our loop-forms are pure numbers, our construction implies that all leading singularities of the correlator are simply linear combinations of the chamber forms~\eqref{eq: ChamberForm}. One can straightforwardly verify that this is indeed the case. For example, we have
\begin{equation}
    \frac{1}{t u \Delta (s-t)}=\frac{1}{2}\sum_{\pm}\pm\left(\sum_{i=3,4,6}\omega_{r_i}^\pm\right)\frac{1}{\langle \! \langle \mathbf{X}_1, \mathbf{X}_2, \mathbf{X}_3, \mathbf{X}_4\rangle \! \rangle^4}.
\end{equation}
Other leading singularities can be similarly represented.

\subsubsection{Chamber-inspired rewriting of the three-loop correlator}

Throughout this paper, we focus on the loop integrand of the four-point correlator as derived from the Correlahedron. Very recently, the correlator integrand has been bootstrapped using $f$-graphs up to  $12$ loops, based on a new and powerful constraint formulated purely at the graphical level~\cite{He:2024cej, Bourjaily:2025iad}. However, it remains an important open question how to obtain the integrated results for all required conformal integrals: although one- and two-loop cases (box, its square, and double-box integrals) are well known, it was not until~\cite{Drummond:2013nda} that all three-loop conformal integrals needed for correlators were determined analytically. While the three-loop ladder, tennis-court (which is equivalent to the ladder), and lower-loop products are relatively easy to compute, two additional integrals known as ``easy" and ``hard" integrals require more work. The main obstacle was that these integrals involve non-trivial leading singularities, and the hard integral also features additional nontrivial {\it symbol} letters (or singularities of the polylogarithmic functions)~\cite{Goncharov:2010jf, Duhr:2011zq, Duhr:2012fh}; the authors of~\cite{Drummond:2013nda} managed to compute these leading singularities and bootstrapped corresponding polylogarithmic functions: the easy integral evaluates to single-valued harmonic polylogarithmic functions (SVHPL) and the hard integral evaluates to single-valued multiple polylogarithmic functions (SVMPL), which has additional symbol letters.

Our geometric approach makes it clear that all these integrals should be reorganized into combinations corresponding to loop forms that have unit leading singularities, multiplied by chamber forms (leading singularities), as given in eq.~\eqref{eq: corr potential}. In this subsection, we take this form seriously and directly compute these unit-leading-singularity integrals at three loops, thereby rewriting the final results of the four-point correlator as pure SVMPL functions multiplied by chamber forms. Naively, the loop integrands for such unit-LS integrals appear more complicated, but our geometric picture guarantees that each of them evaluates to a pure function. We can either perform the loop integrations using {\it e.g.} \texttt{ HyperlogProcedures}~\cite{Schnetz:2013hqa} or bootstrap these functions based on an ansatz and boundary values. It turns out that we only need two independent functions, denoted as ${\cal E}$ and ${\cal H}_a$, to express our $E_c$, $E'$, and $H_c$ integrals:
\begin{equation}
    \begin{split}
        E^{\prime12;34}&:=\frac{1}{ \Delta (u-t)} \left({\cal E}(z, \bar{z}) + {\cal E}(\frac{z}{z-1} , \frac{\bar{z}}{\bar{z}-1}) \right),\\
        E_c^{12;34}&:=\frac{u}{\Delta(u-t)}\left({\cal E}(z, \bar{z}) + 
    \frac{t}{u}{\cal E}(\frac{z}{z-1} , \frac{\bar{z}}{\bar{z}-1})\right)\\
  &=\frac{1}{2\Delta}\left({\cal E}(z, \bar{z}) - {\cal E}(\frac{z}{z-1} , \frac{\bar{z}}{\bar{z}-1}) \right)-\frac{t+u}{2}E^{\prime12;34},\\
  &\phantom{ccccc}H_c^{12;34} +  E^{\prime12;34}:=-\frac{1}{\Delta^2} {\cal H}_a(z, \bar{z}).
    \end{split}
\end{equation}
where recall that $z, \bar{z}$ are defined by $z \bar{z}=U$, 
and permutations of external points $1,2,3,4$ correspond to Möbius transformations of $z, \bar{z}$. Note that in the above equation, up to an overall normalization factor $\frac{1}{x_{1,3}^2 x_{2,4}^2}$, the prefactor of $E'^{12;34}$ and $E_c^{12;34}$ in terms of $z,\bar{z}$ is proportional to $\frac{1}{(z-\bar{z})(z+\bar{z}-z \bar{z})}$, and the prefactor in front of ${\cal H}_a$ is proportional to  $\frac{1}{(z-\bar{z})^2}$, thus, these functions are parity odd and even, respectively.  

Here ${\cal E}$ is a remarkably simple SVHPL function, whose symbol has only four letters $z, \bar{z}, 1-z, 1-\bar{z}$, while ${\cal H}_a$ is more complicated and involve additional letters $z-\bar{z}$ (and its image $1-z \bar{z}$ under permutations).  Before proceeding, we emphasize that it is highly non-trivial and a rather surprising fact that the hard integral $H$ is closely related to the easy integral $E$: note that $H$ contains two parts, the ``even" part with LS proportional to $1/(z-\bar{z})^2$ which is a SVMPL function, ${\cal H}_a$, and the ``odd" part with LS proportional to $1/(z-\bar{z})$; the second part is exactly 
the $E'$ integral, which is given by the symmetrized version of SVHPL function ${\cal E}$, which also determines the easy integral $E$! We find it rather intriguing that the geometry allows us to find this hidden relation between (the odd part of) hard integral and the easy integral. 

In order to present their explicit forms, let us briefly review these functions. The SVHPL functions can be succinctly recorded as ${\cal L}_w$ for a word of $0$ and $1$ (the length of $w$ is the {\it weight} of the function, here $|w|=6$), where the standard notation is related to that in \texttt{HyperlogProcedures} by ${\cal L}_w:=\mathrm{I}_{z,w,0}$, and it is recursively defined as~\cite{Schnetz:2013hqa}:
\begin{equation}
    \begin{aligned}
    &{\cal L}_{a_n, a_{n{-}1}, \cdots, a_1}=\int_{sv} \frac{d z}{z-a_n} {\cal L}_{a_{n{-}1}, \cdots, a_1}, 
    \, a_{n}=0,1; \\
    &{\cal L}_{\emptyset}=1; {\cal L}_0=\log z \bar{z}, \, {\cal L}_1=\log (1-z)(1-\bar{z})
    \end{aligned}
\end{equation}
$\int_{sv}$ is the single-valued integration defined by Eq.~(2.53) in \cite{Schnetz:2013hqa}, which not only is an integration of $z$, but also accounts for the antiholomorphic part in $\bar{z}$ and the boundary condition which requires the integration vanishes when $z$ approaches 0 (regularized at $z=0$). It keeps the result single-valued. This definition can be viewed as the single-valued version of HPL functions,  which can be expressed as sum of products of  standard $G(z)$ and $G(\bar{z})$ functions, from which one can trivially read off the symbol; schematically one writes ${\cal L}_w=\sum_{i,j} c_{i,j}~G_{w_i} (z) G_{w_j} (\bar{z})$, with the standard definition for $G$ (slightly different from the standard definition for HPL in \cite{Remiddi:1999ew}) as:
\begin{equation}
G_{a_{n},a_{n-1},\ldots,a_{1}}(z)=\int_{0}^{z}\frac{\mathrm{d}z^{\prime}}{z^{\prime}-a_{n}}G_{a_{n-1},\ldots,a_{1}} (z), \, a_{n}=0,\pm 1; \quad G_{\emptyset}(z):=1.
\end{equation}
which are understood to be already regularized around $z=0$. For instance, ${\cal L}_{1,0}$ is a weight-2 function given by $G_1(z) G_0(\bar{z}) + G_{1,0}(z) + G_{0,1} (\bar{z})$ which gives the symbol $z\otimes (1-z)$. In this notation, the $L$-loop ladder integral becomes simply
\begin{equation}
\frac{(-1)^{L}}{\Delta}\times \big({\cal L}_{{\scriptscriptstyle \underbrace{0,...,0}_{L}},1,{\scriptscriptstyle \underbrace{0,...,0}_{L-1}}}-{\cal L}_{{\scriptscriptstyle \underbrace{0,...,0}_{L-1}},1,{\scriptscriptstyle \underbrace{0,...,0}_{L}}}\big)\,.   
\end{equation}

After this short review, we can write down the ${\cal E}$ function, which gives $E, E'$ and the ``odd" part of $H$ integrals, and the form is extremely simple:
\begin{eqnarray}\label{eq_E}
{\cal E}(z,\bar{z})&=2 {\cal L}_{0,0,0,1,0,1}-2 {\cal L}_{0,0,1,0,1,0}-{\cal L}_{0,0,1,1,0,1}+{\cal L}_{0,0,1,1,1,0}-2 {\cal L}_{0,1,0,0,0,1}+2 {\cal L}_{0,1,0,1,0,0}-{\cal L}_{0,1,0,1,1,0}\nonumber\\
&+{\cal L}_{0,1,1,0,0,1}+{\cal L}_{0,1,1,0,1,0}-{\cal L}_{0,1,1,1,0,0}+2 {\cal L}_{1,0,0,0,1,0}-{\cal L}_{1,0,0,1,0,1}-{\cal L}_{1,0,0,1,1,0}-2 {\cal L}_{1,0,1,0,0,0}\nonumber\\
&+{\cal L}_{1,0,1,0,0,1}+{\cal L}_{1,0,1,1,0,0}-20 \zeta _5 {\cal L}_1+ 8 \zeta _3 {\cal L}_{0,0,1}-4 \zeta _3 {\cal L}_{0,1,1}-2 \zeta _3 {\cal L}_{1,0,1}\,.
\end{eqnarray}
Although not needed, we also record the symmetrized version (where some cancellations happen), $(z-\bar{z})(z+\bar{z}+ z \bar{z}) E'={\cal E}(z, \bar{z}) + {\cal E}(\frac{z}{z-1} , \frac{\bar{z}}{\bar{z}-1})$ reads
\begin{eqnarray}\label{eq_Hb}
&4 {\cal L}_{0,0,0,1,0,1}-2 {\cal L}_{0,0,0,1,1,1}-4 {\cal L}_{0,0,1,0,1,0}+2 {\cal L}_{0,0,1,0,1,1}-2 {\cal L}_{0,0,1,1,0,1}+2 {\cal L}_{0,0,1,1,1,0}-4 {\cal L}_{0,1,0,0,0,1}\nonumber\\
&+2 {\cal L}_{0,1,0,0,1,1}+4 {\cal L}_{0,1,0,1,0,0}-2 {\cal L}_{0,1,0,1,0,1}-2 {\cal L}_{0,1,0,1,1,0}+{\cal L}_{0,1,0,1,1,1}+2 {\cal L}_{0,1,1,0,0,1}+2 {\cal L}_{0,1,1,0,1,0}\nonumber\\
&-2 {\cal L}_{0,1,1,0,1,1}-2 {\cal L}_{0,1,1,1,0,0}+{\cal L}_{0,1,1,1,0,1}+4 {\cal L}_{1,0,0,0,1,0}-2 {\cal L}_{1,0,0,0,1,1}-2 {\cal L}_{1,0,0,1,0,1}-2 {\cal L}_{1,0,0,1,1,0}\nonumber\\
&+2 {\cal L}_{1,0,0,1,1,1}-4 {\cal L}_{1,0,1,0,0,0}+2 {\cal L}_{1,0,1,0,0,1}+2 {\cal L}_{1,0,1,0,1,0}-{\cal L}_{1,0,1,0,1,1}+2 {\cal L}_{1,0,1,1,0,0}-{\cal L}_{1,0,1,1,1,0}\nonumber\\
&+2 {\cal L}_{1,1,0,0,0,1}-2 {\cal L}_{1,1,0,0,1,0}-2 {\cal L}_{1,1,0,1,0,0}+{\cal L}_{1,1,0,1,0,1}+2 {\cal L}_{1,1,0,1,1,0}-{\cal L}_{1,1,0,1,1,1}+2 {\cal L}_{1,1,1,0,0,0}\nonumber\\&
-2 {\cal L}_{1,1,1,0,0,1}-{\cal L}_{1,1,1,0,1,0}+{\cal L}_{1,1,1,0,1,1}+4 \zeta _3 {\cal L}_{1,0,1}-2 \zeta _3 {\cal L}_{1,1,1} 
\end{eqnarray}

The last function, ${\cal H}_a$, contains functions that are not SVHPL (involving new letters). Adopting the notation from \texttt{HyperlogProcedures}, we can still record these SVMPL functions as ${\bf L}_w$ where the word $w$ now also contains $\bar{z}$ in addition to $0$ and $1$, ${\bf L}_w:=I_{z, w, 0}$ as defined in~\cite{Schnetz:2013hqa, Schnetz:2021ebf}; upon expanding in terms of standard $G$ functions now they involve new letters such as $z-\bar{z}$. ${\cal H}_a(z,\bar{z})$ is given by
\begin{eqnarray}\label{eq_Ha}
& -4 {\cal L}_{0,0,0,1,0,1}+2 {\cal L}_{0,0,0,1,1,1}+4 {\cal L}_{0,0,1,0,1,0}-2 {\cal L}_{0,0,1,0,1,1}+2 {\cal L}_{0,0,1,1,0,1}-2 {\cal L}_{0,0,1,1,1,0}\nonumber\\
&+4 {\cal L}_{0,1,0,0,0,1}+8 {\cal L}_{0,1,0,0,1,0}-10 {\cal L}_{0,1,0,0,1,1}-4 {\cal L}_{0,1,0,1,0,0}+2 {\cal L}_{0,1,0,1,0,1}+2 {\cal L}_{0,1,0,1,1,0}\nonumber\\
&+{\cal L}_{0,1,0,1,1,1}-8 {\cal L}_{0,1,1,0,0,0}+6 {\cal L}_{0,1,1,0,0,1}-2 {\cal L}_{0,1,1,0,1,0}+2 {\cal L}_{0,1,1,1,0,0}-{\cal L}_{0,1,1,1,0,1}\nonumber\\
&+4 {\cal L}_{1,0,0,0,1,0}-6 {\cal L}_{1,0,0,0,1,1}-2 {\cal L}_{1,0,0,1,0,1}+2 {\cal L}_{1,0,0,1,1,0}+2 {\cal L}_{1,0,0,1,1,1}-4 {\cal L}_{1,0,1,0,0,0}\nonumber\\
&+6 {\cal L}_{1,0,1,0,0,1}+2 {\cal L}_{1,0,1,0,1,0}-3 {\cal L}_{1,0,1,0,1,1}-2 {\cal L}_{1,0,1,1,0,0}+2 {\cal L}_{1,0,1,1,0,1}-{\cal L}_{1,0,1,1,1,0}\nonumber\\
&+2 {\cal L}_{1,1,0,0,0,1}-2 {\cal L}_{1,1,0,0,1,0}-2 {\cal L}_{1,1,0,1,0,0}+{\cal L}_{1,1,0,1,0,1}+2 {\cal L}_{1,1,1,0,0,0}-2 {\cal L}_{1,1,1,0,0,1}\nonumber\\
&+{\cal L}_{1,1,1,0,1,0}+20 \zeta _3 {\cal L}_{0,1,1}+16 \zeta _3 {\cal L}_{1,0,1}+12 \zeta _3 {\cal L}_{1,1,0}-8 \zeta _3 {\cal L}_{1,1,1}
\nonumber\\
&+8 {\bf L}_{0,\bar{z},0,0,1,1}-8 {\bf L}_{0,\bar{z},0,1,1,0}-8 {\bf L}_{0,\bar{z},1,0,0,1}+8 {\bf L}_{0,\bar{z},1,1,0,0}+4 {\bf L}_{1,\bar{z},0,0,1,1}-4 {\bf L}_{1,\bar{z},0,1,1,0}-4 {\bf L}_{1,\bar{z},1,0,0,1}\nonumber\\
&+4 {\bf L}_{1,\bar{z},1,1,0,0}+8 {\bf L}_{\bar{z},0,0,0,1,1}+8 {\bf L}_{\bar{z},0,0,1,0,1}-8 {\bf L}_{\bar{z},0,0,1,1,0}-4 {\bf L}_{\bar{z},0,0,1,1,1}-8 {\bf L}_{\bar{z},0,1,0,0,1}-8 {\bf L}_{\bar{z},0,1,0,1,0}\nonumber\\
&+4 {\bf L}_{\bar{z},0,1,0,1,1}+8 {\bf L}_{\bar{z},0,1,1,0,0}-4 {\bf L}_{\bar{z},0,1,1,0,1}+4 {\bf L}_{\bar{z},0,1,1,1,0}-8 {\bf L}_{\bar{z},1,0,0,0,1}-8 {\bf L}_{\bar{z},1,0,0,1,0}+12 {\bf L}_{\bar{z},1,0,0,1,1}\nonumber\\
&+8 {\bf L}_{\bar{z},1,0,1,0,0}-4 {\bf L}_{\bar{z},1,0,1,0,1}-4 {\bf L}_{\bar{z},1,0,1,1,0}+8 {\bf L}_{\bar{z},1,1,0,0,0}-4 {\bf L}_{\bar{z},1,1,0,0,1}+4 {\bf L}_{\bar{z},1,1,0,1,0}-4 {\bf L}_{\bar{z},1,1,1,0,0}\nonumber\\
&-16 {\bf L}_{\bar{z},\bar{z},0,0,1,1}+16 {\bf L}_{\bar{z},\bar{z},0,1,1,0}+16 {\bf L}_{\bar{z},\bar{z},1,0,0,1}-16 {\bf L}_{\bar{z},\bar{z},1,1,0,0}-32 \zeta _3 {\bf L}_{\bar{z},1,1}
\end{eqnarray}

We remark that eqs.~\eqref{eq_E} and~\eqref{eq_Ha}, together with the well-known results for other integrals expressed as products of ladders, provide all the pure functions of three-loop four-point correlators. We have seen that the chamber picture not only predicts all correct LS or prefactors, but also inspires a remarkably compact rewriting of the accompanying integrals which evaluate to pure, SVMPL functions. We expect a similar rewriting for higher-loop correlators as well, though with much more complicated pure functions as we will now see at four loops. 

\section{ Four-loops}\label{sec5}
We now turn to the four loops. The loop-form is constructed in a manner similar to before: one computes the lower-dimensional canonical forms for the cuts and matches them to a basis of local integrands. The geometry of the cuts once again depends on the kinematics, whose chamber structure remains identical to that at three-loop order, {\it i.e.} eq.~\eqref{eq:L=3_chambers}. The explicit loop-form can be organized in a similar pattern as with three-loop case:
\begin{equation}\label{eq: 4Loops}
\Omega^{(4)\pm}_{r_i}= 2\Delta^2 A_{\sigma_3} \pm2\Delta\big(B+ (\sigma_2-\sigma_1)C_{\sigma_2,\sigma_1} + (\sigma_3-\sigma_1)C_{\sigma_3,\sigma_1} + (\sigma_3-\sigma_2) C_{\sigma_3,\sigma_2} \big)\,,   
\end{equation}
where the integrand in each individual block is:
\begin{equation}\label{eq: 4LoopIntAB}
    \begin{split}
        &B:={-}\frac{1}{2} I_{\Theta^{\prime}}^{12;34}+ 2\, \text{perms.}+  I_{S_{}}^{12;34}+ 3\, \text{perms.}+ \big[I_{X^{\prime\prime}}^{12;34}{+} I_{Q_{}}^{12;34} {+} I_{K^{\prime}}^{12;34}+ 5\, \text{perms.}\big]\\
        &\phantom{myff}+  \big[I_{P_{}}^{12;34}- I_{M^{\prime\prime}}^{12;34}+I_{J_{}}^{12;34}+I_{Y^{\prime\prime}}^{12;34}+ I_{N^{\prime}}^{12;34}+ I_{V_{}}^{12;34}+ I_{R^{\prime}}^{12;34}+ 11\, \text{perms.} \big]\\
        &\phantom{myff}+ \big[ I_{P^{\prime\prime}}^{12;34}+I_{Y_{}}^{12;34}+I_{U_{}}^{12;34}+ 23\, \text{perms.} \big],\\
        &A_s:= 8\, I_{F}^{12;34}+ I_{hh^\prime}^{12;34}+\big[ 4\,I_{G}^{12;34}+I_{gL}^{12;34}+I_{hh}^{12;34}+ (1,2){\leftrightarrow}(3,4)\big]+ \big[I_{Z}^{12;34}\phantom{my}\\
        &\phantom{mfffy}+I_{N}^{12;34}+ 7\, \text{perms.}\big|_s \big] +\big[I_{S^{\prime\prime}}^{12;34}+I_{g T}^{12;34}+I_{W}^{12;34}+ 3\, \text{perms.}\big|_s \big] +\big[  I_{K}^{14;23}\\
        &\phantom{mfffy}+ I_{X}^{14;23}+ (1,4){\leftrightarrow}(2,3)+ I^{14;23}_{M} + (3{\leftrightarrow} 4)\big],\\
    \end{split}
\end{equation}
The integrands in $C_{\sigma_i,\sigma_j}$ are further separated into an ``unordered" part ($C^{(1)}_{\sigma_i,\sigma_j}=C^{(1)}_{\sigma_j,\sigma_i}$) and an ordered subset $C^{(2)}_{\sigma_i,\sigma_j}\neq C^{(2)}_{\sigma_j,\sigma_i}$. More explicitly, they are given by:
\begin{eqnarray}\label{eq: 4LoopC}
    \begin{split}
        &C_{t,u}^{(1)}:=\frac{1}{2} I_{\Theta_{}}^{12;34}+[I_{X^\prime}^{12;34}+I_{\Xi}^{12;34}+2 \, I_{M^\prime}^{12;34}+I_{K^{\prime\prime}}^{12;34}+ (1,2){\leftrightarrow}(3,4)\big]\\
        &\phantom{mfffy}+\big[\frac{1}{2} I_{P^\prime}^{12;34}+ I_{Y^\prime}^{12;34}+ \frac{1}{2}I_{S^\prime}^{12;34}+3\, \text{perms.}\big|_s \big]+\big[ I_{O}^{12;34}+7\, \text{perms.}\big|_s \big].\\
        &C_{t,u}^{(2)}:=\big[2I_{R}^{12;43}+ I_{\text{I}}^{12;43}+2I_{U^\prime}^{12;43}+(1,3){\leftrightarrow}(2,4)\big]+ (1,2){\leftrightarrow}(3,4),\\
             &C_{u,t}^{(2)}:=\big[2I_{R}^{12;34}+ I_{\text{I}}^{12;34}+2I_{U^\prime}^{12;34}+(1,3){\leftrightarrow}(2,4)\big]+ (1,2){\leftrightarrow}(3,4).
    \end{split}
\end{eqnarray}
Here, $\text{perms.}\big|_s$ denotes the subset of $P_4$ permutations that preserve  $s:=x_{1,2}^2 x_{3,4}^2$, explicitly: $\{1,2,3,4\}$,  $\{1,2,4,3\}$, $\{2,1,3,4\}$, $\{2,1,4,3\}$, $\{3,4,1,2\}$,  $\{4,3,1,2\}$, $\{3,4,2,1\}$, and $\{4,3,2,1\}$. Explicit expressions for the integrands can be found in Appendix~\ref{sec:four-loop integral basis}. Again, summing over all six chambers $\mathcal{G}^{(4)}(\mathbf{X},Y)=\frac{1}{2}\sum_{i,\pm}\;\omega_i^\pm \Omega_{r_i}^{(4)\pm}$, exactly reproduces the four-loop answer of~\cite{EDEN2012450}, expressed in terms of 32 conformal integrands from three distinct $f$-graph topologies.

Just as three loops, the block pattern in eq.~\eqref{eq: 4Loops} for each chamber reflects the ``diagonalization" of leading singularities. 
However, an important caveat arises: at four loops one begins to encounter elliptic functions, as can be seen from the cuts in eq.~\eqref{eq: Elliptic}. In practice, many local integrands carry ordinary (rational) leading singularities and elliptic cuts. Our results show that  the loop form of these local integrands can be reorganized such that each contributes either to the leading singularities or to the elliptic cuts, the latter given by $I_{G}^{ij;kl}$ in block $A_{\sigma_3}$. Thus, aside from this elliptic integrand, \textit{all remaining terms in each block of eq.~\eqref{eq: 4Loops} are pure functions! }\footnote{It is unsatisfactory that the numerators for the pure integrands $\Theta$ and $\Theta'$ are rather large. We pick one possible choice, as the solution is not unique. The Gram determinant condition may further simplify these expressions.}   

In the following subsection, we give explicit examples of how we engineer local integrands that diagonalize the leading singularities and elliptic cuts. Interestingly, different elliptic cuts appear in different chambers, so that the chamber dissection is also visible from the perspective of the elliptic cuts.


\subsection{Diagonalization of leading singularities and elliptic cut}
We now illustrate in detail how our local integrand representation in eqs.~\eqref{eq: 4LoopIntAB} and~\eqref{eq: 4LoopC} diagonalizes the leading singularities and elliptic cuts. 
At four loops, there are two types of elliptic cuts that can potentially contribute to the correlator. These are cuts where, after localizing the loop variables on 14-cut conditions, the resulting Jacobian factor consists of a square root of a degree-four polynomial in the remaining two variables, which cannot be rationalized. There are two such topologies, which we denote in the following as $\mu_s$- and $\nu_s$-cuts:
\begin{eqnarray}\label{eq: Elliptic}
&&\vcenter{\hbox{\scalebox{1}{
\begin{tikzpicture}[x=0.75pt,y=0.75pt,yscale=-1,xscale=1]


\draw  [line width=0.75] (-30,140) node [anchor=north west][inner sep=0.75pt]  [align=center] {$ \begin{tikzpicture}[x=0.75pt,y=0.75pt,yscale=-1,xscale=1]

\draw  [line width=0.75]  (135.27,102.72) -- (178.71,102.72) -- (178.71,146.16) -- (135.27,146.16) -- cycle ;
\draw  [line width=0.75]  (178.71,146.16) -- (222.16,146.16) -- (222.16,189.61) -- (178.71,189.61) -- cycle ;
\draw  [line width=0.75]   (123.3,90.75) -- (135.27,102.72) ;
\draw   [line width=0.75]  (222.16,189.61) -- (234.13,201.58) ;
\draw   [line width=0.75]  (178.8,85.8) -- (178.71,102.72) ;
\draw   [line width=0.75]  (135.27,189.61) -- (121.13,202.13) ;
\draw   [line width=0.75]  (178.71,189.61) -- (178.47,206.8) ;
\draw  [line width=0.75]  (178.71,102.72) -- (222.16,102.72) -- (222.16,146.16) -- (178.71,146.16) -- cycle ;
\draw  [line width=0.75]  (135.27,146.16) -- (178.71,146.16) -- (178.71,189.61) -- (135.27,189.61) -- cycle ;
\draw   [line width=0.75]  (236.29,90.19) -- (222.16,102.72) ;
\draw [line width=0.75] [color=red  ,draw opacity=1 ]   (157.15,98.16) -- (157.15,106.88) ;
\draw [line width=0.75] [color=red  ,draw opacity=1 ]   (157.75,140.96) -- (157.75,149.68) ;
\draw [line width=0.75] [color=red  ,draw opacity=1 ]   (161,140.96) -- (161,149.68) ;
\draw [line width=0.75]  [color=red  ,draw opacity=1 ]   (158.35,186.16) -- (158.35,194.88) ;
\draw [line width=0.75] [color=red  ,draw opacity=1 ]   (200.35,98.56) -- (200.35,107.28) ;
\draw [line width=0.75] [color=red  ,draw opacity=1 ]   (199.35,141.56) -- (199.35,150.28) ;
\draw [line width=0.75] [color=red  ,draw opacity=1 ]   (203.35,141.56) -- (203.35,150.28) ;
\draw [line width=0.75] [color=red  ,draw opacity=1 ]   (200.35,185.56) -- (200.35,194.28) ;
\draw [line width=0.75] [color=red  ,draw opacity=1 ]   (139.71,126.92) -- (130.99,126.92) ;
\draw [line width=0.75] [color=red  ,draw opacity=1 ]   (139.71,170.92) -- (130.99,170.92) ;
\draw [line width=0.75] [color=red  ,draw opacity=1 ]   (183.71,125.92) -- (174.99,125.92) ;
\draw [line width=0.75] [color=red  ,draw opacity=1 ]   (226.71,125.92) -- (217.99,125.92) ;
\draw [line width=0.75]  [color=red  ,draw opacity=1 ]   (182.71,170.92) -- (173.99,170.92) ;
\draw [line width=0.75]  [color=red  ,draw opacity=1 ]   (226.71,170.92) -- (217.99,170.92) ;

\draw (154.69,119) node [anchor=north west][inner sep=0.75pt]    {$a$};
\draw (195.58,119.45) node [anchor=north west][inner sep=0.75pt]    {$b$};
\draw (154.06,161) node [anchor=north west][inner sep=0.75pt]   [align=left] {$\displaystyle c$};
\draw (194.08,158) node [anchor=north west][inner sep=0.75pt]   [align=left] {$\displaystyle d$};
\draw (151.39,79.98) node [anchor=north west][inner sep=0.75pt]    {$1$};
\draw (195.07,80.72) node [anchor=north west][inner sep=0.75pt]    {$2$};
\draw (194.89,195.14) node [anchor=north west][inner sep=0.75pt]    {$1$};
\draw (115.14,139.37) node [anchor=north west][inner sep=0.75pt]    {$3$};
\draw (231.75,139.13) node [anchor=north west][inner sep=0.75pt]    {$4$};
\draw (152.4,193.72) node [anchor=north west][inner sep=0.75pt]    {$2$};

\end{tikzpicture}
$};

\draw  [line width=0.75] (210,136) node [anchor=north west][inner sep=0.75pt]  [align=center] {$ \begin{tikzpicture}[x=0.75pt,y=0.75pt,yscale=-1,xscale=1]

\draw   [line width=0.75] (135.27,102.72) -- (178.71,102.72) -- (178.71,146.16) -- (135.27,146.16) -- cycle ;
\draw   [line width=0.75] (178.71,146.16) -- (222.16,146.16) -- (222.16,189.61) -- (178.71,189.61) -- cycle ;
\draw    [line width=0.75] (123.3,90.75) -- (135.27,102.72) ;
\draw    [line width=0.75] (222.16,189.61) -- (234.13,201.58) ;
\draw   [line width=0.75]  (178.8,85.8) -- (178.71,102.72) ;
\draw    [line width=0.75] (135.27,189.61) -- (121.13,202.13) ;
\draw    [line width=0.75] (178.71,189.61) -- (178.47,206.8) ;
\draw   [line width=0.75] (178.71,102.72) -- (222.16,102.72) -- (222.16,146.16) -- (178.71,146.16) -- cycle ;
\draw   [line width=0.75] (135.27,146.16) -- (178.71,146.16) -- (178.71,189.61) -- (135.27,189.61) -- cycle ;
\draw    [line width=0.75] (236.29,90.19) -- (222.16,102.72) ;
\draw  [line width=0.75][color=red  ,draw opacity=1 ]   (157.15,98.16) -- (157.15,106.88) ;
\draw [line width=0.75] [color=red  ,draw opacity=1 ]   (157.75,140.96) -- (157.75,149.68) ;
\draw [line width=0.75] [color=red  ,draw opacity=1 ]   (161,140.96) -- (161,149.68) ;

\draw  [line width=0.75][color=red  ,draw opacity=1 ]   (158.35,186.16) -- (158.35,194.88) ;
\draw  [line width=0.75][color=red  ,draw opacity=1 ]   (200.35,98.56) -- (200.35,107.28) ;
\draw [line width=0.75] [color=red  ,draw opacity=1 ]   (199.35,141.56) -- (199.35,150.28) ;
\draw [line width=0.75] [color=red  ,draw opacity=1 ]   (203.35,141.56) -- (203.35,150.28) ;
\draw  [line width=0.75][color=red  ,draw opacity=1 ]   (200.35,185.56) -- (200.35,194.28) ;
\draw  [line width=0.75][color=red  ,draw opacity=1 ]   (139.71,126.92) -- (130.99,126.92) ;
\draw  [line width=0.75] [color=red  ,draw opacity=1 ]   (139.71,170.92) -- (130.99,170.92) ;
\draw  [line width=0.75] [color=red  ,draw opacity=1 ]   (183.71,125.92) -- (174.99,125.92) ;
\draw  [line width=0.75] [color=red  ,draw opacity=1 ]   (226.71,125.92) -- (217.99,125.92) ;
\draw  [line width=0.75] [color=red  ,draw opacity=1 ]   (182.71,170.92) -- (173.99,170.92) ;
\draw  [line width=0.75] [color=red  ,draw opacity=1 ]   (226.71,170.92) -- (217.99,170.92) ;
\draw    [line width=0.75] (239.08,146.17) -- (222.16,146.16) ;
\draw   [line width=0.75]  (135.27,146.16) -- (118.35,146.16) ;

\draw (154.69,119) node [anchor=north west][inner sep=0.75pt]    {$a$};
\draw (195.58,119.45) node [anchor=north west][inner sep=0.75pt]    {$b$};
\draw (154.06,163) node [anchor=north west][inner sep=0.75pt]   [align=left] {$\displaystyle c$};
\draw (194.08,162) node [anchor=north west][inner sep=0.75pt]   [align=left] {$\displaystyle d$};
\draw (111.39,113.98) node [anchor=north west][inner sep=0.75pt]    {$1$};
\draw (150.74,76.05) node [anchor=north west][inner sep=0.75pt]    {$2$};
\draw (234.56,164.14) node [anchor=north west][inner sep=0.75pt]    {$1$};
\draw (233.8,118.71) node [anchor=north west][inner sep=0.75pt]    {$3$};
\draw (194.75,77.46) node [anchor=north west][inner sep=0.75pt]    {$4$};
\draw (196.4,199.38) node [anchor=north west][inner sep=0.75pt]    {$2$};
\draw (113.47,161.04) node [anchor=north west][inner sep=0.75pt]    {$3$};
\draw (152.08,198.46) node [anchor=north west][inner sep=0.75pt]    {$4$};

\end{tikzpicture}
$};

\draw (-90,200) node [anchor=north west][inner sep=0.75pt]    {$\mu_s\text{-cut}:$};

\draw (155,200) node [anchor=north west][inner sep=0.75pt]    {$\nu_s\text{-cut}:$};

\end{tikzpicture}

}
}}
\end{eqnarray}
There are also their permuted images $\mu_{t,u}$ and $\nu_{t,u}$. Choosing the following loop parameterization for the $\mu_s$-cut, 
\begin{equation}\label{rq: 4LoopPara}
    \begin{split}
         \ (A,B) &= \left(Z_{1,1} + x_1 Z_{2,1} - w_1 Z_{2,3},\quad y_1 Z_{2,1} + Z_{1,3} + z_1 Z_{2,3} \right),\\ 
         \ (C,D) &= \left(Z_{1,2} + x_2 Z_{2,2} - w_2 Z_{2,4},\quad y_2 Z_{2,2} + Z_{1,4} + z_2 Z_{2,4} \right),\\
         \ (E,F) &= \left(Z_{1,2} + x_3 Z_{2,2} - w_3 Z_{2,3},\quad y_3 Z_{2,2} + Z_{1,3} + z_3 Z_{2,3} \right),\\
         \ (G,H) &= \left(Z_{1,1} + x_4 Z_{2,1} - w_4 Z_{2,4},\quad y_4 Z_{2,1} + Z_{1,4} + z_4 Z_{2,4} \right),
    \end{split}
\end{equation}
after localizing 14 loop variables, the Jacobian for the $\mu_s$-cut takes the form:
\begin{equation}\label{eq: elliptic J1}
    \begin{split}
    &z_1^{-2}\Big( \big({-}c_1 (c_1 {+} 1) ( c_2{-}1) {+} (c_1 {-} c_2)  ( c_1{-}1) ( 
      c_2{-}1) X + ( c_1{-}1) c_2 ( c_2+1) X^2\big)^2\\
      &\phantom{hghhhhhhhf} {-}4 X (c_1 {+} c_1^2 c_2 (X{-}1 ) {-} c_2 X) (c_1 {+} c_1 c_2^2 (X {-}1 ) {-} 
    c_2 X)\Big)^{{-}\frac{1}{2}} \, ,
    \end{split}
\end{equation}
where we have $X=x_4 / z_1$. After changing variables to $X$ and taking the residue at $z_1 = 0$, we are left with an integral over a degree-four elliptic curve.  Writing the curve as $Y^2=f(X)$, where $f(X)$ is a degree-four polynomial, we can map it to a degree-three form through a modular transformation that sends $(X, Y) = (1, 0) \rightarrow (\infty, \infty)$. After a suitable rescaling of $X$, the curve takes the covariant form:\footnote{Starting from $Y^2 = f(X)$, we perform a modular transformation $(X,Y) \mapsto \left(1 {+} \frac{1}{ X}, \frac{Y}{X^2} \right)$, followed by a shift $X \rightarrow X {+} \frac{c_1^2 (c_2{-}1)^2{-}2 c_1 (c_2{+}1)^2{+}c_2 (c_2{+}10){+}1}{12 (c_1{-}c_2)}$. After rescaling $(X,Y) \rightarrow (a^2 X, a^3 Y)$ with $a = \frac{\sqrt{-(c_1 - 1)(c_2 - 1)}}{(c_1 - c_2)^2}$, and shifting $X \rightarrow X {-} \frac{(c_1{-}1)(c_2{-}1)^2 {+} 4 c_1 c_2}{12 (c_1{-}1)(c_2{-}1)}$, the curve takes the covariant form shown in \eqref{eq: elliptic J1 corvariant}.}
\begin{equation}\label{eq: elliptic J1 corvariant}
    Y^2 = \frac{(s X - t)(4 u X^2 - s X + t)}{s u}=\frac{(U X-V)(4 X^2-U X+V)}{U}\,.
\end{equation}
The equivalence between the above and eq.~\eqref{eq: elliptic J1} can also be verified by comparing the corresponding $j$-invariants, which yields
\begin{equation}
    \frac{\left(U^4-16 U^2 V+16 V^2\right)^3}{U^2 V^4 (U^2-16 V)}\,.
\end{equation}

For the $\nu_s$-cut, adopting the parametrization 
\begin{equation}\label{rq: 4LoopPara2}
    \begin{split}
         \ (A,B) &= \left(Z_{1,1} + x_1 Z_{2,1} - w_1 Z_{2,2},\quad y_1 Z_{2,1} + Z_{1,2} + z_1 Z_{2,2} \right),\\ 
         \ (C,D) &= \left(Z_{1,3} + x_2 Z_{2,3} - w_2 Z_{2,4},\quad y_2 Z_{2,3} + Z_{1,4} + z_2 Z_{2,4} \right),\\
         \ (E,F) &= \left(Z_{1,3} + x_3 Z_{2,3} - w_3 Z_{2,4},\quad y_3 Z_{2,3} + Z_{1,4} + z_3 Z_{2,4} \right),\\
         \ (G,H) &= \left(Z_{1,1} + x_4 Z_{2,1} - w_4 Z_{2,2},\quad y_4 Z_{2,1} + Z_{1,2} + z_4 Z_{2,2} \right),
    \end{split}
\end{equation}
its Jacobian factor is
\begin{equation}\label{eq: elliptic J2}
    \begin{split}
       \Big(4 (c_1{-}1) ( c_2{-}1) ( X{-}1) (c_1 {-} c_2 X) ( c_1{-}1 {+} X {-} c_2 X) (c_1 {-} 
   c_1 c_2 {-} c_2 X {+} c_1 c_2 X)\Big)^{{-}\frac{1}{2}}
    \end{split}
\end{equation}
Upon changing variables via $X=z_2 / x_3$, and performing the same procedure. 
 The covariant expression of the elliptic curve takes the form:
 \begin{equation}
     \begin{split}
         Y^2=\frac{4X (sX+u)(s X+t)}{s^2}=\frac{4 X (U X+1)(U X+V)}{U^2}
     \end{split}
 \end{equation}
with $j$-invariant 
\begin{equation}
    \frac{256 (1-V-V^2)^3}{(-1+V)^2 V^2}.
\end{equation}

As we will see in the next subsection, in our geometric construction only the $\mu$-cut is present, while the $\nu$-cut is nowhere accessible within $\mathbb{T}_4$. Thus to diagonalize the elliptic cut, we need to ensure that the local integrands must have vanishing $\nu$-cut.

Note that since the elliptic cuts are non-maximal, local integrands that contribute to such cuts can at the same time contribute to maximal cuts with rational leading singularities. For example, the $\mu$-cut receives contributions from integrands with the topology of $G$, $R$, $\Theta$, and $W$:

\begin{eqnarray}\label{eq: mu-cut and integral}
&&\vcenter{\hbox{\scalebox{1}{
\begin{tikzpicture}[x=0.75pt,y=0.75pt,yscale=-1,xscale=1]

\draw (-50,150) node [anchor=north west][inner sep=0.75pt]     {$\text{$\mu_s$-cut}:$};





\draw  [line width=0.75] (15,50) node [anchor=north west][inner sep=0.75pt]  [align=center] {$ \hbox{\scalebox{0.75}{\begin{tikzpicture}[x=0.75pt,y=0.75pt,yscale=-1,xscale=1]

\draw  [line width=0.75]  (135.27,102.72) -- (178.71,102.72) -- (178.71,146.16) -- (135.27,146.16) -- cycle ;
\draw  [line width=0.75]  (178.71,146.16) -- (222.16,146.16) -- (222.16,189.61) -- (178.71,189.61) -- cycle ;
\draw  [line width=0.75]   (123.3,90.75) -- (135.27,102.72) ;
\draw   [line width=0.75]  (222.16,189.61) -- (234.13,201.58) ;
\draw   [line width=0.75]  (178.8,85.8) -- (178.71,102.72) ;
\draw   [line width=0.75]  (135.27,189.61) -- (121.13,202.13) ;
\draw   [line width=0.75]  (178.71,189.61) -- (178.47,206.8) ;
\draw  [line width=0.75]  (178.71,102.72) -- (222.16,102.72) -- (222.16,146.16) -- (178.71,146.16) -- cycle ;
\draw  [line width=0.75]  (135.27,146.16) -- (178.71,146.16) -- (178.71,189.61) -- (135.27,189.61) -- cycle ;
\draw   [line width=0.75]  (236.29,90.19) -- (222.16,102.72) ;
\draw (154.69,119) node [anchor=north west][inner sep=0.75pt]    {$a$};
\draw (195.58,119.45) node [anchor=north west][inner sep=0.75pt]    {$b$};
\draw (154.06,161) node [anchor=north west][inner sep=0.75pt]   [align=left] {$\displaystyle c$};
\draw (194.08,158) node [anchor=north west][inner sep=0.75pt]   [align=left] {$\displaystyle d$};
\draw (151.39,79.98) node [anchor=north west][inner sep=0.75pt]    {$1$};
\draw (195.07,80.72) node [anchor=north west][inner sep=0.75pt]    {$2$};
\draw (194.89,195.14) node [anchor=north west][inner sep=0.75pt]    {$1$};
\draw (115.14,139.37) node [anchor=north west][inner sep=0.75pt]    {$3$};
\draw (231.75,139.13) node [anchor=north west][inner sep=0.75pt]    {$4$};
\draw (152.4,193.72) node [anchor=north west][inner sep=0.75pt]    {$2$};

\draw (160,210) node [anchor=north west][inner sep=0.75pt]   [font=\large]  {$I_G^{12;34}$};

\end{tikzpicture}
}}
$};

\draw  [line width=0.75] (190,50) node [anchor=north west][inner sep=0.75pt]  [align=center] {$ \hbox{\scalebox{0.75}{ \begin{tikzpicture}[x=0.75pt,y=0.75pt,yscale=-1,xscale=1]

\draw  [line width=0.75] (246.47,91.92) -- (289.91,91.92) -- (289.91,135.36) -- (246.47,135.36) -- cycle ;
\draw  [line width=0.75] (289.91,135.36) -- (333.36,135.36) -- (333.36,178.81) -- (289.91,178.81) -- cycle ;
\draw   [line width=0.75] (234.5,79.95) -- (246.47,91.92) ;
\draw  [line width=0.75]  (246.47,155.94) -- (268.33,178.81) ;
\draw  [line width=0.75]  (289.91,178.81) -- (268.33,178.81) ;
\draw   [line width=0.75] (246.47,135.36) -- (246.47,155.94) ;
\draw  [line width=0.75]  (311.49,91.92) -- (289.91,91.92) ;
\draw   [line width=0.75] (311.49,91.92) -- (333.36,114.78) ;
\draw   [line width=0.75] (333.36,114.78) -- (333.36,135.36) ;
\draw   [line width=0.75] (333.36,178.81) -- (345.33,190.78) ;
\draw  [line width=0.75]  (317.5,76.5) -- (311.49,91.92) ;
\draw   [line width=0.75] (350,108) -- (333.36,114.78) ;
\draw  [line width=0.75]  (246.47,155.94) -- (230.5,161) ;
\draw   [line width=0.75] (268.33,178.81) -- (262.33,194.23) ;
\draw   [line width=0.75] (246.47,135.36) -- (231.29,135.36) ;

\draw (295.89,106.24) node [anchor=north west][inner sep=0.75pt]    {$b$};
\draw (266.78,150.65) node [anchor=north west][inner sep=0.75pt]    {$c$};
\draw (265.26,107) node [anchor=north west][inner sep=0.75pt]   [align=left] {$\displaystyle a$};
\draw (306.28,150.41) node [anchor=north west][inner sep=0.75pt]   [align=left] {$\displaystyle d$};
\draw (227.59,101.18) node [anchor=north west][inner sep=0.75pt]    {$1$};
\draw (328.6,87.58) node [anchor=north west][inner sep=0.75pt]    {$2$};
\draw (341.6,135.58) node [anchor=north west][inner sep=0.75pt]    {$4$};
\draw (294.09,183.68) node [anchor=north west][inner sep=0.75pt]    {$1$};
\draw (269.17,68.41) node [anchor=north west][inner sep=0.75pt]    {$3$};
\draw (224.61,139.99) node [anchor=north west][inner sep=0.75pt]    {$3$};
\draw (240.6,165.58) node [anchor=north west][inner sep=0.75pt]    {$2$};

\draw (365,130) node [anchor=north west][inner sep=0.75pt]  [font=\large]   {$\text{and $1\leftrightarrow 2,\ 3\leftrightarrow 4$}$};
\draw (365,150) node [anchor=north west][inner sep=0.75pt]  [font=\large]   {$\phantom{myk} (1,3)\leftrightarrow (2,4)$};

\draw (280,200) node [anchor=north west][inner sep=0.75pt]   [font=\large]  {$I_R^{14;23}$};

\end{tikzpicture}}}
$};

\draw  [line width=0.75] (10,174) node [anchor=north west][inner sep=0.75pt]  [align=center] {$ \hbox{\scalebox{0.75}{\begin{tikzpicture}[x=0.75pt,y=0.75pt,yscale=-1,xscale=1]

\draw   [line width=0.75] (313.58,128.41) -- (299.79,153.5) -- (272.21,153.5) -- (258.42,128.41) -- (272.21,103.31) -- (299.79,103.31) -- cycle ;
\draw   [line width=0.75] (258.42,128.41) -- (244.62,153.5) -- (217.04,153.5) -- (203.25,128.41) -- (217.04,103.31) -- (244.62,103.31) -- cycle ;
\draw    [line width=0.75] (258.42,78.22) -- (272.21,103.31) ;
\draw    [line width=0.75] (258.42,78.22) -- (244.62,103.31) ;
\draw    [line width=0.75] (187.25,128.41) -- (203.25,128.41) ;
\draw   [line width=0.75]  (313.58,128.41) -- (329.58,128.41) ;
\draw    [line width=0.75] (299.79,103.31) -- (309.72,93.63) ;
\draw    [line width=0.75] (207.11,163.18) -- (217.04,153.5) ;
\draw    [line width=0.75] (217.04,103.31) -- (201.59,93.63) ;
\draw    [line width=0.75] (315.24,163.18) -- (299.79,153.5) ;
\draw    [line width=0.75] (258.42,64.31) -- (258.42,78.22) ;
\draw    [line width=0.75] (258.42,178.59) -- (258.42,192.5) ;
\draw    [line width=0.75] (244.62,153.5) -- (258.42,178.59) ;
\draw    [line width=0.75] (272.21,153.5) -- (258.42,178.59) ;

\draw (225.09,122) node [anchor=north west][inner sep=0.75pt]    {$c$};
\draw (282.46,119.43) node [anchor=north west][inner sep=0.75pt]    {$b$};
\draw (252.52,97) node [anchor=north west][inner sep=0.75pt]   [align=left] {$\displaystyle a$};
\draw (251.93,145) node [anchor=north west][inner sep=0.75pt]   [align=left] {$\displaystyle d$};
\draw (222.14,75.91) node [anchor=north west][inner sep=0.75pt]    {$1$};
\draw (280.84,163.83) node [anchor=north west][inner sep=0.75pt]    {$1$};
\draw (315.82,138.68) node [anchor=north west][inner sep=0.75pt]    {$2$};
\draw (192.26,102.9) node [anchor=north west][inner sep=0.75pt]    {$2$};
\draw (191.11,137.51) node [anchor=north west][inner sep=0.75pt]    {$3$};
\draw (224.74,163.31) node [anchor=north west][inner sep=0.75pt]    {$4$};
\draw (277.89,78.26) node [anchor=north west][inner sep=0.75pt]    {$3$};
\draw (312.95,102.9) node [anchor=north west][inner sep=0.75pt]    {$4$};

\draw (312.95,102.9) node [anchor=north west][inner sep=0.75pt]    {$4$};

\draw (345.95,120) node [anchor=north west][inner sep=0.75pt]  [font=\large]   {$\text{and $1\leftrightarrow 2$}$};

\draw (245,200) node [anchor=north west][inner sep=0.75pt]   [font=\large]  {$I_W^{12;34}$};

\end{tikzpicture}
}}
$};

\draw  [line width=0.75] (200,178) node [anchor=north west][inner sep=0.75pt]  [align=center] {$ \hbox{\scalebox{0.75}{\begin{tikzpicture}[x=0.75pt,y=0.75pt,yscale=-1,xscale=1]

\draw   [line width=0.75] (303.27,175.42) -- (267.9,175.42) -- (242.9,150.41) -- (242.9,115.05) -- (267.9,90.05) -- (303.27,90.05) -- (328.27,115.05) -- (328.27,150.41) -- cycle ;
\draw   [line width=0.75]  (243.07,132.735) -- (328.11,132.735) ;
\draw   [line width=0.75]  (285.585,89.56) -- (285.585,175.14) ;
\draw   [line width=0.75]  (343.19,115.05) -- (328.27,115.05) ;
\draw  [line width=0.75]   (343.19,150.41) -- (328.27,150.41) ;
\draw   [line width=0.75]  (242.9,115.05) -- (227.98,115.05) ;
\draw  [line width=0.75]   (242.9,150.41) -- (227.98,150.41) ;
\draw   [line width=0.75]  (303.27,90.05) -- (303.27,75.13) ;
\draw   [line width=0.75]  (267.9,90.05) -- (267.9,75.13) ;
\draw   [line width=0.75]  (303.27,190.33) -- (303.27,175.42) ;
\draw   [line width=0.75]  (267.9,190.33) -- (267.9,175.42) ;

\draw (263.37,107.41) node [anchor=north west][inner sep=0.75pt]   [align=left] {$\displaystyle a$};
\draw (298.08,107.41) node [anchor=north west][inner sep=0.75pt]   [align=left] {$\displaystyle c$};
\draw (263.37,144.59) node [anchor=north west][inner sep=0.75pt]   [align=left] {$\displaystyle b$};
\draw (298.08,144.59) node [anchor=north west][inner sep=0.75pt]   [align=left] {$\displaystyle d$};

\draw (245,85) node [anchor=north west][inner sep=0.75pt]   [align=left] {$\displaystyle 1$};
\draw (317,85) node [anchor=north west][inner sep=0.75pt]   [align=left] {$\displaystyle 2$};
\draw (245,165) node [anchor=north west][inner sep=0.75pt]   [align=left] {$\displaystyle 3$};
\draw (317,165) node [anchor=north west][inner sep=0.75pt]   [align=left] {$\displaystyle 4$};

\draw (281,73) node [anchor=north west][inner sep=0.75pt]   [align=left] {$\displaystyle 3$};
\draw (281,180) node [anchor=north west][inner sep=0.75pt]   [align=left] {$\displaystyle 2$};

\draw (281,180) node [anchor=north west][inner sep=0.75pt]   [align=left] {$\displaystyle 2$};

\draw (230,125.05) node [anchor=north west][inner sep=0.75pt]   [align=left] {$\displaystyle 4$};
\draw (332,125.05) node [anchor=north west][inner sep=0.75pt]   [align=left] {$\displaystyle 1$};

\draw (360.95,125) node [anchor=north west][inner sep=0.75pt]  [font=\large]   {$\text{and $3\leftrightarrow 4$}$};

\draw (245,200) node [anchor=north west][inner sep=0.75pt]   [font=\large]  {$I_{\Theta}^{14;23}/ I_{\Theta^\prime}^{14;23}$};

\end{tikzpicture}}}
$};



\end{tikzpicture}

}
}}
\end{eqnarray}
Thus \textit{diagonalization} at four loops requires us to disentangle elliptic and rational leading singularities. Remarkably, this can be done. The explicit numerators for these integrands are engineered so that, apart from $I_G^{12;34}$—which has no other leading singularities—the remaining integrands are set up to vanish on the elliptic cut, while their leading singularities are proportional to the same function.

As an example, consider the conformal integrand $I_W^{12;34}$, which is one among the 32 $f$-graph representations, albeit with a distinct numerator:  
\begin{eqnarray}\label{eq: FourLoopNewNum}
&&\vcenter{\hbox{\scalebox{1}{
\begin{tikzpicture}[x=0.75pt,y=0.75pt,yscale=-1,xscale=1]


\draw  [line width=0.75] (0,130) node [anchor=north west][inner sep=0.75pt]  [align=center] {$ 
\begin{tikzpicture}[x=0.75pt,y=0.75pt,yscale=-1,xscale=1]

\draw   [line width=0.75] (285.83,112.91) -- (272.04,138) -- (244.46,138) -- (230.67,112.91) -- (244.46,87.81) -- (272.04,87.81) -- cycle ;
\draw   [line width=0.75] (230.67,112.91) -- (216.87,138) -- (189.29,138) -- (175.5,112.91) -- (189.29,87.81) -- (216.87,87.81) -- cycle ;
\draw    [line width=0.75] (230.67,62.72) -- (244.46,87.81) ;
\draw    [line width=0.75] (230.67,62.72) -- (216.87,87.81) ;
\draw     [line width=0.75] (159.5,112.91) -- (175.5,112.91) ;
\draw    [line width=0.75] (285.83,112.91) -- (301.83,112.91) ;
\draw     [line width=0.75] (272.04,87.81) -- (281.97,78.13) ;
\draw    [line width=0.75] (179.36,147.68) -- (189.29,138) ;
\draw    [line width=0.75] (189.29,87.81) -- (173.84,78.13) ;
\draw    [line width=0.75] (287.49,147.68) -- (272.04,138) ;
\draw     [line width=0.75](230.67,48.81) -- (230.67,62.72) ;
\draw    [line width=0.75] (230.67,163.09) -- (230.67,177) ;
\draw    [line width=0.75] (216.87,138) -- (230.67,163.09) ;
\draw    [line width=0.75] (244.46,138) -- (230.67,163.09) ;

\draw (197.34,104.56) node [anchor=north west][inner sep=0.75pt]    {$c$};
\draw (254.71,103.93) node [anchor=north west][inner sep=0.75pt]    {$b$};
\draw (224.77,82) node [anchor=north west][inner sep=0.75pt]   [align=left] {$\displaystyle a$};
\draw (224.18,130) node [anchor=north west][inner sep=0.75pt]   [align=left] {$\displaystyle d$};
\draw (194.39,60.41) node [anchor=north west][inner sep=0.75pt]    {$1$};
\draw (253.09,148.33) node [anchor=north west][inner sep=0.75pt]    {$1$};
\draw (288.07,123.18) node [anchor=north west][inner sep=0.75pt]    {$2$};
\draw (164.51,87.4) node [anchor=north west][inner sep=0.75pt]    {$2$};
\draw (163.36,122.01) node [anchor=north west][inner sep=0.75pt]    {$3$};
\draw (198.99,146.06) node [anchor=north west][inner sep=0.75pt]    {$4$};
\draw (250.14,62.76) node [anchor=north west][inner sep=0.75pt]    {$3$};
\draw (285.2,87.4) node [anchor=north west][inner sep=0.75pt]    {$4$};

\end{tikzpicture}
$};

\draw  (-60,180) node [anchor=north west][inner sep=0.75pt]    {$\begin{split}
    W^{12;34}: 
\end{split}$};

\draw  (165,170) node [anchor=north west][inner sep=0.75pt]    {$\begin{split}
     x_{1,3}^2 x_{1,4}^2 x_{b,c}^4 \rightarrow &(x_{b,c}^2 x_{1,3}^2 {-}  x_{b,3}^2 x_{c,1}^2 {-}  x_{b,1}^2 x_{c,3}^2)\\
     \times  &(x_{b,c}^2 x_{1,4}^2 {-}  x_{b,4}^2 x_{c,1}^2 {-}  x_{b,1}^2 x_{c,4}^2).
\end{split}$};

\end{tikzpicture}

}
}}
\end{eqnarray}
The new numerator ensures that the integrand vanishes on the elliptic cut, while for the remaining cuts
\begin{eqnarray}
&&\vcenter{\hbox{\scalebox{1}{
\begin{tikzpicture}[x=0.75pt,y=0.75pt,yscale=-1,xscale=1]


\draw  [line width=0.75]   (90,104) node [anchor=north west][inner sep=0.75pt]  [align=center] {$ \hbox{\scalebox{0.75}{\begin{tikzpicture}[x=0.75pt,y=0.75pt,yscale=-1,xscale=1]

\draw   [line width=0.75] (313.58,128.41) -- (299.79,153.5) -- (272.21,153.5) -- (258.42,128.41) -- (272.21,103.31) -- (299.79,103.31) -- cycle ;
\draw   [line width=0.75] (258.42,128.41) -- (244.62,153.5) -- (217.04,153.5) -- (203.25,128.41) -- (217.04,103.31) -- (244.62,103.31) -- cycle ;
\draw    [line width=0.75] (258.42,78.22) -- (272.21,103.31) ;
\draw    [line width=0.75] (258.42,78.22) -- (244.62,103.31) ;
\draw    [line width=0.75] (187.25,128.41) -- (203.25,128.41) ;
\draw   [line width=0.75]  (313.58,128.41) -- (329.58,128.41) ;
\draw    [line width=0.75] (299.79,103.31) -- (309.72,93.63) ;
\draw    [line width=0.75] (207.11,163.18) -- (217.04,153.5) ;
\draw    [line width=0.75] (217.04,103.31) -- (201.59,93.63) ;
\draw    [line width=0.75] (315.24,163.18) -- (299.79,153.5) ;
\draw    [line width=0.75] (258.42,64.31) -- (258.42,78.22) ;
\draw    [line width=0.75] (258.42,178.59) -- (258.42,192.5) ;
\draw    [line width=0.75] (244.62,153.5) -- (258.42,178.59) ;
\draw    [line width=0.75] (272.21,153.5) -- (258.42,178.59) ;
\draw  [line width=0.75][color=red  ,draw opacity=1 ]   (229.1,106.78) -- (229.1,98.06) ;
\draw  [line width=0.75][color=red  ,draw opacity=1 ]   (230.6,158.28) -- (230.6,149.56) ;
\draw  [line width=0.75][color=red  ,draw opacity=1 ]   (287.1,108.03) -- (287.1,99.31) ;
\draw  [line width=0.75][color=red  ,draw opacity=1 ]   (288.85,158.03) -- (288.85,149.31) ;
\draw  [line width=0.75][color=red  ,draw opacity=1 ]   (205.57,143.1) -- (213.13,138.74) ;
\draw  [line width=0.75][color=red  ,draw opacity=1 ]   (302.57,138.99) -- (310.13,143.35) ;
\draw  [line width=0.75][color=red  ,draw opacity=1 ]   (310.88,113.74) -- (303.32,118.1) ;
\draw  [line width=0.75][color=red  ,draw opacity=1 ]   (205.82,113.49) -- (213.38,117.85) ;
\draw  [line width=0.75][color=red  ,draw opacity=1 ]   (246.82,116.35) -- (253.88,111.75) ;
\draw  [line width=0.75][color=red  ,draw opacity=1 ]   (261.82,142.85) -- (268.88,138.25) ;
\draw  [line width=0.75][color=red  ,draw opacity=1 ]   (262.82,93.6) -- (269.88,89) ;
\draw  [line width=0.75][color=red  ,draw opacity=1 ]   (255.63,93.25) -- (249.13,88.5) ;
\draw  [line width=0.75][color=red  ,draw opacity=1 ]   (270.13,116) -- (263.63,111.25) ;
\draw  [line width=0.75][color=red  ,draw opacity=1 ]   (254.13,144) -- (247.63,139.25) ;
\draw  [line width=0.75][color=red  ,draw opacity=1 ]   (247.82,167.6) -- (254.88,163) ;
\draw  [line width=0.75][color=red  ,draw opacity=1 ]   (270.13,167.5) -- (263.63,162.75) ;

\draw (225.09,122) node [anchor=north west][inner sep=0.75pt]    {$a$};
\draw (282.46,119.43) node [anchor=north west][inner sep=0.75pt]    {$b$};
\draw (252.52,97) node [anchor=north west][inner sep=0.75pt]   [align=left] {$\displaystyle c$};
\draw (251.93,145) node [anchor=north west][inner sep=0.75pt]   [align=left] {$\displaystyle d$};
\draw (222.14,75.91) node [anchor=north west][inner sep=0.75pt]    {$1$};
\draw (280.84,163.83) node [anchor=north west][inner sep=0.75pt]    {$1$};
\draw (315.82,138.68) node [anchor=north west][inner sep=0.75pt]    {$2$};
\draw (192.26,102.9) node [anchor=north west][inner sep=0.75pt]    {$2$};
\draw (191.11,137.51) node [anchor=north west][inner sep=0.75pt]    {$3$};
\draw (224.74,163.31) node [anchor=north west][inner sep=0.75pt]    {$4$};
\draw (277.89,78.26) node [anchor=north west][inner sep=0.75pt]    {$3$};
\draw (312.95,102.9) node [anchor=north west][inner sep=0.75pt]    {$4$};

\end{tikzpicture}
}}
$};

\draw  [line width=0.75] (240,104) node [anchor=north west][inner sep=0.75pt]  [align=center] {$ \hbox{\scalebox{0.75}{\begin{tikzpicture}[x=0.75pt,y=0.75pt,yscale=-1,xscale=1]

\draw   [line width=0.75] (313.58,128.41) -- (299.79,153.5) -- (272.21,153.5) -- (258.42,128.41) -- (272.21,103.31) -- (299.79,103.31) -- cycle ;
\draw   [line width=0.75] (258.42,128.41) -- (244.62,153.5) -- (217.04,153.5) -- (203.25,128.41) -- (217.04,103.31) -- (244.62,103.31) -- cycle ;
\draw    [line width=0.75] (258.42,78.22) -- (272.21,103.31) ;
\draw    [line width=0.75] (258.42,78.22) -- (244.62,103.31) ;
\draw    [line width=0.75] (187.25,128.41) -- (203.25,128.41) ;
\draw   [line width=0.75]  (313.58,128.41) -- (329.58,128.41) ;
\draw    [line width=0.75] (299.79,103.31) -- (309.72,93.63) ;
\draw    [line width=0.75] (207.11,163.18) -- (217.04,153.5) ;
\draw    [line width=0.75] (217.04,103.31) -- (201.59,93.63) ;
\draw    [line width=0.75] (315.24,163.18) -- (299.79,153.5) ;
\draw    [line width=0.75] (258.42,64.31) -- (258.42,78.22) ;
\draw    [line width=0.75] (258.42,178.59) -- (258.42,192.5) ;
\draw    [line width=0.75] (244.62,153.5) -- (258.42,178.59) ;
\draw    [line width=0.75] (272.21,153.5) -- (258.42,178.59) ;
\draw  [line width=0.75][color=red  ,draw opacity=1 ]   (230.6,158.28) -- (230.6,149.56) ;
\draw  [line width=0.75][color=red  ,draw opacity=1 ]   (288.85,158.03) -- (288.85,149.31) ;
\draw  [line width=0.75][color=red  ,draw opacity=1 ]   (205.57,143.1) -- (213.13,138.74) ;
\draw  [line width=0.75][color=red  ,draw opacity=1 ]   (302.57,138.99) -- (310.13,143.35) ;
\draw  [line width=0.75][color=red  ,draw opacity=1 ]   (205.82,113.49) -- (213.38,117.85) ;
\draw  [line width=0.75][color=red  ,draw opacity=1 ]   (246.82,116.35) -- (253.88,111.75) ;
\draw  [line width=0.75][color=red  ,draw opacity=1 ]   (261.82,142.85) -- (268.88,138.25) ;
\draw  [line width=0.75][color=red  ,draw opacity=1 ]   (262.82,93.6) -- (269.88,89) ;
\draw  [line width=0.75][color=red  ,draw opacity=1 ]   (255.63,93.25) -- (249.13,88.5) ;
\draw  [line width=0.75][color=red  ,draw opacity=1 ]   (270.13,116) -- (263.63,111.25) ;
\draw  [line width=0.75][color=red  ,draw opacity=1 ]   (254.13,144) -- (247.63,139.25) ;
\draw  [line width=0.75][color=red  ,draw opacity=1 ]   (247.82,167.6) -- (254.88,163) ;
\draw  [line width=0.75][color=red  ,draw opacity=1 ]   (270.13,167.5) -- (263.63,162.75) ;
\draw [line width=0.75][color=red  ,draw opacity=1 ]   (263.57,145.85) -- (270.63,141.25) ;
\draw [line width=0.75][color=red  ,draw opacity=1 ]   (252.63,146.75) -- (246.13,142) ;
\draw [color=red  ,draw opacity=1 ]   (268.88,118.5) -- (262.38,113.75) ;

\draw (225.09,122) node [anchor=north west][inner sep=0.75pt]    {$a$};
\draw (282.46,119.43) node [anchor=north west][inner sep=0.75pt]    {$b$};
\draw (252.52,97) node [anchor=north west][inner sep=0.75pt]   [align=left] {$\displaystyle c$};
\draw (251.93,145) node [anchor=north west][inner sep=0.75pt]   [align=left] {$\displaystyle d$};
\draw (222.14,75.91) node [anchor=north west][inner sep=0.75pt]    {$1$};
\draw (280.84,163.83) node [anchor=north west][inner sep=0.75pt]    {$1$};
\draw (315.82,138.68) node [anchor=north west][inner sep=0.75pt]    {$2$};
\draw (192.26,102.9) node [anchor=north west][inner sep=0.75pt]    {$2$};
\draw (191.11,137.51) node [anchor=north west][inner sep=0.75pt]    {$3$};
\draw (224.74,163.31) node [anchor=north west][inner sep=0.75pt]    {$4$};
\draw (277.89,78.26) node [anchor=north west][inner sep=0.75pt]    {$3$};
\draw (312.95,102.9) node [anchor=north west][inner sep=0.75pt]    {$4$};

\end{tikzpicture}
}}
$};

\draw  [line width=0.75] (90,210) node [anchor=north west][inner sep=0.75pt]  [align=center] {$ \hbox{\scalebox{0.75}{\begin{tikzpicture}[x=0.75pt,y=0.75pt,yscale=-1,xscale=1]

\draw   [line width=0.75] (313.58,128.41) -- (299.79,153.5) -- (272.21,153.5) -- (258.42,128.41) -- (272.21,103.31) -- (299.79,103.31) -- cycle ;
\draw   [line width=0.75] (258.42,128.41) -- (244.62,153.5) -- (217.04,153.5) -- (203.25,128.41) -- (217.04,103.31) -- (244.62,103.31) -- cycle ;
\draw    [line width=0.75] (258.42,78.22) -- (272.21,103.31) ;
\draw    [line width=0.75] (258.42,78.22) -- (244.62,103.31) ;
\draw    [line width=0.75] (187.25,128.41) -- (203.25,128.41) ;
\draw   [line width=0.75]  (313.58,128.41) -- (329.58,128.41) ;
\draw    [line width=0.75] (299.79,103.31) -- (309.72,93.63) ;
\draw    [line width=0.75] (207.11,163.18) -- (217.04,153.5) ;
\draw    [line width=0.75] (217.04,103.31) -- (201.59,93.63) ;
\draw    [line width=0.75] (315.24,163.18) -- (299.79,153.5) ;
\draw    [line width=0.75] (258.42,64.31) -- (258.42,78.22) ;
\draw    [line width=0.75] (258.42,178.59) -- (258.42,192.5) ;
\draw    [line width=0.75] (244.62,153.5) -- (258.42,178.59) ;
\draw    [line width=0.75] (272.21,153.5) -- (258.42,178.59) ;
\draw  [line width=0.75][color=red  ,draw opacity=1 ]   (229.1,106.78) -- (229.1,98.06) ;
\draw  [line width=0.75][color=red  ,draw opacity=1 ]   (230.6,158.28) -- (230.6,149.56) ;
\draw  [line width=0.75][color=red  ,draw opacity=1 ]   (287.1,108.03) -- (287.1,99.31) ;
\draw  [line width=0.75][color=red  ,draw opacity=1 ]   (302.57,138.99) -- (310.13,143.35) ;
\draw  [line width=0.75][color=red  ,draw opacity=1 ]   (310.88,113.74) -- (303.32,118.1) ;
\draw  [line width=0.75][color=red  ,draw opacity=1 ]   (205.82,113.49) -- (213.38,117.85) ;
\draw  [line width=0.75][color=red  ,draw opacity=1 ]   (246.82,116.35) -- (253.88,111.75) ;
\draw  [line width=0.75][color=red  ,draw opacity=1 ]   (261.82,142.85) -- (268.88,138.25) ;
\draw  [line width=0.75][color=red  ,draw opacity=1 ]   (262.82,93.6) -- (269.88,89) ;
\draw  [line width=0.75][color=red  ,draw opacity=1 ]   (255.63,93.25) -- (249.13,88.5) ;
\draw  [line width=0.75][color=red  ,draw opacity=1 ]   (270.13,116) -- (263.63,111.25) ;
\draw  [line width=0.75][color=red  ,draw opacity=1 ]   (254.13,144) -- (247.63,139.25) ;
\draw  [line width=0.75][color=red  ,draw opacity=1 ]   (247.82,167.6) -- (254.88,163) ;
\draw  [line width=0.75][color=red  ,draw opacity=1 ]   (270.13,167.5) -- (263.63,162.75) ;
\draw [line width=0.75][color=red  ,draw opacity=1 ]   (252.63,146.75) -- (246.13,142) ;
\draw [color=red  ,draw opacity=1 ]   (268.88,118.5) -- (262.38,113.75) ;

\draw (225.09,122) node [anchor=north west][inner sep=0.75pt]    {$a$};
\draw (282.46,119.43) node [anchor=north west][inner sep=0.75pt]    {$b$};
\draw (252.52,97) node [anchor=north west][inner sep=0.75pt]   [align=left] {$\displaystyle c$};
\draw (251.93,145) node [anchor=north west][inner sep=0.75pt]   [align=left] {$\displaystyle d$};
\draw (222.14,75.91) node [anchor=north west][inner sep=0.75pt]    {$1$};
\draw (280.84,163.83) node [anchor=north west][inner sep=0.75pt]    {$1$};
\draw (315.82,138.68) node [anchor=north west][inner sep=0.75pt]    {$2$};
\draw (192.26,102.9) node [anchor=north west][inner sep=0.75pt]    {$2$};
\draw (191.11,137.51) node [anchor=north west][inner sep=0.75pt]    {$3$};
\draw (224.74,163.31) node [anchor=north west][inner sep=0.75pt]    {$4$};
\draw (277.89,78.26) node [anchor=north west][inner sep=0.75pt]    {$3$};
\draw (312.95,102.9) node [anchor=north west][inner sep=0.75pt]    {$4$};

\end{tikzpicture}
}}
$};

\draw  [line width=0.75] (240,210) node [anchor=north west][inner sep=0.75pt]  [align=center] {$ \hbox{\scalebox{0.75}{\begin{tikzpicture}[x=0.75pt,y=0.75pt,yscale=-1,xscale=1]

\draw   [line width=0.75] (313.58,128.41) -- (299.79,153.5) -- (272.21,153.5) -- (258.42,128.41) -- (272.21,103.31) -- (299.79,103.31) -- cycle ;
\draw   [line width=0.75] (258.42,128.41) -- (244.62,153.5) -- (217.04,153.5) -- (203.25,128.41) -- (217.04,103.31) -- (244.62,103.31) -- cycle ;
\draw    [line width=0.75] (258.42,78.22) -- (272.21,103.31) ;
\draw    [line width=0.75] (258.42,78.22) -- (244.62,103.31) ;
\draw    [line width=0.75] (187.25,128.41) -- (203.25,128.41) ;
\draw   [line width=0.75]  (313.58,128.41) -- (329.58,128.41) ;
\draw    [line width=0.75] (299.79,103.31) -- (309.72,93.63) ;
\draw    [line width=0.75] (207.11,163.18) -- (217.04,153.5) ;
\draw    [line width=0.75] (217.04,103.31) -- (201.59,93.63) ;
\draw    [line width=0.75] (315.24,163.18) -- (299.79,153.5) ;
\draw    [line width=0.75] (258.42,64.31) -- (258.42,78.22) ;
\draw    [line width=0.75] (258.42,178.59) -- (258.42,192.5) ;
\draw    [line width=0.75] (244.62,153.5) -- (258.42,178.59) ;
\draw    [line width=0.75] (272.21,153.5) -- (258.42,178.59) ;
\draw  [line width=0.75][color=red  ,draw opacity=1 ]   (229.1,106.78) -- (229.1,98.06) ;
\draw  [line width=0.75][color=red  ,draw opacity=1 ]   (287.1,108.03) -- (287.1,99.31) ;
\draw  [line width=0.75][color=red  ,draw opacity=1 ]   (205.57,143.1) -- (213.13,138.74) ;
\draw  [line width=0.75][color=red  ,draw opacity=1 ]   (302.57,138.99) -- (310.13,143.35) ;
\draw  [line width=0.75][color=red  ,draw opacity=1 ]   (310.88,113.74) -- (303.32,118.1) ;
\draw  [line width=0.75][color=red  ,draw opacity=1 ]   (205.82,113.49) -- (213.38,117.85) ;
\draw  [line width=0.75][color=red  ,draw opacity=1 ]   (246.82,116.35) -- (253.88,111.75) ;
\draw  [line width=0.75][color=red  ,draw opacity=1 ]   (261.82,142.85) -- (268.88,138.25) ;
\draw  [line width=0.75][color=red  ,draw opacity=1 ]   (262.82,93.6) -- (269.88,89) ;
\draw  [line width=0.75][color=red  ,draw opacity=1 ]   (255.63,93.25) -- (249.13,88.5) ;
\draw  [line width=0.75][color=red  ,draw opacity=1 ]   (270.13,116) -- (263.63,111.25) ;
\draw  [line width=0.75][color=red  ,draw opacity=1 ]   (254.13,144) -- (247.63,139.25) ;
\draw [line width=0.75][color=red  ,draw opacity=1 ]   (252.63,146.75) -- (246.13,142) ;
\draw [color=red  ,draw opacity=1 ]   (268.88,118.5) -- (262.38,113.75) ;

\draw  [line width=0.75][color=red  ,draw opacity=1 ]   (247.82,167.6) -- (254.88,163) ;
\draw  [line width=0.75][color=red  ,draw opacity=1 ]   (270.13,167.5) -- (263.63,162.75) ;

\draw (225.09,122) node [anchor=north west][inner sep=0.75pt]    {$a$};
\draw (282.46,119.43) node [anchor=north west][inner sep=0.75pt]    {$b$};
\draw (252.52,97) node [anchor=north west][inner sep=0.75pt]   [align=left] {$\displaystyle c$};
\draw (251.93,145) node [anchor=north west][inner sep=0.75pt]   [align=left] {$\displaystyle d$};
\draw (222.14,75.91) node [anchor=north west][inner sep=0.75pt]    {$1$};
\draw (280.84,163.83) node [anchor=north west][inner sep=0.75pt]    {$1$};
\draw (315.82,138.68) node [anchor=north west][inner sep=0.75pt]    {$2$};
\draw (192.26,102.9) node [anchor=north west][inner sep=0.75pt]    {$2$};
\draw (191.11,137.51) node [anchor=north west][inner sep=0.75pt]    {$3$};
\draw (224.74,163.31) node [anchor=north west][inner sep=0.75pt]    {$4$};
\draw (277.89,78.26) node [anchor=north west][inner sep=0.75pt]    {$3$};
\draw (312.95,102.9) node [anchor=north west][inner sep=0.75pt]    {$4$};

\end{tikzpicture}
}}
$};



\draw (200,170) node [anchor=north west][inner sep=0.75pt]    {$,$};
\draw (350,170) node [anchor=north west][inner sep=0.75pt]    {$,$};
\draw (200,280) node [anchor=north west][inner sep=0.75pt]    {$,$};
\draw (350,280) node [anchor=north west][inner sep=0.75pt]    {$.$};

\end{tikzpicture}

}
}}
\end{eqnarray}
all leading singularities vanish except the first, yielding the leading singularity $1/\Delta^2$. This is to be compared with the original $f$-graph numerators, whose leading singularities associated with these cuts (listed in order) are given by 
\begin{equation} \frac{1}{\Delta^2}, \quad \frac{t}{\Delta s (s-t)}, \quad \frac{1}{\Delta (s-t)}, \quad \text{and} \quad \frac{1}{\Delta (s-u)}. 
\end{equation}
Thus we see that the numerator in eq.~\eqref{eq: FourLoopNewNum} not only removes the elliptic $\mu_s$-cut contribution but also leads to a pure integrand. On the other hand, integrands $I_{\Xi}^{12;34}$, $I_{\Theta}^{12;34}$ and  $I_{\Theta^\prime}^{12;34}$ can potentially contribute to the $\nu_s$-cut:
\begin{eqnarray}
&&\vcenter{\hbox{\scalebox{1}{
\begin{tikzpicture}[x=0.75pt,y=0.75pt,yscale=-1,xscale=1]

\draw (-50,220) node [anchor=north west][inner sep=0.75pt]     {$\text{$\nu_s$-cut}:$};





\draw  [line width=0.75] (10,174) node [anchor=north west][inner sep=0.75pt]  [align=center] {$ \hbox{\scalebox{0.75}{\begin{tikzpicture}[x=0.75pt,y=0.75pt,yscale=-1,xscale=1]

\draw [line width=0.75]  (246.47,91.92) -- (289.91,91.92) -- (289.91,135.36) -- (246.47,135.36) -- cycle ;
\draw  [line width=0.75] (289.91,135.36) -- (333.36,135.36) -- (333.36,178.81) -- (289.91,178.81) -- cycle ;
\draw  [line width=0.75]  (234.5,79.95) -- (246.47,91.92) ;
\draw  [line width=0.75]  (246.47,155.94) -- (268.33,178.81) ;
\draw  [line width=0.75]  (289.91,178.81) -- (268.33,178.81) ;
\draw  [line width=0.75]  (246.47,135.36) -- (246.47,155.94) ;
\draw  [line width=0.75]  (311.49,91.92) -- (289.91,91.92) ;
\draw  [line width=0.75]  (311.49,91.92) -- (333.36,114.78) ;
\draw  [line width=0.75]  (333.36,114.78) -- (333.36,135.36) ;
\draw   [line width=0.75] (333.36,178.81) -- (345.33,190.78) ;
\draw   [line width=0.75] (317.5,76.5) -- (311.49,91.92) ;
\draw   [line width=0.75] (350,108) -- (333.36,114.78) ;
\draw  [line width=0.75]  (246.47,155.94) -- (230.5,161) ;
\draw   [line width=0.75] (268.33,178.81) -- (262.33,194.23) ;
\draw  [line width=0.75]  (348.53,135.36) -- (333.36,135.36) ;
\draw   [line width=0.75] (246.47,135.36) -- (231.29,135.36) ;

\draw (265.89,104.24) node [anchor=north west][inner sep=0.75pt]    {$a$};
\draw (306.78,108.65) node [anchor=north west][inner sep=0.75pt]    {$b$};
\draw (265.26,146.7) node [anchor=north west][inner sep=0.75pt]   [align=left] {$\displaystyle c$};
\draw (305.28,145.41) node [anchor=north west][inner sep=0.75pt]   [align=left] {$\displaystyle d$};
\draw (227.59,101.18) node [anchor=north west][inner sep=0.75pt]    {$1$};
\draw (273.6,72.58) node [anchor=north west][inner sep=0.75pt]    {$2$};
\draw (341.6,150.58) node [anchor=north west][inner sep=0.75pt]    {$2$};
\draw (294.09,183.68) node [anchor=north west][inner sep=0.75pt]    {$1$};
\draw (325.67,83.91) node [anchor=north west][inner sep=0.75pt]    {$3$};
\draw (243.17,168.41) node [anchor=north west][inner sep=0.75pt]    {$3$};
\draw (224.61,139.99) node [anchor=north west][inner sep=0.75pt]    {$4$};
\draw (343.11,113.99) node [anchor=north west][inner sep=0.75pt]    {$4$};

\draw (280,210) node [anchor=north west][inner sep=0.75pt]   [font=\large]  {$I_{\Xi}^{12;34}$};

\end{tikzpicture}
}}
$};

\draw  [line width=0.75] (200,178) node [anchor=north west][inner sep=0.75pt]  [align=center] {$ \hbox{\scalebox{0.75}{\begin{tikzpicture}[x=0.75pt,y=0.75pt,yscale=-1,xscale=1]

\draw   [line width=0.75] (303.27,175.42) -- (267.9,175.42) -- (242.9,150.41) -- (242.9,115.05) -- (267.9,90.05) -- (303.27,90.05) -- (328.27,115.05) -- (328.27,150.41) -- cycle ;
\draw   [line width=0.75]  (243.07,132.735) -- (328.11,132.735) ;
\draw   [line width=0.75]  (285.585,89.56) -- (285.585,175.14) ;
\draw   [line width=0.75]  (343.19,115.05) -- (328.27,115.05) ;
\draw  [line width=0.75]   (343.19,150.41) -- (328.27,150.41) ;
\draw   [line width=0.75]  (242.9,115.05) -- (227.98,115.05) ;
\draw  [line width=0.75]   (242.9,150.41) -- (227.98,150.41) ;
\draw   [line width=0.75]  (303.27,90.05) -- (303.27,75.13) ;
\draw   [line width=0.75]  (267.9,90.05) -- (267.9,75.13) ;
\draw   [line width=0.75]  (303.27,190.33) -- (303.27,175.42) ;
\draw   [line width=0.75]  (267.9,190.33) -- (267.9,175.42) ;

\draw (263.37,107.41) node [anchor=north west][inner sep=0.75pt]   [align=left] {$\displaystyle a$};
\draw (298.08,107.41) node [anchor=north west][inner sep=0.75pt]   [align=left] {$\displaystyle b$};
\draw (263.37,144.59) node [anchor=north west][inner sep=0.75pt]   [align=left] {$\displaystyle c$};
\draw (298.08,144.59) node [anchor=north west][inner sep=0.75pt]   [align=left] {$\displaystyle d$};

\draw (245,85) node [anchor=north west][inner sep=0.75pt]   [align=left] {$\displaystyle 1$};
\draw (317,85) node [anchor=north west][inner sep=0.75pt]   [align=left] {$\displaystyle 4$};
\draw (245,165) node [anchor=north west][inner sep=0.75pt]   [align=left] {$\displaystyle 3$};
\draw (317,165) node [anchor=north west][inner sep=0.75pt]   [align=left] {$\displaystyle 2$};

\draw (281,73) node [anchor=north west][inner sep=0.75pt]   [align=left] {$\displaystyle 3$};
\draw (281,180) node [anchor=north west][inner sep=0.75pt]   [align=left] {$\displaystyle 4$};

\draw (281,180) node [anchor=north west][inner sep=0.75pt]   [align=left] {$\displaystyle 4$};

\draw (230,125.05) node [anchor=north west][inner sep=0.75pt]   [align=left] {$\displaystyle 2$};
\draw (332,125.05) node [anchor=north west][inner sep=0.75pt]   [align=left] {$\displaystyle 1$};


\draw (245,200) node [anchor=north west][inner sep=0.75pt]   [font=\large]  {$I_{\Theta}^{12;34}/I_{\Theta^\prime}^{12;34}$};

\end{tikzpicture}}}
$};



\end{tikzpicture}

}
}}
\end{eqnarray}
However, their numerators are engineered such that they vanish on the $\nu_s$-cut.

The remaining integrands in eqs.~\eqref{eq: 4LoopIntAB} and~\eqref{eq: 4LoopC} are dressed with numerators such that each contains only one leading singularity. For example, consider the numerator for the following topology:
\begin{eqnarray}\label{eq: TopologyS1234}
&&\vcenter{\hbox{\scalebox{1}{
\begin{tikzpicture}[x=0.75pt,y=0.75pt,yscale=-1,xscale=1]


\draw  [line width=0.75] (100,100) node [anchor=north west][inner sep=0.75pt]  [align=center] {$ \begin{tikzpicture}[x=0.75pt,y=0.75pt,yscale=-1,xscale=1]

\draw  [line width=0.75] (307.23,186.05) -- (295.18,206.91) -- (271.09,206.91) -- (259.05,186.05) -- (271.09,165.18) -- (295.18,165.18) -- cycle ;
\draw  [line width=0.75] (238.15,173.89) -- (250.2,153.02) -- (271.06,165.07) -- (259.02,185.94) -- cycle ;
\draw  [line width=0.75] (316.05,153.13) -- (328.1,174) -- (307.23,186.05) -- (295.18,165.18) -- cycle ;
\draw  [line width=0.75] (271.09,206.91) -- (295.19,206.91) -- (295.19,231.01) -- (271.09,231.01) -- cycle ;
\draw  [line width=0.75]  (295.19,231.01) -- (304.39,240.2) ;
\draw   [line width=0.75] (271.09,231.01) -- (261.9,240.2) ;
\draw  [line width=0.75]  (319.42,140.56) -- (316.05,153.13) ;
\draw  [line width=0.75]  (340.66,177.36) -- (328.1,174) ;
\draw   [line width=0.75] (225.58,177.26) -- (238.15,173.89) ;
\draw  [line width=0.75]  (246.83,140.46) -- (250.2,153.02) ;

\draw (173,186.05) node [anchor=north west][inner sep=0.75pt]    {$S^{12,34} :$};
\draw (227.02,146.99) node [anchor=north west][inner sep=0.75pt]   [align=left] {$\displaystyle 2$};
\draw (280.13,137.05) node [anchor=north west][inner sep=0.75pt]   [align=left] {$\displaystyle 1$};
\draw (278.76,180.51) node [anchor=north west][inner sep=0.75pt]   [align=left] {$\displaystyle a$};
\draw (249.85,160.78) node [anchor=north west][inner sep=0.75pt]   [align=left] {$\displaystyle b$};
\draw (307.46,163.78) node [anchor=north west][inner sep=0.75pt]   [align=left] {$\displaystyle c$};
\draw (278.7,210.76) node [anchor=north west][inner sep=0.75pt]   [align=left] {$\displaystyle d$};
\draw (314.03,201.64) node [anchor=north west][inner sep=0.75pt]   [align=left] {$\displaystyle 4$};
\draw (242.67,196.91) node [anchor=north west][inner sep=0.75pt]   [align=left] {$\displaystyle 3$};
\draw (333.88,146.04) node [anchor=north west][inner sep=0.75pt]   [align=left] {$\displaystyle 2$};
\draw (276.89,236.75) node [anchor=north west][inner sep=0.75pt]   [align=left] {$\displaystyle 2$};

\end{tikzpicture}$};



\end{tikzpicture}

}
}}
\end{eqnarray}
From this topology and its permuted image, we  define a set of seven integrands with distinct numerators:
\begin{eqnarray}
    &&I_S^{12;34}:  n_{S}^{12;34} {=} (\textcolor{blue}{x_{1,3}^{2} x_{1,4}^{2} x_{3,4}^{2}  x_{a,2}^{4}}{+} 3\, x_{1,4}^2   x_{2,3}^2 x_{a,1}^2 (x_{2,4}^2 x_{a,3}^2 {-}x_{3,4}^2 x_{a,2}^2))/6 {+}\text{perms}(1,3,4), \nonumber\\
  &&I_{S^{\prime}}^{12;34},\, I_{S^{\prime}}^{12;43},\, I_{S^{\prime}}^{32;41}:\,   {n}_{S}^{\prime\,ij;kl} = x_{a,l}^2 (x_{i,j}^{2}  x_{a,k}^{2} {-}x_{j,k}^{2}  x_{a,i}^{2} {-} x_{i,k}^{2}  x_{a,j}^{2}),\\
  &&I_{S^{\prime\prime}}^{12;34},\, I_{S^{\prime\prime}}^{32;14},\, I_{S^{\prime\prime}}^{42;13}:\,  {n}_{S}^{\prime\prime\,ij;kl} = x_{i,j}^{2} x_{a,k}^{2}  x_{a,l}^{2}\nonumber.
\end{eqnarray}
where the action of $\text{perms}(1,3,4)$ is as follows:
after dividing  $n_{S}^{12;34}$ by the propagators in eq.~\eqref{eq: TopologyS1234}, we apply the permutation among labels $(1,3,4)$. 
The topology in eq.~\eqref{eq: TopologyS1234} contains the following maximal cuts: 
\begin{eqnarray}
&&\vcenter{\hbox{\scalebox{1}{
\begin{tikzpicture}[x=0.75pt,y=0.75pt,yscale=-1,xscale=1]


\draw  [line width=0.75]   (70,126) node [anchor=north west][inner sep=0.75pt]  [align=center] {$ \hbox{\scalebox{1.0}{\begin{tikzpicture}[x=0.75pt,y=0.75pt,yscale=-1,xscale=1]

\draw  [line width=0.75]  (205.9,56.91) -- (229.99,56.91) -- (229.99,81.01) -- (205.9,81.01) -- cycle ;
\draw  [line width=0.75]  (254.09,56.91) -- (278.19,56.91) -- (278.19,81.01) -- (254.09,81.01) -- cycle ;
\draw  [line width=0.75]  (229.99,81.01) -- (254.09,81.01) -- (254.09,105.11) -- (229.99,105.11) -- cycle ;
\draw   [line width=0.75]  (254.09,105.11) -- (263.29,114.3) ;
\draw   [line width=0.75]  (229.99,105.11) -- (220.8,114.3) ;
\draw  [line width=0.75]  (229.99,56.91) -- (254.09,56.91) -- (254.09,81.01) -- (229.99,81.01) -- cycle ;
\draw   [line width=0.75]  (196.7,47.72) -- (205.9,56.91) ;
\draw  [line width=0.75]   (278.19,81.01) -- (287.38,90.21) ;
\draw   [line width=0.75]  (205.9,81.01) -- (196.7,90.21) ;
\draw    [line width=0.75]  (287.38,47.72) -- (278.19,56.91) ;

\draw  [line width=0.75]  [color=red  ,draw opacity=1 ]   (217.5,61.25) -- (217.5,50.64) ;
\draw  [line width=0.75]  [color=red  ,draw opacity=1 ]   (217.5,85.25) -- (217.5,74.64) ;
\draw  [line width=0.75]  [color=red  ,draw opacity=1 ]   (241.5,110.25) -- (241.5,99.64) ;
\draw  [line width=0.75]  [color=red  ,draw opacity=1 ]   (241.5,62.25) -- (241.5,51.64) ;
\draw  [line width=0.75] [color=red  ,draw opacity=1 ]   (264.5,63.25) -- (264.5,52.64) ;
\draw  [line width=0.75]  [color=red  ,draw opacity=1 ]   (265.5,87.25) -- (265.5,76.64) ;
\draw  [line width=0.75]  [color=red  ,draw opacity=1 ]   (239.5,87.25) -- (239.5,76.64) ;
\draw  [line width=0.75]  [color=red  ,draw opacity=1 ]   (242.5,87.25) -- (242.5,76.64) ;
\draw  [line width=0.75]  [color=red  ,draw opacity=1 ]   (273.19,69.94) -- (283.81,69.94) ;
\draw  [line width=0.75]  [color=red  ,draw opacity=1 ]   (249.64,92.94) -- (260.25,92.94) ;
\draw  [line width=0.75] [color=red  ,draw opacity=1 ]   (224.64,92.94) -- (235.25,92.94) ;
\draw  [line width=0.75]  [color=red  ,draw opacity=1 ]   (199.64,69.94) -- (210.25,69.94) ;
\draw  [line width=0.75] [color=red  ,draw opacity=1 ]   (225.64,67.94) -- (236.25,67.94) ;
\draw  [line width=0.75] [color=red  ,draw opacity=1 ]   (249.64,67.94) -- (260.25,67.94) ;
\draw  [line width=0.75] [color=red  ,draw opacity=1 ]   (249.64,70.94) -- (260.25,70.94) ;
\draw  [line width=0.75] [color=red  ,draw opacity=1 ]   (225.64,70.94) -- (236.25,70.94) ;

\draw (189,63) node [anchor=north west][inner sep=0.75pt]   [align=left] {$\displaystyle 2$};
\draw (237.83,37) node [anchor=north west][inner sep=0.75pt]   [align=left] {$\displaystyle 1$};
\draw (236.86,64) node [anchor=north west][inner sep=0.75pt]   [align=left] {$\displaystyle a$};
\draw (212.35,62) node [anchor=north west][inner sep=0.75pt]   [align=left] {$\displaystyle b$};
\draw (262.76,64) node [anchor=north west][inner sep=0.75pt]   [align=left] {$\displaystyle c$};
\draw (237.2,86) node [anchor=north west][inner sep=0.75pt]   [align=left] {$\displaystyle d$};
\draw (262.53,85.34) node [anchor=north west][inner sep=0.75pt]   [align=left] {4};
\draw (211.97,86) node [anchor=north west][inner sep=0.75pt]   [align=left] {$\displaystyle 3$};
\draw (284.78,63) node [anchor=north west][inner sep=0.75pt]   [align=left] {$\displaystyle 2$};
\draw (237,111.25) node [anchor=north west][inner sep=0.75pt]   [align=left] {$\displaystyle 2$};

\end{tikzpicture}
}}
$};

\draw  [line width=0.75] (240,104) node [anchor=north west][inner sep=0.75pt]  [align=center] {$ \hbox{\scalebox{1.0}{\begin{tikzpicture}[x=0.75pt,y=0.75pt,yscale=-1,xscale=1]

\draw  [line width=0.75] (363.9,60.11) -- (387.99,60.11) -- (387.99,84.21) -- (363.9,84.21) -- cycle ;
\draw  [line width=0.75] (387.99,36.02) -- (412.09,36.02) -- (412.09,60.11) -- (387.99,60.11) -- cycle ;
\draw  [line width=0.75] (387.99,84.21) -- (412.09,84.21) -- (412.09,108.31) -- (387.99,108.31) -- cycle ;
\draw   [line width=0.75] (412.09,108.31) -- (421.29,117.5) ;
\draw   [line width=0.75] (387.99,108.31) -- (378.8,117.5) ;
\draw  [line width=0.75] (387.99,60.11) -- (412.09,60.11) -- (412.09,84.21) -- (387.99,84.21) -- cycle ;
\draw   [line width=0.75] (354.7,50.92) -- (363.9,60.11) ;
\draw   [line width=0.75] (378.8,26.82) -- (387.99,36.02) ;
\draw   [line width=0.75] (363.9,84.21) -- (354.7,93.41) ;
\draw   [line width=0.75] (421.29,26.82) -- (412.09,36.02) ;
\draw [line width=0.75] [color=red  ,draw opacity=1 ]   (375.25,65.25) -- (375.25,54.64) ;
\draw [line width=0.75][color=red  ,draw opacity=1 ]   (375.25,89.25) -- (375.25,78.64) ;
\draw [line width=0.75][color=red  ,draw opacity=1 ]   (397.25,90.25) -- (397.25,79.64) ;
\draw [line width=0.75][color=red  ,draw opacity=1 ]   (401.25,90.25) -- (401.25,79.64) ;
\draw [line width=0.75][color=red  ,draw opacity=1 ]   (401.25,66.25) -- (401.25,55.64) ;
\draw [line width=0.75][color=red  ,draw opacity=1 ]   (397.25,66.25) -- (397.25,55.64) ;
\draw [line width=0.75][color=red  ,draw opacity=1 ]   (400.25,42.25) -- (400.25,31.64) ;
\draw [line width=0.75][color=red  ,draw opacity=1 ]   (401.25,114.25) -- (401.25,103.64) ;
\draw [line width=0.75][color=red  ,draw opacity=1 ]   (382.94,47.69) -- (393.56,47.69) ;
\draw [line width=0.75][color=red  ,draw opacity=1 ]   (406.94,48.69) -- (417.56,48.69) ;
\draw [line width=0.75][color=red  ,draw opacity=1 ]   (407.94,72.69) -- (418.56,72.69) ;
\draw [line width=0.75][color=red  ,draw opacity=1 ]   (407.94,96.69) -- (418.56,96.69) ;
\draw [line width=0.75] [color=red  ,draw opacity=1 ]   (383.94,96.69) -- (394.56,96.69) ;
\draw [line width=0.75][color=red  ,draw opacity=1 ]   (383.94,73.69) -- (394.56,73.69) ;
\draw [line width=0.75][color=red  ,draw opacity=1 ]   (383.94,70.69) -- (394.56,70.69) ;
\draw [line width=0.75] [color=red  ,draw opacity=1 ]   (358.94,72.69) -- (369.56,72.69) ;

\draw (348.72,66) node [anchor=north west][inner sep=0.75pt]   [align=left] {$\displaystyle 2$};
\draw (369.83,38.35) node [anchor=north west][inner sep=0.75pt]   [align=left] {$\displaystyle 1$};
\draw (394.86,68) node [anchor=north west][inner sep=0.75pt]   [align=left] {$\displaystyle a$};
\draw (370.35,63) node [anchor=north west][inner sep=0.75pt]   [align=left] {$\displaystyle b$};
\draw (394.76,37.6) node [anchor=north west][inner sep=0.75pt]   [align=left] {$\displaystyle c$};
\draw (395.2,88.46) node [anchor=north west][inner sep=0.75pt]   [align=left] {$\displaystyle d$};
\draw (418.53,66) node [anchor=north west][inner sep=0.75pt]   [align=left] {4};
\draw (369.97,85.81) node [anchor=north west][inner sep=0.75pt]   [align=left] {$\displaystyle 3$};
\draw (396,19) node [anchor=north west][inner sep=0.75pt]   [align=left] {$\displaystyle 2$};
\draw (396,114.45) node [anchor=north west][inner sep=0.75pt]   [align=left] {$\displaystyle 2$};

\end{tikzpicture}
}}
$};

\draw  [line width=0.75] (390,104) node [anchor=north west][inner sep=0.75pt]  [align=center] {$ \hbox{\scalebox{1.0}{
\begin{tikzpicture}[x=0.75pt,y=0.75pt,yscale=-1,xscale=1]

\draw    [line width=0.75] (518.49,65.77) -- (542.59,65.77) -- (542.59,89.87) -- (518.49,89.87) -- cycle ;
\draw   [line width=0.75]   (494.39,41.67) -- (518.49,41.67) -- (518.49,65.77) -- (494.39,65.77) -- cycle ;
\draw   [line width=0.75]  (494.39,89.87) -- (518.49,89.87) -- (518.49,113.96) -- (494.39,113.96) -- cycle ;
\draw    [line width=0.75]  (518.49,113.96) -- (527.69,123.16) ;
\draw    [line width=0.75]  (494.39,113.96) -- (485.2,123.16) ;
\draw   [line width=0.75]  (494.39,65.77) -- (518.49,65.77) -- (518.49,89.87) -- (494.39,89.87) -- cycle ;
\draw    [line width=0.75]  (542.59,89.87) -- (551.78,99.06) ;
\draw    [line width=0.75]  (485.2,32.48) -- (494.39,41.67) ;
\draw    [line width=0.75]  (551.78,56.57) -- (542.59,65.77) ;
\draw    [line width=0.75]  (527.69,32.48) -- (518.49,41.67) ;
\draw  [line width=0.75] [color=red  ,draw opacity=1 ]   (489.69,54.44) -- (500.31,54.44) ;
\draw  [line width=0.75] [color=red  ,draw opacity=1 ]   (489.69,77.44) -- (500.31,77.44) ;
\draw  [line width=0.75] [color=red  ,draw opacity=1 ]   (489.69,103.44) -- (500.31,103.44) ;
\draw  [line width=0.75] [color=red  ,draw opacity=1 ]   (514.69,103.44) -- (525.31,103.44) ;
\draw  [line width=0.75] [color=red  ,draw opacity=1 ]   (538.69,78.44) -- (549.31,78.44) ;
\draw  [line width=0.75] [color=red  ,draw opacity=1 ]   (514.89,80.44) -- (525.5,80.44) ;
\draw  [line width=0.75] [color=red  ,draw opacity=1 ]   (514.89,77.44) -- (525.5,77.44) ;
\draw  [line width=0.75] [color=red  ,draw opacity=1 ]   (513.89,54.44) -- (524.5,54.44) ;
\draw  [line width=0.75] [color=red  ,draw opacity=1 ]   (530.19,60.14) -- (530.19,70.75) ;
\draw  [line width=0.75] [color=red  ,draw opacity=1 ]   (531.19,85.14) -- (531.19,95.75) ;
\draw  [line width=0.75] [color=red  ,draw opacity=1 ]   (504.19,85.14) -- (504.19,95.75) ;
\draw  [line width=0.75] [color=red  ,draw opacity=1 ]   (507.19,85.14) -- (507.19,95.75) ;
\draw  [line width=0.75] [color=red  ,draw opacity=1 ]   (504.19,61.14) -- (504.19,71.75) ;
\draw  [line width=0.75] [color=red  ,draw opacity=1 ]   (507.19,61.14) -- (507.19,71.75) ;
\draw  [line width=0.75] [color=red  ,draw opacity=1 ]   (506.19,36.14) -- (506.19,46.75) ;
\draw  [line width=0.75] [color=red  ,draw opacity=1 ]   (507.19,110.14) -- (507.19,120.75) ;

\draw (547.52,68.75) node [anchor=north west][inner sep=0.75pt]   [align=left] {$\displaystyle 2$};
\draw (527.43,42.41) node [anchor=north west][inner sep=0.75pt]   [align=left] {$\displaystyle 1$};
\draw (501.26,72) node [anchor=north west][inner sep=0.75pt]   [align=left] {$\displaystyle a$};
\draw (501.15,44.53) node [anchor=north west][inner sep=0.75pt]   [align=left] {$\displaystyle b$};
\draw (526.76,72) node [anchor=north west][inner sep=0.75pt]   [align=left] {$\displaystyle c$};
\draw (501.6,94.11) node [anchor=north west][inner sep=0.75pt]   [align=left] {$\displaystyle d$};
\draw (523.73,94.6) node [anchor=north west][inner sep=0.75pt]   [align=left] {4};
\draw (471.57,65.87) node [anchor=north west][inner sep=0.75pt]   [align=left] {$\displaystyle 3$};
\draw (501,24) node [anchor=north west][inner sep=0.75pt]   [align=left] {$\displaystyle 2$};
\draw (501,120.11) node [anchor=north west][inner sep=0.75pt]   [align=left] {$\displaystyle 2$};

\end{tikzpicture}
}}
$};

\draw  [line width=0.75] (70,230) node [anchor=north west][inner sep=0.75pt]  [align=center] {$ \hbox{\scalebox{1.0}{\begin{tikzpicture}[x=0.75pt,y=0.75pt,yscale=-1,xscale=1]

\draw   [line width=0.75] (268.15,210.95) -- (245.91,220.85) -- (229.62,202.76) -- (241.79,181.68) -- (265.6,186.74) -- cycle ;
\draw    [line width=0.75](208.85,190.64) -- (220.9,169.77) -- (241.76,181.82) -- (229.72,202.69) -- cycle ;
\draw   [line width=0.75] (289.55,184.05) -- (292.24,208) -- (268.29,210.68) -- (265.6,186.74) -- cycle ;
\draw    [line width=0.75] (245.91,220.85) -- (267.98,211.17) -- (277.65,233.24) -- (255.59,242.92) -- cycle ;
\draw    [line width=0.75] (277.65,233.24) -- (286.85,242.44) ;
\draw    [line width=0.75] (255.59,242.92) -- (254.95,254.5) ;
\draw   [line width=0.75]  (299.7,177.5) -- (289.55,184.05) ;
\draw    [line width=0.75] (304.8,211.36) -- (292.24,208) ;
\draw    [line width=0.75] (196.28,194.01) -- (208.85,190.64) ;
\draw    [line width=0.75] (217.53,157.21) -- (220.9,169.77) ;
\draw  [line width=0.75] [color=red  ,draw opacity=1 ]   (232.36,169.72) -- (227.03,178.89) ;
\draw  [line width=0.75][color=red  ,draw opacity=1 ]   (220.36,191.72) -- (215.03,200.89) ;
\draw  [line width=0.75] [color=red  ,draw opacity=1 ]   (219.28,182.72) -- (210.11,177.39) ;
\draw  [line width=0.75] [color=red  ,draw opacity=1 ]   (240.28,197.72) -- (231.11,192.39) ;

\draw  [line width=0.75] [color=red  ,draw opacity=1 ]   (278.05,179.79) -- (279.18,190.33) ;
\draw  [line width=0.75] [color=red  ,draw opacity=1 ]   (280.55,203.29) -- (281.68,213.83) ;
\draw  [line width=0.75] [color=red  ,draw opacity=1 ]   (296.39,195.99) -- (285.85,197.13) ;
\draw  [line width=0.75] [color=red  ,draw opacity=1 ]   (272.39,196.24) -- (261.85,197.38) ;
\draw  [line width=0.75] [color=red  ,draw opacity=1 ]   (272.39,199.24) -- (261.85,200.38) ;
\draw  [line width=0.75] [color=red  ,draw opacity=1 ]   (253.1,188.62) -- (255,180) ;

\draw  [line width=0.75] [color=red  ,draw opacity=1 ]   (261,220) -- (257.48,211.21) ;
\draw  [line width=0.75] [color=red  ,draw opacity=1 ]   (269,242) -- (265.48,233.21) ;
\draw  [line width=0.75] [color=red  ,draw opacity=1 ]   (268.6,224.37) -- (277.38,220.85) ;
\draw  [line width=0.75] [color=red  ,draw opacity=1 ]   (245.6,234.37) -- (254.38,230.85) ;
\draw  [line width=0.75] [color=red  ,draw opacity=1 ]   (235.25,215.25) -- (241.75,208.5) ;

\draw  [line width=0.75] [color=red  ,draw opacity=1 ]   (259,221) -- (255.48,212.21) ;

\draw (197.72,163.74) node [anchor=north west][inner sep=0.75pt]   [align=left] {$\displaystyle 2$};
\draw (250.83,158) node [anchor=north west][inner sep=0.75pt]   [align=left] {$\displaystyle 1$};
\draw (246,194) node [anchor=north west][inner sep=0.75pt]   [align=left] {$\displaystyle a$};
\draw (220.55,177.53) node [anchor=north west][inner sep=0.75pt]   [align=left] {$\displaystyle b$};
\draw (275.16,192) node [anchor=north west][inner sep=0.75pt]   [align=left] {$\displaystyle c$};
\draw (256.65,219) node [anchor=north west][inner sep=0.75pt]   [align=left] {$\displaystyle d$};
\draw (284.73,218.39) node [anchor=north west][inner sep=0.75pt]   [align=left] {$\displaystyle 4$};
\draw (213.37,213.66) node [anchor=north west][inner sep=0.75pt]   [align=left] {$\displaystyle 3$};
\draw (300,185.79) node [anchor=north west][inner sep=0.75pt]   [align=left] {$\displaystyle 2$};
\draw (265.59,243.5) node [anchor=north west][inner sep=0.75pt]   [align=left] {$\displaystyle 2$};

\end{tikzpicture}
}}
$};

\draw  [line width=0.75] (230,230) node [anchor=north west][inner sep=0.75pt]  [align=center] {$ \hbox{\scalebox{1.0}{\begin{tikzpicture}[x=0.75pt,y=0.75pt,yscale=-1,xscale=1]

\draw  [line width=0.75] (374.64,211.37) -- (376.7,187.12) -- (400.4,181.58) -- (413,202.41) -- (397.07,220.83) -- cycle ;
\draw  [line width=0.75] (421.04,169.23) -- (433.5,189.85) -- (412.88,202.32) -- (400.42,181.7) -- cycle ;
\draw  [line width=0.75] (387.87,243.1) -- (365.6,233.9) -- (374.8,211.63) -- (397.07,220.83) -- cycle ;
\draw  [line width=0.75] (376.7,187.12) -- (374.52,211.11) -- (350.53,208.94) -- (352.7,184.95) -- cycle ;
\draw   [line width=0.75] (350.53,208.94) -- (338.03,212.56) ;
\draw  [line width=0.75]  (352.7,184.95) -- (342.86,178.8) ;
\draw  [line width=0.75]  (388.71,255.15) -- (387.87,243.1) ;
\draw   [line width=0.75] (356.59,243.28) -- (365.6,233.9) ;
\draw  [line width=0.75]  (424.15,156.6) -- (421.04,169.23) ;
\draw  [line width=0.75]  (446.13,192.97) -- (433.5,189.85) ;

\draw [line width=0.75] [color=red  ,draw opacity=1 ]   (428.12,199.24) -- (422.18,190.45) ;
\draw [line width=0.75] [color=red  ,draw opacity=1 ]   (413.52,178.88) -- (407.57,170.09) ;
\draw [line width=0.75] [color=red  ,draw opacity=1 ]   (422.09,181.81) -- (430.88,175.87) ;
\draw [line width=0.75] [color=red  ,draw opacity=1 ]   (401.46,195.21) -- (410.25,189.27) ;
\draw [line width=0.75] [color=red  ,draw opacity=1 ]   (396.75,234.5) -- (388.47,230.97) ;
\draw [line width=0.75] [color=red  ,draw opacity=1 ]   (373.25,225.75) -- (364.97,222.51) ;
\draw [line width=0.75] [color=red  ,draw opacity=1 ]   (374.65,243.5) -- (378.21,233.52) ;
\draw [line width=0.75] [color=red  ,draw opacity=1 ]   (384.85,221.78) -- (388.4,211.79) ;
\draw [line width=0.75] [color=red  ,draw opacity=1 ]   (382.15,220.47) -- (385.1,212.16) -- (385.25,211.75) ;
\draw [line width=0.75] [color=red  ,draw opacity=1 ]   (400.1,207.71) -- (407.03,213.17) ;
\draw [line width=0.75] [color=red  ,draw opacity=1 ]   (371.14,197.2) -- (380.59,197.85) ;
\draw [line width=0.75] [color=red  ,draw opacity=1 ]   (347.1,196.85) -- (356.55,197.5) ;
\draw [line width=0.75] [color=red  ,draw opacity=1 ]   (361.16,206.15) -- (360.52,215.59) ;
\draw [line width=0.75] [color=red  ,draw opacity=1 ]   (365.14,181.09) -- (364.5,190.53) ;
\draw [line width=0.75] [color=red  ,draw opacity=1 ]   (385.85,182.07) -- (389.11,190.86) ;
\draw [line width=0.75] [color=red  ,draw opacity=1 ]   (371.14,200.2) -- (380.59,200.85) ;

\draw (433.93,163.94) node [anchor=north west][inner sep=0.75pt]   [align=left] {$\displaystyle 2$};
\draw (417.01,215.26) node [anchor=north west][inner sep=0.75pt]   [align=left] {$\displaystyle 4$};
\draw (387,194) node [anchor=north west][inner sep=0.75pt]   [align=left] {$\displaystyle a$};
\draw (410.84,180) node [anchor=north west][inner sep=0.75pt]   [align=left] {$\displaystyle c$};
\draw (376.77,220) node [anchor=north west][inner sep=0.75pt]   [align=left] {$\displaystyle d$};
\draw (358.84,189.86) node [anchor=north west][inner sep=0.75pt]   [align=left] {$\displaystyle b$};
\draw (346.02,215.28) node [anchor=north west][inner sep=0.75pt]   [align=left] {$\displaystyle 3$};
\draw (382.15,156) node [anchor=north west][inner sep=0.75pt]   [align=left] {$\displaystyle 1$};
\draw (367.45,241.74) node [anchor=north west][inner sep=0.75pt]   [align=left] {$\displaystyle 2$};
\draw (331.32,185.4) node [anchor=north west][inner sep=0.75pt]   [align=left] {$\displaystyle 2$};

\end{tikzpicture}
}}
$};

\draw  [line width=0.75] (390,230) node [anchor=north west][inner sep=0.75pt]  [align=center] {$ \hbox{\scalebox{1.0}{\begin{tikzpicture}[x=0.75pt,y=0.75pt,yscale=-1,xscale=1]

\draw  [line width=0.75]  (525.74,185.29) -- (545.28,199.81) -- (537.49,222.88) -- (513.15,222.61) -- (505.89,199.37) -- cycle ;
\draw  [line width=0.75]  (537.12,246.93) -- (513.02,246.66) -- (513.29,222.56) -- (537.39,222.83) -- cycle ;
\draw  [line width=0.75]  (491.81,179.82) -- (511.36,165.74) -- (525.44,185.29) -- (505.89,199.37) -- cycle ;
\draw  [line width=0.75]  (545.28,199.81) -- (526.02,185.33) -- (540.51,166.07) -- (559.76,180.56) -- cycle ;
\draw   [line width=0.75]  (540.51,166.07) -- (544.02,153.55) ;
\draw   [line width=0.75]  (559.76,180.56) -- (570.17,175.44) ;
\draw   [line width=0.75]  (481.12,174.19) -- (491.81,179.82) ;
\draw   [line width=0.75]  (508.13,153.14) -- (511.36,165.74) ;
\draw   [line width=0.75]  (546.21,256.23) -- (537.12,246.93) ;
\draw   [line width=0.75]  (503.72,255.75) -- (513.02,246.66) ;

\draw [line width=0.75]  [color=red  ,draw opacity=1 ]   (525.44,243) -- (525.44,251.75) ;
\draw [line width=0.75] [color=red  ,draw opacity=1 ]   (525.44,218) -- (525.44,226.75) ;
\draw [line width=0.75]  [color=red  ,draw opacity=1 ]   (517.32,235.13) -- (508.57,235.13) ;
\draw [line width=0.75] [color=red  ,draw opacity=1 ]   (542.32,235.13) -- (533.57,235.13) ;
\draw [line width=0.75] [color=red  ,draw opacity=1 ]   (501.82,186.34) -- (495.07,191.91) ;
\draw [line width=0.75] [color=red  ,draw opacity=1 ]   (520.82,172.34) -- (514.07,177.91) ;
\draw [line width=0.75]  [color=red  ,draw opacity=1 ]   (504.23,175.5) -- (498.66,168.75) ;
\draw [line width=0.75]  [color=red  ,draw opacity=1 ]   (518.23,196.5) -- (512.66,189.75) ;
\draw [line width=0.75]  [color=red  ,draw opacity=1 ]   (520.23,195.5) -- (514.66,188.75) ;
\draw [line width=0.75]  [color=red  ,draw opacity=1 ]   (537.48,178.25) -- (531,173) ;
\draw [line width=0.75]  [color=red  ,draw opacity=1 ]   (556.48,192.25) -- (550,187) ;
\draw [line width=0.75]  [color=red  ,draw opacity=1 ]   (547.37,176.11) -- (552,169.25) ;
\draw [line width=0.75]  [color=red  ,draw opacity=1 ]   (531.37,194.11) -- (536,187.25) ;
\draw [line width=0.75]  [color=red  ,draw opacity=1 ]   (533.37,196.11) -- (538,189.25) ;
\draw [line width=0.75]  [color=red  ,draw opacity=1 ]   (537.5,209.5) -- (544.5,213.75) ;
\draw [line width=0.75] [color=red  ,draw opacity=1 ]   (506,210.5) -- (514.25,206.75) ;

\draw (518.78,251.61) node [anchor=north west][inner sep=0.75pt]   [align=left] {$\displaystyle 2$};
\draw (484.07,210.19) node [anchor=north west][inner sep=0.75pt]   [align=left] {$\displaystyle 3$};
\draw (520.57,198) node [anchor=north west][inner sep=0.75pt]   [align=left] {$\displaystyle a$};
\draw (519.6,226.95) node [anchor=north west][inner sep=0.75pt]   [align=left] {$\displaystyle d$};
\draw (504.11,175.28) node [anchor=north west][inner sep=0.75pt]   [align=left] {$\displaystyle b$};
\draw (538.3,176) node [anchor=north west][inner sep=0.75pt]   [align=left] {$\displaystyle c$};
\draw (521.88,150.6) node [anchor=north west][inner sep=0.75pt]   [align=left] {$\displaystyle 1$};
\draw (554.24,212.67) node [anchor=north west][inner sep=0.75pt]   [align=left] {$\displaystyle 4$};
\draw (487.67,154.11) node [anchor=north west][inner sep=0.75pt]   [align=left] {$\displaystyle 2$};
\draw (554.59,153.09) node [anchor=north west][inner sep=0.75pt]   [align=left] {$\displaystyle 2$};

\end{tikzpicture}
}}
$};

\draw  [line width=0.75] (80,350) node [anchor=north west][inner sep=0.75pt]  [align=center] {$ \hbox{\scalebox{1.0}{\begin{tikzpicture}[x=0.75pt,y=0.75pt,yscale=-1,xscale=1]

\draw  [line width=0.75] (278.53,324.95) -- (266.48,345.81) -- (242.39,345.81) -- (230.35,324.95) -- (242.39,304.08) -- (266.48,304.08) -- cycle ;
\draw  [line width=0.75] (209.45,312.79) -- (221.5,291.92) -- (242.36,303.97) -- (230.32,324.84) -- cycle ;
\draw  [line width=0.75] (287.35,292.03) -- (299.4,312.9) -- (278.53,324.95) -- (266.48,304.08) -- cycle ;
\draw  [line width=0.75] (242.39,345.81) -- (266.49,345.81) -- (266.49,369.91) -- (242.39,369.91) -- cycle ;
\draw  [line width=0.75]  (266.49,369.91) -- (275.69,379.1) ;
\draw   [line width=0.75] (242.39,369.91) -- (233.2,379.1) ;
\draw   [line width=0.75] (290.72,279.46) -- (287.35,292.03) ;
\draw   [line width=0.75] (311.96,316.26) -- (299.4,312.9) ;
\draw   [line width=0.75] (196.88,316.16) -- (209.45,312.79) ;
\draw  [line width=0.75]  (218.13,279.36) -- (221.5,291.92) ;

\draw [line width=0.75][color=red  ,draw opacity=1 ]   (254.75,374.25) -- (254.5,365.75) ;
\draw [line width=0.75][color=red  ,draw opacity=1 ]   (254.75,350.25) -- (254.5,341.75) ;
\draw [line width=0.75][color=red  ,draw opacity=1 ]   (254.75,308.25) -- (254.5,299.75) ;
\draw [line width=0.75][color=red  ,draw opacity=1 ]   (261.38,357.94) -- (269.88,357.88) ;
\draw [line width=0.75][color=red  ,draw opacity=1 ]   (237.38,357.94) -- (245.88,357.88) ;
\draw [line width=0.75][color=red  ,draw opacity=1 ]   (211.25,300.5) -- (219.65,304.89) ;
\draw [line width=0.75][color=red  ,draw opacity=1 ]   (233.25,313.5) -- (241.65,317.89) ;
\draw [line width=0.75][color=red  ,draw opacity=1 ]   (221.65,314.49) -- (217.26,322.9) ;
\draw [line width=0.75][color=red  ,draw opacity=1 ]   (233.65,293.49) -- (229.26,301.9) ;
\draw [line width=0.75][color=red  ,draw opacity=1 ]   (276.75,312) -- (268.26,316.9) ;
\draw [line width=0.75][color=red  ,draw opacity=1 ]   (298.75,301) -- (290.26,305.9) ;
\draw [line width=0.75][color=red  ,draw opacity=1 ]   (292.95,322.69) -- (288.06,314.2) ;
\draw [line width=0.75][color=red  ,draw opacity=1 ]   (279.95,301.69) -- (275.06,293.2) ;
\draw [line width=0.75][color=red  ,draw opacity=1 ]   (231.26,337.9) -- (239.75,333) ;
\draw [line width=0.75][color=red  ,draw opacity=1 ]   (268.25,333.5) -- (276.65,337.89) ;

\draw (198.32,285.89) node [anchor=north west][inner sep=0.75pt]   [align=left] {$\displaystyle 2$};
\draw (251.43,275.95) node [anchor=north west][inner sep=0.75pt]   [align=left] {$\displaystyle 1$};
\draw (250.06,317) node [anchor=north west][inner sep=0.75pt]   [align=left] {$\displaystyle a$};
\draw (221.15,299.68) node [anchor=north west][inner sep=0.75pt]   [align=left] {$\displaystyle b$};
\draw (278.76,302) node [anchor=north west][inner sep=0.75pt]   [align=left] {$\displaystyle c$};
\draw (250,349.66) node [anchor=north west][inner sep=0.75pt]   [align=left] {$\displaystyle d$};
\draw (285.33,340.54) node [anchor=north west][inner sep=0.75pt]   [align=left] {4};
\draw (213.97,335.81) node [anchor=north west][inner sep=0.75pt]   [align=left] {$\displaystyle 3$};
\draw (305.18,284.94) node [anchor=north west][inner sep=0.75pt]   [align=left] {$\displaystyle 2$};
\draw (248.19,375.65) node [anchor=north west][inner sep=0.75pt]   [align=left] {$\displaystyle 2$};

\end{tikzpicture}
}}
$};



\draw (200,180) node [anchor=north west][inner sep=0.75pt]    {$,$};
\draw (350,180) node [anchor=north west][inner sep=0.75pt]    {$,$};
\draw (490,180) node [anchor=north west][inner sep=0.75pt]    {$,$};
\draw (200,300) node [anchor=north west][inner sep=0.75pt]    {$,$};
\draw (350,300) node [anchor=north west][inner sep=0.75pt]    {$,$};
\draw (490,300) node [anchor=north west][inner sep=0.75pt]    {$,$};
\draw (200,420) node [anchor=north west][inner sep=0.75pt]    {$.$};

\end{tikzpicture}
}
}}
\end{eqnarray}
For the $f$-graph integrand, the leading singularities associated with these cuts are, in chronological order:
\begin{equation}
    \frac{s}{\Delta^2},\ \frac{t}{\Delta^2},\ \frac{u}{\Delta^2},\ \frac{s}{(s{-}t)\Delta} \Big/ \frac{t}{(s{-}t)\Delta},\ \frac{s}{(s{-}u)\Delta} \Big/ \frac{u}{(s{-}u)\Delta},\  \frac{t}{(t-u)\Delta} \Big/  \frac{u}{(t-u)\Delta}, \ \frac{1}{\Delta} .
\end{equation}
Here, the leading singularity expressed as $A/B$ indicates that the result depends on the order in which the residues are taken on the two Jacobian factors in the mutual cut.

This should be compared with our local integrand basis, where each yields a single leading singularity across all non-vanishing cuts:
\begin{equation}
    \begin{split}
     &\Delta:\  I_S^{12;34}, \\
        &(s-t)\Delta : I_{S^\prime}^{12,34},\ \Delta (s-u): I_{S^\prime}^{12,43},\ \Delta (s-u): I_{S^\prime}^{32,41},\\
        &\Delta^2:\ I_{S^{\prime\prime}}^{12;34}, I_{S^{\prime\prime}}^{32;14}, I_{S^{\prime\prime}}^{42;13}\,.
    \end{split}
\end{equation}

\subsection{Elliptic cuts and  Chambers}\label{sec: elliptic geo}

Now that we have introduced the local integrand responsible for the elliptic cut, 
we turn to analyze in which regions of $\mathbb{T}_4$ the elliptic cut is accessible. 

To proceed, let us go upstairs and consider the three-dimensional thirteen-cut geometry
\begin{eqnarray}
&&\vcenter{\hbox{\scalebox{1}{
\begin{tikzpicture}[x=0.75pt,y=0.75pt,yscale=-1,xscale=1]


\draw  [line width=0.75] (-10,140) node [anchor=north west][inner sep=0.75pt]  [align=center] {$ \begin{tikzpicture}[x=0.75pt,y=0.75pt,yscale=-1,xscale=1]

\draw  [line width=0.75]  (135.27,102.72) -- (178.71,102.72) -- (178.71,146.16) -- (135.27,146.16) -- cycle ;
\draw  [line width=0.75]  (178.71,146.16) -- (222.16,146.16) -- (222.16,189.61) -- (178.71,189.61) -- cycle ;
\draw  [line width=0.75]   (123.3,90.75) -- (135.27,102.72) ;
\draw   [line width=0.75]  (222.16,189.61) -- (234.13,201.58) ;
\draw   [line width=0.75]  (178.8,85.8) -- (178.71,102.72) ;
\draw   [line width=0.75]  (135.27,189.61) -- (121.13,202.13) ;
\draw   [line width=0.75]  (178.71,189.61) -- (178.47,206.8) ;
\draw  [line width=0.75]  (178.71,102.72) -- (222.16,102.72) -- (222.16,146.16) -- (178.71,146.16) -- cycle ;
\draw  [line width=0.75]  (135.27,146.16) -- (178.71,146.16) -- (178.71,189.61) -- (135.27,189.61) -- cycle ;
\draw   [line width=0.75]  (236.29,90.19) -- (222.16,102.72) ;
\draw [line width=0.75] [color=red  ,draw opacity=1 ]   (157.15,98.16) -- (157.15,106.88) ;
\draw [line width=0.75] [color=red  ,draw opacity=1 ]   (157.75,140.96) -- (157.75,149.68) ;
\draw [line width=0.75] [color=red  ,draw opacity=1 ]   (161,140.96) -- (161,149.68) ;
\draw [line width=0.75]  [color=red  ,draw opacity=1 ]   (158.35,186.16) -- (158.35,194.88) ;
\draw [line width=0.75] [color=red  ,draw opacity=1 ]   (200.35,98.56) -- (200.35,107.28) ;
\draw [line width=0.75] [color=red  ,draw opacity=1 ]   (200.35,141.56) -- (200.35,150.28) ;
\draw [line width=0.75] [color=red  ,draw opacity=1 ]   (200.35,185.56) -- (200.35,194.28) ;
\draw [line width=0.75] [color=red  ,draw opacity=1 ]   (139.71,126.92) -- (130.99,126.92) ;
\draw [line width=0.75] [color=red  ,draw opacity=1 ]   (139.71,170.92) -- (130.99,170.92) ;
\draw [line width=0.75] [color=red  ,draw opacity=1 ]   (183.71,125.92) -- (174.99,125.92) ;
\draw [line width=0.75] [color=red  ,draw opacity=1 ]   (226.71,125.92) -- (217.99,125.92) ;
\draw [line width=0.75]  [color=red  ,draw opacity=1 ]   (182.71,170.92) -- (173.99,170.92) ;
\draw [line width=0.75]  [color=red  ,draw opacity=1 ]   (226.71,170.92) -- (217.99,170.92) ;

\draw (154.69,115.04) node [anchor=north west][inner sep=0.75pt]    {$a$};
\draw (195.58,119.45) node [anchor=north west][inner sep=0.75pt]    {$b$};
\draw (154.06,157.5) node [anchor=north west][inner sep=0.75pt]   [align=left] {$\displaystyle c$};
\draw (194.08,156.21) node [anchor=north west][inner sep=0.75pt]   [align=left] {$\displaystyle d$};
\draw (151.39,79.98) node [anchor=north west][inner sep=0.75pt]    {$1$};
\draw (195.07,80.72) node [anchor=north west][inner sep=0.75pt]    {$2$};
\draw (194.89,195.14) node [anchor=north west][inner sep=0.75pt]    {$1$};
\draw (115.14,139.37) node [anchor=north west][inner sep=0.75pt]    {$3$};
\draw (231.75,139.13) node [anchor=north west][inner sep=0.75pt]    {$4$};
\draw (152.4,193.72) node [anchor=north west][inner sep=0.75pt]    {$2$};

\end{tikzpicture}
$};

\draw  [line width=0.75] (18,150) node [anchor=north west][inner sep=0.75pt]  [align=center] {$ \hbox{\scalebox{0.6}{\begin{tikzpicture}[x=0.75pt,y=0.75pt,yscale=-1,xscale=1]


\draw (93.5,66) node [anchor=north west][inner sep=0.75pt]  [font=\large,color=blue  ,opacity=1 ] [align=left] {$\displaystyle 1$};

\end{tikzpicture}

}}
$};

\draw  [line width=0.75] (-5,182) node [anchor=north west][inner sep=0.75pt]  [align=center] {$ \hbox{\scalebox{0.6}{\begin{tikzpicture}[x=0.75pt,y=0.75pt,yscale=-1,xscale=1]


\draw (93.5,66) node [anchor=north west][inner sep=0.75pt]  [font=\large,color=blue  ,opacity=1 ] [align=left] {$\displaystyle 2$};

\end{tikzpicture}

}}
$};

\draw  [line width=0.75] (80,147) node [anchor=north west][inner sep=0.75pt]  [align=center] {$ \hbox{\scalebox{0.6}{\begin{tikzpicture}[x=0.75pt,y=0.75pt,yscale=-1,xscale=1]


\draw (93.5,66) node [anchor=north west][inner sep=0.75pt]  [font=\large,color=blue  ,opacity=1 ] [align=left] {$\displaystyle 3$};

\end{tikzpicture}

}}
$};

\draw  [line width=0.75] (105,180) node [anchor=north west][inner sep=0.75pt]  [align=center] {$ \hbox{\scalebox{0.6}{\begin{tikzpicture}[x=0.75pt,y=0.75pt,yscale=-1,xscale=1]


\draw (93.5,66) node [anchor=north west][inner sep=0.75pt]  [font=\large,color=blue  ,opacity=1 ] [align=left] {$\displaystyle 4$};

\end{tikzpicture}

}}
$};

\draw  [line width=0.75] (15,253) node [anchor=north west][inner sep=0.75pt]  [align=center] {$ \hbox{\scalebox{0.6}{\begin{tikzpicture}[x=0.75pt,y=0.75pt,yscale=-1,xscale=1]


\draw (93.5,66) node [anchor=north west][inner sep=0.75pt]  [font=\large,color=blue  ,opacity=1 ] [align=left] {$\displaystyle 5$};

\end{tikzpicture}

}}
$};

\draw  [line width=0.75] (-5,225) node [anchor=north west][inner sep=0.75pt]  [align=center] {$ \hbox{\scalebox{0.6}{\begin{tikzpicture}[x=0.75pt,y=0.75pt,yscale=-1,xscale=1]


\draw (93.5,66) node [anchor=north west][inner sep=0.75pt]  [font=\large,color=blue  ,opacity=1 ] [align=left] {$\displaystyle 6$};

\end{tikzpicture}

}}
$};

\draw  [line width=0.75] (80,253) node [anchor=north west][inner sep=0.75pt]  [align=center] {$ \hbox{\scalebox{0.6}{\begin{tikzpicture}[x=0.75pt,y=0.75pt,yscale=-1,xscale=1]


\draw (93.5,66) node [anchor=north west][inner sep=0.75pt]  [font=\large,color=blue  ,opacity=1 ] [align=left] {$\displaystyle 7$};

\end{tikzpicture}

}}
$};

\draw  [line width=0.75] (105,225) node [anchor=north west][inner sep=0.75pt]  [align=center] {$ \hbox{\scalebox{0.6}{\begin{tikzpicture}[x=0.75pt,y=0.75pt,yscale=-1,xscale=1]


\draw (93.5,66) node [anchor=north west][inner sep=0.75pt]  [font=\large,color=blue  ,opacity=1 ] [align=left] {$\displaystyle 8$};

\end{tikzpicture}

}}
$};

\draw  [line width=0.75] (40,180) node [anchor=north west][inner sep=0.75pt]  [align=center] {$ \hbox{\scalebox{0.6}{\begin{tikzpicture}[x=0.75pt,y=0.75pt,yscale=-1,xscale=1]


\draw (93.5,66) node [anchor=north west][inner sep=0.75pt]  [font=\large,color=blue  ,opacity=1 ] [align=left] {$\displaystyle 9$};

\end{tikzpicture}

}}
$};

\draw  [line width=0.75] (35,225) node [anchor=north west][inner sep=0.75pt]  [align=center] {$ \hbox{\scalebox{0.6}{\begin{tikzpicture}[x=0.75pt,y=0.75pt,yscale=-1,xscale=1]


\draw (90,68) node [anchor=north west][inner sep=0.75pt]  [font=\large,color=blue  ,opacity=1 ] [align=left] {$\displaystyle 10$};

\end{tikzpicture}

}}
$};

\draw  [line width=0.75] (70,193) node [anchor=north west][inner sep=0.75pt]  [align=center] {$ \hbox{\scalebox{0.6}{\begin{tikzpicture}[x=0.75pt,y=0.75pt,yscale=-1,xscale=1]


\draw (90,68) node [anchor=north west][inner sep=0.75pt]  [font=\large,color=blue  ,opacity=1 ] [align=left] {$\displaystyle 11$};

\end{tikzpicture}

}}
$};

\draw  [line width=0.75] (18,193) node [anchor=north west][inner sep=0.75pt]  [align=center] {$ \hbox{\scalebox{0.6}{\begin{tikzpicture}[x=0.75pt,y=0.75pt,yscale=-1,xscale=1]


\draw (90,68) node [anchor=north west][inner sep=0.75pt]  [font=\large,color=blue  ,opacity=1 ] [align=left] {$\displaystyle 12$};

\end{tikzpicture}

}}
$};

\draw  [line width=0.75] (38,210) node [anchor=north west][inner sep=0.75pt]  [align=center] {$ \hbox{\scalebox{0.6}{\begin{tikzpicture}[x=0.75pt,y=0.75pt,yscale=-1,xscale=1]


\draw (90,68) node [anchor=north west][inner sep=0.75pt]  [font=\large,color=blue  ,opacity=1 ] [align=left] {$\displaystyle 13$};

\end{tikzpicture}

}}
$};



\end{tikzpicture}

}
}}
\end{eqnarray}
In the above, we have denoted the sequence in which the cut conditions are imposed. In terms of the parameterization in eq.~\eqref{rq: 4LoopPara}, this sequence corresponds to 
\begin{equation}
    (w_1,y_1, w_2,y_2,w_3,y_3,w_4,y_4,x_1,z_4,x_2,z_3,x_3 )\,.
\end{equation}
This is the predecessor of the elliptic $\mu_s$-cut discussed earlier. Starting from this cut geometry, we can probe the elliptic boundary in a controlled fashion. 
It receives contributions from the local integrands $I_G^{12;34}$, $I_R^{23;14}$, and $I_R^{13;24}$, while all other integrands in eq.~\eqref{eq: mu-cut and integral} are absent on this 13-cut. As we will see, the elliptic $\mu_s$-cut  emerges from this geometry only in chambers $r_4$ and $r_6$, {\it i.e.} when $\text{max}(s,t,u)= s$.

First, for chambers with  $\text{min}(s,t,u)= s$ (namely $r_1$ and $r_2$), the cut is never positive, and hence the geometry is empty in this region. Next consider the geometry in chambers $r_5, r_3$, for the net-positive sector (and positive branch $\Delta>0$), the three-dimensional geometry in $z_1, z_2$ and $x_4$, 
\begin{align}
    r_5\;(u<s<t):&\quad  z_1<0\land \frac{c_1^2 (1-c_2)z_1}{c_2^2 (1-c_1)}<x_4<z_1\land \frac{c_1 z_1}{c_2}<z_2<\frac{c_2 (1-c_1)x_4}{c_1 (1-c_2)}\\
     r_3\; (t<s<u):&\quad z_1>0\land \frac{c_2 z_1}{c_1}<x_4<\frac{(1-c_1)z_1}{1-c_2}\land x_4<z_2\nonumber\\
       &\quad \phantom{kf} \land z_2<\frac{z_1((c_2-1)(1+c_2-c_1)c_1 x_4-c_2 (c_1-1)z_1)}{(c_2-1)c_1 x_4+c_2 (c_2-c_1-1)(c_1-1)z_1}.
\end{align}
For the net-negative orientation sector, it simply coincides with swapping all $``>"$ and $``<"$ inequalities in the above.
Since all of the inequalities are rational, one can straightforwardly construct the canonical form for this geometry.

Finally, for chambers $r_4$ and $r_6$ (where $\text{max}(s,t,u)= s$), the inequalities cannot be fully rationalized-reduced as previously. Since the complete expression is lengthy, we focus here on the net-positive orientation and the region with  $z_1>0$. In this region part of the inequalities can be fully rationalized while some cannot. 
For the non-rational (elliptic) part, the geometry depends on the sign of:
\begin{align}
    &\mathfrak{p}{:=} \left(z_2-\frac{z_1 \left((c_1{-}1) c_2 (c_2{+}1) X^2+(c_1{-}1) (c_2{-}1)(c_1{-}c_2) X {-}c_1 (c_1{+}1) (c_2{-}1)\right)}{2 c_1 \left(c_2^2 (X-1)+1\right)-2 c_2 X}\right)^2\nonumber\\
        &\phantom{dd}{-}\frac{z_1^2}{4 \left(c_1 c_2^2 (X{-}1)+c_1{-}c_2 X\right)^2} \Big( {-}4 X (c_1 {+} c_1^2 c_2 (X{-}1 ) {-} c_2 X) (c_1 {+} c_1 c_2^2 (X {-}1 ) {-} 
    c_2 X)\nonumber \\
        & \phantom{dddddddd}{+}\big({-}c_1 (c_1 {+} 1) ( c_2{-}1) {+} (c_1 {-} c_2)  ( c_1{-}1) ( 
      c_2{-}1) X + ( c_1{-}1) c_2 ( c_2+1) X^2\big)^2\Big)
\end{align}
with $X=x_4/z_1$. For example, when $\mathfrak{p}>0$, the regions are given by
\begin{align}
     &z_1>0\, \land \left (x_4{-} \frac{c_1 z_1}{c_2} \right)^2 + \frac{1}{c_2}  \left( \frac{c_1 z_1 } {c_2}\right)^2>0\, \land \, \Bigg[\left(\frac{c_2(c_1-1)x_4}{c_1 (c_2-1)}<z_2< \frac{c_1 z_1}{c_2 }\right) \nonumber\\
        &\phantom{khhhf}\lor\,\left(    z_1<x_4<\frac{c_1 (c_2^2-1)z_1}{c_2 (c_1 c_2 -1)}\, \land\, \frac{c_2(c_1-1)x_4}{c_1 (c_2-1)}<z_2\right)\Bigg],\\
        \lor\, &z_1>0\, \land    x_4<0\land z_2<x_4\,, \nonumber
\end{align}
For $\mathfrak{p}<0$ , we instead find:
\begin{equation}
    \begin{split}
        &z_1>0\, \land x_4>\frac{c_1(c_2^2-1)z_1}{c_2 ( c_1c_2-1)}\, \land \Bigg[\left( \frac{c_2 (c_1-1)  x_4}{c_1 (c_2-1)}<z_2 \right) \lor \left(  z_2<\frac{c_1 z_1}{c_2}\right)\Bigg].
    \end{split}
\end{equation}
Since the geometry depends on the sign of $\mathfrak{p}$, the condition $\mathfrak{p}=0$ defines a boundary. Using this, we can localize $z_2$ as
\begin{equation}
    \begin{split}
        &z_2=\frac{z_1 \left((c_1{-}1) c_2 (c_2{+}1) X^2+(c_1{-}1) (c_2{-}1)(c_1{-}c_2) X {-}c_1 (c_1{+}1) (c_2{-}1)\right)}{2 c_1 \left(c_2^2 (X-1)+1\right)-2 c_2 X}\\
        &\phantom{dd}{+}\frac{z_1}{2 \left(c_1 c_2^2 (X{-}1)+c_1{-}c_2 X\right)} \Big( {-}4 X (c_1 {+} c_1^2 c_2 (X{-}1 ) {-} c_2 X) (c_1 {+} c_1 c_2^2 (X {-}1 ) {-} 
    c_2 X)\\
        & \phantom{dddddddd}{+}\big({-}c_1 (c_1 {+} 1) ( c_2{-}1) {+} (c_1 {-} c_2)  ( c_1{-}1) ( 
      c_2{-}1) X + ( c_1{-}1) c_2 ( c_2+1) X^2\big)^2\Big)^{\frac{1}{2}}
    \end{split}
\end{equation}
Note that the square-root factor coincides precisely with the Jacobian factor of the elliptic cut~\eqref{eq: elliptic J1}. \textit{Thus, the elliptic-$\mu_s$ cut arises only in the region where $\text{max}(s,t,u) = s$, {\it i.e.} chambers $r_4$ and $r_6$}. Similarly, the elliptic cuts $\mu_t$ and $\mu_u$ reside in chambers $(r_2, r_5)$ and $(r_1, r_3)$, respectively.

Since the positivity condition for the elliptic $\mu_s$-cut is $\text{max}(s,t,u) = s$, this suggests that it belongs to the block $A_s$, which carries the universal prefactor $\Delta^2$.
As the forms associated with the rational parts of the geometry can be explicitly computed and matched to local integrands, we can assign the elliptic integrand responsible for reproducing the $\mu_s$-cut, {\it i.e.} $I_G^{12;34}$, a factor of $c\,\Delta^2$, where $c$ is a proportionality constant. By matching to known results, this constant is fixed to be $4$. Since the elliptic $\mu_s$-cut imposes two mutual cuts, it encounters internal boundaries twice, which we conjecture to be responsible for this overall multiplicity.

Similarly, we can use the same strategy to another elliptic cut $\nu_s$ by considering the $12$-cut geometry:
\begin{eqnarray}
&&\vcenter{\hbox{\scalebox{1}{
\begin{tikzpicture}[x=0.75pt,y=0.75pt,yscale=-1,xscale=1]


\draw  [line width=0.75] (-10,140) node [anchor=north west][inner sep=0.75pt]  [align=center] {$ \begin{tikzpicture}[x=0.75pt,y=0.75pt,yscale=-1,xscale=1]

\draw   [line width=0.75] (135.27,102.72) -- (178.71,102.72) -- (178.71,146.16) -- (135.27,146.16) -- cycle ;
\draw   [line width=0.75] (178.71,146.16) -- (222.16,146.16) -- (222.16,189.61) -- (178.71,189.61) -- cycle ;
\draw    [line width=0.75] (123.3,90.75) -- (135.27,102.72) ;
\draw    [line width=0.75] (222.16,189.61) -- (234.13,201.58) ;
\draw   [line width=0.75]  (178.8,85.8) -- (178.71,102.72) ;
\draw    [line width=0.75] (135.27,189.61) -- (121.13,202.13) ;
\draw    [line width=0.75] (178.71,189.61) -- (178.47,206.8) ;
\draw   [line width=0.75] (178.71,102.72) -- (222.16,102.72) -- (222.16,146.16) -- (178.71,146.16) -- cycle ;
\draw   [line width=0.75] (135.27,146.16) -- (178.71,146.16) -- (178.71,189.61) -- (135.27,189.61) -- cycle ;
\draw    [line width=0.75] (236.29,90.19) -- (222.16,102.72) ;
\draw  [line width=0.75][color=red  ,draw opacity=1 ]   (157.15,98.16) -- (157.15,106.88) ;
\draw  [line width=0.75][color=red  ,draw opacity=1 ]   (158.75,140.96) -- (158.75,149.68) ;
\draw  [line width=0.75][color=red  ,draw opacity=1 ]   (158.35,186.16) -- (158.35,194.88) ;
\draw  [line width=0.75][color=red  ,draw opacity=1 ]   (200.35,98.56) -- (200.35,107.28) ;
\draw  [line width=0.75][color=red  ,draw opacity=1 ]   (200.35,141.56) -- (200.35,150.28) ;
\draw  [line width=0.75][color=red  ,draw opacity=1 ]   (200.35,185.56) -- (200.35,194.28) ;
\draw  [line width=0.75][color=red  ,draw opacity=1 ]   (139.71,126.92) -- (130.99,126.92) ;
\draw  [line width=0.75] [color=red  ,draw opacity=1 ]   (139.71,170.92) -- (130.99,170.92) ;
\draw  [line width=0.75] [color=red  ,draw opacity=1 ]   (183.71,125.92) -- (174.99,125.92) ;
\draw  [line width=0.75] [color=red  ,draw opacity=1 ]   (226.71,125.92) -- (217.99,125.92) ;
\draw  [line width=0.75] [color=red  ,draw opacity=1 ]   (182.71,170.92) -- (173.99,170.92) ;
\draw  [line width=0.75] [color=red  ,draw opacity=1 ]   (226.71,170.92) -- (217.99,170.92) ;
\draw    [line width=0.75] (239.08,146.17) -- (222.16,146.16) ;
\draw   [line width=0.75]  (135.27,146.16) -- (118.35,146.16) ;

\draw (152,119.04) node [anchor=north west][inner sep=0.75pt]    {$a$};
\draw (195.58,119.45) node [anchor=north west][inner sep=0.75pt]    {$b$};
\draw (152,164.5) node [anchor=north west][inner sep=0.75pt]   [align=left] {$ c$};
\draw (194.08,161.5) node [anchor=north west][inner sep=0.75pt]   [align=left] {$ d$};
\draw (111.39,113.98) node [anchor=north west][inner sep=0.75pt]    {$1$};
\draw (150.74,76.05) node [anchor=north west][inner sep=0.75pt]    {$2$};
\draw (234.56,164.14) node [anchor=north west][inner sep=0.75pt]    {$1$};
\draw (233.8,118.71) node [anchor=north west][inner sep=0.75pt]    {$3$};
\draw (194.75,77.46) node [anchor=north west][inner sep=0.75pt]    {$4$};
\draw (196.4,199.38) node [anchor=north west][inner sep=0.75pt]    {$2$};
\draw (113.47,161.04) node [anchor=north west][inner sep=0.75pt]    {$3$};
\draw (152.08,198.46) node [anchor=north west][inner sep=0.75pt]    {$4$};

\end{tikzpicture}
$};

\draw  [line width=0.75] (0,180) node [anchor=north west][inner sep=0.75pt]  [align=center] {$ \hbox{\scalebox{0.6}{\begin{tikzpicture}[x=0.75pt,y=0.75pt,yscale=-1,xscale=1]


\draw (93.5,68) node [anchor=north west][inner sep=0.75pt]  [font=\large,color=blue  ,opacity=1 ] [align=left] {$\displaystyle 1$};

\end{tikzpicture}

}}
$};

\draw  [line width=0.75] (32,152) node [anchor=north west][inner sep=0.75pt]  [align=center] {$ \hbox{\scalebox{0.6}{\begin{tikzpicture}[x=0.75pt,y=0.75pt,yscale=-1,xscale=1]


\draw (93.5,68) node [anchor=north west][inner sep=0.75pt]  [font=\large,color=blue  ,opacity=1 ] [align=left] {$\displaystyle 2$};

\end{tikzpicture}

}}
$};

\draw  [line width=0.75] (107,180) node [anchor=north west][inner sep=0.75pt]  [align=center] {$ \hbox{\scalebox{0.6}{\begin{tikzpicture}[x=0.75pt,y=0.75pt,yscale=-1,xscale=1]


\draw (93.5,68) node [anchor=north west][inner sep=0.75pt]  [font=\large,color=blue  ,opacity=1 ] [align=left] {$\displaystyle 3$};

\end{tikzpicture}

}}
$};

\draw  [line width=0.75] (75,152) node [anchor=north west][inner sep=0.75pt]  [align=center] {$ \hbox{\scalebox{0.6}{\begin{tikzpicture}[x=0.75pt,y=0.75pt,yscale=-1,xscale=1]


\draw (93.5,68) node [anchor=north west][inner sep=0.75pt]  [font=\large,color=blue  ,opacity=1 ] [align=left] {$\displaystyle 4$};

\end{tikzpicture}

}}
$};

\draw  [line width=0.75] (0,232) node [anchor=north west][inner sep=0.75pt]  [align=center] {$ \hbox{\scalebox{0.6}{\begin{tikzpicture}[x=0.75pt,y=0.75pt,yscale=-1,xscale=1]


\draw (93.5,68) node [anchor=north west][inner sep=0.75pt]  [font=\large,color=blue  ,opacity=1 ] [align=left] {$\displaystyle 5$};

\end{tikzpicture}

}}
$};

\draw  [line width=0.75] (32,240) node [anchor=north west][inner sep=0.75pt]  [align=center] {$ \hbox{\scalebox{0.6}{\begin{tikzpicture}[x=0.75pt,y=0.75pt,yscale=-1,xscale=1]


\draw (93.5,68) node [anchor=north west][inner sep=0.75pt]  [font=\large,color=blue  ,opacity=1 ] [align=left] {$\displaystyle 6$};

\end{tikzpicture}

}}
$};

\draw  [line width=0.75] (107,232) node [anchor=north west][inner sep=0.75pt]  [align=center] {$ \hbox{\scalebox{0.6}{\begin{tikzpicture}[x=0.75pt,y=0.75pt,yscale=-1,xscale=1]


\draw (93.5,68) node [anchor=north west][inner sep=0.75pt]  [font=\large,color=blue  ,opacity=1 ] [align=left] {$\displaystyle 7$};

\end{tikzpicture}

}}
$};

\draw  [line width=0.75] (75,240) node [anchor=north west][inner sep=0.75pt]  [align=center] {$ \hbox{\scalebox{0.6}{\begin{tikzpicture}[x=0.75pt,y=0.75pt,yscale=-1,xscale=1]


\draw (93.5,68) node [anchor=north west][inner sep=0.75pt]  [font=\large,color=blue  ,opacity=1 ] [align=left] {$\displaystyle 8$};

\end{tikzpicture}

}}
$};

\draw  [line width=0.75] (46,188) node [anchor=north west][inner sep=0.75pt]  [align=center] {$ \hbox{\scalebox{0.6}{\begin{tikzpicture}[x=0.75pt,y=0.75pt,yscale=-1,xscale=1]


\draw (93.5,68) node [anchor=north west][inner sep=0.75pt]  [font=\large,color=blue  ,opacity=1 ] [align=left] {$\displaystyle 9$};

\end{tikzpicture}

}}
$};

\draw  [line width=0.75] (43,231) node [anchor=north west][inner sep=0.75pt]  [align=center] {$ \hbox{\scalebox{0.6}{\begin{tikzpicture}[x=0.75pt,y=0.75pt,yscale=-1,xscale=1]


\draw (93.5,68) node [anchor=north west][inner sep=0.75pt]  [font=\large,color=blue  ,opacity=1 ] [align=left] {$\displaystyle 10$};

\end{tikzpicture}

}}
$};

\draw  [line width=0.75] (32,198) node [anchor=north west][inner sep=0.75pt]  [align=center] {$ \hbox{\scalebox{0.6}{\begin{tikzpicture}[x=0.75pt,y=0.75pt,yscale=-1,xscale=1]


\draw (93.5,68) node [anchor=north west][inner sep=0.75pt]  [font=\large,color=blue  ,opacity=1 ] [align=left] {$\displaystyle 11$};

\end{tikzpicture}

}}
$};

\draw  [line width=0.75] (74,198) node [anchor=north west][inner sep=0.75pt]  [align=center] {$ \hbox{\scalebox{0.6}{\begin{tikzpicture}[x=0.75pt,y=0.75pt,yscale=-1,xscale=1]


\draw (93.5,68) node [anchor=north west][inner sep=0.75pt]  [font=\large,color=blue  ,opacity=1 ] [align=left] {$\displaystyle 12$};

\end{tikzpicture}

}}
$};



\end{tikzpicture}

}
}}
\end{eqnarray}
and solve the cuts in the following sequential order
\begin{equation}
    (w_1,y_1, w_2,y_2,w_3,y_3,w_4,y_4,x_1,x_4,x_2,z_3 )\,.
\end{equation}
 As it turns out all regions, the inequality can be linearly reduced. The elliptic cut $\nu_s$ does not exist.

\subsection{Elliptic cuts and Leading singularities: chambers vs. $f$-graphs}
It is instructive to compare the elliptic cuts and leading singularities that appear in $f$-graphs with those of the chamber loop-forms. Since both sum to the same result, the relevant comparison is between individual $f$-graphs and the local integrand in the loop-forms. As emphasized in previous sections, the loop-form for each chamber is diagonalized; hence, each set of integrands possesses only one leading singularity or one elliptic cut. In contrast, individual $f$-graphs can exhibit multiple singularities, some of which are even spurious.

\paragraph{Elliptic cut:} Let us consider the elliptic $\mu_s$-cut.  Four conformal integrands generated from the $f$-graphs contribute to this cut:\footnote{The labels $I_{14}$ and $I_{30}$ denote the numbering of the 32 conformal integrands obtained from the $f$-graph and carry no additional meaning.
The same applies to $I_{7}$, $I_{10}$, and $I_{13}$ mentioned later in the text. }

\begin{eqnarray}
&&\vcenter{\hbox{\scalebox{1}{
\begin{tikzpicture}[x=0.75pt,y=0.75pt,yscale=-1,xscale=1]





\draw  [line width=0.75] (-30,174) node [anchor=north west][inner sep=0.75pt]  [align=center] {$ \hbox{\scalebox{0.75}{\begin{tikzpicture}[x=0.75pt,y=0.75pt,yscale=-1,xscale=1]

\draw   [line width=0.75] (313.58,128.41) -- (299.79,153.5) -- (272.21,153.5) -- (258.42,128.41) -- (272.21,103.31) -- (299.79,103.31) -- cycle ;
\draw   [line width=0.75] (258.42,128.41) -- (244.62,153.5) -- (217.04,153.5) -- (203.25,128.41) -- (217.04,103.31) -- (244.62,103.31) -- cycle ;
\draw    [line width=0.75] (258.42,78.22) -- (272.21,103.31) ;
\draw    [line width=0.75] (258.42,78.22) -- (244.62,103.31) ;
\draw    [line width=0.75] (187.25,128.41) -- (203.25,128.41) ;
\draw   [line width=0.75]  (313.58,128.41) -- (329.58,128.41) ;
\draw    [line width=0.75] (299.79,103.31) -- (309.72,93.63) ;
\draw    [line width=0.75] (207.11,163.18) -- (217.04,153.5) ;
\draw    [line width=0.75] (217.04,103.31) -- (201.59,93.63) ;
\draw    [line width=0.75] (315.24,163.18) -- (299.79,153.5) ;
\draw    [line width=0.75] (258.42,64.31) -- (258.42,78.22) ;
\draw    [line width=0.75] (258.42,178.59) -- (258.42,192.5) ;
\draw    [line width=0.75] (244.62,153.5) -- (258.42,178.59) ;
\draw    [line width=0.75] (272.21,153.5) -- (258.42,178.59) ;

\draw (225.09,122) node [anchor=north west][inner sep=0.75pt]    {$c$};
\draw (282.46,119.43) node [anchor=north west][inner sep=0.75pt]    {$b$};
\draw (252.52,97) node [anchor=north west][inner sep=0.75pt]   [align=left] {$\displaystyle a$};
\draw (251.93,145) node [anchor=north west][inner sep=0.75pt]   [align=left] {$\displaystyle d$};
\draw (222.14,75.91) node [anchor=north west][inner sep=0.75pt]    {$1$};
\draw (280.84,163.83) node [anchor=north west][inner sep=0.75pt]    {$1$};
\draw (315.82,138.68) node [anchor=north west][inner sep=0.75pt]    {$2$};
\draw (192.26,102.9) node [anchor=north west][inner sep=0.75pt]    {$2$};
\draw (191.11,137.51) node [anchor=north west][inner sep=0.75pt]    {$3$};
\draw (224.74,163.31) node [anchor=north west][inner sep=0.75pt]    {$4$};
\draw (277.89,78.26) node [anchor=north west][inner sep=0.75pt]    {$3$};
\draw (312.95,102.9) node [anchor=north west][inner sep=0.75pt]    {$4$};

\draw (312.95,102.9) node [anchor=north west][inner sep=0.75pt]    {$4$};



\end{tikzpicture}
}}
$};

\draw (-70,106.5) node [anchor=north west][inner sep=0.75pt]   [align=left] {$\displaystyle I_{14}:=$};
\draw (80,106.5) node [anchor=north west][inner sep=0.75pt]   [align=left] {$\displaystyle \times \phantom{ j} x_{1,2}^2 x_{1,3}^2 x_{2,4}^2 x_{3,4}^2 x_{a,d}^2 x_{b,c}^2\ ,$};
\draw (-80,216.5) node [anchor=north west][inner sep=0.75pt]   [align=left] {$\displaystyle I_{30}:=$};
\draw (90,216.5) node [anchor=north west][inner sep=0.75pt]   [align=left] {$\displaystyle \times \phantom{ j} x_{1,2}^2 x_{3,4}^2 x_{1,3}^2 x_{1,4}^2 x_{b,c}^4\ ,$};

\draw (-20,290) node [anchor=north west][inner sep=0.75pt]   [align=left] {$\displaystyle I_{14}^\prime:=\left.I_{14}\right|_{3\leftrightarrow4}, \quad I_{30}^\prime:=\left.I_{30}\right|_{1\leftrightarrow2 } \,.$};

\draw  [line width=0.75] (-20,68) node [anchor=north west][inner sep=0.75pt]  [align=center] {$ \hbox{\scalebox{0.75}{\begin{tikzpicture}[x=0.75pt,y=0.75pt,yscale=-1,xscale=1]

\draw   [line width=0.75] (303.27,175.42) -- (267.9,175.42) -- (242.9,150.41) -- (242.9,115.05) -- (267.9,90.05) -- (303.27,90.05) -- (328.27,115.05) -- (328.27,150.41) -- cycle ;
\draw   [line width=0.75]  (243.07,132.735) -- (328.11,132.735) ;
\draw   [line width=0.75]  (285.585,89.56) -- (285.585,175.14) ;
\draw   [line width=0.75]  (343.19,115.05) -- (328.27,115.05) ;
\draw  [line width=0.75]   (343.19,150.41) -- (328.27,150.41) ;
\draw   [line width=0.75]  (242.9,115.05) -- (227.98,115.05) ;
\draw  [line width=0.75]   (242.9,150.41) -- (227.98,150.41) ;
\draw   [line width=0.75]  (303.27,90.05) -- (303.27,75.13) ;
\draw   [line width=0.75]  (267.9,90.05) -- (267.9,75.13) ;
\draw   [line width=0.75]  (303.27,190.33) -- (303.27,175.42) ;
\draw   [line width=0.75]  (267.9,190.33) -- (267.9,175.42) ;

\draw (263.37,107.41) node [anchor=north west][inner sep=0.75pt]   [align=left] {$\displaystyle a$};
\draw (298.08,107.41) node [anchor=north west][inner sep=0.75pt]   [align=left] {$\displaystyle c$};
\draw (263.37,144.59) node [anchor=north west][inner sep=0.75pt]   [align=left] {$\displaystyle b$};
\draw (298.08,144.59) node [anchor=north west][inner sep=0.75pt]   [align=left] {$\displaystyle d$};

\draw (245,85) node [anchor=north west][inner sep=0.75pt]   [align=left] {$\displaystyle 1$};
\draw (317,85) node [anchor=north west][inner sep=0.75pt]   [align=left] {$\displaystyle 2$};
\draw (245,165) node [anchor=north west][inner sep=0.75pt]   [align=left] {$\displaystyle 3$};
\draw (317,165) node [anchor=north west][inner sep=0.75pt]   [align=left] {$\displaystyle 4$};

\draw (281,73) node [anchor=north west][inner sep=0.75pt]   [align=left] {$\displaystyle 3$};
\draw (281,180) node [anchor=north west][inner sep=0.75pt]   [align=left] {$\displaystyle 2$};

\draw (281,180) node [anchor=north west][inner sep=0.75pt]   [align=left] {$\displaystyle 2$};

\draw (230,125.05) node [anchor=north west][inner sep=0.75pt]   [align=left] {$\displaystyle 4$};
\draw (332,125.05) node [anchor=north west][inner sep=0.75pt]   [align=left] {$\displaystyle 1$};



\end{tikzpicture}}}
$};



\end{tikzpicture}

}
}}
\end{eqnarray}
Their propagator structures coincide with those of our $W$ and $\Theta$ topologies. Note that $I_{30}$ contains two elliptic cuts, namely the $\mu_s$-cut and $\mu_u$-cut.  Each $f$-graph integrand yields the same functional form on the elliptic cut, with the final one-dimensional integral given by
\begin{equation}
    \begin{split}
      4 \times \frac{1}{s t u}  \frac{  s}{\Delta \sqrt{(t- s X)(4uX^2-s  X+t)}} d X\,.
    \end{split}
\end{equation}
From the chamber basis, the $\mu_s$-cut arises from a single integrand, $I_G^{12;34}$, which is present in chambers $r_4$ and $r_6$. Summing over these two chambers and their branches yields complete agreement:
\begin{equation}
    \begin{split}
        &\frac{1}{2}\left(\sum_{\pm}\sum_{i=4,6}\omega_{r_i}^\pm \, \Delta^2 \right)\frac{1}{ \Delta }\frac{4 }{\sqrt{( s X-t)(4 u X^2-s X+t)}} d X\Bigg|_{\langle \! \langle \mathbf{X}_1, \mathbf{X}_2, \mathbf{X}_3, \mathbf{X}_4\rangle \! \rangle^4\rightarrow 1}\\
        =& \frac{1}{t u \Delta} \frac{4 }{\sqrt{(s X-t)(4uX^2-s X+t)}} d X .
    \end{split}
\end{equation}
For the $\nu_s$-cut, the integrand in the $f$-graphs contributing to this cut is
\begin{equation}
    I_{14}^{\prime\prime}:=\left.I_{14}\right|_{2\leftrightarrow4}\   \Rightarrow \  \left.I_{14}^{\prime\prime}\right|_{\nu_s}=0.
\end{equation}
Hence, both the $f$-graph basis and the chamber basis vanish on the $\nu_s$-cut.

\paragraph{ Leading singularities:}
We can also compute directly the possible leading singularities of the 32 four-loop conformal integrands arising from the three $f$-graphs (see~\cite{He:2025vqt}).  Below we briefly summarize the results, which agree with our chamber analysis.\footnote{We thank Xuhang Jiang for performing the detailed analysis of all possible LS of the $32$ conformal integrands and for sharing the unpublished results with us.} We find that all leading singularities can be reproduced as linear combinations of the chamber forms. However, individual integrands often produce leading singularities that do not lie within the space of chamber forms; we refer to these as \textit{spurious} leading singularities. As expected, they cancel in the sum. In the following, we present a few examples. 

Consider the cut:
\begin{eqnarray}\label{eq: spurious cut 2}
&&\vcenter{\hbox{\scalebox{1}{
\begin{tikzpicture}[x=0.75pt,y=0.75pt,yscale=-1,xscale=1]

\draw  [line width=0.75]   (224.62,158.65) -- (239.3,173.33) ;
\draw  [line width=0.75]   (368.98,205.75) -- (383.66,220.43) ;
\draw  [line width=0.75]   (383.66,158.65) -- (368.98,173.33) ;
\draw  [line width=0.75]   (239.3,205.75) -- (224.62,220.43) ;
\draw  [line width=0.75]  (239.3,173.33) -- (271.72,173.33) -- (271.72,205.75) -- (239.3,205.75) -- cycle ;
\draw  [line width=0.75]  (271.72,173.33) -- (304.14,173.33) -- (304.14,205.75) -- (271.72,205.75) -- cycle ;
\draw  [line width=0.75]  (304.14,173.33) -- (336.56,173.33) -- (336.56,205.75) -- (304.14,205.75) -- cycle ;
\draw  [line width=0.75]  (336.56,173.33) -- (368.98,173.33) -- (368.98,205.75) -- (336.56,205.75) -- cycle ;
\draw   [line width=0.75]  (271.72,156.5) -- (271.72,173.33) ;
\draw   [line width=0.75]  (304.14,205.75) -- (304.14,222.59) ;
\draw   [line width=0.75]  (336.56,156.5) -- (336.56,173.33) ;
\draw [line width=0.75] [color=red  ,draw opacity=1 ]   (253.5,168) -- (253.5,177.67) ;
\draw [line width=0.75] [color=red  ,draw opacity=1 ]   (286.83,168.67) -- (286.83,178.33) ;
\draw [line width=0.75] [color=red  ,draw opacity=1 ]   (318.83,168.67) -- (318.83,178.33) ;
\draw [line width=0.75] [color=red  ,draw opacity=1 ]   (351.83,168.67) -- (351.83,178.33) ;
\draw [line width=0.75] [color=red  ,draw opacity=1 ]   (351.83,201) -- (351.83,210.67) ;
\draw [line width=0.75]  [color=red  ,draw opacity=1 ]   (319.17,201.33) -- (319.17,211) ;
\draw [line width=0.75] [color=red  ,draw opacity=1 ]   (286.5,202) -- (286.5,211.67) ;
\draw [line width=0.75] [color=red  ,draw opacity=1 ]   (253.5,201) -- (253.5,210.67) ;
\draw [line width=0.75] [color=red  ,draw opacity=1 ]   (244.67,190.17) -- (235,190.17) ;
\draw [line width=0.75] [color=red  ,draw opacity=1 ]   (276.67,189.17) -- (267,189.17) ;
\draw [line width=0.75] [color=red  ,draw opacity=1 ]   (276.67,192.17) -- (267,192.17) ;

\draw [line width=0.75] [color=red  ,draw opacity=1 ]   (373.67,190.17) -- (364,190.17) ;
\draw [line width=0.75] [color=red  ,draw opacity=1 ]   (308.67,188.17) -- (299,188.17) ;
\draw [line width=0.75] [color=red  ,draw opacity=1 ]   (308.67,191.17) -- (299,191.17) ;

\draw [line width=0.75] [color=red  ,draw opacity=1 ]   (341.67,188.17) -- (332,188.17) ;
\draw [line width=0.75] [color=red  ,draw opacity=1 ]   (341.67,191.17) -- (332,191.17) ;

\draw (221.37,181.74) node [anchor=north west][inner sep=0.75pt]   [align=left] {$\displaystyle 4$};
\draw (249,184.41) node [anchor=north west][inner sep=0.75pt]   [align=left] {$\displaystyle a$};
\draw (284.05,182) node [anchor=north west][inner sep=0.75pt]   [align=left] {$\displaystyle b$};
\draw (315.12,185) node [anchor=north west][inner sep=0.75pt]   [align=left] {$\displaystyle c$};
\draw (347.79,181.53) node [anchor=north west][inner sep=0.75pt]   [align=left] {$\displaystyle d$};
\draw (248.37,154.17) node [anchor=north west][inner sep=0.75pt]   [align=left] {$\displaystyle 1$};
\draw (345.53,154.17) node [anchor=north west][inner sep=0.75pt]   [align=left] {$\displaystyle 2$};
\draw (331.2,210.17) node [anchor=north west][inner sep=0.75pt]   [align=left] {$\displaystyle 1$};
\draw (298.87,154.17) node [anchor=north west][inner sep=0.75pt]   [align=left] {$\displaystyle 3$};
\draw (375.87,181.24) node [anchor=north west][inner sep=0.75pt]   [align=left] {$\displaystyle 4$};
\draw (265.2,210.91) node [anchor=north west][inner sep=0.75pt]   [align=left] {$\displaystyle 2$};

\end{tikzpicture}

}
}}
\end{eqnarray}
The following integrands from the $f$-graph contribute to it:
\begin{eqnarray}
&&\vcenter{\hbox{\scalebox{1}{
\begin{tikzpicture}[x=0.75pt,y=0.75pt,yscale=-1,xscale=1]





\draw (-35,217) node [anchor=north west][inner sep=0.75pt]   [align=left] {$\displaystyle I_{7}:=$};

\draw (-35,310.5) node [anchor=north west][inner sep=0.75pt]   [align=left] {$\displaystyle I_{13}:=$};

\draw (135,213.5) node [anchor=north west][inner sep=0.75pt]   [align=left] {$\displaystyle \times \phantom{j} x_{1,2}^2 x_{1,3}^2  x_{3,4}^2 x_{c,2}^2 x_{a,d}^2$};

\draw (145,307.5) node [anchor=north west][inner sep=0.75pt]   [align=left] {$\displaystyle \times \phantom{j} x_{1,2}^4 x_{3,4}^2 x_{1,3}^2 x_{b,d}^2$};

\draw (20,370) node [anchor=north west][inner sep=0.75pt]   [align=left] {$\displaystyle I_{7}^\prime:=\left.I_{7}\right|_{1\leftrightarrow2}, \quad I_{13}^\prime:=\left.I_{13}\right|_{1\leftrightarrow2 } \,.$};

\draw  [line width=0.75] (10,174) node [anchor=north west][inner sep=0.75pt]  [align=center] {$ \hbox{\scalebox{0.75}{\begin{tikzpicture}[x=0.75pt,y=0.75pt,yscale=-1,xscale=1]

\draw  [line width=0.75] (100,100) node [anchor=north west][inner sep=0.75pt]  [align=center] {$ \begin{tikzpicture}[x=0.75pt,y=0.75pt,yscale=-1,xscale=1]

\draw  [line width=0.75]  (362.83,235.28) -- (374.66,247.12) ;
\draw   [line width=0.75] (253.28,235.33) -- (241.45,247.17) ;
\draw  [line width=0.75] (318.05,145.72) -- (362.83,145.72) -- (362.83,190.5) -- (318.05,190.5) -- cycle ;
\draw   [line width=0.75] (241.45,133.88) -- (253.28,145.72) ;
\draw  [line width=0.75]  (374.66,133.88) -- (362.83,145.72) ;
\draw  [line width=0.75] (318.05,190.5) -- (362.83,190.5) -- (362.83,235.28) -- (318.05,235.28) -- cycle ;
\draw [line width=0.75]  (285.67,145.72) -- (318.05,145.72) -- (318.05,235.33) -- (285.67,235.33) -- cycle ;
\draw  [line width=0.75] (253.28,145.72) -- (285.67,145.72) -- (285.67,235.33) -- (253.28,235.33) -- cycle ;
\draw  [line width=0.75]  (362.83,170.5) -- (380.78,170.5) ;
\draw  [line width=0.75]  (235.45,190.5) -- (253.28,190.5) ;

\draw (235,160.5) node [anchor=north west][inner sep=0.75pt]   [align=left] {$\displaystyle 2$};
\draw (235,210.5) node [anchor=north west][inner sep=0.75pt]   [align=left] {$\displaystyle 4$};
\draw (265.06,183.29) node [anchor=north west][inner sep=0.75pt]   [align=left] {$\displaystyle d$};
\draw (299.09,185) node [anchor=north west][inner sep=0.75pt]   [align=left] {$\displaystyle c$};
\draw (333.72,162.65) node [anchor=north west][inner sep=0.75pt]   [align=left] {$\displaystyle a$};
\draw (334.85,206.48) node [anchor=north west][inner sep=0.75pt]   [align=left] {$\displaystyle b$};
\draw (302.18,126.24) node [anchor=north west][inner sep=0.75pt]   [align=left] {$\displaystyle 1$};
\draw (302.19,238.16) node [anchor=north west][inner sep=0.75pt]   [align=left] {$\displaystyle 3$};
\draw (370.78,147.69) node [anchor=north west][inner sep=0.75pt]   [align=left] {$\displaystyle 4$};
\draw (370.78,195) node [anchor=north west][inner sep=0.75pt]   [align=left] {$\displaystyle 2$};

\end{tikzpicture}$};

\end{tikzpicture}
}}
$};

\draw  [line width=0.75] (10,288) node [anchor=north west][inner sep=0.75pt]  [align=center] {$ \hbox{\scalebox{0.75}{\begin{tikzpicture}[x=0.75pt,y=0.75pt,yscale=-1,xscale=1]

\draw  [line width=0.75] (100,100) node [anchor=north west][inner sep=0.75pt]  [align=center] {$ \begin{tikzpicture}[x=0.75pt,y=0.75pt,yscale=-1,xscale=1]

\draw   [line width=0.75] (224.62,158.65) -- (239.3,173.33) ;
\draw  [line width=0.75]  (368.98,205.75) -- (383.66,220.43) ;
\draw  [line width=0.75]  (383.66,158.65) -- (368.98,173.33) ;
\draw  [line width=0.75]  (239.3,205.75) -- (224.62,220.43) ;
\draw  [line width=0.75] (239.3,173.33) -- (271.72,173.33) -- (271.72,205.75) -- (239.3,205.75) -- cycle ;
\draw [line width=0.75]  (271.72,173.33) -- (304.14,173.33) -- (304.14,205.75) -- (271.72,205.75) -- cycle ;
\draw  [line width=0.75] (304.14,173.33) -- (336.56,173.33) -- (336.56,205.75) -- (304.14,205.75) -- cycle ;
\draw  [line width=0.75] (336.56,173.33) -- (368.98,173.33) -- (368.98,205.75) -- (336.56,205.75) -- cycle ;
\draw  [line width=0.75]  (353.56,205.75) -- (353.56,222.59) ;
\draw   [line width=0.75]  (288.72,205.75) -- (288.72,222.59) ;

\draw (300,153.74) node [anchor=north west][inner sep=0.75pt]   [align=left] {$\displaystyle 1$};
\draw (249,185.53) node [anchor=north west][inner sep=0.75pt]   [align=left] {$\displaystyle a$};
\draw (284.05,183.53) node [anchor=north west][inner sep=0.75pt]   [align=left] {$\displaystyle b$};
\draw (315.12,185.53) node [anchor=north west][inner sep=0.75pt]   [align=left] {$\displaystyle c$};
\draw (347.79,183.53) node [anchor=north west][inner sep=0.75pt]   [align=left] {$\displaystyle d$};
\draw (260.7,208.5) node [anchor=north west][inner sep=0.75pt]   [align=left] {$\displaystyle 2$};
\draw (315,208.5) node [anchor=north west][inner sep=0.75pt]   [align=left] {$\displaystyle 3$};
\draw (357,208.5) node [anchor=north west][inner sep=0.75pt]   [align=left] {$\displaystyle 2$};
\draw (375.87,181.24) node [anchor=north west][inner sep=0.75pt]   [align=left] {$\displaystyle 4$};
\draw (225.53,181.24) node [anchor=north west][inner sep=0.75pt]   [align=left] {$\displaystyle 4$};

\end{tikzpicture}$};

\end{tikzpicture}}}
$};



\end{tikzpicture}

}
}}
\end{eqnarray}
Their propagator structures coincide with those of our $Z$ and $N$ topologies.

Let us show how the cuts are performed for these two integrands separately.
For $I_{7}$, we first cut propagators $x_{b,2}^{2},\, x_{b,3}^2, \, x_{a,b}^{2},\, x_{b,c}^2$ and arrive at
    \begin{equation}
        \frac{x_{1,2}^2 x_{1,3}^2 x_{3,4}^2 x_{c,2}^2   x_{a,d}^2}{x_{a,1}^2 x_{a,2}^2 x_{a,4}^2  x_{c,1}^2 x_{c,3}^2   x_{d,1}^2  x_{d,2}^2 x_{3,d}^2
  x_{d,4}^2 x_{a,c}^2  x_{c,d}^2  \lambda_{2,3,a,c}}\,,
    \end{equation}
where we have introduced  Gram-determinant Jacobians, $\lambda_{i,j,k,l}^2:={\rm det} |x^2_{a,b}|_{a,b=i,j,k,l}$. Next, cutting $x_c$ on $x_{c,1}^{2},x_{c,3}^{2},x_{c,d}^{2},\lambda_{2,3,a,c}$ where $\lambda_{2,3,a,c}$ simplifies to $x_{c,2}^{2}x_{a,3}^{2}-x_{2,3}^{2}x_{a,c}^{2}$. The remaining integrand is
    \begin{equation}
        \frac{x_{1,2}^2 x_{1,3}^2 x_{2,3}^2 x_{3,4}^2  x_{a,d}^2}{  x_{a,1}^2 x_{a,2}^2 x_{a,3}^2 x_{a,4}^2   x_{d,1}^2 x_{d,2}^2 x_{d,3}^2 x_{d,4}^2  \lambda_{1,3,d,\{a2\}}}.
    \end{equation}
where we have used $x_{c,2}^{2}x_{a,3}^{2}-x_{2,3}^{2}x_{a,c}^{2}=0$ to eliminate $x_{a,c}^{2}$ in the denominator, and $\lambda_{1,3,d,\{a2\}}$ is the Jacobian corresponding to this cut. 
    By cutting $x_{d,1}^2, x_{d,2}^{2},x_{d,4}^{2},\lambda_{1,3,d,\{a2\}}$ where $\lambda_{1,3,d,\{a2\}}$ simplifies to
$(x_{2,3}^{2}x_{a,1}^{2}-x_{1,2}^{2}x_{a,3}^{2})x_{d,3}^{2}-x_{2,3}^{2}x_{1,3}^{2}x_{a,d}^{2}$ the integrand becomes 
    \begin{equation}
        \frac{x_{1,2}^2 x_{3,4}^2(x_{2,3}^{2}x_{a,1}^{2}-x_{1,2}^{2}x_{a,3}^{2})}{x_{a,1}^2  x_{a,2}^2 x_{a,3}^2 x_{a,4}^2  \lambda_{1,2,4,\{a3\}} }\,,
    \end{equation}
    where $\lambda_{1,2,4,\{a3\}}$ is the Jacobian for this cut of $x_d$. Then after the last cut for $x_a$: $x_{a,1}^{2}$, $x_{a,2}^{2}$, $x_{a,4}^{2}$,  $\lambda_{1,2,4,\{a3\}}$ where $\lambda_{1,2,4\{a3\}}$ simplifies $x_{1,2}^{2}x_{a,3}^{2}(x_{2,3}^{2}x_{1,4}^{2}+x_{2,4}^{2}x_{1,3}^{2}-x_{1,2}^{2}x_{3,4}^{2})$ we obtain    \begin{equation}\label{eq:I7final}
        I^{final}_{7}=-\frac{s}{t+u-s }\frac{1}{{ \Delta}}.
    \end{equation}
Similarly we perform the computation for $I_{13}$, first with the following cuts:
    \begin{equation}
    \text{cut: } x_{a,1}^2,\, x_{a,2}^2,\, x_{a,4}^2,\,   x_{a,b}^2,\, x_{c,1}^2,\,  x_{c,3}^2,\, x_{c,d}^2,\, x_{b,c}^2,
    \end{equation}
which produces two Jacobians $\lambda_{1,2,4,b}$ and $\lambda_{243b}$; 
next we cut $x_{b,2}^{2},x_{b,3}^{2},\lambda_{1,2,4,b},\lambda_{2,3,4,b}$ where the two Jacobians simplify to $\lambda_{1,2,4,b}=x_{1,2}^{2}x_{b,4}^{2}-x_{2,4}^{2}x_{b,1}^{2}$,  $\lambda_{1,3,b,d}=x_{b,1}^{2}x_{d,3}^{2}-x_{1,3}^{2}x_{b,d}^{2}$, and the integrand becomes
\begin{equation}
        \frac{x_{1,2}^4 x_{3,4}^2 }{x_{d,2}^2 x_{d,4}^2 x_{d,1}^2 \lambda_{2,3\{14\}\{d1\}}}.
    \end{equation}
Finally we cut 
$x_{d,1}^{2},x_{d,2}^{2},x_{d,4}^{2},\lambda_{2,3,\{14\},\{d1\}}$ where $\lambda_{2,3,\{14\},\{d1\}}$ simplifies
thus we obtain  
\begin{equation}\label{eq:I13final}
        I^{final}_{13}=\frac{s}{t+u-s }\frac{1}{{ \Delta}}.
    \end{equation}
Note that this \textit{spurious} leading singularity cannot be constructed from linear combinations of chamber forms. As we will see, their contributions cancel in the end. Performing the same analysis for the remaining permuted integrands, we find that under the cut~\eqref{eq: spurious cut 2}
\begin{equation}
    -I_{7}\big|_{\eqref{eq: spurious cut 2}}=-I_{7}^\prime\big|_{\eqref{eq: spurious cut 2}}=I_{13}\big|_{\eqref{eq: spurious cut 2}}=I_{13}^\prime\big|_{\eqref{eq: spurious cut 2}}=\frac{s}{t+u-s }\frac{1}{{ \Delta}}.
\end{equation}
Therefore, the net contribution vanishes
\begin{equation}
    I_{7}\big|_{\eqref{eq: spurious cut 2}}+I_{7}^\prime\big|_{\eqref{eq: spurious cut 2}}+I_{13}\big|_{\eqref{eq: spurious cut 2}}+I_{13}^\prime\big|_{\eqref{eq: spurious cut 2}}=0.
\end{equation}

As another example, consider the cut:
\begin{eqnarray}\label{eq: spurious cut 3}
&&\vcenter{\hbox{\scalebox{1}{
\begin{tikzpicture}[x=0.75pt,y=0.75pt,yscale=-1,xscale=1]


\draw  [line width=0.75] (320,250) node [anchor=north west][inner sep=0.75pt]  [align=center] {$\begin{tikzpicture}[x=0.75pt,y=0.75pt,yscale=-1,xscale=1]

\draw    [line width=0.75] (224.62,158.65) -- (239.3,173.33) ;
\draw    [line width=0.75] (368.98,205.75) -- (383.66,220.43) ;
\draw    [line width=0.75] (383.66,158.65) -- (368.98,173.33) ;
\draw    [line width=0.75] (239.3,205.75) -- (224.62,220.43) ;
\draw   [line width=0.75] (239.3,173.33) -- (271.72,173.33) -- (271.72,205.75) -- (239.3,205.75) -- cycle ;
\draw   [line width=0.75] (271.72,173.33) -- (304.14,173.33) -- (304.14,205.75) -- (271.72,205.75) -- cycle ;
\draw   [line width=0.75] (304.14,173.33) -- (336.56,173.33) -- (336.56,205.75) -- (304.14,205.75) -- cycle ;
\draw   [line width=0.75] (336.56,173.33) -- (368.98,173.33) -- (368.98,205.75) -- (336.56,205.75) -- cycle ;
\draw    [line width=0.75] (304.14,156.5) -- (304.14,173.33) ;
\draw    [line width=0.75] (271.72,205.75) -- (271.72,222.59) ;
\draw    [line width=0.75] (304.14,205.75) -- (304.14,222.59) ;
\draw    [line width=0.75] (336.56,205.75) -- (336.56,222.59) ;

\draw  [line width=0.75][color=red  ,draw opacity=1 ]   (253.5,168) -- (253.5,177.67) ;
\draw  [line width=0.75][color=red  ,draw opacity=1 ]   (286.83,168.67) -- (286.83,178.33) ;
\draw  [line width=0.75][color=red  ,draw opacity=1 ]   (318.83,168.67) -- (318.83,178.33) ;
\draw  [line width=0.75][color=red  ,draw opacity=1 ]   (351.83,168.67) -- (351.83,178.33) ;
\draw  [line width=0.75][color=red  ,draw opacity=1 ]   (351.83,201) -- (351.83,210.67) ;
\draw  [line width=0.75][color=red  ,draw opacity=1 ]   (319.17,201.33) -- (319.17,211) ;
\draw  [line width=0.75][color=red  ,draw opacity=1 ]   (286.5,202) -- (286.5,211.67) ;
\draw  [line width=0.75][color=red  ,draw opacity=1 ]   (253.5,201) -- (253.5,210.67) ;
\draw  [line width=0.75][color=red  ,draw opacity=1 ]   (244.67,190.17) -- (235,190.17) ;
\draw  [line width=0.75][color=red  ,draw opacity=1 ]   (276.67,189.17) -- (267,189.17) ;
\draw  [line width=0.75][color=red  ,draw opacity=1 ]   (276.67,192.17) -- (267,192.17) ;

\draw  [line width=0.75][color=red  ,draw opacity=1 ]   (373.67,190.17) -- (364,190.17) ;
\draw  [line width=0.75][color=red  ,draw opacity=1 ]   (308.67,188.17) -- (299,188.17) ;
\draw  [line width=0.75][color=red  ,draw opacity=1 ]   (308.67,191.17) -- (299,191.17) ;

\draw  [line width=0.75][color=red  ,draw opacity=1 ]   (341.67,188.17) -- (332,188.17) ;
\draw  [line width=0.75][color=red  ,draw opacity=1 ]   (341.67,191.17) -- (332,191.17) ;

\draw (331.53,155) node [anchor=north west][inner sep=0.75pt]   [align=left] {$\displaystyle 1$};
\draw (267.37,155) node [anchor=north west][inner sep=0.75pt]   [align=left] {$\displaystyle 2$};
\draw (284,213) node [anchor=north west][inner sep=0.75pt]   [align=left] {$\displaystyle 4$};
\draw (249,183.41) node [anchor=north west][inner sep=0.75pt]   [align=left] {$\displaystyle d$};
\draw (284.05,183.25) node [anchor=north west][inner sep=0.75pt]   [align=left] {$\displaystyle b$};
\draw (315.12,185.86) node [anchor=north west][inner sep=0.75pt]   [align=left] {$\displaystyle a$};
\draw (347.79,185.53) node [anchor=north west][inner sep=0.75pt]   [align=left] {$\displaystyle c$};
\draw (250.7,213) node [anchor=north west][inner sep=0.75pt]   [align=left] {$\displaystyle 3$};
\draw (315.2,213) node [anchor=north west][inner sep=0.75pt]   [align=left] {$\displaystyle 3$};
\draw (347.87,213) node [anchor=north west][inner sep=0.75pt]   [align=left] {$\displaystyle 4$};
\draw (375.87,183) node [anchor=north west][inner sep=0.75pt]   [align=left] {$\displaystyle 2$};
\draw (222,183) node [anchor=north west][inner sep=0.75pt]   [align=left] {$\displaystyle 1$};

\end{tikzpicture}
$};

\end{tikzpicture}

}
}}
\end{eqnarray}
The two integrands contributing to this cut are $I_{14}^\prime$ and
\begin{eqnarray}
&&\vcenter{\hbox{\scalebox{1}{
\begin{tikzpicture}[x=0.75pt,y=0.75pt,yscale=-1,xscale=1]





\draw (-35,195) node [anchor=north west][inner sep=0.75pt]   [align=left] {$\displaystyle I_{10}=$};
\draw (140,191) node [anchor=north west][inner sep=0.75pt]   [align=left] {$\displaystyle \times\phantom{ j} x_{1,4}^2 x_{2,3}^2 x_{3,4}^2 x_{a,2}^2 x_{b,1}^2$\,.};

\draw  [line width=0.75] (10,174) node [anchor=north west][inner sep=0.75pt]  [align=center] {$ \hbox{\scalebox{0.75}{\begin{tikzpicture}[x=0.75pt,y=0.75pt,yscale=-1,xscale=1]

\draw   [line width=0.75] (224.62,158.65) -- (239.3,173.33) ;
\draw  [line width=0.75]  (368.98,205.75) -- (383.66,220.43) ;
\draw  [line width=0.75]  (383.66,158.65) -- (368.98,173.33) ;
\draw  [line width=0.75]  (239.3,205.75) -- (224.62,220.43) ;
\draw  [line width=0.75] (239.3,173.33) -- (271.72,173.33) -- (271.72,205.75) -- (239.3,205.75) -- cycle ;
\draw [line width=0.75]  (271.72,173.33) -- (304.14,173.33) -- (304.14,205.75) -- (271.72,205.75) -- cycle ;
\draw  [line width=0.75] (304.14,173.33) -- (336.56,173.33) -- (336.56,205.75) -- (304.14,205.75) -- cycle ;
\draw  [line width=0.75] (336.56,173.33) -- (368.98,173.33) -- (368.98,205.75) -- (336.56,205.75) -- cycle ;
\draw  [line width=0.75]  (320.56,156.5) -- (320.56,173.33) ;
\draw   [line width=0.75]  (288.72,205.75) -- (288.72,222.59) ;

\draw (287.37,153.74) node [anchor=north west][inner sep=0.75pt]   [align=left] {$\displaystyle 3$};
\draw (249,183.53) node [anchor=north west][inner sep=0.75pt]   [align=left] {$\displaystyle d$};
\draw (284.05,183.53) node [anchor=north west][inner sep=0.75pt]   [align=left] {$\displaystyle b$};
\draw (315.12,186.53) node [anchor=north west][inner sep=0.75pt]   [align=left] {$\displaystyle a$};
\draw (347.79,186.53) node [anchor=north west][inner sep=0.75pt]   [align=left] {$\displaystyle c$};
\draw (260.7,208.5) node [anchor=north west][inner sep=0.75pt]   [align=left] {$\displaystyle 2$};
\draw (341.53,153.41) node [anchor=north west][inner sep=0.75pt]   [align=left] {$\displaystyle 1$};
\draw (327.2,208.5) node [anchor=north west][inner sep=0.75pt]   [align=left] {$\displaystyle 4$};
\draw (375.87,181.24) node [anchor=north west][inner sep=0.75pt]   [align=left] {$\displaystyle 2$};
\draw (225.53,182.91) node [anchor=north west][inner sep=0.75pt]   [align=left] {$\displaystyle 1$};

\end{tikzpicture}
}}
$};



\end{tikzpicture}

}
}}
\end{eqnarray}
Following the same procedure as before, we obtain a spurious leading singularity with the opposite sign
\begin{equation}
    \pm \frac{s}{\Delta \sqrt{(s-t)^2- 4 t u}}
\end{equation}
Thus, these terms cancel in the total sum 
\begin{equation}
    I_{10}\big|_{\eqref{eq: spurious cut 3}}+I_{14}^\prime\big|_{\eqref{eq: spurious cut 3}}=0.
\end{equation}

\section{Conclusion and Discussions}

In this paper, we have examined the chamber dissection of loop geometries of the Correlahedron, which reproduces the integrand of the four-point stress–energy correlator in planar SYM—at least up to four loops—as a sum of universal chamber forms (or leading singularities) multiplied by loop forms (or loop integrands) whose leading singularities are unity. We have shown that the four-loop case exhibits precisely the same chamber forms as those appearing at lower loop orders, suggesting that these chamber forms constitute the all-loop leading singularities of the correlator integrand.

Remarkably, the loop form associated with each chamber is expressible in terms of local integrands that are diagonal in the space of leading singularities. More concretely, the components of each chamber loop form organize into blocks, each carrying only one nontrivial leading singularity. This structure implies that the integrated answer is naturally expressed in terms of pure functions, as we explicitly demonstrate at three loops. The four-loop case exhibits additional richness due to the emergence of elliptic integrals. Nonetheless, we find that this diagonalization persists even in the presence of elliptic cuts: every integrand carries either a single leading singularity or an elliptic cut. Equivalently, the four-loop result decomposes cleanly into contributions expected to integrate to pure functions and contributions that are purely elliptic, already visible at the level of local integrands. Furthermore, the elliptic behavior is not universal across all kinematic configurations: each elliptic cut appears only in a subset of chambers. Although the absence of a definition for canonical forms with elliptic boundaries prevents us from determining the precise kinematic normalization of the three elliptic integrands ($I_G^{12;34}$ and its two images under $2\leftrightarrow 3$ and $2\leftrightarrow 4$), the fact that the elliptic cut resides in the $A_{\sigma_i}$ block of eq.~\eqref{eq: ChamberBlocks} suggests that it should be normalized by $\Delta^2$. Understanding the geometric origin of this factor is an interesting open question.

There are several natural directions for further research. We outline a few of them below.

\paragraph{Evaluating or bootstrapping the chamber integrals}
Given that the four-loop result is now organized into integrands expected to integrate to pure functions, these integrals present an attractive target for direct evaluation or bootstrap methods. In particular, the integrals with unit leading singularities are linear combinations of well-studied conformal integrals, including the 32 $f$-graph integrals and additional ones (analogous to $E'$ at three loops). Many such integrals can be evaluated directly (see~\cite{He:2025vqt}), and one can verify that they indeed assemble into weight-8 SVMPL pure functions. A significant subset, however, remains unknown because their kinematic limits and boundary behaviors are not yet systematically understood. Completing the evaluation of these pure functions—especially the elliptic ones—thus remains an important task for future work.

Alternatively, a compelling approach is to bootstrap the correlator’s final answer directly, representing it as chamber forms multiplied by an ansatz of pure functions and constraining this ansatz using physical data, thereby avoiding explicit computation of each individual integral. Even at three loops, it would be illuminating to attempt such a bootstrap without first determining the explicit forms of ${\cal E}$ and ${\cal H}_a$. A more refined understanding of the analytic structure and kinematic limits of these correlators and their pure integrals would greatly facilitate this program.

\paragraph{Interpretation of chambers from positroid cells}
The fact that the chamber structure at four loops coincides with that at three loops indicates that this pattern likely persists at all loop orders. A similar phenomenon occurs in the Amplituhedron, where chambers correspond to intersections of cells in the positive Grassmannian. At N$^k$MHV, each chamber is associated with the overlap of $4k$-dimensional cells of $G_{\geq0}(k,n)$. Because only finitely many such cells exist, the number of chambers naturally saturates at sufficiently high loop order. It would be highly interesting to determine whether our chambers admit a similar interpretation in terms of an underlying combinatorial geometry.

\paragraph{Extending Correlahedron geometry beyond the maximal sector}
For higher-point correlators,
\begin{align}
G_n=G_{n,0}+G_{n,1}+\cdots+G_{n,n-4},
\end{align}
where each $G_{n,k}$ is homogeneous of degree $4k$ in the Grassmann variables. The current formulation of the Correlahedron, along with its loop-level construction, captures only the top-degree component $G_{n,n-4}$, corresponding to the maximal sector. A natural and important question is how to extend this geometric framework to the lower-degree sectors.
Following the proposal of~\cite{Eden:2017fow}, one may define the geometry associated with fixed Grassmann degree $k$ as
\begin{equation}\Big\{  Y\in \operatorname{Gr}(n+k,n+k+4),X_i\in  \operatorname{Gr}(2,n+k+4):\  \langle Y X_i X_j\rangle>0  \Big\}.
\end{equation}

A natural starting point is the next-to-maximal sector ($k = n{-}5$). The first nontrivial example occurs for the six-point NMHV case,\footnote{The $G_{n,0}$ components at tree level are special and do not follow the same structure; they take the form $G_{n,0}=\prod_{i=1}^n \frac{y_{i,i+1}^2}{x_{i,i+1}^2}+ (S_n\text{-perm})$, interpretable as a sum over cyclic $n$-gons and their permutations. Therefore, the tree-level six-point NMHV case is the appropriate first example.} whose kinematics is embedded in an 11-dimensional space. Note that in this construction, superconformal Ward identities are trivialized to Schouten identities, see eq.~(4.17) in~\cite{Eden:2017fow}. However, the associated positive geometry lives in $Y\in \operatorname{Gr}(7,11)$ which is 28-dimensional, rendering a direct computation of its canonical form currently impractical. To make progress, one needs to understand how to utilize additional symmetries of the system, such as GL(2) redundancy of the bi-twistors, to effectively reduce the dimension. 

\section*{Acknowledgement}
We thank Dmitry Chicherin, James Drummond, Paul Heslop, Xuhang Jiang,   Jiahao Liu, Yichao Tang,  Kai Yan, and Chi Zhang for their enlightening discussions that contributed to this work.  In particular, Xuhang Jiang provided substantial assistance. The authors
would also like to thank HU  for hosting the conference ``New Ways to Higher Points"  where the results of this paper were discussed. The work of S. He has been supported by the National Natural Science Foundation of China under Grant No. 12225510, 12447101, 12247103, and by the New Cornerstone Science Foundation through the XPLORER PRIZE. 
Y.-t. Huang is supported by the Taiwan Ministry of Science and Technology Grant No. 112-2628-M-002-003-MY3 and
114-2923-M-002-011-MY5.
C.-K. Kuo is funded by the European Union (ERC, 101118787). Views and opinions expressed are, however, those of the author(s) only and do not necessarily reflect those of the European Union or the European Research Council Executive Agency. Neither the European Union nor the granting authority can be held responsible for them.

\appendix

\section{Three-loop chamber form}\label{sec:3loop original form}
Here we present the chamber loop-form for the region $s>t>u$, organized in terms of the standard easy and hard integrands, which exhibit multiple leading singularities. The result reads:
\begin{equation}
    \begin{split}
         & \Delta\, T^{12;34}+\Delta\,  E^{12;34}_c + 11\, \text{perms.} + \Delta\,  L^{12;34}  + 5\, \text{perms.}+\Delta^2(gh^{12;34} {+}  gh^{34;12}) \\
         +&\Delta^2\big[  \,  H_c^{14;23}+H_c^{23;14}+H_c^{13;24}+H_c^{24;13}\big] +    (t-3u)\Delta [{E^{\prime}}^{12;34} + {E^{\prime}}^{34;12}]
        \\
          + &(s-3u -\Delta )\Delta [{E^{\prime}}^{14;23} + {E^{\prime}}^{23;14}]+ (s-3t-\Delta )\Delta [{E^{\prime}}^{13;24} + {E^{\prime}}^{24;13}]  .
    \end{split}
\end{equation}

\section{Four-loop integrands basis  }\label{sec:four-loop integral basis}
This appendix defines the four-loop conformal integrands used for the chamber loop integrand of $\mathcal{N}=4$ SYM correlator. For simplicity, we only explicitly present the numerators of these integrands, and the denominators can be easily inferred from the corresponding diagram. For instance,  the integral $I_G^{12;34}$  (shown below) is
\begin{equation}
    \begin{split}
        I_G^{12;34}:=&\int_{a,b,c,d} \frac{n_{G}^{12;34}}{x_{a,1}^2 x_{a,3}^2 x_{b,2}^2 x_{b,4}^2 x_{c,2}^2 x_{c,3}^2 x_{d,1}^2 x_{d,4}^2 x_{a,b}^2  x_{a,c}^2 x_{b,d}^2  x_{c,d}^2 }+\text{perms}(a,b,c,d)
\end{split}
\end{equation}
In some cases, the numerator is accompanied by an external permutation, such as $n_{P}^{12;34}$ with $3\leftrightarrow 4$. In these cases, the permutation acts on the full integrand, meaning that the numerator is first divided by its corresponding denominator before applying the external label permutation.
Explicitly, for $I_P^{12;34}$, we have
\begin{equation}
    \begin{split}
        I_P^{12;34} :=\int_{a,b,c,d} & \frac{n_{P}^{12;34}}{x_{a,2}^2 x_{a,4}^2 x_{b,1}^2  x_{c,1}^2  x_{c,2}^2  x_{c,3}^2  x_{d,1}^2  x_{d,3}^2  x_{d,4}^2  x_{a,b}^2  x_{a,c}^2 x_{a,d}^2 x_{b,c}^2 x_{b,d}^2 } +3\leftrightarrow 4\\
        &+\text{perms}(a,b,c,d)
    \end{split}
\end{equation}
We have performed multiple cut checks for the expected purity of these integrands.
In the following, we present the conjectured pure integrand basis.

There are three  ladder topologies
\begin{eqnarray}
&&\vcenter{\hbox{\scalebox{1}{
\begin{tikzpicture}[x=0.75pt,y=0.75pt,yscale=-1,xscale=1]

\draw (-5.97,235) node [anchor=north west][inner sep=0.75pt]    {$N^{12;34}$};
\draw (169,235) node [anchor=north west][inner sep=0.75pt]    {$O^{12;34} $};
\draw (340,235) node [anchor=north west][inner sep=0.75pt]    {${K}^{12;34} $};

\draw  [line width=0.75] (-50,174) node [anchor=north west][inner sep=0.75pt]  [align=center] {$ \hbox{\scalebox{0.75}{\begin{tikzpicture}[x=0.75pt,y=0.75pt,yscale=-1,xscale=1]

\draw  [line width=0.75] (100,100) node [anchor=north west][inner sep=0.75pt]  [align=center] {$ \begin{tikzpicture}[x=0.75pt,y=0.75pt,yscale=-1,xscale=1]

\draw   [line width=0.75] (224.62,158.65) -- (239.3,173.33) ;
\draw  [line width=0.75]  (368.98,205.75) -- (383.66,220.43) ;
\draw  [line width=0.75]  (383.66,158.65) -- (368.98,173.33) ;
\draw  [line width=0.75]  (239.3,205.75) -- (224.62,220.43) ;
\draw  [line width=0.75] (239.3,173.33) -- (271.72,173.33) -- (271.72,205.75) -- (239.3,205.75) -- cycle ;
\draw [line width=0.75]  (271.72,173.33) -- (304.14,173.33) -- (304.14,205.75) -- (271.72,205.75) -- cycle ;
\draw  [line width=0.75] (304.14,173.33) -- (336.56,173.33) -- (336.56,205.75) -- (304.14,205.75) -- cycle ;
\draw  [line width=0.75] (336.56,173.33) -- (368.98,173.33) -- (368.98,205.75) -- (336.56,205.75) -- cycle ;
\draw  [line width=0.75]  (353.56,205.75) -- (353.56,222.59) ;
\draw   [line width=0.75]  (288.72,205.75) -- (288.72,222.59) ;

\draw (300,153.74) node [anchor=north west][inner sep=0.75pt]   [align=left] {$\displaystyle 1$};
\draw (249,183.53) node [anchor=north west][inner sep=0.75pt]   [align=left] {$\displaystyle a$};
\draw (284.05,183.53) node [anchor=north west][inner sep=0.75pt]   [align=left] {$\displaystyle b$};
\draw (315.12,183.53) node [anchor=north west][inner sep=0.75pt]   [align=left] {$\displaystyle c$};
\draw (347.79,183.53) node [anchor=north west][inner sep=0.75pt]   [align=left] {$\displaystyle d$};
\draw (260.7,208.5) node [anchor=north west][inner sep=0.75pt]   [align=left] {$\displaystyle 2$};
\draw (315,208.5) node [anchor=north west][inner sep=0.75pt]   [align=left] {$\displaystyle 3$};
\draw (357,208.5) node [anchor=north west][inner sep=0.75pt]   [align=left] {$\displaystyle 2$};
\draw (375.87,181.24) node [anchor=north west][inner sep=0.75pt]   [align=left] {$\displaystyle 4$};
\draw (225.53,181.24) node [anchor=north west][inner sep=0.75pt]   [align=left] {$\displaystyle 4$};

\end{tikzpicture}$};

\end{tikzpicture}
}}
$};

\draw  [line width=0.75] (120,174) node [anchor=north west][inner sep=0.75pt]  [align=center] {$ \hbox{\scalebox{0.75}{\begin{tikzpicture}[x=0.75pt,y=0.75pt,yscale=-1,xscale=1]

\draw  [line width=0.75] (80,100) node [anchor=north west][inner sep=0.75pt]  [align=center] {$ \begin{tikzpicture}[x=0.75pt,y=0.75pt,yscale=-1,xscale=1]

\draw   [line width=0.75] (224.62,158.65) -- (239.3,173.33) ;
\draw  [line width=0.75]  (368.98,205.75) -- (383.66,220.43) ;
\draw  [line width=0.75]  (383.66,158.65) -- (368.98,173.33) ;
\draw  [line width=0.75]  (239.3,205.75) -- (224.62,220.43) ;
\draw  [line width=0.75] (239.3,173.33) -- (271.72,173.33) -- (271.72,205.75) -- (239.3,205.75) -- cycle ;
\draw [line width=0.75]  (271.72,173.33) -- (304.14,173.33) -- (304.14,205.75) -- (271.72,205.75) -- cycle ;
\draw  [line width=0.75] (304.14,173.33) -- (336.56,173.33) -- (336.56,205.75) -- (304.14,205.75) -- cycle ;
\draw  [line width=0.75] (336.56,173.33) -- (368.98,173.33) -- (368.98,205.75) -- (336.56,205.75) -- cycle ;
\draw  [line width=0.75]  (336.56,156.5) -- (336.56,173.33) ;
\draw   [line width=0.75]  (304.14,205.75) -- (304.14,222.59) ;

\draw (287.37,153.74) node [anchor=north west][inner sep=0.75pt]   [align=left] {$\displaystyle 1$};
\draw (249,183.53) node [anchor=north west][inner sep=0.75pt]   [align=left] {$\displaystyle a$};
\draw (284.05,183.53) node [anchor=north west][inner sep=0.75pt]   [align=left] {$\displaystyle b$};
\draw (315.12,183.53) node [anchor=north west][inner sep=0.75pt]   [align=left] {$\displaystyle c$};
\draw (347.79,183.53) node [anchor=north west][inner sep=0.75pt]   [align=left] {$\displaystyle d$};
\draw (260.7,208.5) node [anchor=north west][inner sep=0.75pt]   [align=left] {$\displaystyle 4$};
\draw (341.53,153.41) node [anchor=north west][inner sep=0.75pt]   [align=left] {$\displaystyle 3$};
\draw (327.2,208.5) node [anchor=north west][inner sep=0.75pt]   [align=left] {$\displaystyle 2$};
\draw (375.87,181.24) node [anchor=north west][inner sep=0.75pt]   [align=left] {$\displaystyle 4$};
\draw (225.53,182.91) node [anchor=north west][inner sep=0.75pt]   [align=left] {$\displaystyle 3$};

\end{tikzpicture}$};

\end{tikzpicture}
}}
$};

\draw  [line width=0.75] (280,174) node [anchor=north west][inner sep=0.75pt]  [align=center] {$ \hbox{\scalebox{0.75}{\begin{tikzpicture}[x=0.75pt,y=0.75pt,yscale=-1,xscale=1]

\draw  [line width=0.75] (100,100) node [anchor=north west][inner sep=0.75pt]  [align=center] {$ \begin{tikzpicture}[x=0.75pt,y=0.75pt,yscale=-1,xscale=1]

\draw  [line width=0.75]  (244.74,157.1) -- (257.09,169.44) ;
\draw  [line width=0.75]  (364.28,214.32) -- (376.62,226.67) ;
\draw   [line width=0.75] (376.48,157.58) -- (364.14,169.92) ;
\draw  [line width=0.75]  (257.01,214.32) -- (244.66,226.66) ;
\draw [line width=0.75]  (283.55,178.49) -- (310.81,178.49) -- (310.81,205.75) -- (283.55,205.75) -- cycle ;
\draw  [line width=0.75] (310.81,178.49) -- (338.07,178.49) -- (338.07,205.75) -- (310.81,205.75) -- cycle ;
\draw [line width=0.75]  (257.09,169.44) -- (283.45,178.06) -- (283.4,205.79) -- (257.01,214.32) -- (240.74,191.85) -- cycle ;
\draw  [line width=0.75] (364.14,169.92) -- (380.34,192.07) -- (364.28,214.32) -- (338.16,205.92) -- (338.07,178.49) -- cycle ;
\draw  [line width=0.75]  (397.6,192.07) -- (380.34,192.07) ;
\draw  [line width=0.75]  (240.74,192.07) -- (223.48,192.07) ;

\draw (306.29,155.06) node [anchor=north west][inner sep=0.75pt]   [align=left] {$\displaystyle 2$};
\draw (306.23,212.28) node [anchor=north west][inner sep=0.75pt]   [align=left] {$\displaystyle 1$};
\draw (258.27,185.47) node [anchor=north west][inner sep=0.75pt]   [align=left] {$\displaystyle a$};
\draw (293.12,185.47) node [anchor=north west][inner sep=0.75pt]   [align=left] {$\displaystyle b$};
\draw (319.25,185.47) node [anchor=north west][inner sep=0.75pt]   [align=left] {$\displaystyle c$};
\draw (351.42,185.47) node [anchor=north west][inner sep=0.75pt]   [align=left] {$\displaystyle d$};
\draw (227.88,201.5) node [anchor=north west][inner sep=0.75pt]   [align=left] {$\displaystyle 3$};
\draw (382.78,165.08) node [anchor=north west][inner sep=0.75pt]   [align=left] {$\displaystyle 4$};
\draw (229.98,165.08) node [anchor=north west][inner sep=0.75pt]   [align=left] {$\displaystyle 4$};
\draw (382.28,201.5) node [anchor=north west][inner sep=0.75pt]   [align=left] {$\displaystyle 3$};

\end{tikzpicture}$};

\end{tikzpicture}
}}
$};



\end{tikzpicture}

}
}}
\end{eqnarray}

\begin{itemize}
    \item $N^{12;34}$ with two distinct numerators:
    \begin{equation}
    \begin{split}
    &n_{N}^{12;34} := x_{1,2}^{2} (x_{b,d}^{2} x_{1,3}^{2}{-}x_{b,3}^2 x_{d,1}^2{-}x_{b,1}^2 x_{d,3}^2), \\
    &n_{N}^{\prime\,12;34} := x_{1,2}^{2} x_{1,3}^{2}  x_{1,4}^{2} x_{b,3}^{2} x_{d,2}^{2}.
\end{split}
\end{equation}
\item $O^{12;34}$ with  numerator: 
\begin{equation}
    n_{O_{}}^{12;34} := x_{1,4}^{2}.
\end{equation}
\item $K^{12;34}$ with three distinct numerators:
\begin{equation}
    \begin{split}
    &n_{K}^{12;34} := x_{1,2}^{2}  (x_{1,2}^{2} 
 x_{a,d}^{2}{-} x_{a,1}^{2} 
 x_{d,2}^{2}{-} x_{a,2}^{2} 
 x_{d,1}^{2}), \\ &n_{K}^{\prime\,12;34} := x_{1,2}^{6}  x_{a,3}^{2} x_{d,4}^{2}, \\  &n_{K_{}}^{\prime\prime\,12;34} := x_{1,2}^{2} x_{a,1}^{2}  x_{d,2}^{2}.
\end{split}
\end{equation}
\end{itemize}

There are four  field-like topologies
\begin{eqnarray}
&&\vcenter{\hbox{\scalebox{1}{
\begin{tikzpicture}[x=0.75pt,y=0.75pt,yscale=-1,xscale=1]

\draw (-5.97,275) node [anchor=north west][inner sep=0.75pt]    {$U^{12;34}$};
\draw (169,275) node [anchor=north west][inner sep=0.75pt]    {$\text{I}^{12;34} $};
\draw (-5.97,405) node [anchor=north west][inner sep=0.75pt]    {$Z^{12;34} $};
\draw (165,405) node [anchor=north west][inner sep=0.75pt]    {$Y^{12;34} $};

\draw  [line width=0.75] (-50,174) node [anchor=north west][inner sep=0.75pt]  [align=center] {$ \hbox{\scalebox{0.75}{\begin{tikzpicture}[x=0.75pt,y=0.75pt,yscale=-1,xscale=1]

\draw  [line width=0.75]  (362.83,235.28) -- (374.66,247.12) ;
\draw   [line width=0.75] (253.28,235.33) -- (241.45,247.17) ;
\draw  [line width=0.75] (318.05,145.72) -- (362.83,145.72) -- (362.83,190.5) -- (318.05,190.5) -- cycle ;
\draw   [line width=0.75] (241.45,133.88) -- (253.28,145.72) ;
\draw  [line width=0.75]  (374.66,133.88) -- (362.83,145.72) ;
\draw  [line width=0.75] (318.05,190.5) -- (362.83,190.5) -- (362.83,235.28) -- (318.05,235.28) -- cycle ;
\draw [line width=0.75]  (285.67,145.72) -- (318.05,145.72) -- (318.05,235.33) -- (285.67,235.33) -- cycle ;
\draw  [line width=0.75] (253.28,145.72) -- (285.67,145.72) -- (285.67,235.33) -- (253.28,235.33) -- cycle ;
\draw  [line width=0.75]  (300.78,235.83) -- (300.78,250.5) ;

\draw (232,182.69) node [anchor=north west][inner sep=0.75pt]   [align=left] {$\displaystyle 1$};
\draw (265.06,183.29) node [anchor=north west][inner sep=0.75pt]   [align=left] {$\displaystyle a$};
\draw (299.09,183) node [anchor=north west][inner sep=0.75pt]   [align=left] {$\displaystyle b$};
\draw (333.72,162.65) node [anchor=north west][inner sep=0.75pt]   [align=left] {$\displaystyle c$};
\draw (334.85,206.48) node [anchor=north west][inner sep=0.75pt]   [align=left] {$\displaystyle d$};
\draw (330.69,237.85) node [anchor=north west][inner sep=0.75pt]   [align=left] {$\displaystyle 4$};
\draw (302.18,126.24) node [anchor=north west][inner sep=0.75pt]   [align=left] {$\displaystyle 2$};
\draw (266.19,238.16) node [anchor=north west][inner sep=0.75pt]   [align=left] {$\displaystyle 3$};
\draw (370.78,182.69) node [anchor=north west][inner sep=0.75pt]   [align=left] {$\displaystyle 1$};

\end{tikzpicture}
}}
$};

\draw  [line width=0.75] (120,174) node [anchor=north west][inner sep=0.75pt]  [align=center] {$ \hbox{\scalebox{0.75}{ \begin{tikzpicture}[x=0.75pt,y=0.75pt,yscale=-1,xscale=1]

\draw  [line width=0.75]  (362.83,235.28) -- (374.66,247.12) ;
\draw   [line width=0.75] (253.28,235.33) -- (241.45,247.17) ;
\draw  [line width=0.75] (318.05,145.72) -- (362.83,145.72) -- (362.83,190.5) -- (318.05,190.5) -- cycle ;
\draw   [line width=0.75] (241.45,133.88) -- (253.28,145.72) ;
\draw  [line width=0.75]  (374.66,133.88) -- (362.83,145.72) ;
\draw  [line width=0.75] (318.05,190.5) -- (362.83,190.5) -- (362.83,235.28) -- (318.05,235.28) -- cycle ;
\draw [line width=0.75]  (285.67,145.72) -- (318.05,145.72) -- (318.05,235.33) -- (285.67,235.33) -- cycle ;
\draw  [line width=0.75] (253.28,145.72) -- (285.67,145.72) -- (285.67,235.33) -- (253.28,235.33) -- cycle ;
\draw  [line width=0.75]  (362.83,170.5) -- (380.78,170.5) ;
\draw  [line width=0.75]  (285.67,130.5) -- (285.67,145.5) ;

\draw (265,126.24) node [anchor=north west][inner sep=0.75pt]   [align=left] {$\displaystyle 3$};
\draw (235,180.5) node [anchor=north west][inner sep=0.75pt]   [align=left] {$\displaystyle 4$};
\draw (265.06,183.29) node [anchor=north west][inner sep=0.75pt]   [align=left] {$\displaystyle a$};
\draw (299.09,183) node [anchor=north west][inner sep=0.75pt]   [align=left] {$\displaystyle b$};
\draw (333.72,162.65) node [anchor=north west][inner sep=0.75pt]   [align=left] {$\displaystyle c$};
\draw (334.85,206.48) node [anchor=north west][inner sep=0.75pt]   [align=left] {$\displaystyle d$};
\draw (302.18,126.24) node [anchor=north west][inner sep=0.75pt]   [align=left] {$\displaystyle 1$};
\draw (302.19,238.16) node [anchor=north west][inner sep=0.75pt]   [align=left] {$\displaystyle 2$};
\draw (370.78,147.69) node [anchor=north west][inner sep=0.75pt]   [align=left] {$\displaystyle 4$};
\draw (370.78,195) node [anchor=north west][inner sep=0.75pt]   [align=left] {$\displaystyle 3$};

\end{tikzpicture}
}}
$};

\draw  [line width=0.75] (-50,300) node [anchor=north west][inner sep=0.75pt]  [align=center] {$ \hbox{\scalebox{0.75}{\begin{tikzpicture}[x=0.75pt,y=0.75pt,yscale=-1,xscale=1]

\draw  [line width=0.75] (100,100) node [anchor=north west][inner sep=0.75pt]  [align=center] {$ \begin{tikzpicture}[x=0.75pt,y=0.75pt,yscale=-1,xscale=1]

\draw  [line width=0.75]  (362.83,235.28) -- (374.66,247.12) ;
\draw   [line width=0.75] (253.28,235.33) -- (241.45,247.17) ;
\draw  [line width=0.75] (318.05,145.72) -- (362.83,145.72) -- (362.83,190.5) -- (318.05,190.5) -- cycle ;
\draw   [line width=0.75] (241.45,133.88) -- (253.28,145.72) ;
\draw  [line width=0.75]  (374.66,133.88) -- (362.83,145.72) ;
\draw  [line width=0.75] (318.05,190.5) -- (362.83,190.5) -- (362.83,235.28) -- (318.05,235.28) -- cycle ;
\draw [line width=0.75]  (285.67,145.72) -- (318.05,145.72) -- (318.05,235.33) -- (285.67,235.33) -- cycle ;
\draw  [line width=0.75] (253.28,145.72) -- (285.67,145.72) -- (285.67,235.33) -- (253.28,235.33) -- cycle ;
\draw  [line width=0.75]  (362.83,170.5) -- (380.78,170.5) ;
\draw  [line width=0.75]  (235.45,190.5) -- (253.28,190.5) ;

\draw (235,160.5) node [anchor=north west][inner sep=0.75pt]   [align=left] {$\displaystyle 2$};
\draw (235,210.5) node [anchor=north west][inner sep=0.75pt]   [align=left] {$\displaystyle 4$};
\draw (265.06,183.29) node [anchor=north west][inner sep=0.75pt]   [align=left] {$\displaystyle a$};
\draw (299.09,183) node [anchor=north west][inner sep=0.75pt]   [align=left] {$\displaystyle b$};
\draw (333.72,162.65) node [anchor=north west][inner sep=0.75pt]   [align=left] {$\displaystyle c$};
\draw (334.85,206.48) node [anchor=north west][inner sep=0.75pt]   [align=left] {$\displaystyle d$};
\draw (302.18,126.24) node [anchor=north west][inner sep=0.75pt]   [align=left] {$\displaystyle 1$};
\draw (302.19,238.16) node [anchor=north west][inner sep=0.75pt]   [align=left] {$\displaystyle 3$};
\draw (370.78,147.69) node [anchor=north west][inner sep=0.75pt]   [align=left] {$\displaystyle 4$};
\draw (370.78,195) node [anchor=north west][inner sep=0.75pt]   [align=left] {$\displaystyle 2$};

\end{tikzpicture}$};

\end{tikzpicture}
}}
$};

\draw  [line width=0.75] (120,300) node [anchor=north west][inner sep=0.75pt]  [align=center] {$ \hbox{\scalebox{0.75}{\begin{tikzpicture}[x=0.75pt,y=0.75pt,yscale=-1,xscale=1]

\draw  [line width=0.75] (100,100) node [anchor=north west][inner sep=0.75pt]  [align=center] {$ \begin{tikzpicture}[x=0.75pt,y=0.75pt,yscale=-1,xscale=1]

\draw  [line width=0.75]  (362.83,235.28) -- (374.66,247.12) ;
\draw   [line width=0.75] (253.28,235.33) -- (241.45,247.17) ;
\draw  [line width=0.75] (318.05,145.72) -- (362.83,145.72) -- (362.83,190.5) -- (318.05,190.5) -- cycle ;
\draw   [line width=0.75] (241.45,133.88) -- (253.28,145.72) ;
\draw  [line width=0.75]  (374.66,133.88) -- (362.83,145.72) ;
\draw  [line width=0.75] (318.05,190.5) -- (362.83,190.5) -- (362.83,235.28) -- (318.05,235.28) -- cycle ;
\draw [line width=0.75]  (285.67,145.72) -- (318.05,145.72) -- (318.05,235.33) -- (285.67,235.33) -- cycle ;
\draw  [line width=0.75] (253.28,145.72) -- (285.67,145.72) -- (285.67,235.33) -- (253.28,235.33) -- cycle ;
\draw  [line width=0.75]  (362.83,170.5) -- (380.78,170.5) ;
\draw  [line width=0.75]  (235.45,190.5) -- (253.28,190.5) ;

\draw (235,160.5) node [anchor=north west][inner sep=0.75pt]   [align=left] {$\displaystyle 2$};
\draw (235,210.5) node [anchor=north west][inner sep=0.75pt]   [align=left] {$\displaystyle 4$};
\draw (265.06,183.29) node [anchor=north west][inner sep=0.75pt]   [align=left] {$\displaystyle a$};
\draw (299.09,183) node [anchor=north west][inner sep=0.75pt]   [align=left] {$\displaystyle b$};
\draw (333.72,162.65) node [anchor=north west][inner sep=0.75pt]   [align=left] {$\displaystyle c$};
\draw (334.85,206.48) node [anchor=north west][inner sep=0.75pt]   [align=left] {$\displaystyle d$};
\draw (302.18,126.24) node [anchor=north west][inner sep=0.75pt]   [align=left] {$\displaystyle 1$};
\draw (302.19,238.16) node [anchor=north west][inner sep=0.75pt]   [align=left] {$\displaystyle 3$};
\draw (370.78,147.69) node [anchor=north west][inner sep=0.75pt]   [align=left] {$\displaystyle 4$};
\draw (370.78,195) node [anchor=north west][inner sep=0.75pt]   [align=left] {$\displaystyle 2$};

\end{tikzpicture}$};

\end{tikzpicture}
}}
$};



\end{tikzpicture}

}
}}
\end{eqnarray}

\begin{itemize}
    \item $U^{12;34}$ with two distinct numerators:
    \begin{equation}
    \begin{split}
    &n_{U}^{12;34} :=  x_{2,4}^{2} x_{b,1}^{2} (x_{2,3}^{2} x_{b,1}^{2} {-}x_{1,3}^{2}x_{b,2}^{2}),\\
    & n_{U}^{\prime\,12;34} := x_{b,1}^{2} x_{b,2}^2.
\end{split}
\end{equation}
\item ${\text{I}}^{12;34}$ with  numerator: \begin{equation}
    n_{\text{I}}^{12;34} :=  x_{2,3}^{2} x_{b,c}^{2}  {+} x_{b,2}^{2} x_{c,3}^{2} {-}x_{b,3}^{2} x_{c,2}^{2}.
\end{equation}
\item ${Z}^{12;34}$ with  numerator: 
\begin{equation}
    \begin{split}
    n_{Z}^{12;34} :=& x_{1,3}^{2} x_{b,2}^{2} x_{a,c}^2 {+} x_{2,3}^2 x_{a,1}^2 x_{b,c}^2{-} x_{1,3}^2 x_{a,2}^2 x_{b,c}^2 {-}x_{a,1}^2 x_{b,2}^2 x_{c,3}^2,\\
    &  {+}x_{a,1}^2 x_{b,3}^2 x_{c,2}^2{-}x_{a,3}^2 x_{b,2}^2 x_{c,1}^2{+}x_{1,2}^2x_{a,3}^2 x_{b,c}^2.
\end{split}
\end{equation}
\item  $Y^{12;34}$ with three distinct numerators:
\begin{equation}
    \begin{split}
    &n_{Y}^{12;34} := x_{2,4}^{2} x_{d,1}^{2}  (x_{1,4}^2 x_{b,3}^2{-}x_{1,3}^2 x_{b,4}^2) , \\
    &n_{Y}^{\prime\,12;34} := x_{d,1}^{2} x_{b,4}^{2},\\
    &n_{Y}^{\prime\prime\,12;34} := x_{1,4}^{4} x_{b,2}^{2} x_{d,3}^{2}.  \phantom{my long teddddcccdfff}
\end{split}
\end{equation}
\end{itemize}

There are six  wheel topologies
\begin{eqnarray}
&&\vcenter{\hbox{\scalebox{1}{
\begin{tikzpicture}[x=0.75pt,y=0.75pt,yscale=-1,xscale=1]

\draw (-5.97,275) node [anchor=north west][inner sep=0.75pt]    {$F^{12;34}$};
\draw (159,275) node [anchor=north west][inner sep=0.75pt]    {$G^{12;34} $};
\draw (329,275) node [anchor=north west][inner sep=0.75pt]    {$R^{12;34} $};
\draw (-5.97,405) node [anchor=north west][inner sep=0.75pt]    {$\Xi^{12;34} $};
\draw (155,405) node [anchor=north west][inner sep=0.75pt]    {$M^{12;34} $};
\draw (329,405) node [anchor=north west][inner sep=0.75pt]    {$J^{12;34} $};

\draw  [line width=0.75] (-50,174) node [anchor=north west][inner sep=0.75pt]  [align=center] {$ \hbox{\scalebox{0.75}{\begin{tikzpicture}[x=0.75pt,y=0.75pt,yscale=-1,xscale=1]

\draw   [line width=0.75] (246.47,91.92) -- (289.91,91.92) -- (289.91,135.36) -- (246.47,135.36) -- cycle ;
\draw   [line width=0.75] (289.91,135.36) -- (333.36,135.36) -- (333.36,178.81) -- (289.91,178.81) -- cycle ;
\draw   [line width=0.75]  (234.5,79.95) -- (246.47,91.92) ;
\draw   [line width=0.75]  (333.36,178.81) -- (345.33,190.78) ;
\draw   [line width=0.75]  (246.47,178.81) -- (232.33,191.33) ;
\draw   [line width=0.75] (289.91,91.92) -- (333.36,91.92) -- (333.36,135.36) -- (289.91,135.36) -- cycle ;
\draw   [line width=0.75] (246.47,135.36) -- (289.91,135.36) -- (289.91,178.81) -- (246.47,178.81) -- cycle ;
\draw    [line width=0.75] (347.49,79.39) -- (333.36,91.92) ;
\draw (265.89,104.24) node [anchor=north west][inner sep=0.75pt]    {$a$};
\draw (306.78,108.65) node [anchor=north west][inner sep=0.75pt]    {$b$};
\draw (265.26,146.7) node [anchor=north west][inner sep=0.75pt]   [align=left] {$\displaystyle c$};
\draw (305.28,145.41) node [anchor=north west][inner sep=0.75pt]   [align=left] {$\displaystyle d$};
\draw (283.27,69.92) node [anchor=north west][inner sep=0.75pt]    {$2$};
\draw (284.09,187.34) node [anchor=north west][inner sep=0.75pt]    {$1$};
\draw (226.34,128.57) node [anchor=north west][inner sep=0.75pt]    {$3$};
\draw (342.95,128.33) node [anchor=north west][inner sep=0.75pt]    {$4$};

\end{tikzpicture}
}}
$};

\draw  [line width=0.75] (120,174) node [anchor=north west][inner sep=0.75pt]  [align=center] {$ \hbox{\scalebox{0.75}{ \begin{tikzpicture}[x=0.75pt,y=0.75pt,yscale=-1,xscale=1]

\draw  [line width=0.75] (246.47,91.92) -- (289.91,91.92) -- (289.91,135.36) -- (246.47,135.36) -- cycle ;
\draw  [line width=0.75]  (289.91,135.36) -- (333.36,135.36) -- (333.36,178.81) -- (289.91,178.81) -- cycle ;
\draw   [line width=0.75] (234.5,79.95) -- (246.47,91.92) ;
\draw    [line width=0.75] (333.36,178.81) -- (345.33,190.78) ;
\draw   [line width=0.75]  (290,75) -- (289.91,91.92) ;
\draw   [line width=0.75]  (246.47,178.81) -- (232.33,191.33) ;
\draw   [line width=0.75]  (289.91,178.81) -- (289.67,196) ;
\draw  [line width=0.75] (289.91,91.92) -- (333.36,91.92) -- (333.36,135.36) -- (289.91,135.36) -- cycle ;
\draw  [line width=0.75]  (246.47,135.36) -- (289.91,135.36) -- (289.91,178.81) -- (246.47,178.81) -- cycle ;
\draw   [line width=0.75]  (347.49,79.39) -- (333.36,91.92) ;

\draw (265.89,104.24) node [anchor=north west][inner sep=0.75pt]    {$a$};
\draw (306.78,108.65) node [anchor=north west][inner sep=0.75pt]    {$b$};
\draw (265.26,146.7) node [anchor=north west][inner sep=0.75pt]   [align=left] {$\displaystyle c$};
\draw (305.28,145.41) node [anchor=north west][inner sep=0.75pt]   [align=left] {$\displaystyle d$};
\draw (264.59,72.18) node [anchor=north west][inner sep=0.75pt]    {$1$};
\draw (307.27,71.92) node [anchor=north west][inner sep=0.75pt]    {$2$};
\draw (306.09,183.34) node [anchor=north west][inner sep=0.75pt]    {$1$};
\draw (226.34,128.57) node [anchor=north west][inner sep=0.75pt]    {$3$};
\draw (342.95,128.33) node [anchor=north west][inner sep=0.75pt]    {$4$};
\draw (263.6,180.92) node [anchor=north west][inner sep=0.75pt]    {$2$};

\end{tikzpicture}
}}
$};

\draw  [line width=0.75] (290,174) node [anchor=north west][inner sep=0.75pt]  [align=center] {$ \hbox{\scalebox{0.75}{ \begin{tikzpicture}[x=0.75pt,y=0.75pt,yscale=-1,xscale=1]

\draw  [line width=0.75] (246.47,91.92) -- (289.91,91.92) -- (289.91,135.36) -- (246.47,135.36) -- cycle ;
\draw  [line width=0.75] (289.91,135.36) -- (333.36,135.36) -- (333.36,178.81) -- (289.91,178.81) -- cycle ;
\draw   [line width=0.75] (234.5,79.95) -- (246.47,91.92) ;
\draw  [line width=0.75]  (246.47,155.94) -- (268.33,178.81) ;
\draw  [line width=0.75]  (289.91,178.81) -- (268.33,178.81) ;
\draw   [line width=0.75] (246.47,135.36) -- (246.47,155.94) ;
\draw  [line width=0.75]  (311.49,91.92) -- (289.91,91.92) ;
\draw   [line width=0.75] (311.49,91.92) -- (333.36,114.78) ;
\draw   [line width=0.75] (333.36,114.78) -- (333.36,135.36) ;
\draw   [line width=0.75] (333.36,178.81) -- (345.33,190.78) ;
\draw  [line width=0.75]  (317.5,76.5) -- (311.49,91.92) ;
\draw   [line width=0.75] (350,108) -- (333.36,114.78) ;
\draw  [line width=0.75]  (246.47,155.94) -- (230.5,161) ;
\draw   [line width=0.75] (268.33,178.81) -- (262.33,194.23) ;
\draw   [line width=0.75] (246.47,135.36) -- (231.29,135.36) ;

\draw (295.89,106.24) node [anchor=north west][inner sep=0.75pt]    {$b$};
\draw (266.78,145.65) node [anchor=north west][inner sep=0.75pt]    {$c$};
\draw (265.26,101.7) node [anchor=north west][inner sep=0.75pt]   [align=left] {$\displaystyle a$};
\draw (306.28,150.41) node [anchor=north west][inner sep=0.75pt]   [align=left] {$\displaystyle d$};
\draw (227.59,101.18) node [anchor=north west][inner sep=0.75pt]    {$1$};
\draw (328.6,87.58) node [anchor=north west][inner sep=0.75pt]    {$3$};
\draw (341.6,135.58) node [anchor=north west][inner sep=0.75pt]    {$2$};
\draw (294.09,183.68) node [anchor=north west][inner sep=0.75pt]    {$1$};
\draw (269.17,68.41) node [anchor=north west][inner sep=0.75pt]    {$4$};
\draw (224.61,139.99) node [anchor=north west][inner sep=0.75pt]    {$4$};
\draw (240.6,165.58) node [anchor=north west][inner sep=0.75pt]    {$3$};

\end{tikzpicture}
}}
$};

\draw  [line width=0.75] (-50,300) node [anchor=north west][inner sep=0.75pt]  [align=center] {$ \hbox{\scalebox{0.75}{\begin{tikzpicture}[x=0.75pt,y=0.75pt,yscale=-1,xscale=1]

\draw  [line width=0.75] (100,100) node [anchor=north west][inner sep=0.75pt]  [align=center] {$ \begin{tikzpicture}[x=0.75pt,y=0.75pt,yscale=-1,xscale=1]

\draw [line width=0.75]  (246.47,91.92) -- (289.91,91.92) -- (289.91,135.36) -- (246.47,135.36) -- cycle ;
\draw  [line width=0.75] (289.91,135.36) -- (333.36,135.36) -- (333.36,178.81) -- (289.91,178.81) -- cycle ;
\draw  [line width=0.75]  (234.5,79.95) -- (246.47,91.92) ;
\draw  [line width=0.75]  (246.47,155.94) -- (268.33,178.81) ;
\draw  [line width=0.75]  (289.91,178.81) -- (268.33,178.81) ;
\draw  [line width=0.75]  (246.47,135.36) -- (246.47,155.94) ;
\draw  [line width=0.75]  (311.49,91.92) -- (289.91,91.92) ;
\draw  [line width=0.75]  (311.49,91.92) -- (333.36,114.78) ;
\draw  [line width=0.75]  (333.36,114.78) -- (333.36,135.36) ;
\draw   [line width=0.75] (333.36,178.81) -- (345.33,190.78) ;
\draw   [line width=0.75] (317.5,76.5) -- (311.49,91.92) ;
\draw   [line width=0.75] (350,108) -- (333.36,114.78) ;
\draw  [line width=0.75]  (246.47,155.94) -- (230.5,161) ;
\draw   [line width=0.75] (268.33,178.81) -- (262.33,194.23) ;
\draw  [line width=0.75]  (348.53,135.36) -- (333.36,135.36) ;
\draw   [line width=0.75] (246.47,135.36) -- (231.29,135.36) ;

\draw (265.89,104.24) node [anchor=north west][inner sep=0.75pt]    {$a$};
\draw (306.78,108.65) node [anchor=north west][inner sep=0.75pt]    {$b$};
\draw (265.26,146.7) node [anchor=north west][inner sep=0.75pt]   [align=left] {$\displaystyle c$};
\draw (305.28,145.41) node [anchor=north west][inner sep=0.75pt]   [align=left] {$\displaystyle d$};
\draw (227.59,101.18) node [anchor=north west][inner sep=0.75pt]    {$1$};
\draw (273.6,72.58) node [anchor=north west][inner sep=0.75pt]    {$2$};
\draw (341.6,150.58) node [anchor=north west][inner sep=0.75pt]    {$2$};
\draw (294.09,183.68) node [anchor=north west][inner sep=0.75pt]    {$1$};
\draw (325.67,83.91) node [anchor=north west][inner sep=0.75pt]    {$3$};
\draw (243.17,168.41) node [anchor=north west][inner sep=0.75pt]    {$3$};
\draw (224.61,139.99) node [anchor=north west][inner sep=0.75pt]    {$4$};
\draw (343.11,113.99) node [anchor=north west][inner sep=0.75pt]    {$4$};

\end{tikzpicture}$};

\end{tikzpicture}
}}
$};

\draw  [line width=0.75] (120,300) node [anchor=north west][inner sep=0.75pt]  [align=center] {$ \hbox{\scalebox{0.75}{\begin{tikzpicture}[x=0.75pt,y=0.75pt,yscale=-1,xscale=1]

\draw  [line width=0.75] (100,100) node [anchor=north west][inner sep=0.75pt]  [align=center] {$ \begin{tikzpicture}[x=0.75pt,y=0.75pt,yscale=-1,xscale=1]

\draw  [line width=0.75]  (342.19,150.41) -- (327.27,150.41) ;
\draw  [line width=0.75]  (242.9,115.05) -- (227.98,115.05) ;
\draw  [line width=0.75]  (267.9,91.05) -- (267.9,76.13) ;
\draw  [line width=0.75]  (303.27,190.33) -- (303.27,175.42) ;
\draw [line width=0.75]  (243.33,91.74) -- (285.29,91.74) -- (285.29,133.7) -- (243.33,133.7) -- cycle ;
\draw  [line width=0.75] (285.29,91.74) -- (327.24,91.74) -- (327.24,133.7) -- (285.29,133.7) -- cycle ;
\draw  [line width=0.75] (243.33,133.7) -- (285.29,133.7) -- (285.29,175.65) -- (243.33,175.65) -- cycle ;
\draw [line width=0.75]  (285.29,133.7) -- (327.24,133.7) -- (327.24,175.65) -- (285.29,175.65) -- cycle ;
\draw   [line width=0.75] (327.24,91.74) -- (337.79,81.2) ;
\draw  [line width=0.75]  (232.79,186.2) -- (243.33,175.65) ;
\draw  [line width=0.75]  (243.33,91.74) -- (232.79,81.2) ;
\draw  [line width=0.75]  (337.79,186.2) -- (327.24,175.65) ;

\draw (263.37,107.41) node [anchor=north west][inner sep=0.75pt]   [align=left] {$\displaystyle a$};
\draw (298.08,107.41) node [anchor=north west][inner sep=0.75pt]   [align=left] {$\displaystyle b$};
\draw (263.37,146.16) node [anchor=north west][inner sep=0.75pt]   [align=left] {$\displaystyle c$};
\draw (298.08,146.16) node [anchor=north west][inner sep=0.75pt]   [align=left] {$\displaystyle d$};
\draw (295.7,72.41) node [anchor=north west][inner sep=0.75pt]   [align=left] {$\displaystyle 4$};
\draw (250.37,72.41) node [anchor=north west][inner sep=0.75pt]   [align=left] {$\displaystyle 3$};
\draw (336.03,108.67) node [anchor=north west][inner sep=0.75pt]   [align=left] {$\displaystyle 3$};
\draw (335.7,157) node [anchor=north west][inner sep=0.75pt]   [align=left] {$\displaystyle 4$};
\draw (260.03,179.67) node [anchor=north west][inner sep=0.75pt]   [align=left] {$\displaystyle 2$};
\draw (311.7,179.33) node [anchor=north west][inner sep=0.75pt]   [align=left] {$\displaystyle 1$};
\draw (224.03,90) node [anchor=north west][inner sep=0.75pt]   [align=left] {$\displaystyle 2$};
\draw (223.03,141) node [anchor=north west][inner sep=0.75pt]   [align=left] {$\displaystyle 1$};

\end{tikzpicture}$};

\end{tikzpicture}
}}
$};

\draw  [line width=0.75] (290,300) node [anchor=north west][inner sep=0.75pt]  [align=center] {$ \hbox{\scalebox{0.75}{\begin{tikzpicture}[x=0.75pt,y=0.75pt,yscale=-1,xscale=1]

\draw  [line width=0.75] (100,100) node [anchor=north west][inner sep=0.75pt]  [align=center] {$ \begin{tikzpicture}[x=0.75pt,y=0.75pt,yscale=-1,xscale=1]

\draw  [line width=0.75]  (342.19,115.05) -- (327.27,115.05) ;
\draw  [line width=0.75]  (242.9,115.05) -- (227.98,115.05) ;
\draw  [line width=0.75]  (303.27,190.33) -- (303.27,175.42) ;
\draw [line width=0.75]  (243.33,91.74) -- (285.29,91.74) -- (285.29,133.7) -- (243.33,133.7) -- cycle ;
\draw  [line width=0.75] (285.29,91.74) -- (327.24,91.74) -- (327.24,133.7) -- (285.29,133.7) -- cycle ;
\draw  [line width=0.75] (243.33,133.7) -- (285.29,133.7) -- (285.29,175.65) -- (243.33,175.65) -- cycle ;
\draw [line width=0.75]  (285.29,133.7) -- (327.24,133.7) -- (327.24,175.65) -- (285.29,175.65) -- cycle ;
\draw   [line width=0.75] (327.24,91.74) -- (337.79,81.2) ;
\draw  [line width=0.75]  (232.79,186.2) -- (243.33,175.65) ;
\draw  [line width=0.75]  (243.33,91.74) -- (232.79,81.2) ;
\draw  [line width=0.75]  (337.79,186.2) -- (327.24,175.65) ;

\draw (263.37,107.41) node [anchor=north west][inner sep=0.75pt]   [align=left] {$\displaystyle a$};
\draw (298.08,107.41) node [anchor=north west][inner sep=0.75pt]   [align=left] {$\displaystyle b$};
\draw (263.37,146.16) node [anchor=north west][inner sep=0.75pt]   [align=left] {$\displaystyle c$};
\draw (298.08,146.16) node [anchor=north west][inner sep=0.75pt]   [align=left] {$\displaystyle d$};
\draw (285,72.41) node [anchor=north west][inner sep=0.75pt]   [align=left] {$\displaystyle 4$};
\draw (336.03,90) node [anchor=north west][inner sep=0.75pt]   [align=left] {$\displaystyle 2$};
\draw (335.7,140) node [anchor=north west][inner sep=0.75pt]   [align=left] {$\displaystyle 3$};
\draw (265,179.67) node [anchor=north west][inner sep=0.75pt]   [align=left] {$\displaystyle 2$};
\draw (310,179.33) node [anchor=north west][inner sep=0.75pt]   [align=left] {$\displaystyle 1$};
\draw (224.03,95) node [anchor=north west][inner sep=0.75pt]   [align=left] {$\displaystyle 2$};
\draw (224.03,138) node [anchor=north west][inner sep=0.75pt]   [align=left] {$\displaystyle 1$};

\end{tikzpicture}$};

\end{tikzpicture}
}}
$};



\end{tikzpicture}

}
}}
\end{eqnarray}

\begin{itemize}
    \item $F^{12;34}$ with one numerators:
    \begin{equation}
    \begin{split}
    &n_{F}^{12;34} := 1.
\end{split}
\end{equation}
 \item $G^{12;34}$ with one numerators:
    \begin{equation}
    \begin{split}
    &n_{G}^{12;34} := 1.
\end{split}
\end{equation}
\item $R^{12;34}$ with two distinct numerators:
    \begin{equation}
    \begin{split}
    &n_{R_{ }}^{12;34}:= x_{b,c}^2 x_{1,4}^2{-} x_{b,1}^2 x_{c,4}^2{-} x_{b,4}^2 x_{c,1}^2\\
    &n_{R}^{\prime\,12;34}:= x_{b,1}^2 x_{c,4}^2 x_{1,3}^2 x_{2,4}^2 
\end{split}
\end{equation}
\item $\Xi^{12;34}$ with one  numerator:
    \begin{equation}
    \begin{split}
    &n_{ \Xi}^{12;34} :=x_{1,2}^{2} x_{a,b}^{2}  {+} x_{a,2}^2 x_{b,1}^2 {-} x_{a,1}^2 x_{b,2}^2 
\end{split}
\end{equation}
\item $M^{12;34}$ with three distinct numerators:
    \begin{equation}
    \begin{split}
    &n_{M }^{12;34} :=(x_{a,d}^{2} x_{1,2}^{2} {-} x_{a,1}^{2} x_{d,2}^{2} {-} x_{a,2}^{2} x_{d,1}^{2})\times (x_{a,d}^{2} x_{3,4}^{2} {-} x_{a,3}^{2} x_{d,4}^{2} {-} x_{a,4}^{2} x_{d,3}^{2})\\ 
    & n_{M }^{\prime\,12;34} :=x_{a,3}^{2} x_{d,4}^{2}  (x_{a,d}^2 x_{1,2}^2 {-}x_{a,1}^2 x_{d,2}^2 {-}x_{a,2}^2 x_{d,1}^2) \\
    &n_{M}^{\prime\prime\,12;34}:= x_{a,2}^{2} x_{a,3}^{2} x_{d,1}^{2} x_{d,4}^{2} (x_{1,2}^{2}x_{3,4}^{2}{+}x_{1,3}^{2}x_{2,4}^{2}  {-}x_{1,4}^{2}x_{2,3}^{2})/2 {-}x_{1,2}^2 x_{1,3}^2 x_{3,4}^2  x_{a,2}^2 x_{d,4}^2 x_{a,d}^2 . 
\end{split}
\end{equation}
\item $J^{12;34}$ with one  numerator:
    \begin{equation}
    \begin{split}
    &n_{J}^{12;34} := x_{1,2}^{2} x_{2,3}^{2}  x_{a,d}^{2} (x_{b,1}^2 x_{2,4}^2{-}x_{b,2}^2 x_{1,4}^2) \\
    &\quad  {+} x_{1,2}^2 x_{2,3}^2 x_{a,b}^2 (x_{1,2}^2 x_{d,4}^2 {-} x_{1,4}^2 x_{d,2}^2  {-} x_{2,4}^2 x_{d,1}^2 ) 
\end{split}
\end{equation}
\item $Q^{12;34}$ with one  numerator:
\begin{equation}
    \begin{split}
    n_{Q}^{12;34} &:= ({-}x_{1,2}^2 x_{3,4}^2{-}x_{1,4}^2 x_{2,3}^2{+}x_{1,3}^2 x_{2,4}^2)x_{b,c}^2\\
    & \quad {+} x_{1,2}^2  x_{b,3}^2 x_{c,4}^2{+} x_{3,4}^2  x_{b,1}^2 x_{c,2}^2 
\end{split}
\end{equation}
\end{itemize}

There are three  playground-like topologies
\begin{eqnarray}
&&\vcenter{\hbox{\scalebox{1}{
\begin{tikzpicture}[x=0.75pt,y=0.75pt,yscale=-1,xscale=1]

\draw (-5.97,265) node [anchor=north west][inner sep=0.75pt]    {$V^{12;34}$};
\draw (159,265) node [anchor=north west][inner sep=0.75pt]    {$P^{12;34} $};
\draw (329,265) node [anchor=north west][inner sep=0.75pt]    {$Q^{12;34} $};

\draw  [line width=0.75] (-50,174) node [anchor=north west][inner sep=0.75pt]  [align=center] {$ \hbox{\scalebox{0.75}{ \begin{tikzpicture}[x=0.75pt,y=0.75pt,yscale=-1,xscale=1]

\draw  [line width=0.75] (253.55,186.22) -- (284.55,186.22) -- (284.55,217.21) -- (253.55,217.21) -- cycle ;
\draw  [line width=0.75]  (346.54,217.21) -- (358.37,229.04) ;
\draw  [line width=0.75]  (253.55,217.21) -- (241.72,229.04) ;
\draw [line width=0.75]  (284.55,186.22) -- (315.55,186.22) -- (315.55,217.21) -- (284.55,217.21) -- cycle ;
\draw  [line width=0.75] (315.55,186.22) -- (346.54,186.22) -- (346.54,217.21) -- (315.55,217.21) -- cycle ;
\draw  [line width=0.75] (253.55,155.46) -- (346.48,155.46) -- (346.48,186.34) -- (253.55,186.34) -- cycle ;
\draw  [line width=0.75]  (241.72,143.63) -- (253.55,155.46) ;
\draw  [line width=0.75]  (358.31,143.63) -- (346.48,155.46) ;

\draw (294.78,223.69) node [anchor=north west][inner sep=0.75pt]   [align=left] {$\displaystyle 1$};
\draw (294.56,162.29) node [anchor=north west][inner sep=0.75pt]   [align=left] {$\displaystyle a$};
\draw (265.09,193.5) node [anchor=north west][inner sep=0.75pt]   [align=left] {$\displaystyle b$};
\draw (295.72,194) node [anchor=north west][inner sep=0.75pt]   [align=left] {$\displaystyle c$};
\draw (324.85,193.98) node [anchor=north west][inner sep=0.75pt]   [align=left] {$\displaystyle d$};
\draw (295.69,133.35) node [anchor=north west][inner sep=0.75pt]   [align=left] {$\displaystyle 4$};
\draw (355.68,178.24) node [anchor=north west][inner sep=0.75pt]   [align=left] {$\displaystyle 3$};
\draw (232.19,178.66) node [anchor=north west][inner sep=0.75pt]   [align=left] {$\displaystyle 2$};

\end{tikzpicture}
}}
$};

\draw  [line width=0.75] (120,174) node [anchor=north west][inner sep=0.75pt]  [align=center] {$ \hbox{\scalebox{0.75}{ \begin{tikzpicture}[x=0.75pt,y=0.75pt,yscale=-1,xscale=1]

\draw  [line width=0.75] (245.5,157.46) -- (279.69,157.46) -- (279.69,222.5) -- (245.5,222.5) -- cycle ;
\draw  [line width=0.75] (279.69,157.46) -- (312.11,157.46) -- (312.11,189.88) -- (279.69,189.88) -- cycle ;
\draw  [line width=0.75] (279.69,189.88) -- (312.11,189.88) -- (312.11,222.3) -- (279.69,222.3) -- cycle ;
\draw  [line width=0.75] (312.11,157.26) -- (346.3,157.26) -- (346.3,222.3) -- (312.11,222.3) -- cycle ;

\draw  [line width=0.75]  (230.82,142.78) -- (245.5,157.46) ;
\draw  [line width=0.75]  (346.3,222.3) -- (360.98,236.98) ;
\draw  [line width=0.75]  (360.98,142.58) -- (346.3,157.26) ;
\draw  [line width=0.75]  (245.5,222.5) -- (230.82,237.18) ;
\draw  [line width=0.75]  (296.14,156.8) -- (296.14,136.04) ;

\draw (263.7,135.74) node [anchor=north west][inner sep=0.75pt]   [align=left] {$\displaystyle 3$};
\draw (322.2,137.67) node [anchor=north west][inner sep=0.75pt]   [align=left] {$\displaystyle 4$};
\draw (225.03,181.33) node [anchor=north west][inner sep=0.75pt]   [align=left] {$\displaystyle 2$};
\draw (291,168) node [anchor=north west][inner sep=0.75pt]   [align=left] {$\displaystyle a$};
\draw (291,199.25) node [anchor=north west][inner sep=0.75pt]   [align=left] {$\displaystyle b$};
\draw (259.62,182.86) node [anchor=north west][inner sep=0.75pt]   [align=left] {$\displaystyle c$};
\draw (324.62,182.86) node [anchor=north west][inner sep=0.75pt]   [align=left] {$\displaystyle d$};
\draw (355.03,181.83) node [anchor=north west][inner sep=0.75pt]   [align=left] {$\displaystyle 2$};
\draw (290.53,230.24) node [anchor=north west][inner sep=0.75pt]   [align=left] {$\displaystyle 1$};

\end{tikzpicture}
}}
$};

\draw  [line width=0.75] (290,174) node [anchor=north west][inner sep=0.75pt]  [align=center] {$ \hbox{\scalebox{0.75}{  \begin{tikzpicture}[x=0.75pt,y=0.75pt,yscale=-1,xscale=1]

\draw  [line width=0.75] (244.26,158.02) -- (276.68,158.02) -- (276.68,190.44) -- (244.26,190.44) -- cycle ;
\draw  [line width=0.75]  (229.58,143.34) -- (244.26,158.02) ;
\draw   [line width=0.75] (341.72,222.75) -- (356.4,237.43) ;
\draw  [line width=0.75]  (356.4,143.34) -- (341.72,158.02) ;
\draw [line width=0.75]   (244.26,222.65) -- (229.58,237.33) ;
\draw  [line width=0.75] (341.72,158.02) -- (341.72,190.33) -- (276.68,190.33) -- (276.68,158.02) -- cycle ;
\draw  [line width=0.75] (309.3,190.33) -- (341.72,190.33) -- (341.72,222.75) -- (309.3,222.75) -- cycle ;
\draw [line width=0.75]  (309.3,190.33) -- (309.3,222.65) -- (244.26,222.65) -- (244.26,190.33) -- cycle ;

\draw (288.37,131.74) node [anchor=north west][inner sep=0.75pt]   [align=left] {$\displaystyle 2$};
\draw (356.87,185) node [anchor=north west][inner sep=0.75pt]   [align=left] {$\displaystyle 4$};
\draw (257,167.41) node [anchor=north west][inner sep=0.75pt]   [align=left] {$\displaystyle b$};
\draw (305.55,167.25) node [anchor=north west][inner sep=0.75pt]   [align=left] {$\displaystyle c$};
\draw (271.62,199.53) node [anchor=north west][inner sep=0.75pt]   [align=left] {$\displaystyle a$};
\draw (319.29,199.53) node [anchor=north west][inner sep=0.75pt]   [align=left] {$\displaystyle d$};
\draw (221.7,182.5) node [anchor=north west][inner sep=0.75pt]   [align=left] {$\displaystyle 3$};
\draw (290.53,228.91) node [anchor=north west][inner sep=0.75pt]   [align=left] {$\displaystyle 1$};

\end{tikzpicture}
}}
$};



\end{tikzpicture}

}
}}
\end{eqnarray}
\begin{itemize}
    \item $V^{12;34}$ with one numerators:
    \begin{equation}
    \begin{split}
    &n_{V}^{12;34} := x_{2,3}^{2} x_{a,1}^{4}.
\end{split}
\end{equation}
\item  $P^{12;34}$ with three distinct numerators:
\begin{equation}
\begin{split}
    &n_{P}^{12;34} := x_{a,b}^{2} x_{1,4}^{2} (x_{1,3}^{2} x_{c,2}^{2} {-} x_{2,3}^{2} x_{c,1}^{2})/2  {+} x_{a,c}^2 x_{b,1}^2 (x_{1,4}^2 x_{2,3}^2 {+} x_{1,3}^2 x_{2,4}^2 )/2
    \\ &\phantom{jjjjjjjj}{-}x_{1,3}^2 x_{1,4}^2x_{b,2}^2 x_{a,c}^2,\\
    & n_{P}^{\prime\,12;34} := x_{a,c}^2 x_{d,1}^2 {+} x_{a,d}^2 x_{c,1}^2 {-} x_{c,d}^2 x_{a,1}^2,\\
    & n_{P}^{\prime\prime\,12;34} := x_{1,3}^2 x_{a,1}^2 x_{c,d}^2 x_{d,2}^2.
\end{split}    
\end{equation}
where  $n_P^{12;34}$ with symmetry $3\leftrightarrow4$.
\item $Q^{12;34}$ with one numerators:
    \begin{equation}
    \begin{split}
    n_{Q}^{12;34} &:= ({-}x_{1,2}^2 x_{3,4}^2{-}x_{1,4}^2 x_{2,3}^2{+}x_{1,3}^2 x_{2,4}^2)x_{b,c}^2
  {+} x_{1,2}^2  x_{b,3}^2 x_{c,4}^2{+} x_{3,4}^2  x_{b,1}^2 x_{c,2}^2 
\end{split}
\end{equation}
\end{itemize}

The following four topologies are the product of the  lower loop
\begin{eqnarray}
&&\vcenter{\hbox{\scalebox{1}{
\begin{tikzpicture}[x=0.75pt,y=0.75pt,yscale=-1,xscale=1]

\draw (-5.97,265) node [anchor=north west][inner sep=0.75pt]    {$gT^{12;34}$};
\draw (200,265) node [anchor=north west][inner sep=0.75pt]    {$hh^{12;34} $};
\draw (-5.97,385) node [anchor=north west][inner sep=0.75pt]   {$hh^{\prime12;34} $};
\draw (200,385) node [anchor=north west][inner sep=0.75pt]    {$gL^{12;34} $};

\draw  [line width=0.75] (-90,174) node [anchor=north west][inner sep=0.75pt]  [align=center] {$ \hbox{\scalebox{0.75}{\begin{tikzpicture}[x=0.75pt,y=0.75pt,yscale=-1,xscale=1]

\draw   [line width=0.75] (327.78,223.67) -- (336.92,232.81) ;
\draw  [line width=0.75]  (258.59,224.14) -- (249.45,233.28) ;
\draw [line width=0.75]  (258.3,181.45) -- (258.06,146.87) -- (292.64,146.63) -- (292.87,181.21) -- cycle ;
\draw  [line width=0.75] (292.87,181.21) -- (292.64,146.63) -- (327.22,146.4) -- (327.45,180.98) -- cycle ;
\draw  [line width=0.75] (258.59,224.14) -- (258.3,181.45) -- (327.49,180.98) -- (327.78,223.67) -- cycle ;

\draw   [line width=0.75] (248.93,137.73) -- (258.06,146.87) ;
\draw  [line width=0.75]  (336.35,137.26) -- (327.22,146.4) ;
\draw  [line width=0.75] (404.47,207.46) -- (404.23,172.89) -- (438.81,172.65) -- (439.04,207.23) -- cycle ;
\draw  [line width=0.75]  (447.94,163.51) -- (438.81,172.65) ;
\draw  [line width=0.75]  (395.09,163.75) -- (404.23,172.89) ;
\draw  [line width=0.75]  (439.04,207.23) -- (448.18,216.36) ;
\draw   [line width=0.75] (404.47,207.46) -- (395.33,216.6) ;

\draw (238.77,179.87) node [anchor=north west][inner sep=0.75pt]   [align=left] {$\displaystyle 1$};
\draw (287.24,195) node [anchor=north west][inner sep=0.75pt]   [align=left] {$\displaystyle a$};
\draw (270.67,155.4) node [anchor=north west][inner sep=0.75pt]   [align=left] {$\displaystyle b$};
\draw (304.74,157.52) node [anchor=north west][inner sep=0.75pt]   [align=left] {$\displaystyle c$};
\draw (416.43,183.19) node [anchor=north west][inner sep=0.75pt]   [align=left] {$\displaystyle d$};
\draw (288.11,128.18) node [anchor=north west][inner sep=0.75pt]   [align=left] {$\displaystyle 3$};
\draw (445.65,184.48) node [anchor=north west][inner sep=0.75pt]   [align=left] {$\displaystyle 2$};
\draw (335.99,177.92) node [anchor=north west][inner sep=0.75pt]   [align=left] {$\displaystyle 2$};
\draw (286.95,228.62) node [anchor=north west][inner sep=0.75pt]   [align=left] {$\displaystyle 4$};
\draw (362.61,180.74) node [anchor=north west][inner sep=0.75pt]  [font=\normalsize]  {$\times $};
\draw (390.51,182.96) node [anchor=north west][inner sep=0.75pt]   [align=left] {$\displaystyle 1$};
\draw (414.37,155.97) node [anchor=north west][inner sep=0.75pt]   [align=left] {$\displaystyle 3$};
\draw (416.3,210.63) node [anchor=north west][inner sep=0.75pt]   [align=left] {$\displaystyle 4$};

\end{tikzpicture}
}}
$};

\draw  [line width=0.75] (120,194) node [anchor=north west][inner sep=0.75pt]  [align=center] {$ \hbox{\scalebox{0.75}{ \begin{tikzpicture}[x=0.75pt,y=0.75pt,yscale=-1,xscale=1]

\draw  [line width=0.75]  (326.78,201.67) -- (335.92,210.81) ;
\draw  [line width=0.75]  (257.59,202.14) -- (248.45,211.28) ;
\draw [line width=0.75]  (257.3,202.45) -- (257.06,167.87) -- (291.64,167.63) -- (291.87,202.21) -- cycle ;
\draw  [line width=0.75] (291.87,202.21) -- (291.64,167.63) -- (326.22,167.4) -- (326.45,201.98) -- cycle ;
\draw   [line width=0.75] (247.93,158.73) -- (257.06,167.87) ;
\draw  [line width=0.75]  (335.35,158.26) -- (326.22,167.4) ;
\draw  [line width=0.75]  (464.78,200.67) -- (473.92,209.81) ;
\draw  [line width=0.75]  (395.59,201.14) -- (386.45,210.28) ;
\draw  [line width=0.75] (395.3,201.45) -- (395.06,166.87) -- (429.64,166.63) -- (429.87,201.21) -- cycle ;
\draw  [line width=0.75] (429.87,201.21) -- (429.64,166.63) -- (464.22,166.4) -- (464.45,200.98) -- cycle ;
\draw  [line width=0.75]  (385.93,157.73) -- (395.06,166.87) ;
\draw   [line width=0.75] (473.35,157.26) -- (464.22,166.4) ;

\draw (239.77,177.87) node [anchor=north west][inner sep=0.75pt]   [align=left] {$\displaystyle 3$};
\draw (269.67,179) node [anchor=north west][inner sep=0.75pt]   [align=left] {$\displaystyle a$};
\draw (303.74,176.52) node [anchor=north west][inner sep=0.75pt]   [align=left] {$\displaystyle b$};
\draw (287.11,149.18) node [anchor=north west][inner sep=0.75pt]   [align=left] {$\displaystyle 2$};
\draw (337.99,178.92) node [anchor=north west][inner sep=0.75pt]   [align=left] {$\displaystyle 4$};
\draw (287.95,206.62) node [anchor=north west][inner sep=0.75pt]   [align=left] {$\displaystyle 1$};
\draw (355.61,180.74) node [anchor=north west][inner sep=0.75pt]  [font=\normalsize]  {$\times $};
\draw (377.77,176.87) node [anchor=north west][inner sep=0.75pt]   [align=left] {$\displaystyle 3$};
\draw (407.67,179.4) node [anchor=north west][inner sep=0.75pt]   [align=left] {$\displaystyle c$};
\draw (441.74,175.52) node [anchor=north west][inner sep=0.75pt]   [align=left] {$\displaystyle d$};
\draw (425.11,148.18) node [anchor=north west][inner sep=0.75pt]   [align=left] {$\displaystyle 2$};
\draw (475.99,177.92) node [anchor=north west][inner sep=0.75pt]   [align=left] {$\displaystyle 4$};
\draw (425.95,205.62) node [anchor=north west][inner sep=0.75pt]   [align=left] {$\displaystyle 1$};

\end{tikzpicture}
}}
$};

\draw  [line width=0.75] (-90,290) node [anchor=north west][inner sep=0.75pt]  [align=center] {$ \hbox{\scalebox{0.75}{\begin{tikzpicture}[x=0.75pt,y=0.75pt,yscale=-1,xscale=1]

\draw  [line width=0.75] (100,100) node [anchor=north west][inner sep=0.75pt]  [align=center] {$ \begin{tikzpicture}[x=0.75pt,y=0.75pt,yscale=-1,xscale=1]

\draw   [line width=0.75] (326.78,201.67) -- (335.92,210.81) ;
\draw   [line width=0.75] (257.59,202.14) -- (248.45,211.28) ;
\draw  [line width=0.75] (257.3,202.45) -- (257.06,167.87) -- (291.64,167.63) -- (291.87,202.21) -- cycle ;
\draw  [line width=0.75] (291.87,202.21) -- (291.64,167.63) -- (326.22,167.4) -- (326.45,201.98) -- cycle ;
\draw   [line width=0.75] (247.93,158.73) -- (257.06,167.87) ;
\draw   [line width=0.75] (335.35,158.26) -- (326.22,167.4) ;
\draw   [line width=0.75] (447.78,224.67) -- (456.92,233.81) ;
\draw   [line width=0.75] (412.87,225.21) -- (403.74,234.35) ;
\draw  [line width=0.75] (412.64,190.63) -- (412.4,156.06) -- (446.98,155.82) -- (447.22,190.4) -- cycle ;
\draw  [line width=0.75] (412.87,225.21) -- (412.64,190.63) -- (447.22,190.4) -- (447.45,224.98) -- cycle ;
\draw   [line width=0.75] (403.27,146.92) -- (412.4,156.06) ;
\draw   [line width=0.75] (456.12,146.68) -- (446.98,155.82) ;

\draw (239.77,177.87) node [anchor=north west][inner sep=0.75pt]   [align=left] {$\displaystyle 3$};
\draw (269.67,176.4) node [anchor=north west][inner sep=0.75pt]   [align=left] {$\displaystyle a$};
\draw (303.74,176.52) node [anchor=north west][inner sep=0.75pt]   [align=left] {$\displaystyle b$};
\draw (287.11,149.18) node [anchor=north west][inner sep=0.75pt]   [align=left] {$\displaystyle 2$};
\draw (337.99,178.92) node [anchor=north west][inner sep=0.75pt]   [align=left] {$\displaystyle 4$};
\draw (287.95,206.62) node [anchor=north west][inner sep=0.75pt]   [align=left] {$\displaystyle 1$};
\draw (360.61,180.74) node [anchor=north west][inner sep=0.75pt]  [font=\normalsize]  {$\times $};
\draw (391.77,179.87) node [anchor=north west][inner sep=0.75pt]   [align=left] {$\displaystyle 3$};
\draw (423.17,163.4) node [anchor=north west][inner sep=0.75pt]   [align=left] {$\displaystyle c$};
\draw (424.74,199.52) node [anchor=north west][inner sep=0.75pt]   [align=left] {$\displaystyle d$};
\draw (423.11,131.18) node [anchor=north west][inner sep=0.75pt]   [align=left] {$\displaystyle 2$};
\draw (460.99,182.42) node [anchor=north west][inner sep=0.75pt]   [align=left] {$\displaystyle 4$};
\draw (423.95,228.12) node [anchor=north west][inner sep=0.75pt]   [align=left] {$\displaystyle 1$};

\end{tikzpicture}$};

\end{tikzpicture}
}}
$};

\draw  [line width=0.75] (120,303) node [anchor=north west][inner sep=0.75pt]  [align=center] {$ \hbox{\scalebox{0.75}{\begin{tikzpicture}[x=0.75pt,y=0.75pt,yscale=-1,xscale=1]

\draw  [line width=0.75] (100,100) node [anchor=north west][inner sep=0.75pt]  [align=center] {$ \begin{tikzpicture}[x=0.75pt,y=0.75pt,yscale=-1,xscale=1]

\draw   [line width=0.75] (361.03,201.74) -- (370.17,210.88) ;
\draw   [line width=0.75] (257.59,202.14) -- (248.45,211.28) ;
\draw  [line width=0.75] (257.3,202.45) -- (257.06,167.87) -- (291.64,167.63) -- (291.87,202.21) -- cycle ;
\draw [line width=0.75]  (291.87,202.21) -- (291.64,167.63) -- (326.22,167.4) -- (326.45,201.98) -- cycle ;
\draw   [line width=0.75] (247.93,158.73) -- (257.06,167.87) ;
\draw  [line width=0.75]  (369.93,158.03) -- (360.79,167.16) ;
\draw  [line width=0.75]  (464.78,200.67) -- (473.92,209.81) ;
\draw  [line width=0.75]  (429.87,201.21) -- (420.74,210.35) ;
\draw [line width=0.75]  (326.45,201.98) -- (326.22,167.4) -- (360.79,167.16) -- (361.03,201.74) -- cycle ;
\draw  [line width=0.75] (429.87,201.21) -- (429.64,166.63) -- (464.22,166.4) -- (464.45,200.98) -- cycle ;
\draw  [line width=0.75]  (420.5,157.5) -- (429.64,166.63) ;
\draw   [line width=0.75] (473.35,157.26) -- (464.22,166.4) ;

\draw (239.77,177.87) node [anchor=north west][inner sep=0.75pt]   [align=left] {$\displaystyle 3$};
\draw (269.67,176.4) node [anchor=north west][inner sep=0.75pt]   [align=left] {$\displaystyle a$};
\draw (303.74,176.52) node [anchor=north west][inner sep=0.75pt]   [align=left] {$\displaystyle b$};
\draw (302.11,148.18) node [anchor=north west][inner sep=0.75pt]   [align=left] {$\displaystyle 2$};
\draw (364.99,177.92) node [anchor=north west][inner sep=0.75pt]   [align=left] {$\displaystyle 4$};
\draw (301.95,206.62) node [anchor=north west][inner sep=0.75pt]   [align=left] {$\displaystyle 1$};
\draw (386.61,177.74) node [anchor=north west][inner sep=0.75pt]  [font=\normalsize]  {$\times $};
\draw (412.77,176.87) node [anchor=north west][inner sep=0.75pt]   [align=left] {$\displaystyle 3$};
\draw (339.67,175.4) node [anchor=north west][inner sep=0.75pt]   [align=left] {$\displaystyle c$};
\draw (441.74,175.52) node [anchor=north west][inner sep=0.75pt]   [align=left] {$\displaystyle d$};
\draw (442.11,148.18) node [anchor=north west][inner sep=0.75pt]   [align=left] {$\displaystyle 2$};
\draw (473.99,177.92) node [anchor=north west][inner sep=0.75pt]   [align=left] {$\displaystyle 4$};
\draw (440.95,207.62) node [anchor=north west][inner sep=0.75pt]   [align=left] {$\displaystyle 1$};

\end{tikzpicture}$};

\end{tikzpicture}
}}
$};



\end{tikzpicture}

}
}}
\end{eqnarray}

\begin{itemize}
    \item $gT^{12;34}$ with one numerators:
    \begin{equation}
    \begin{split}
    &n_{gT}^{12;34} := x_{1,2}^2 x_{a,3}^2-x_{1,3}^2 
\end{split}
\end{equation}
\item $hh^{12;34}$ with one numerators:
    \begin{equation}
    \begin{split}
    &n_{hh}^{12;34} := x_{1,2}^4.
\end{split}
\end{equation}
\item $hh^{\prime12;34}$ with one numerators:
    \begin{equation}
    \begin{split}
    &n_{hh^\prime}^{12;34} := x_{1,2}^2 x_{3,4}^2 .
\end{split}
\end{equation}
\item $gL^{12;34}$ with one numerators:
    \begin{equation}
    \begin{split}
    &n_{L\times g}^{12;34} := x_{1,2}^4 . 
\end{split}
\end{equation}
\end{itemize}

\noindent
The following are the remaining topologies.
\begin{eqnarray}
&&\vcenter{\hbox{\scalebox{1}{
\begin{tikzpicture}[x=0.75pt,y=0.75pt,yscale=-1,xscale=1]

\draw (-40,275) node [anchor=north west][inner sep=0.75pt]    {$S^{12;34}$};
\draw (180,275) node [anchor=north west][inner sep=0.75pt]    {$X^{12;34} $};
\draw (-40,415) node [anchor=north west][inner sep=0.75pt]   {$W^{12;34} $};
\draw (180,415) node [anchor=north west][inner sep=0.75pt]    {$\Theta^{12;34} $};

\draw  [line width=0.75] (-70,184) node [anchor=north west][inner sep=0.75pt]  [align=center] {$ \hbox{\scalebox{0.75}{\begin{tikzpicture}[x=0.75pt,y=0.75pt,yscale=-1,xscale=1]

\draw  [line width=0.75] (307.23,186.05) -- (295.18,206.91) -- (271.09,206.91) -- (259.05,186.05) -- (271.09,165.18) -- (295.18,165.18) -- cycle ;
\draw  [line width=0.75] (238.15,173.89) -- (250.2,153.02) -- (271.06,165.07) -- (259.02,185.94) -- cycle ;
\draw  [line width=0.75] (316.05,153.13) -- (328.1,174) -- (307.23,186.05) -- (295.18,165.18) -- cycle ;
\draw  [line width=0.75] (271.09,206.91) -- (295.19,206.91) -- (295.19,231.01) -- (271.09,231.01) -- cycle ;
\draw  [line width=0.75]  (295.19,231.01) -- (304.39,240.2) ;
\draw   [line width=0.75] (271.09,231.01) -- (261.9,240.2) ;
\draw  [line width=0.75]  (319.42,140.56) -- (316.05,153.13) ;
\draw  [line width=0.75]  (340.66,177.36) -- (328.1,174) ;
\draw   [line width=0.75] (225.58,177.26) -- (238.15,173.89) ;
\draw  [line width=0.75]  (246.83,140.46) -- (250.2,153.02) ;

\draw (227.02,146.99) node [anchor=north west][inner sep=0.75pt]   [align=left] {$\displaystyle 2$};
\draw (280.13,137.05) node [anchor=north west][inner sep=0.75pt]   [align=left] {$\displaystyle 1$};
\draw (278.76,180.51) node [anchor=north west][inner sep=0.75pt]   [align=left] {$\displaystyle a$};
\draw (249.85,160.78) node [anchor=north west][inner sep=0.75pt]   [align=left] {$\displaystyle b$};
\draw (307.46,163.78) node [anchor=north west][inner sep=0.75pt]   [align=left] {$\displaystyle c$};
\draw (278.7,210.76) node [anchor=north west][inner sep=0.75pt]   [align=left] {$\displaystyle d$};
\draw (314.03,201.64) node [anchor=north west][inner sep=0.75pt]   [align=left] {$\displaystyle 4$};
\draw (242.67,196.91) node [anchor=north west][inner sep=0.75pt]   [align=left] {$\displaystyle 3$};
\draw (333.88,146.04) node [anchor=north west][inner sep=0.75pt]   [align=left] {$\displaystyle 2$};
\draw (276.89,236.75) node [anchor=north west][inner sep=0.75pt]   [align=left] {$\displaystyle 2$};

\end{tikzpicture}
}}
$};

\draw  [line width=0.75] (120,194) node [anchor=north west][inner sep=0.75pt]  [align=center] {$ \hbox{\scalebox{0.75}{ \begin{tikzpicture}[x=0.75pt,y=0.75pt,yscale=-1,xscale=1]

\draw  [line width=0.75] (261.27,152.17) -- (301.62,165.36) -- (301.55,207.82) -- (261.15,220.87) -- (236.25,186.48) -- cycle ;
\draw [line width=0.75]  (301.62,165.36) -- (322.79,165.36) -- (322.79,186.53) -- (301.62,186.53) -- cycle ;
\draw  [line width=0.75] (301.62,186.53) -- (322.79,186.53) -- (322.79,207.7) -- (301.62,207.7) -- cycle ;
\draw [line width=0.75]  (362.99,152) -- (388.05,186.26) -- (363.21,220.69) -- (322.79,207.7) -- (322.65,165.24) -- cycle ;
\draw  [line width=0.75]  (236.25,186.48) -- (221.34,186.48) ;
\draw  [line width=0.75]  (402.97,186.26) -- (388.05,186.26) ;
\draw  [line width=0.75]  (362.99,152) -- (367.73,137.07) ;
\draw  [line width=0.75]  (261.15,220.87) -- (256.4,235.79) ;
\draw  [line width=0.75]  (370.93,235.47) -- (363.21,220.69) ;
\draw  [line width=0.75]  (261.27,152.17) -- (251.33,137.47) ;

\draw (265.5,180.25) node [anchor=north west][inner sep=0.75pt]   [align=left] {$\displaystyle a$};
\draw (348.22,180.25) node [anchor=north west][inner sep=0.75pt]   [align=left] {$\displaystyle b$};
\draw (306.62,170.36) node [anchor=north west][inner sep=0.75pt]   [align=left] {$\displaystyle c$};
\draw (306.62,189.36) node [anchor=north west][inner sep=0.75pt]   [align=left] {$\displaystyle d$};

\draw (308.03,141.74) node [anchor=north west][inner sep=0.75pt]   [align=left] {$\displaystyle 1$};
\draw (307.7,213.74) node [anchor=north west][inner sep=0.75pt]   [align=left] {$\displaystyle 2$};
\draw (226.7,154.67) node [anchor=north west][inner sep=0.75pt]   [align=left] {$\displaystyle 4$};
\draw (387.37,157.33) node [anchor=north west][inner sep=0.75pt]   [align=left] {$\displaystyle 4$};
\draw (225.37,206) node [anchor=north west][inner sep=0.75pt]   [align=left] {$\displaystyle 3$};
\draw (384.03,213.33) node [anchor=north west][inner sep=0.75pt]   [align=left] {$\displaystyle 3$};

\end{tikzpicture}
}}
$};

\draw  [line width=0.75] (-90,305) node [anchor=north west][inner sep=0.75pt]  [align=center] {$ \hbox{\scalebox{0.75}{\begin{tikzpicture}[x=0.75pt,y=0.75pt,yscale=-1,xscale=1]

\draw  [line width=0.75] (100,100) node [anchor=north west][inner sep=0.75pt]  [align=center] {$  \begin{tikzpicture}[x=0.75pt,y=0.75pt,yscale=-1,xscale=1]

\draw  [line width=0.75] (302.04,114.65) -- (286.04,143.17) -- (254.05,143.17) -- (238.05,114.65) -- (254.05,86.13) -- (286.04,86.13) -- cycle ;
\draw   [line width=0.75] (238.05,114.65) -- (222.05,143.17) -- (190.06,143.17) -- (174.06,114.65) -- (190.06,86.13) -- (222.05,86.13) -- cycle ;
\draw   [line width=0.75] (222.05,143.17) -- (239.97,167.99) ;
\draw  [line width=0.75]  (236.13,61.31) -- (254.05,86.13) ;
\draw   [line width=0.75] (236.13,61.31) -- (222.05,86.13) ;
\draw   [line width=0.75] (254.05,143.17) -- (239.97,167.99) ;
\draw   [line width=0.75] (155.5,114.65) -- (174.06,114.65) ;
\draw   [line width=0.75] (302.04,114.65) -- (320.6,114.65) ;
\draw  [line width=0.75]  (286.04,86.13) -- (297.56,75.14) ;
\draw   [line width=0.75] (178.54,154.16) -- (190.06,143.17) ;
\draw  [line width=0.75]  (190.06,86.13) -- (172.14,75.14) ;
\draw   [line width=0.75] (303.96,154.16) -- (286.04,143.17) ;
\draw   [line width=0.75] (236.13,45.5) -- (236.13,61.31) ;
\draw  [line width=0.75]  (239.97,167.99) -- (239.97,183.8) ;

\draw (199.6,102.7) node [anchor=north west][inner sep=0.75pt]    {$a$};
\draw (266.9,108.23) node [anchor=north west][inner sep=0.75pt]    {$b$};
\draw (233.09,80.72) node [anchor=north west][inner sep=0.75pt]   [align=left] {$\displaystyle c$};
\draw (231.41,132.85) node [anchor=north west][inner sep=0.75pt]   [align=left] {$\displaystyle d$};
\draw (196.93,56.03) node [anchor=north west][inner sep=0.75pt]    {$1$};
\draw (264.27,158.7) node [anchor=north west][inner sep=0.75pt]    {$1$};
\draw (305.6,127.37) node [anchor=north west][inner sep=0.75pt]    {$2$};
\draw (162.27,86.7) node [anchor=north west][inner sep=0.75pt]    {$2$};
\draw (160.93,126.03) node [anchor=north west][inner sep=0.75pt]    {$3$};
\draw (202.27,153.37) node [anchor=north west][inner sep=0.75pt]    {$4$};
\draw (261.6,58.7) node [anchor=north west][inner sep=0.75pt]    {$3$};
\draw (302.27,86.7) node [anchor=north west][inner sep=0.75pt]    {$4$};

\end{tikzpicture}$};

\end{tikzpicture}
}}
$};

\draw  [line width=0.75] (150,313) node [anchor=north west][inner sep=0.75pt]  [align=center] {$ \hbox{\scalebox{0.75}{\begin{tikzpicture}[x=0.75pt,y=0.75pt,yscale=-1,xscale=1]

\draw  [line width=0.75] (100,100) node [anchor=north west][inner sep=0.75pt]  [align=center] {$  \begin{tikzpicture}[x=0.75pt,y=0.75pt,yscale=-1,xscale=1]

\draw   [line width=0.75] (303.27,175.42) -- (267.9,175.42) -- (242.9,150.41) -- (242.9,115.05) -- (267.9,90.05) -- (303.27,90.05) -- (328.27,115.05) -- (328.27,150.41) -- cycle ;
\draw   [line width=0.75]  (243.07,132.735) -- (328.11,132.735) ;
\draw   [line width=0.75]  (285.585,89.56) -- (285.585,175.14) ;
\draw   [line width=0.75]  (343.19,115.05) -- (328.27,115.05) ;
\draw  [line width=0.75]   (343.19,150.41) -- (328.27,150.41) ;
\draw   [line width=0.75]  (242.9,115.05) -- (227.98,115.05) ;
\draw  [line width=0.75]   (242.9,150.41) -- (227.98,150.41) ;
\draw   [line width=0.75]  (303.27,90.05) -- (303.27,75.13) ;
\draw   [line width=0.75]  (267.9,90.05) -- (267.9,75.13) ;
\draw   [line width=0.75]  (303.27,190.33) -- (303.27,175.42) ;
\draw   [line width=0.75]  (267.9,190.33) -- (267.9,175.42) ;

\draw (263.37,107.41) node [anchor=north west][inner sep=0.75pt]   [align=left] {$\displaystyle a$};
\draw (298.08,107.41) node [anchor=north west][inner sep=0.75pt]   [align=left] {$\displaystyle b$};
\draw (263.37,144.59) node [anchor=north west][inner sep=0.75pt]   [align=left] {$\displaystyle c$};
\draw (298.08,144.59) node [anchor=north west][inner sep=0.75pt]   [align=left] {$\displaystyle d$};

\draw (245,85) node [anchor=north west][inner sep=0.75pt]   [align=left] {$\displaystyle 1$};
\draw (317,85) node [anchor=north west][inner sep=0.75pt]   [align=left] {$\displaystyle 4$};
\draw (245,165) node [anchor=north west][inner sep=0.75pt]   [align=left] {$\displaystyle 3$};
\draw (317,165) node [anchor=north west][inner sep=0.75pt]   [align=left] {$\displaystyle 2$};

\draw (281,73) node [anchor=north west][inner sep=0.75pt]   [align=left] {$\displaystyle 3$};
\draw (281,180) node [anchor=north west][inner sep=0.75pt]   [align=left] {$\displaystyle 4$};

\draw (281,180) node [anchor=north west][inner sep=0.75pt]   [align=left] {$\displaystyle 4$};

\draw (230,125.05) node [anchor=north west][inner sep=0.75pt]   [align=left] {$\displaystyle 2$};
\draw (332,125.05) node [anchor=north west][inner sep=0.75pt]   [align=left] {$\displaystyle 1$};

\end{tikzpicture}$};

\end{tikzpicture}
}}
$};

\end{tikzpicture}

}
}}
\end{eqnarray}

\begin{itemize}
    \item $S^{12;34}$ with three  distinct numerators:
    \begin{equation}
    \begin{split}
    &n_{S}^{12;34} := (x_{1,3}^{2} x_{1,4}^{2} x_{3,4}^{2}  x_{a,2}^{4}{+} 3\, x_{14}^2   x_{2,3}^2 x_{a,1}^2 (x_{2,4}^2 x_{a,3}^2 {-}x_{3,4}^2 x_{a,2}^2))/6,  \\
    & {n}_{S_2}^{\prime\,12;34} := x_{a,4}^2 (x_{1,2}^{2}  x_{a,3}^{2} {-}x_{2,3}^{2}  x_{a,1}^{2} {-} x_{1,3}^{2}  x_{a,2}^{2}),\\
    &n_{S}^{\prime\prime\,12;34} := x_{1,2}^{2} x_{a,3}^{2}  x_{a,4}^{2}.
\end{split}
\end{equation}
where  $n_S^{12,34}$ with symmetry $\text{perms}\,(1,3,4)$
\item $X^{12;34}$ with three  distinct numerators:
    \begin{equation}
    \begin{split}
    &n_{X}^{12;34} := x_{a,b}^2 ( x_{a,b}^2  x_{1,2}^2 {-} x_{a,1}^2  x_{b,2}^2{-} x_{a,2}^2  x_{b,1}^2), \\ 
    &n_{X}^{\prime\,12;34} := x_{a,1}^{2} x_{b,2}^{2}  x_{a,b}^{2},\\ 
    &n_{X}^{\prime\prime\,12;34} :=x_{1,2}^{2} x_{a,3}^{2} x_{b,4}^{2}  ( x_{a,b}^{2} x_{1,2}^{2} {-}x_{a,1}^{2} x_{b,2}^{2} {-} x_{a,2}^{2} x_{b,1}^{2}).
\end{split}
\end{equation}
\item $W^{12;34}$ with one   numerator:
    \begin{equation}
    \begin{split}
    &n_W^{12;34} := (x_{a,b}^2 x_{1,3}^2 {-} x_{a,1}^2 x_{b,3}^2 {-} x_{a,3}^2 x_{b,1}^2) \times  (x_{a,b}^2 x_{1,4}^2 {-} x_{a,1}^2 x_{b,4}^2 {-} x_{a,4}^2 x_{b,1}^2).
\end{split}
\end{equation}
\item $\Theta^{12;34}$ with two distinct   numerators, which is the most complicated topology at four-loop:
    \begin{equation}
    \begin{split}
        &n_{\Theta}^{12;34} :=\frac{1}{16} ({-}x_{a,4}^2 x_{b,2}^2 x_{c,1}^2 x_{d,3}^2 {+} x_{a,3}^2 x_{b,1}^2 x_{c,2}^2 x_{d,4}^2 {-} x_{1,2}^2 x_{3,4}^2 x_{a,d}^2 x_{b,c}^2 )\\
        & \phantom{my lj} {+}\frac{1}{8}({-}x_{a,2}^2 x_{b,4}^2 x_{c,3}^2 x_{d,1}^2 {-} x_{a,1}^2 x_{b,4}^2 x_{c,3}^2 x_{d,2}^2 {+} x_{1,2}^2 x_{a,4}^2 x_{d,3}^2 x_{b,c}^2 {+} x_{1,4}^2 x_{2,3}^2 x_{a,d}^2 x_{b,c}^2\\
        &\phantom{my ljkkkk}{-} x_{1,4}^2 x_{2,3}^2 x_{a,b}^2 x_{c,d}^2 {-} x_{1,3}^2 x_{2,4}^2 x_{a,b}^2 x_{c,d}^2 {+} x_{1,2}^2 x_{3,4}^2 x_{a,b}^2 x_{c,d}^2 )\\
        &\phantom{my lj}{+}\frac{1}{4} (x_{a,2}^2 x_{b,4}^2 x_{c,1}^2 x_{d,3}^2 {+} x_{a,1}^2 x_{b,3}^2 x_{c,2}^2 x_{d,4}^2 {-} x_{2,4}^2 x_{c,3}^2 x_{d,1}^2 x_{a,b}^2 {+} x_{2,4}^2 x_{c,1}^2 x_{d,3}^2 x_{a,b}^2 \\
        &\phantom{my ljfff} {+} x_{1,3}^2 x_{c,4}^2 x_{d,2}^2 x_{a,b}^2 {-} x_{1,3}^2 x_{c,2}^2 x_{d,4}^2 x_{a,b}^2 {+} x_{3,4}^2 x_{a,2}^2 x_{d,1}^2 x_{b,c}^2 ) \\
        &\phantom{my lj} {+} \frac{1}{2} (x_{a,2}^2 x_{b,3}^2 x_{c,1}^2 x_{d,4}^2 {+} x_{2,3}^2 x_{c,4}^2 x_{d,1}^2 x_{a,b}^2 {-} x_{1,2}^2 x_{c,4}^2 x_{d,3}^2 x_{a,b}^2 {+} x_{2,3}^2 x_{c,1}^2 x_{d,4}^2 x_{a,b}^2 \\
        &\phantom{my ljjjjjj} {-} x_{1,2}^2 x_{c,3}^2 x_{d,4}^2 x_{a,b}^2 {-} x_{1,4}^2 x_{a,2}^2 x_{d,3}^2 x_{b,c}^2 {-} x_{1,3}^2 x_{a,2}^2 x_{d,4}^2 x_{b,c}^2),    \end{split}
\end{equation}

\begin{equation}
    \begin{split}
        &n_{\Theta}^{\prime\,12;34} :=\frac{1}{8} ({-}x_{1,4}^2 x_{2,3}^2 x_{a,4}^2 x_{b,2}^2 x_{c,1}^2 x_{d,3}^2 {-} x_{1,4}^2 x_{2,3}^2 x_{a,2}^2 x_{b,4}^2 x_{c,3}^2 x_{d,1}^2{-} x_{1,3}^2 x_{2,4}^2 x_{a,2}^2 x_{b,4}^2 x_{c,3}^2 x_{d,1}^2  \\
        &\phantom{my ljkkk} {-}x_{1,4}^2 x_{2,3}^2 x_{a,3}^2 x_{b,1}^2 x_{c,2}^2 x_{d,4}^2 {+} x_{1,4}^4 x_{2,3}^4 x_{a,d}^2 x_{b,c}^2{-} x_{1,4}^4 x_{2,3}^4 x_{a,c}^2 x_{b,d}^2 {-} x_{1,3}^4 x_{2,4}^4 x_{a,c}^2 x_{b,d}^2  \\
        &\phantom{my ljkkk} {-}x_{1,4}^2 x_{2,3}^2 x_{1,3}^2 x_{2,4}^2 x_{a,d}^2 x_{b,c}^2 {-} x_{1,2}^2 x_{3,4}^2 x_{1,4}^2 x_{2,3}^2 x_{a,d}^2 x_{b,c}^2  {+} x_{1,2}^2 x_{3,4}^2 x_{1,4}^2 x_{2,3}^2 x_{a,c}^2 x_{b,d}^2 \\
        &\phantom{my ljkkk} {+} x_{1,2}^2 x_{3,4}^2 x_{1,3}^2 x_{2,4}^2 x_{a,c}^2 x_{b,d}^2){+}\frac{5}{4} x_{1,2}^2 x_{3,4}^2 x_{1,4}^2 x_{2,3}^2 x_{b,4}^2 x_{c,3}^2 x_{a,d}^2 {+}\frac{3}{4} ( {-}x_{1,4}^2 x_{2,3}^2 x_{a,3}^2 \\
        &\phantom{mykk} x_{b,4}^2 x_{c,2}^2 x_{d,1}^2 {-} x_{1,4}^2 x_{2,3}^2 x_{a,1}^2 x_{b,4}^2 x_{c,3}^2 x_{d,2}^2 {+} x_{1,4}^2 x_{2,3}^2 x_{2,4}^2 x_{c,3}^2 x_{d,1}^2 x_{a,b}^2 {-}  x_{1,4}^2 x_{2,3}^2 x_{2,4}^2 x_{c,1}^2 \\
        &\phantom{mykk}x_{d,3}^2 x_{a,b}^2 ){+}\frac{1}{4} (x_{1,4}^2 x_{2,3}^2 x_{a,4}^2 x_{b,3}^2 x_{c,1}^2 x_{d,2}^2{+} x_{1,3}^2 x_{2,4}^2 x_{a,4}^2 x_{b,3}^2 x_{c,1}^2 x_{d,2}^2 {+} x_{1,4}^2 x_{2,3}^2 x_{a,4}^2 x_{b,1}^2 \\
        &\phantom{mykk}x_{c,2}^2 x_{d,3}^2 {+}  x_{1,3}^2 x_{2,4}^2 x_{a,3}^2 x_{b,4}^2 x_{c,2}^2 x_{d,1}^2 {-} x_{1,2}^2 x_{3,4}^2 x_{a,1}^2 x_{b,4}^2 x_{c,3}^2 x_{d,2}^2 {-} x_{1,3}^2 x_{2,4}^4 x_{c,3}^2 x_{d,1}^2 x_{a,b}^2 \\
        &\phantom{mykk} {+}x_{1,3}^2 x_{2,4}^4 x_{c,1}^2 x_{d,3}^2 x_{a,b}^2 {+} x_{1,3}^2 x_{1,4}^2 x_{2,3}^2 x_{c,4}^2 x_{d,2}^2 x_{a,b}^2 {+} x_{1,3}^4 x_{2,4}^2 x_{c,4}^2 x_{d,2}^2 x_{a,b}^2 {-} x_{1,3}^2 x_{1,4}^2 x_{2,3}^2 \\
        &\phantom{mykk} x_{c,2}^2 x_{d,4}^2 x_{a,b}^2 {-} x_{1,3}^4 x_{2,4}^2 x_{c,2}^2 x_{d,4}^2 x_{a,b}^2 {+} x_{1,4}^2 x_{2,3}^2 x_{3,4}^2 x_{b,2}^2 x_{c,1}^2 x_{a,d}^2 {-} x_{1,4}^2 x_{2,3}^2 x_{3,4}^2 x_{b,1}^2 x_{c,2}^2 \\
        &\phantom{mykk} x_{a,d}^2 {+} x_{1,2}^2 x_{1,4}^2 x_{2,3}^2 x_{b,4}^2 x_{c,3}^2 x_{a,d}^2 {+}  x_{1,4}^2 x_{2,3}^2 x_{1,3}^2 x_{2,4}^2 x_{a,c}^2 x_{b,d}^2   ){+}\frac{3}{2} ({-} x_{1,3}^2 x_{2,4}^2 x_{a,4}^2 x_{b,3}^2 x_{c,2}^2 \\
        &\phantom{mykk}  x_{d,1}^2 {+} x_{1,3}^2 x_{1,4}^2 x_{2,3}^2 x_{b,3}^2 x_{c,2}^2 x_{a,d}^2  ) {+} \frac{1}{2} (x_{1,4}^2 x_{2,3}^2 x_{a,4}^2 x_{b,3}^2 x_{c,2}^2 x_{d,1}^2 {-} x_{1,4}^2 x_{2,3}^2  x_{3,4}^2 x_{c,2}^2 x_{d,1}^2  \\
        &\phantom{mykk} x_{a,b}^2 {-} x_{1,3}^2 x_{2,4}^2 x_{3,4}^2 x_{c,2}^2 x_{d,1}^2 x_{a,b}^2 {-} x_{1,3}^2 x_{2,4}^2 x_{3,4}^2 x_{c,2}^2 x_{d,1}^2 x_{a,b}^2 {-} x_{1,4}^2 x_{2,3}^2 x_{3,4}^2 x_{c,1}^2 x_{d,2}^2 x_{a,b}^2\\
        &\phantom{mykk}  {-} x_{1,3}^2 x_{2,4}^2 x_{3,4}^2 x_{c,1}^2 x_{d,2}^2 x_{a,b}^2{+} x_{1,4}^4 x_{2,3}^2 x_{c,3}^2 x_{d,2}^2 x_{a,b}^2 {+} x_{1,3}^2 x_{2,4}^2 x_{1,4}^2 x_{c,3}^2 x_{d,2}^2 x_{a,b}^2{+} x_{1,4}^4 x_{2,3}^2 \\
        &\phantom{mykk} x_{c,2}^2 x_{d,3}^2 x_{a,b}^2 {-} x_{1,3}^2 x_{2,4}^2 x_{1,4}^2 x_{c,2}^2 x_{d,3}^2 x_{a,b}^2 {-} x_{1,4}^2 x_{2,3}^2 x_{2,4}^2 x_{b,3}^2 x_{c,1}^2 x_{a,d}^2 {-} x_{1,3}^2 x_{2,4}^4 x_{b,3}^2 x_{c,1}^2 x_{a,d}^2  \\
        &\phantom{mykk}{-} x_{1,3}^2 x_{2,4}^4 x_{b,3}^2 x_{c,1}^2 x_{a,d}^2 {-} x_{1,4}^4 x_{2,3}^2 x_{b,3}^2 x_{c,2}^2 x_{a,d}^2 ) 
    \end{split}
\end{equation}
where $n_{\Theta}^{12;34}$ and $n_{\Theta}^{\prime\,12;34}$ with symmetry $\{1,2,3,4\}$,  $\{1,2,4,3\}$, $\{2,1,3,4\}$, $\{2,1,4,3\}$, $\{3,4,1,2\}$,  $\{4,3,1,2\}$, $\{3,4,2,1\}$, and $\{4,3,2,1\}$.

\end{itemize}

\bibliographystyle{utphys}
\bibliography{bib}

\end{document}